\documentclass[a4paper,11pt]{article}
\pdfoutput=1
\usepackage[utf8]{inputenc}
\usepackage{jheppub}
\usepackage{empheq}
\usepackage{comment}
\usepackage{bm}
\usepackage[dvipsnames]{xcolor}
\usepackage{tikz}
\usepackage{titlesec}

\allowdisplaybreaks[1]

\DeclareMathOperator{\Ei}{Ei}
\DeclareMathOperator{\Li}{Li}

\DeclareMathOperator{\K}{\mathcal{K}}
\DeclareMathOperator{\II}{\mathcal{I}}
\DeclareMathOperator{\G}{\mathcal{G}}

\DeclareMathOperator{\Div}{Div}

\renewcommand{\[}{\begin{equation}}
\renewcommand{\]}{\end{equation}}
\newcommand{\I}{\mathrm{i}}
\newcommand{\D}{\mathrm{d}}

\newcommand{\Z}{\mathbb{Z}}

\renewcommand{\O}{\mathcal{O}}
\newcommand{\PLp}{\mathcal{D}^{(+)}}  
\newcommand{\PLm}{\mathcal{D}^{(-)}}  
\newcommand{\op}{\O^{[2]}}
\newcommand{\Op}{\O^{[3]}}
\newcommand{\src}{\phi^{[1]}}
\newcommand{\Src}{\phi^{[0]}}
\newcommand{\f}{\Phi^{[2]}}
\newcommand{\F}{\Phi^{[3]}}
\newcommand{\act}{\mathfrak{a}}
\newcommand{\<}{\langle}
\renewcommand{\>}{\rangle}
\newcommand{\lla}{\langle \! \langle}
\newcommand{\rra}{\rangle \! \rangle}

\newcommand{\bs}[1]{\boldsymbol{#1}}
\newcommand{\reg}[1]{\hat{#1}}
\newcommand{\dreg}{\reg{d}}
\newcommand{\Dreg}{\reg{\Delta}}
\newcommand{\areg}{\reg{\alpha}}
\newcommand{\breg}{\reg{\beta}}
\newcommand{\Kreg}{\reg{\K}}
\newcommand{\Ireg}{\reg{\II}}
\newcommand{\Greg}{\reg{\G}}

\newcommand{\nn}{\nonumber}
\renewcommand{\a}{\alpha}

\newcommand{\ep}{\epsilon}
\newcommand{\z}{\zeta}

\newcommand{\ino}{i}
\newcommand{\ireg}{\hat{i}}
\newcommand{\idiv}{i^{\text{div}}}
\newcommand{\ifin}{i^{\text{fin}}}
\newcommand{\iren}{i^{\text{ren}}}

\newcommand{\ict}{i^{\text{ct}}}
\newcommand{\icts}[1]{i^{\text{ct}({#1})}}
\newcommand{\ilog}{i^{\text{log}}}

\newcommand{\jno}{j}
\newcommand{\jreg}{\hat{j}}
\newcommand{\jdiv}{j^{\text{div}}}

\newcommand{\jfin}{j^{\text{fin}}}

\newcommand{\s}[2]{\sigma_{(#1) #2}}
\newcommand{\m}[1]{l_{#1 -}}
\newcommand{\p}[1]{l_{#1 +}}

\newcommand{\nicebox}{\fboxsep=10pt\fbox}

\titleformat*{\section}{\LARGE\bfseries}
\titleformat*{\subsection}{\Large\bfseries}
\titleformat*{\subsubsection}{\large\bfseries}
\titlespacing*{\section}{0pt}{28pt}{18pt}
\titlespacing*{\subsection}{0pt}{24pt}{14pt}
\titlespacing*{\subsubsection}{0pt}{20pt}{6pt}

\newcommand{\triR}[5]{
\draw [fill=black] (-2.121 + #1, -2.121 + #2) circle [radius=0.1];
\draw [fill=black] (-2.121 + #1, 2.121 + #2) circle [radius=0.1];
\draw [fill=black] (-1 + #1, 0 + #2) circle [radius=0.1];
\draw [fill=black] (1.121 + #1, 0 + #2) circle [radius=0.1];
\draw (-2.121 + #1, -2.121 + #2) -- (-1 + #1, 0 + #2) -- (-2.121 + #1, 2.121 + #2);
\draw (-1 + #1, 0 + #2) -- (1.121 + #1,0 + #2);
\node [right] at (-1.7 + #1, 1.4 + #2) {#3};
\node [right] at (-1.7 + #1, -1.4 + #2) {#4};
\node [above] at (0.2 + #1, 0 + #2) {#5};
}

\newcommand{\triL}[5]{
\draw [fill=black] (2.121 + #1, -2.121 + #2) circle [radius=0.1];
\draw [fill=black] (2.121 + #1, 2.121 + #2) circle [radius=0.1];
\draw [fill=black] (1 + #1, 0 + #2) circle [radius=0.1];
\draw [fill=black] (-1.121 + #1, 0 + #2) circle [radius=0.1];
\draw (2.121 + #1, -2.121 + #2) -- (1 + #1, 0 + #2) -- (2.121 + #1, 2.121 + #2);
\draw (1 + #1, 0 + #2) -- (-1.121 + #1,0 + #2);
\node [left] at (1.7 + #1, -1.4 + #2) {#3};
\node [left] at (1.7 + #1, 1.4 + #2) {#4};
\node [above] at (-0.2 + #1, 0 + #2) {#5};
}

\newtheorem{fact}{Fact}

\begin{document}

\title{\LARGE A handbook of holographic 4-point functions}

\author[a]{Adam Bzowski,}
\affiliation[a]{Faculty of Physics, 
	University of Warsaw, 
	Pasteura 5,
	02-093 Warsaw, 
	Poland}

\author[b]{Paul McFadden}
\affiliation[b]{School of Mathematics, 
	Statistics \& Physics, Newcastle University, 
	Newcastle NE1 7RU, U.K.} 

\author[c]{and Kostas Skenderis.}
\affiliation[c]{STAG Research Center \& Mathematical Sciences, 
	University of Southampton,
	Highfield, \\
	Southampton 
	SO17 1BJ, U.K.}

\emailAdd{a.bzowski@physics.uoc.gr}
\emailAdd{paul.l.mcfadden@newcastle.ac.uk}
\emailAdd{k.skenderis@soton.ac.uk}

\abstract{
We present a comprehensive discussion of tree-level  holographic $4$-point functions of scalar operators in momentum space. We show that each individual Witten diagram satisfies the conformal Ward identities on its own and is thus a valid conformal correlator.  When the $\beta = \Delta - d/2$ are half-integral,  with $\Delta$  the dimensions of the operators and $d$ the spacetime dimension, the Witten diagrams can be evaluated in closed form and we present explicit formulae for the case  $d=3$ and $\Delta=2,3$.  These correlators require renormalization, which we carry out explicitly, and lead to new conformal anomalies and beta functions. Correlators of operators of different dimension may be linked via weight-shifting operators, which allow new correlators to be generated from given `seed' correlators. We present a new  derivation of weight-shifting operators 
 in momentum space 
 and uncover several subtleties associated with their use:  such operators map
 exchange diagrams  to a linear combination of exchange and contact diagrams, and special care must be taken when renormalization is required.   
}

\maketitle

\section{Introduction} 

Witten diagrams have become an indispensable tool in analyzing CFTs at strong coupling. Yet explicit closed formulae in momentum space are rare.
This is even more the case when the correlators require renormalization. It is the purpose of this paper to show how to obtain such formulae for tree-level holographic $4$-point functions of operators of dimension $\Delta$, with $\Delta$  such that the combinations $\beta = \Delta - d/2$ are half-integral, where $d$ is the spacetime dimension. 

The simplest cases, which are still of considerable interest, are those of 4-point functions of scalar operators of dimension $\Delta=2,3$ in three-dimensional CFTs. For these cases, we discuss and present explicit formulae for all holographic CFT $4$-point functions in momentum space that can be constructed  from tree-level contact and exchange Witten diagrams, with exchanged scalar operators of dimension $\Delta=2,3$.  The methodology we discuss extends to other operators and spacetime dimensions (provided the $\beta$ are half-integral)  and the Mathematica file we supply may be adapted to such cases.
 
 Our focus on $4$-point functions of scalar operators with dimensions $\Delta=2,3$ in three spacetime dimensions is motivated in part by recent works on 
the cosmological bootstrap \cite{Arkani-Hamed:2015bza, Arkani-Hamed:2018kmz, Baumann:2019oyu, Baumann:2020dch, Baumann:2021fxj,Sleight:2019mgd,Sleight:2019hfp,Sleight:2020obc,Sleight:2021plv,DiPietro:2021sjt, Meltzer:2021zin}, in which late-time correlators in four-dimensional de Sitter spacetime are constructed by solving three-dimensional conformal Ward identities.  These operators are dual to conformally coupled and massless bulk scalars, with the latter modelling the inflaton.
The same approach was also  explored at the level of the $3$-point function  in the earlier holographic cosmology literature \cite{Maldacena:2011nz,Bzowski:2012ih, Mata:2012bx, McFadden:2013ria, Anninos:2014lwa}. The correlation functions of these operators require renormalization and
a key question is whether the divergences affecting the corresponding $4$-point correlators in anti-de Sitter spacetime, as studied here, also play a role in de Sitter.   This question will be addressed in a follow-up paper \cite{toappear} using the results for renormalized AdS/CFT correlators developed here.   

Witten diagrams \cite{Witten:1998qj} were historically computed in position space, see \cite{Freedman:1998tz, Liu:1998ty, DHoker:1999kzh,Arutyunov:2000py, Dolan:2004iy} for a sample of computations from the early days of AdS/CFT.
However, the correlators we discuss here contain divergences as tabulated in Table \ref{degdivtable}, and one of the central aims of this work is to provide a complete, yet readable, account  of their renormalization.  The extraction of divergences is best performed in momentum space  \cite{Bzowski:2015pba,Bzowski:2017poo,Bzowski:2018fql}, and as such we work throughout in momentum space.  The momentum-space results are also  of current interest for many applications 
including inflationary cosmology.  
As the Fourier transform is well-defined after renormalization, our results can in principle also be transformed back to position space.

The framework we use for CFT in momentum space was introduced in \cite{Bzowski:2013sza}. The general structure of 4-point functions in momentum space was given in \cite{Bzowski:2019kwd, Bzowski:2020kfw} and a number of special cases were discussed 
in \cite{Arkani-Hamed:2018kmz, Isono:2018rrb, Albayrak:2018tam, Albayrak:2019asr, Maglio:2019grh, Isono:2019wex, Coriano:2019nkw, Albayrak:2020isk}.\footnote{For a sample of other recent developments in momentum-space CFT, see \cite{Anand:2019lkt, 
Gillioz:2020wgw,Jain:2021wyn,
Jain:2021vrv, 
Armstrong:2020woi,Albayrak:2020fyp,Gomez:2021qfd,Coriano:2021nvn,Coriano:2020ees}.}
Our hope is that this handbook of renormalized holographic $4$-point functions, which we believe is the most comprehensive 
of its kind, will be of use to a wide range of researchers.  To this end, 
we provide full documentation including a set of accompanying Mathematica notebooks. 
Researchers interested in the 
renormalization of CFT correlators 
will find a complete discussion of tree-level $4$-point functions, while those seeking specific results need only consult the relevant listings or notebooks.

\begin{table}[t]
\begin{center}
\begin{tabular}{|c|c|c|c|} \hline
External operator & Contact  & $s$-channel exchange & $s$-channel exchange \\[-2ex] 
dimensions & diagram & with $\Delta_x = 2$ & with $\Delta_x = 3$ \\ \hline
$[22,22]$ & 0 & 0 & 0 \\ \hline
$[32,22]$ & 1 & 2 & 1 \\ \hline
$[33,22]$ & 1 & 1 & 2 \\ \hline
$[32,32]$ & 1 & 2 & 1 \\ \hline
$[32,33]$ & 1 & 2 & 2 \\ \hline
$[33,33]$ & 1 & 1 & 2 \\ \hline
\end{tabular}
\caption{Degree of divergence $n$ for Witten diagrams dimensionally regulated according to the scheme \eqref{special}. Each diagram diverges as $\ep^{-n}$ for $\ep\rightarrow 0$, and is  labelled by the dimensions $[\Delta_1\Delta_2,\Delta_3\Delta_4]$  of the external operators 
  and that of the exchanged operator $\Delta_x$ where present. 
  \label{degdivtable}}
\end{center}
\end{table}

Aside from their applications, an additional reason for focusing on correlators with $\Delta=2,3$ in $d=3$ is the relative simplicity of their construction: the corresponding Witten diagrams feature propagators involving Bessel functions with half-integer indices.  The latter reduce to elementary functions enabling all diagrams to be easily evaluated in terms of dilogarithms.  It is desirable to preserve this simplicity in the regulated theory, and we show this is possible by regulating the spacetime and operator dimensions as
\begin{align}\label{special}
 d \longmapsto \dreg = d + 2 \ep, \qquad \Delta_j \longmapsto \Dreg_j = \Delta_j + \ep, 
\end{align}
where $j$ runs over all external and exchanged (`$x$') operators $j = 1,2,3,4,x$ and $\ep$ is the regulator.
This `half-integer' scheme preserves the indices on all Bessel functions, which are given by $\Delta_j-d/2 = \Dreg_j-\dreg/2$, and is sufficient to regulate all the divergences we encounter here.
Renormalization can then be achieved by constructing suitable counterterms directly within the boundary CFT.

For completeness, we also explain how to change to a fully general dimensional regularization scheme
\begin{align} \label{genreg}
 d \longmapsto \dreg = d + 2 u \ep, \qquad \Delta_j \longmapsto \Dreg_j = \Delta_j + (u + v_j) \ep,
\end{align}
where  $u$ and $v_j$, for $j = 1,2,3,4,x$, are real constants  parametrizing the  scheme. This can be achieved by expanding the Bessel functions in their indices, for which several results are available.
An alternative to both the above approaches would be to employ holographic renormalization \cite{Skenderis:2002wp}, however in practice the resulting cut-off radial integrals  are less convenient to evaluate than those encountered in dimensional regularization.

Usually in quantum field theory individual Feynman diagrams are not physically meaningful and do not satisfy Ward identities on their own.  Instead, one must sum all relevant diagrams to construct observable quantities, such as correlators, that satisfy the relevant Ward identities.  
However, an interesting general feature of  tree-level Witten diagrams associated with correlators of scalar operators is that they {\it individually} satisfy the conformal Ward identities.\footnote{We emphasize that this property does {\it not} hold for Witten diagrams associated with correlators of non-abelian symmetry currents and/or the energy-momentum tensor. In such cases the corresponding bulk action involves a (bulk) gauge symmetry and summing over all Witten diagrams is necessary in order to get a physically meaningful answer for the correlator.}
This follows from that the fact that there is a choice of  bulk AdS Lagrangian such that any given Witten diagram is the only diagram contributing to a corresponding correlator. We illustrate this in the cases we analyze, and we expect this to be a general feature. The existence of a bulk Lagrangian implies that the diagram must satisfy the conformal Ward identities, as these follow from the bulk isometries, and also that we should be able to renormalize at the level of individual Witten diagrams.
This does not necessarily mean however that there exists an actual CFT with the chosen  spectrum 
and interaction.  Where such a CFT does not exist (for example, via bootstrap arguments) then the corresponding bulk AdS Lagrangian would be in the swampland. 

An interesting approach to the construction of correlators is to use weight-shifting operators. These operators were introduced in position space in \cite{Karateev:2017jgd} (see also \cite{Costa:2018mcg}), and by construction, map CFT correlators of operators of given conformal dimensions to those of operators with different (shifted) conformal dimensions.  One may thus start from `seed' correlators and obtain others using weight-shifting operators. These were applied in momentum space in \cite{Arkani-Hamed:2018kmz,Baumann:2019oyu} and we further develop and analyze this approach here. 
There are however a number of subtleties that need to be taken into account when using this method. First, the application of weight-shifting operators to exchange diagrams generally produces a linear combination of shifted exchange and contact diagrams, rather than a pure exchange diagram.   So if one wants to obtain exchange diagrams using weight-shifting operators one needs to know first all the contact diagrams.  Second, care is needed when renormalization is required: renormalized correlators obey inhomogeneous conformal Ward identities containing anomalies and beta functions, whereas these operators only connect solutions of the homogeneous conformal Ward identities.  In particular, one cannot construct a renormalized correlator, which depends on the RG scale $\mu$, by acting with a ($\mu$-independent) weight-shifting operator on a finite correlator, since the latter has no $\mu$-dependence. 
One may apply weight-shifting operators at the level of the regulated theory, but only a subset of them yield useful relations as sometimes weight-shifting operators map regulated exchange diagrams to shifted contact diagrams only.

For the cases we analyze, one may use weight-shifting operators to reduce all exchange diagrams to a single diagram per exchanged scalar ($\Delta_x=2$  and $\Delta_x=3$), plus all contact diagrams. One may further link the two master exchange integrals via a different relation, so that at the end all exchange integrals may be reduced to the exchange diagram with all external dimensions, and the dimension of the exchanged scalar, equal to three (plus all contact diagrams).
However, while all obstacles have been avoided and the regulated amplitudes  constructed, the resulting scheme is little simpler than evaluating the diagrams directly, at least for the cases we analyze.

The layout of this paper is as follows.   In the next section we present the objects of interest: the regulated Witten diagrams.  Section \ref{sec:reg_amp} proceeds to evaluate these diagrams in the `half-integer' scheme \eqref{special}, after which they are renormalized as described in Section \ref{sec:renormalization}.  Section \ref{sec:schemechange} discusses how to change to the general regularization scheme \eqref{genreg}, and provides a complete list of such results. The discussion of the change of scheme is the most complicated part of this paper. Readers interested only in results may skip this part.
The action of weight-shifting operators is discussed in Section \ref{sec:weightshift}.  We present a simple momentum-space derivation of these operators based on the shadow transform, then give a new formula for their action on exchange diagrams before discussing their limitations when applied to renormalized correlators.
Section \ref{sec:Math} summarises the Mathematica packages accompanying this paper, which contain a complete record of all our results.  We discuss the outlook in Section \ref{sec:outlook}.  The paper contains four appendices. In Appendix \ref{sec:conventions} we summarize our conventions and definitions for momenta.  Appendix \ref{CWIapp} lists the  conformal Ward identities and 
Appendix \ref{sec:form} various useful mathematical results. Finally, in Appendix \ref{appD} we present  additional results on weight-shifting operators.

\section{Definitions for regulated amplitudes} \label{sec:defs}

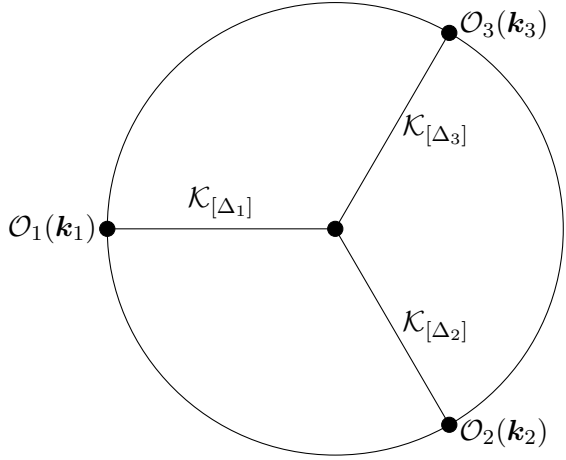
\begin{figure}[t]
\begin{tikzpicture}[scale=1.0]
\draw (0,0) circle [radius=3];
\draw [fill=black] (-3,0) circle [radius=0.1];
\draw [fill=black] (1.5,2.598) circle [radius=0.1];
\draw [fill=black] (1.5,-2.598) circle [radius=0.1];
\draw [fill=black] ( 0, 0) circle [radius=0.1];
\draw (-3,0) -- (0,0) -- (1.5,2.598);
\draw (0,0) -- (1.5,-2.598);
\node [left] at (-3,0) {$\O_1(\bs{k}_1)$}; 
\node [right] at (1.5,2.7) {$\O_3(\bs{k}_3)$};
\node [right] at (1.5,-2.7) {$\O_2(\bs{k}_2)$};
\node [above] at (-1.5,0) {$\K_{[\Delta_1]}$};
\node [right] at (0.75,1.299) {$\K_{[\Delta_3]}$};
\node [right] at (0.75,-1.299) {$\K_{[\Delta_2]}$};
\end{tikzpicture}
\centering
\caption{Witten diagram representing the 3-point amplitude $\ino_{[\Delta_1 \Delta_2 \Delta_3]}$ in \eqref{amp3},  with bulk-to-boundary propagators as given in  \eqref{KPropagator}. \label{fig:intro3pt}}
\end{figure}

Our goal is to calculate the \emph{amplitudes}, \textit{i.e.}, the $3$- and $4$-point \emph{Witten diagrams}, as presented in Figures \ref{fig:intro3pt} and \ref{fig:intro4pt}. As usual, external and internal lines correspond to bulk-to-boundary and bulk-to-bulk propagators in pure Euclidean AdS, while each interior point requires integration over its radial position.  All our conventions for QFT and the momenta are summarized in Appendix \ref{sec:conventions}.
At this point, we will not make reference to any  specific bulk action: rather, we will simply focus on the individual amplitudes defined by the expressions below, postponing consideration of their relation to the bulk action.

\begin{itemize}
\item The scalar bulk-to-boundary propagator is
\begin{align} \label{KPropagator}
	\K_{d, \Delta}(z, k) = \frac{k^{\Delta - \frac{d}{2}} z^{\frac{d}{2}} K_{\Delta - \frac{d}{2}}(k z)}{2^{\Delta - \frac{d}{2} - 1} \Gamma \left( \Delta - \frac{d}{2} \right)}
\end{align}
while the scalar bulk-to-bulk propagator is
\begin{align} \label{GPropagator}
	\G_{d, \Delta}(z, k; \z) = \left\{ \begin{array}{ll}
		(z \z)^{\frac{d}{2}} I_{\Delta - \frac{d}{2}}(k z) K_{\Delta - \frac{d}{2}}(k \z) & \text{ for } z < \z, \\
		(z \z)^{\frac{d}{2}} K_{\Delta - \frac{d}{2}}(k z) I_{\Delta - \frac{d}{2}}(k \z) & \text{ for } z > \z.
	\end{array} \right.	
\end{align}
where $I_{\beta}$ and $K_{\beta}$ are the modified Bessel functions.

To avoid clutter, we will use $\Kreg_{[\Delta]}$ and $\Greg_{[\Delta]}$ to denote the propagators for the regulated parameters $\dreg$ and $\Dreg$ defined in \eqref{genreg}, leaving the specific scheme implicit.

\item For convenience,  regulated 2-point amplitudes are normalized so as  to match the holographic 2-point functions:
\begin{empheq}[box=\nicebox]{align} \label{amp2}
\ireg_{[\Delta \Delta]}(k) = (2 \Dreg - \dreg) \times\text{coefficient of } z^{\Dreg} \text{ in } \Kreg_{[\Delta]}(z, k).
\end{empheq}
All non-diagonal 2-point amplitudes $\ireg_{[\Delta \Delta']}$ with $\Delta \neq \Delta'$ vanish.

\item We define the regulated 3-point amplitudes as
\begin{empheq}[box=\nicebox]{align} \label{amp3}
\ireg_{[\Delta_1 \Delta_2 \Delta_3]}(k_1, k_2, k_3) = \int_0^{\infty} \D z \, z^{-\dreg-1} \, \Kreg_{[\Delta_1]}(z, k_1) \Kreg_{[\Delta_2]}(z, k_2) \Kreg_{[\Delta_3]}(z, k_3).
\end{empheq}
The corresponding Witten diagram is presented in Figure \ref{fig:intro3pt}.

\item We use $\ino_{[\Delta_1 \Delta_2 \Delta_3 \Delta_4]}$ to  denote the contact diagram with four external scalars of dimensions $\Delta_1, \Delta_2, \Delta_3, \Delta_4$, as presented in Figure \ref{fig:intro4pt}. The regulated expression is
\begin{empheq}[box=\nicebox]{align} \label{amp4c}
& \ireg_{[\Delta_1 \Delta_2 \Delta_3 \Delta_4]}(k_1, k_2, k_3, k_4)  \nn\\
& \qquad = \int_0^\infty \D z \, z^{-\dreg-1} \Kreg_{[\Delta_1]}(z, k_1) \Kreg_{[\Delta_2]}(z, k_2) \Kreg_{[\Delta_3]}(z, k_3) \Kreg_{[\Delta_4]}(z, k_4).
\end{empheq}

\begin{figure}[t]
\begin{tikzpicture}[scale=1.0]
\draw (0,0) circle [radius=3];
\draw [fill=black] (-2.121,-2.121) circle [radius=0.1];
\draw [fill=black] (-2.121, 2.121) circle [radius=0.1];
\draw [fill=black] ( 2.121,-2.121) circle [radius=0.1];
\draw [fill=black] ( 2.121, 2.121) circle [radius=0.1];
\draw [fill=black] ( 0, 0) circle [radius=0.1];
\draw (-2.121,-2.121) -- ( 2.121, 2.121);
\draw ( 2.121,-2.121) -- (-2.121, 2.121);	
\node [left] at (-2.121, 2.2) {$\O_1(\bs{k}_1)$}; 
\node [left] at (-2.121,-2.2) {$\O_2(\bs{k}_2)$}; 	
\node [right] at ( 2.121,-2.2) {$\O_3(\bs{k}_3)$}; 
\node [right] at ( 2.121, 2.2) {$\O_4(\bs{k}_4)$}; 	
\node [above] at (-0.9, 1.06) {$\K_{[\Delta_1]}$};
\node [above] at (-1.3,-1.06) {$\K_{[\Delta_2]}$};
\node [above] at ( 1.2,-1.06) {$\K_{[\Delta_3]}$};
\node [above] at ( 0.8, 1.06) {$\K_{[\Delta_4]}$};
\end{tikzpicture}
\qquad
\begin{tikzpicture}[scale=1.0]
\draw (0,0) circle [radius=3];
\draw [fill=black] (-2.121,-2.121) circle [radius=0.1];
\draw [fill=black] (-2.121, 2.121) circle [radius=0.1];
\draw [fill=black] ( 2.121,-2.121) circle [radius=0.1];
\draw [fill=black] ( 2.121, 2.121) circle [radius=0.1];
\draw [fill=black] (-1, 0) circle [radius=0.1];
\draw [fill=black] ( 1, 0) circle [radius=0.1];
\draw (-2.121,-2.121) -- (-1,0) -- (-2.121, 2.121);
\draw ( 2.121, 2.121) -- ( 1,0) -- ( 2.121,-2.121);
\draw (-1,0) -- (1,0);
\node [left] at (-2.121, 2.2) {$\O_1(\bs{k}_1)$}; 
\node [left] at (-2.121,-2.2) {$\O_2(\bs{k}_2)$}; 	
\node [right] at ( 2.121,-2.2) {$\O_3(\bs{k}_3)$}; 
\node [right] at ( 2.121, 2.2) {$\O_4(\bs{k}_4)$}; 	
\node [right] at (-1.5, 1.2) {$\K_{[\Delta_1]}$};
\node [right] at (-1.5, -1.2) {$\K_{[\Delta_2]}$};
\node [left] at ( 1.5, -1.2) {$\K_{[\Delta_3]}$};
\node [left] at ( 1.5, 1.2) {$\K_{[\Delta_4]}$};
\node [above] at (0,0) {$\G_{[\Delta_x]}$};
\end{tikzpicture}
\centering
\caption{Witten diagrams representing the contact and exchange 4-point amplitudes $\ino_{[\Delta_1 \Delta_2 \Delta_3 \Delta_4]}$ and $\ino_{[\Delta_1 \Delta_2, \Delta_3 \Delta_4 x \Delta_x]}$  given in \eqref{amp4c} and \eqref{amp4x}.\label{fig:intro4pt}}
\end{figure}
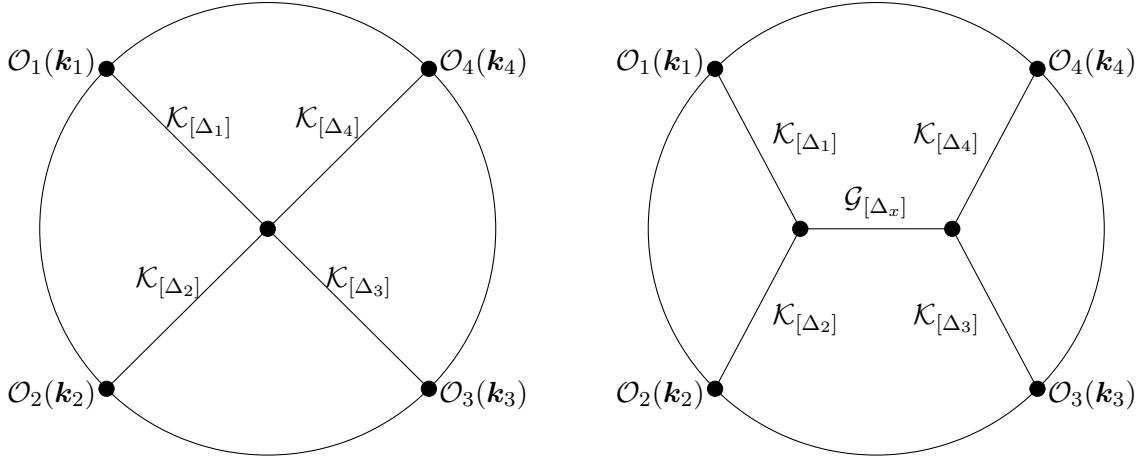

\item We use $\ino_{[\Delta_1 \Delta_2, \Delta_3 \Delta_4 x \Delta_x]}$ to denote the exchange diagram $\Delta_1 \Delta_2 \mapsto \Delta_3 \Delta_4$ with the exchange scalar of dimension $\Delta_x$, as shown in Figure \ref{fig:intro4pt}. The regulated expression is
\begin{empheq}[box=\nicebox]{align} \label{amp4x}
& \ireg_{[\Delta_1 \Delta_2, \Delta_3 \Delta_4 x \Delta_x]}(k_1, k_2, k_3, k_4, s)  \nn\\
& \qquad = \int_0^\infty \D z \, z^{-\dreg-1} \Kreg_{[\Delta_1]}(z, k_1) \Kreg_{[\Delta_2]}(z, k_2) \times\nn\\
& \qquad\qquad \times \int_0^\infty \D \z \, \z^{-\dreg-1} \Greg_{[\Delta_x]}(z, s; \z) \Kreg_{[\Delta_3]}(\z, k_3) \Kreg_{[\Delta_4]}(\z, k_4).
\end{empheq}

\item Many of the regularized amplitudes defined above exhibit divergences as $\ep\rightarrow 0$.  The 3-point amplitudes $\ireg_{[222]}, \ireg_{[322]}, \ireg_{[332]}, \ireg_{[333]}$ all have a single pole $\ep^{-1}$.  For the 4-point amplitudes, the degrees of divergence are summarized  in Table \ref{degdivtable}.

\end{itemize}

\section{Regulated amplitudes in the half-integer scheme} \label{sec:reg_amp}

The idea of the half-integer scheme  \eqref{special} is that all the Bessel functions appearing in the propagators for operators of dimensions $\Delta = 2$ and $\Delta = 3$ reduce to elementary functions. Explicitly, the bulk-to-boundary propagators \eqref{KPropagator} are
\begin{align}
\Kreg_{[2]}(z, k) = \K_{3 + 2 \ep, 2 + \ep}(z, k) & = z^{1 + \ep} e^{-k z}, \\
\Kreg_{[3]}(z, k) = \K_{3 + 2 \ep, 3 + \ep}(z, k) & = z^{\ep} e^{-k z} (1 + k z)
\end{align}
while the bulk-to-bulk propagators \eqref{GPropagator} reduce to 
\begin{align}
	\Greg_{[2]}(z, k; \z) & = \G_{3 + 2 \ep, 2 + \ep}(z, k; \z)  \nn\\
	& = \left\{ \begin{array}{ll}
		\frac{(z \z)^{1 + \ep}}{k} e^{-k \z} \sinh(k z) & \text{ for } z < \z, \\
		\frac{(z \z)^{1 + \ep}}{k} e^{-k z} \sinh(k \z) & \text{ for } z > \z,
	\end{array} \right.	\\[1ex]
	\Greg_{[3]}(z, k; \z) & = \G_{3 + 2 \ep, 3 + \ep}(z, k; \z)  \nn\\
	& = \left\{ \begin{array}{ll}
		\frac{(z \z)^{\ep}}{k^3} e^{-k \z} (1 + k \z) \left( k z \cosh(k z) - \sinh(k z) \right) & \text{ for } z < \z, \\
		\frac{(z \z)^{\ep}}{k^3} e^{-k z} (1 + k z) \left( z \z \cosh(k \z) - \sinh(k \z) \right) & \text{ for } z > \z.
	\end{array} \right.
\end{align}
The radial integrals in the 3- and 4-point contact diagrams \eqref{amp3} and \eqref{amp4c}  can then be evaluated in terms of gamma functions and rational functions. 

For  exchange diagrams, the inner integral in \eqref{amp4x} is expressible in terms of incomplete gamma functions and rational functions. Depending on which end of the integration is fixed, we have
\begin{align}
\gamma(\alpha, x) & = \int_0^x \D t \, e^{-t} t^{\alpha - 1}, & \Gamma(\alpha, x) & = \int_x^{\infty} \D t \, e^{-t} t^{\alpha - 1}.
\end{align}
To calculate the outer integral in \eqref{amp4x} it then suffices to use the following integrals,
\begin{align}
\int_0^{\infty} \D x \, x^{\mu-1} e^{-\beta x} \gamma(\nu, \alpha x) & = \frac{\alpha^\nu \Gamma(\mu + \nu)}{\nu(\alpha + \beta)^{\mu+\nu}} \, {}_2 F_1 \left( 1, \mu + \nu; \nu + 1; \frac{\alpha}{\alpha + \beta}  \right), \\
\int_0^{\infty} \D x \, x^{\mu-1} e^{-\beta x} \Gamma(\nu, \alpha x) & = \frac{\alpha^\nu \Gamma(\mu + \nu)}{\mu(\alpha + \beta)^{\mu+\nu}} \, {}_2 F_1 \left( 1, \mu + \nu; \mu + 1; \frac{\beta}{\alpha + \beta}  \right).
\end{align}
Since for a specific amplitude the variables $\mu$ and $\nu$ depend on the regulator $\ep$, one must series expand the hypergeometric function with respect to its parameters. In all cases the parameters reduce to integer values plus terms of order $O(\ep)$ allowing use of the excellent Mathematica package \verb|HypExp|  \cite{Huber:2005yg}.

In the remainder of this section, we list the $2$-, $3$- and $4$-point  amplitudes defined by expressions \eqref{amp2}, \eqref{amp3}, \eqref{amp4c} and \eqref{amp4x}, all regulated in the half-integer  scheme \eqref{special}. 

\subsection{2-point amplitudes} \label{sec:2pt}

The 2-point amplitudes are
\begin{align}
\ireg_{[22]}  = -k, \qquad \qquad\qquad  \ireg_{[33]}  = k^3, \label{reg2}
\end{align}
where the  normalization matches that of the holographic 2-point functions.
In the half-integer scheme these amplitudes are finite and independent of the regulator $\ep$.

\subsection{3-point amplitudes}

For later use, we will need the regulated 3-point amplitudes expanded to order $\epsilon$. Alternatively, we can simply keep the  full solutions evaluated to all orders in the regulator:
\begin{align}
\ireg_{[222]} & = \Div_1(k_t), \label{ireg222} \\
\ireg_{[322]} & = - \Div_1(k_t) \left[ (k_2 + k_3) + \frac{\ep k_t}{1 - \ep} \right], \\
\ireg_{[332]} & = \frac{1}{2} \Div_1(k_t) \left[ (k_3^2 - k_1^2 - k_2^2) + \frac{\ep k_t}{(1 - \ep)(2 - \ep)} \left( (3 - \ep) k_3 - (1 - \ep) (k_1 + k_2) \right) \right], \label{ireg332} \\
\ireg_{[333]} & = \frac{1}{3} \Div_1(k_t) \Big[ (k_1^3 + k_2^3 + k_3^3) + \frac{\ep k_t}{(1 - \ep)(3 - \ep)} \left( (4 - \ep) (k_1^2 + k_2^2 + k_3^2) \right.\nn\\
& \qquad\qquad\qquad\qquad \left. - (1 - \ep) (k_1 k_2 + k_1 k_3 + k_2 k_3) \right) \Big]. \label{ireg333}
\end{align}
Here and throughout, we use the conventions listed in appendix \ref{sec:conventions}. In particular, $\Div_a(k)= \Gamma(a \ep) k^{-a \ep}$ as given in  \eqref{def:div} and $k_t$ is the total 3-point momentum magnitude as defined in \eqref{total}.
Note that all these 3-point amplitudes exhibit a single pole in the regulator.

\subsection{Contact diagrams}

The regulated contact 4-point amplitudes valid to all orders in $\ep$ are 
\begin{align}
\ireg_{[2222]} & = \frac{2 \ep}{k_T} \Div_2(k_T), \\
\ireg_{[3222]} & = \Div_2(k_T) \left[ 1 + \frac{2 \ep \, k_1}{k_T} \right], \\
\label{idiv3322beta}
\ireg_{[3322]} & =  \Div_2(k_T) \left[  \s{1}{12} - \frac{k_T}{1 - 2 \ep} + \frac{2 \ep \, \s{2}{12}}{k_T} \right], \\
\ireg_{[3332]} & = \Div_2(k_T) \left[ \s{2}{123} - \frac{k_T \s{1}{123}}{1 - 2 \ep} + \frac{k_T^2}{2 (1 - \ep)(1 - 2 \ep)} + \frac{2 \ep \, \s{3}{123}}{k_T} \right], \\
\ireg_{[3333]} & = \Div_2(k_T) \left[ \s{3}{1234} - \frac{k_T \s{2}{1234}}{1 - 2 \ep} + \frac{k_T^3}{(1 - 2 \ep)(3 - 2 \ep)} + \frac{2 \ep \, \s{4}{1234}}{k_T} \right],
\end{align}
where $k_T$ is the total 4-point momentum magnitude  \eqref{total} and $\sigma$ denotes various symmetric polynomials defined in \eqref{def:sigma}. Note that the amplitude $\ireg_{[2222]}$ is finite, $\ireg_{[2222]} = \frac{1}{k_T} + O(\ep)$, while all the remaining amplitudes are linearly divergent.

\subsection{Exchange diagrams}

We split the results for exchange diagrams  into two parts,
\begin{align}
\ireg_{[\Delta_1 \Delta_2, \Delta_3 \Delta_4 x \Delta_x]} = \idiv_{[\Delta_1 \Delta_2, \Delta_3 \Delta_4 x \Delta_x]} + \ifin_{[\Delta_1 \Delta_2, \Delta_3 \Delta_4 x \Delta_x]},
\end{align}
where $\idiv$ contains both the divergences and logarithmic terms, while the finite part $\ifin$ contains the remainder of the amplitude. This split is non-unique, but is  convenient for organizing the amplitude and will be particularly useful in the context of renormalization and change of scheme as we will discuss later.

\subsubsection{Exchange diagrams with {\bs{$\Delta_x = 2$}}} \label{sec:reg2}

For $\Delta_x=2$, we find: 

\begin{itemize}
\item $\ino_{[22,22x2]}$ is finite,
\begin{align}\label{simplest}
\idiv_{[22,22x2]} & = 0 + O(\ep), \\
\ifin_{[22,22x2]} & = - \frac{1}{2 s} \PLp,
\end{align}
where $s$ is the usual Mandelstam variable \eqref{def:stu} and $\PLp$ is defined in \eqref{defD+}.

\item $\ino_{[32,22x2]}$ exhibits a double pole,
\begin{align}
\idiv_{[32,22x2]} & = \frac{1}{2} \Div_1^2(\p{34}) + \Div_2(\p{34}) + O(\ep), \label{i3222x2div} \\[1ex]
\ifin_{[32,22x2]} & =  \frac{k_2}{2 s} \PLp + \frac{1}{2} \PLm  + \frac{1}{4} \left[ 2 - \log^2 \left( \frac{\p{34}}{\p{12}} \right) \right],
\end{align}
where $\PLm$ is defined in \eqref{defD-} and $l_{ij+}$ is defined in \eqref{mij}.

\item $\ino_{[33,22x2]}$ exhibits a single pole,
\begin{align}\label{idiv3322x2beta}
\idiv_{[33,22x2]} & = \frac{k_3 + k_4}{2} \Div_2(k_T) + O(\ep), \\[1ex]
\ifin_{[33,22x2]} & = \frac{k_1^2 + k_2^2 - s^2}{4 s} \PLp + \frac{k_1 + k_2}{2} \left[ \log \left( \frac{\p{34}}{k_T} \right) + 1 \right] + \frac{7}{8} (k_3 + k_4).
\end{align}

\item $\ino_{[32,32x2]}$ exhibits a double pole,
\begin{align}
\idiv_{[32,32x2]} & = - \frac{k_2 + s}{2} \left( \Div_1^2(\p{12}) + 6 \Div_2(\p{12}) \right) - \frac{k_4 + s}{2} \left( \Div_1^2(\p{34}) + 6 \Div_2(\p{34}) \right) \nn\\
& \qquad\qquad + (s - 2 k_1) \Div_2(\p{12}) + (s - 2 k_3) \Div_2(\p{34}) + O(\ep), \\[1ex]
\ifin_{[32,32x2]} & = \frac{s^2 - k_2 k_4}{2 s} \PLp + \frac{k_2 - k_4}{2} \PLm + \frac{k_2 + k_4 + 2s}{4} \log^2 \left( \frac{\p{12}}{\p{34}} \right) \nn\\
& \qquad\qquad + (k_3 + k_4) \log \left( \frac{k_T}{\p{34}} \right) + (k_1 + k_2) \log \left( \frac{k_T}{\p{12}} \right) \nn\\
& \qquad\qquad - 3 k_T - \frac{k_2 + k_4}{2} - 3 s.
\end{align}

\item $\ino_{[32,33x2]}$ exhibits a double pole,
\begin{align}
\idiv_{[32,33x2]} & = \frac{s^2 - k_3^2 - k_4^2}{4} \Div_1^2(\p{34}) + \left[ s (k_3 + k_4) - ( k_3^2 + k_3 k_4 + k_4^2) \right.\nn\\
& \qquad\qquad\qquad\qquad \left. + \frac{5 s^2 + k_1^2 - k_2^2}{4} \right] \Div_2(\p{34}) + O(\ep), \\[1ex]
\ifin_{[32,33x2]} & = \frac{s^2 - k_3^2 - k_4^2}{4} \left[ \frac{k_2}{s} \PLp + \PLm \right] - \frac{s^2 - k_3^2 - k_4^2}{8} \log^2 \left( \frac{\p{34}}{\p{12}} \right) \nn\\
& \qquad\qquad + \frac{1}{4} \left[ k_1^2 - k_2^2 + (2s-k_3-k_4)(k_3 + k_4) \right] \log \left( \frac{\p{34}}{\p{12}} \right) \nn\\
& \qquad\qquad + (s + k_2)(k_3 + k_4) \log \left( \frac{k_T}{\p{12}} \right)\nn\\
& \qquad\qquad + \frac{1}{4} \left[ k_1^2 - (k_4 + k_3 - k_2)^2 + 4 s (k_3 + k_4) \right] \log \left( \frac{\p{12}}{k_T} \right) \nn\\
& \qquad\qquad + \frac{1}{16} \left[ 17 s^2 + 5 k_1^2 - 4 k_1 k_2 - 9 k_2^2 + 4 (5 s + k_1 - k_2) \s{1}{34}
\right. \nn\\& \qquad\qquad \qquad\quad\left.
 - 12 \s{1}{34}^2 + 8 \s{2}{34} \right],
\end{align}	
where $\s{m}{J}$ is defined in \eqref{def:sigma}. 

\item $\ino_{[33,33x2]}$ exhibits a single pole,
\begin{align}
\idiv_{[33,33x2]} & = - \frac{k_1^3 + k_2^3 + k_3^3 + k_4^3}{6} \Div_2(k_T) + O(\ep), \\[1ex]
\ifin_{[33,33x2]} & = - \frac{(s^2 - k_1^2 - k_2^2)(s^2 - k_3^2 - k_4^2)}{8 s} \PLp \nn\\
& \qquad\qquad + \frac{1}{4} (s^2 - k_1^2 - k_2^2)(k_3 + k_4) \log \left( \frac{\p{12}}{k_T}\right) \nn\\
& \qquad\qquad + \frac{1}{4} (s^2 - k_3^2 - k_4^2)(k_1 + k_2) \log \left( \frac{\p{34}}{k_T}\right) \nn\\
& \qquad\qquad + \frac{1}{72} \left( 18 s^2 \s{1}{1234} - 25 \s{1}{1234}^3 + 63 \s{1}{1234} \s{2}{1234} - 27 \s{3}{1234} \right) \nn\\
& \qquad\qquad - \frac{s}{4} (k_1 + k_2)(k_3 + k_4).
\end{align}

\end{itemize}

\subsubsection{Exchange diagrams with {\bs{$\Delta_x = 3$}}} \label{sec:reg3}

For $\Delta_x=3$, we find:

\begin{itemize}
\item $\ino_{[22,22x3]}$ is finite,
\begin{align}
\idiv_{[22,22x3]} & = 0 + O(\ep), \\[1ex]
\ifin_{[22,22x3]} & = \frac{(k_1 + k_2)(k_3 + k_4)}{2 s^3} \PLp - \frac{k_1 + k_2}{s^2} \log \left( \frac{\p{12}}{k_T} \right) - \frac{k_3 + k_4}{s^2} \log \left( \frac{\p{34}}{k_T} \right) + \frac{1}{s}.
\end{align}
\item $\ino_{[32,22x3]}$ exhibits a single pole,
\begin{align}
\idiv_{[32,22x3]} & = \frac{1}{2} \Div_2(\p{12}) + O(\ep), \\[1ex]
\ifin_{[32,22x3]} & = \frac{(s^2 + k_1^2 - k_2^2)(k_3 + k_4)}{4 s^3} \PLp \nn\\
& \qquad\qquad + \frac{k_1^2 - k_2^2}{2 s^2} \log \left( \frac{k_T}{\p{12}} \right) + \frac{(k_1 - k_2)(k_3 + k_4)}{2 s^2} \log \left( \frac{k_T}{\p{34}} \right) \nn\\
& \qquad\qquad + \frac{3 s + 4 (k_1 - k_2)}{8 s}.
\end{align}
\item $\ino_{[33,22x3]}$ exhibits a double pole,
\begin{align}
\idiv_{[33,22x3]} & = - \frac{k_3 + k_4}{6} \Div_1^2(\p{34}) - \frac{6 s + 7(k_3 + k_4)}{9} \Div_2(\p{34}) + O(\ep), \\[1ex]
\ifin_{[33,22x3]} & = - \frac{k_3 + k_4}{6} \left[ \frac{k_1^3 + k_2^3}{s^3} \PLp + \PLm \right] \nn\\
& \qquad\qquad + \frac{k_3 + k_4}{12} \log^2 \left( \frac{\p{12}}{\p{34}} \right) + \frac{s^3 + k_1^3 + k_2^3}{3 s^2} \log \left( \frac{\p{12}}{\p{34}} \right) \nn\\
& \qquad\qquad - \frac{k_T(k_1^2 - k_1 k_2 + k_2^2)}{3 s^2} \log \left( \frac{k_T}{\p{34}} \right) \nn\\
& \qquad\qquad - \frac{1}{3} \left[ \frac{k_1^2 - k_1 k_2 + k_2^2}{s} + (k_1 + k_2) + \frac{7}{3} s + \frac{43}{18} (k_3 + k_4) \right].
\end{align}
\item $\ino_{[32,32x3]}$ exhibits a single pole,
\begin{align}
\idiv_{[32,32x3]} & = - \frac{k_2 + k_4}{2} \Div_2(k_T) + O(\ep), \\[1ex]
\ifin_{[32,32x3]} & = \frac{(s^2 + k_1^2 - k_2^2)(s^2 + k_3^2 - k_4^2)}{8 s^3} \PLp \nn\\
& \qquad\qquad + \frac{(s^2 + k_1^2 - k_2^2)(k_3 - k_4)}{4 s^2} \log \left( \frac{k_T}{\p{12}} \right) \nn\\
& \qquad\qquad + \frac{(s^2 + k_3^2 - k_4^2)(k_1 - k_2)}{4 s^2} \log \left( \frac{k_T}{\p{34}} \right) \nn\\
& \qquad\qquad + \frac{1}{8 s} \left[ 2 (k_1 - k_2)(k_3 - k_4) - 2 s (k_1 + k_3) - 5 s (k_2 + k_4) \right].
\end{align}
\item $\ino_{[32,33x3]}$ exhibits a double pole,
\begin{align}
\idiv_{[32,33x3]} & = - \frac{s^2 + k_1^2 - k_2^2}{12} \Div_1^2(\p{12}) \nn\\
& \qquad\qquad - \frac{1}{36} \left[ 11 s^2 + 12 s(k_1 - k_2) + 4(2 k_1 - 5 k_2) (k_1 + k_2) \right.\nn\\
& \qquad\qquad\qquad\qquad \left. + 9 (k_3^2 + k_4^2) \right] \Div_2(\p{12}) + O(\ep), 
\end{align}
\begin{align}
\ifin_{[32,33x3]} & = \frac{(s^2 + k_1^2 - k_2^2)}{12} \left[ - \frac{k_3^3 + k_4^3}{s^3} \PLp + \PLm \right] \nn\\
& \qquad\qquad + \frac{s^2 + k_1^2 - k_2^2}{24} \log^2 \left( \frac{\p{12}}{\p{34}} \right) - \frac{(k_1 - k_2)(s^3 + k_3^3 + k_4^3)}{6 s^2} \log \left( \frac{\p{12}}{\p{34}} \right) \nn\\
& \qquad\qquad - \frac{k_T}{12} \left[ k_1 + k_2 - k_3 - k_4 + \frac{2(k_1 - k_2)(k_3^2 - k_3 k_4 + k_4^2)}{s^2} \right] \log \left( \frac{k_T}{\p{12}} \right) \nn\\
& \qquad\qquad - \frac{(k_1 - k_2)(k_3^2 - k_3 k_4 + k_4^2)}{6 s} + \frac{1}{432} \left[ -52 k_1^2 + 292 k_2^2 + 108 k_2 (k_3 + k_4) \right.\nn\\[1ex]
& \qquad\qquad\qquad\qquad \left. - 9 (11 k_3^2 + 4 k_3 k_4 + 11 k_4^2) + 12 k_1 (20k_2 - 3 k_3 - 3 k_4) \right] \nn\\[1ex]
& \qquad\qquad - \frac{s(11 k_1 - 17 k_2)}{36} - \frac{85 s^2}{432}.
\end{align}
\item $\ino_{[33,33x3]}$ exhibits a double pole,
\begin{align}
\idiv_{[33,33x3]} & = \frac{s^3 + k_1^3 + k_2^3}{18} \Div_1^2(\p{12}) + \frac{s^3 + k_3^3 + k_4^3}{18} \Div_1^2(\p{34}) \nn\\
& \qquad\qquad + \frac{1}{9} \left[ 3 \s{1}{12s}^3 - 7 \s{1}{12s} \s{2}{12s} + \s{3}{12s} - \frac{s^3}{3} \right] \Div_2(\p{12}) \nn\\
& \qquad\qquad + \frac{1}{9} \left[ 3 \s{1}{34s}^3 - 7 \s{1}{34s} \s{2}{34s} + \s{3}{34s} - \frac{s^3}{3} \right] \Div_2(\p{34}) + O(\ep),
\end{align}
\begin{align}
\ifin_{[33,33x3]} & = - \frac{2 s^3 + k_1^3 + k_2^3 + k_3^3 + k_4^3}{36} \log^2 \left( \frac{\p{12}}{\p{34}} \right) \nn\\
& \qquad\qquad + \frac{1}{9} \left[ s^2 (k_1 + k_2) + s ( k_1^2 - k_1 k_2 + k_2^2 - k_3^2 + k_3 k_4 - k_4^2 ) \right.\nn\\[0.5ex]
& \qquad\qquad\qquad\qquad \left. + \frac{(k_1 + k_2)^3}{3} - \frac{(k_1^3 + k_2^3)(k_3^2 - k_3 k_4 + k_4^2)}{s^2} \right] \log \left( \frac{\p{12}}{k_T} \right) \nn\\[0.5ex]
& \qquad\qquad + \frac{1}{9} \left[ s^2 (k_3 + k_4) + s ( k_3^2 - k_3 k_4 + k_4^2 - k_1^2 + k_1 k_2 - k_2^2 ) \right.\nn\\[0.5ex]
& \qquad\qquad\qquad\qquad \left. + \frac{(k_3 + k_4)^3}{3} - \frac{(k_3^3 + k_4^3)(k_1^2 - k_1 k_2 + k_2^2)}{s^2} \right] \log \left( \frac{\p{34}}{k_T} \right) \nn\\[0.5ex]
& \qquad\qquad + \frac{(k_1^3 + k_2^3)(k_3^3 + k_4^3) - s^6}{18 s^3} \PLp + \frac{k_3^3 + k_4^3 - k_1^3 - k_2^3}{18} \PLm \nn\\[1ex]
& \qquad\qquad + \frac{1}{27} \Big[ 14 s^3 + 11 s^2 k_T + s \left( 8 (k_1^2 + k_2^2 + k_3^2 + k_4^2) - 5 (k_1 k_2 + k_3 k_4) \right)  \nn\\[1ex]
& \qquad\qquad\qquad\qquad + \frac{1}{6} \left( 61 \s{1}{1234}^3 - 129 \s{1}{1234} \s{2}{1234} + 15 \s{3}{1234} \right) \nn\\[1ex]
& \qquad\qquad\qquad\qquad + 5 \left( 2 \s{1}{34} \s{2}{12} + 2 \s{1}{12} \s{2}{34} - \s{1}{12} \s{1}{34} k_T \right) \nn\\[1ex]
& \qquad\qquad\qquad\qquad  + \frac{3 (k_1^2 - k_1 k_2 + k_2^2)(k_3^2 - k_3 k_4 + k_4^2)}{s} \Big].
\end{align}
\end{itemize}

\subsection{OPE limit of exchange diagrams}

An interesting limit of the $s$-channel exchange diagram is
\[
s\rightarrow 0, \qquad k_1\approx k_2, \qquad k_3\approx k_4
\]
which probes configurations where $x_{12}^2\rightarrow 0$ and $x_{34}^2\rightarrow 0$ in position space  \cite{Arkani-Hamed:2015bza, Freedman:1998bj, Suyama:2007bg,  Assassi:2012zq}.  Applying the operator product expansion
\[
\O_{\Delta_i}(\bs{x}_i)\O_{\Delta_j}(\bs{x}_j) \sim \frac{C_{ijx}}{x_{12}^{\Delta_i+\Delta_j-\Delta_x}}\O_{\Delta_x}(\bs{x}_i) +\ldots
\]
to the pairs $(i,j) = (1,2)$ and $(3,4)$ yields the 4-point function
$
x_{12}^{\Delta_x-\Delta_1-\Delta_2}x_{34}^{\Delta_x-\Delta_3-\Delta_4}x_{13}^{-2\Delta_x}
$
which corresponds in momentum space  to
\[\label{naiveOPE}
i_{[\Delta_1\Delta_2,\Delta_3\Delta_4x\Delta_x]}\sim 
k_1^{\Delta_1+\Delta_2-\Delta_x-d}k_3^{\Delta_3+\Delta_4-\Delta_x-d}s^{2\Delta_x-d}.
\]
Naively, one might think this argument gives the leading behaviour in the limit as $s\rightarrow 0$. 
In fact, however, this expectation is only  correct for cases where $\Delta_x < d/2$, and not for those studied here  where $\Delta_x>d/2$. This in  exact agreement with the  results in \cite{Bzowski:2014qja} where the same subtlety was discussed for 3-point functions.

To illustrate this, consider for example the OPE limit of the simplest exchange diagram \eqref{simplest}, where all $\Delta_j=\Delta_x=2$. 
This corresponds to evaluating
\[
\lim_{s\rightarrow 0} \Big[-\frac{1}{2s}\Big(\Li_2\Big(\frac{k_1-s/2}{k_1+k_3}\Big)+\Li_2\Big(\frac{k_3-s/2}{k_1+k_3}\Big)+\log\Big(\frac{k_1-s/2}{k_1+k_3}\Big)\log\Big(\frac{k_3-s/2}{k_1+k_3}\Big)-\frac{\pi^2}{6}\Big)\Big].
\]
The limit is of the form $0/0$ since the numerator vanishes for $s=0$
by the identity 
\[
0 = \Li_2(x)+\Li_2(1-x)+\log x\log(1-x) - \frac{\pi^2}{6}
\]
setting $x=k_1/(k_1+k_3)$.  Since  $0<x<1$ all logarithms and dilogarithms are evaluated on their principal branches. 
Using l'H{\^o}pital's rule, we then find 
\[\label{slimeasy}
\ino_{[22,22x2]} \rightarrow -\frac{1}{2k_3}\log\Big(\frac{k_1}{k_1+k_3}\Big)-\frac{1}{2k_1}\log\Big(\frac{k_3}{k_1+k_3}\Big)
\]
in contradiction with \eqref{naiveOPE}.

The subleading nature of the OPE behaviour \eqref{naiveOPE} for $\Delta_x>d/2$ 
can clearly be seen from the $s\rightarrow 0$ limit of the bulk-bulk propagator \eqref{GPropagator}.  For $z<\zeta$ we have
\[\label{Gslim}
\lim_{s\rightarrow 0}
	\G_{d, \Delta}(z, s; \z) =\frac{z^{\Delta_x}\zeta^{d-\Delta_x}}{2\Delta_x-d}\Big(1+O(s^2)\Big) + s^{2\Delta_x-d}\Big(C z^{\Delta_x}\zeta^{\Delta_x}+O(s^2)\Big)
 \]
 where $C= 2^{d-2\Delta_x-1}\Gamma(d/2-\Delta_x)/\Gamma(\Delta_x-d/2+1)$, and for $z>\zeta$ the same expression holds with $z\leftrightarrow\zeta$.  As $s\rightarrow 0$, therefore, the term proportional to $s^{2\Delta_x-d}$ only dominates for $\Delta_x<d/2$.  In this case, the leading behaviour  is symmetric under interchanging $z\leftrightarrow\zeta$ and the exchange diagram factorizes as 
\begin{align}\label{OPEofxdia}
\ino_{[\Delta_1\Delta_2,\Delta_3\Delta_4x\Delta_x]}&\rightarrow
C s^{2\Delta_x-d}\int_0^\infty \frac{\D z}{z^{d+1}}z^{\Delta_x}\K_{d, \Delta_1}(z, k_1) \K_{d, \Delta_2}(z, k_1) \nn\\&\qquad\qquad\times \int_0^\infty \frac{\D \zeta}{\zeta^{d+1}}\zeta^{\Delta_x}\K_{d, \Delta_3}(\zeta, k_3) \K_{d, \Delta_4}(\zeta, k_3) \nn\\[2ex]&\quad
\sim s^{2\Delta_x-d} k_1^{\Delta_1+\Delta_2-\Delta_x-d}k_3^{\Delta_3+\Delta_4-\Delta_x-d}
\end{align}
consistent with the OPE behaviour \eqref{naiveOPE}.
For $\Delta_x>d/2$,  however,  the leading behaviour of the bulk-bulk propagator corresponds instead to the first term in \eqref{Gslim}.
The $s$-dependence is then subleading invalidating the naive OPE expectation \eqref{naiveOPE} as we saw above.  
As the leading behaviour of the bulk-bulk propagator is no longer symmetric under $z\leftrightarrow\zeta$, we cannot simply factorize the exchange diagram into a product of 3-point functions as in \eqref{OPEofxdia}.  Instead, we have an inner $\zeta$-integral over $0\le \zeta < z$ using \eqref{Gslim} with $z\leftrightarrow\zeta$, and an outer $\zeta$-integral over $z < \zeta<\infty$ using \eqref{Gslim},
giving rise to the result \eqref{slimeasy}.

In conclusion, care must  be taken when applying OPE arguments to ascertain the leading behaviour of momentum-space correlators.  
The application of the OPE  for understanding the conformal block structure of momentum-space correlators is an interesting open direction 
(see, {\it e.g.,} \cite{Gillioz:2020wgw} for recent work).

\section{Renormalization}  \label{sec:renormalization}

While the amplitudes listed in the previous section are naturally associated with AdS Witten diagrams,
we do not necessarily have to invoke any form of holography. In principle, it is sufficient simply to  write down the expressions \eqref{amp2}-\eqref{amp4x} and verify that they solve the regulated conformal Ward identities (CWIs). 
This check is performed explicitly in the Mathematica notebook \verb|RegulatedAmplitudes.nb| using the Ward identities listed in Appendix \ref{CWIapp}. $\,$   Since every amplitude 
satisfies the  Ward identities, each  potentially represents a valid correlation function.

However, the fact that a given  expression satisfies the CWIs does not by itself imply that a CFT with such a 4-point function actually 
exists. Other constraints, for example unitarity or the bootstrap, could  rule out the existence of such a CFT.  Nevertheless, we will show in Section \ref{sec:every} that, for each amplitude listed in the previous section,  a holographic CFT can be constructed
for which the corresponding correlation function is equal to this amplitude, {\it i.e.}, a bulk action exists such that the Witten diagram is the entire correlator. If it turns out that a CFT with the desired spectrum and correlators does not exist, then the corresponding bulk theory would be in the swampland.

Next, we will discuss the renormalization  of the regulated amplitudes listed in the previous section.  In a textbook approach, it is usually the correlation functions themselves that are renormalized.  However, as every regulated amplitude here is  equal to a valid regulated correlator in some CFT, every amplitude must be individually renormalizable. This means, for example, that its divergences must be of an appropriately local form, so that eventually they can be removed by a suitable local counterterm. In particular, the divergences of an $n$-point function must take the form of lower-point functions of the same operators, plus possible additional ultralocal pieces.  Our central focus will therefore be on renormalizing individual Witten diagrams. We will be able to write down counterterm actions and counterterm contributions that can be used in the renormalization procedure once the theory is specified. The renormalized amplitudes are listed in Section \ref{sec:ren_amp}.

As an illustration of the complete renormalization procedure, we will consider two holographic theories which we call   \emph{symmetric} and \emph{asymmetric}. The asymmetric theory, defined in Section \ref{sec:every}, contains five operators of dimension two or three  and is the least symmetric theory in the sense that there are no non-vanishing 4-point functions containing identical operators.
Even when some operators have the same dimension, the correlators are not symmetric under exchange of these operators as they are not identical.
The symmetric theory, on the other hand, contains only two operators, one of dimension two and one of dimension three, and there are non-vanishing 4-point function containing identical operators. These correlators are symmetric under exchange of the operators of the same dimension. The dual bulk theory contains bulk scalars  corresponding to the boundary operators (five and two scalars, respectively) plus suitable $3$- and $4$-point interaction terms. In Sections \ref{sec:asym_ren} and \ref{sec:symmetric}, we  carry out the complete renormalization procedure for these two theories. For the symmetric theory, we further derive the beta functions and anomalies and verify the Callan-Symanzik equation.

From a holographic perspective, the renormalization procedure we adopt is a hybrid one: boundary correlation functions are derived from bulk amplitudes at generic dimensions $d$ and $\Delta_j$, and the subsequent renormalization is performed entirely within the boundary theory.  Thus, instead of regulating the bulk theory by cutting off the radial direction as per holographic renormalization, we employ dimensional regularization in the boundary theory as, \textit{e.g.}, in \cite{Schwimmer:2000cu,Schwimmer:2003eq}. The resulting counterterms are an integral part of the bulk theory (dimensionally regularized as in \cite{Bzowski:2016kni,Bzowski:2019xri})  as the counterterms are  also uniquely determined by requiring the bulk variational problem to be well-posed \cite{Papadimitriou:2005ii}. 

\subsection{Conventions and definitions}

In this section, we use the following conventions and definitions:
\begin{itemize}
\item The counterterm contribution $\ict$ is added to the regulated amplitude $\ireg$, to yield the finite renormalized amplitude $\iren$
\begin{align}
\iren = \lim_{\epsilon \rightarrow 0} \left[ \ireg + \ict \right].
\end{align}
\item For the exchange amplitudes, we will list only their \emph{logarithmic terms}  defined by
\begin{align}
	\ilog_{[\Delta_1 \Delta_2, \Delta_3 \Delta_4 x \Delta_x]} = \lim_{\epsilon \rightarrow 0} \left[ \idiv_{[\Delta_1 \Delta_2, \Delta_3, \Delta_4 x \Delta_x]} + \ict_{[\Delta_1 \Delta_2, \Delta_3 \Delta_4 x \Delta_x]} \right].
\end{align}
The full amplitude then reads
\begin{align}
	\iren_{[\Delta_1 \Delta_2, \Delta_3 \Delta_4 x \Delta_x]} = \ilog_{[\Delta_1 \Delta_2, \Delta_3 \Delta_4 x \Delta_x]} + \ifin_{[\Delta_1 \Delta_2, \Delta_3 \Delta_4 x \Delta_x]},
\end{align}
where the finite parts $\ifin$ are as listed in Sections \ref{sec:reg2} and \ref{sec:reg3}.
\item As previously, the Euclidean $\text{AdS}_4$ metric in Poincar\'{e} coordinates is
\begin{align}
	\D s^2 = \frac{\D z^2 + \D \bs{x}^2}{z^2}.
\end{align}

The AdS/CFT correspondence in the half-integer regularization scheme then gives
\begin{align}
	& W[\phi_j] = - S_{\text{on-shell}}[\phi_j], && \< \O \>_{s, \text{reg}} = - (2 \Delta - d) \phi_{(\Delta + \ep)}. \label{1pt}
\end{align}
In particular, the combination $2 \Delta - d$ remains unchanged in the half-integer scheme. 

For every operator $\O_j$,  we denote its source by $\phi_j$. If the operator has dimension $\Delta_j$, its source has  dimension $d - \Delta_j$ and thus $\O_j^{[\Delta_j]}$ couples to $\phi_j^{[d - \Delta_j]}$. In particular $\O^{[2]}$ is sourced by $\phi^{[1]}$, while $\O^{[3]}$ is sourced by $\phi^{[0]}$.
\item All counterterms appear with the renormalization scale $\mu$, and multiplied by constants
that we collectively denote as $\mathfrak{a}$. These constants are series in $\ep$,
\begin{align} \label{a_exp}
\mathfrak{a} = 1 + \ep \, \mathfrak{a}^{(1)} + \ep^2 \, \mathfrak{a}^{(2)} + O(\ep^3),
\end{align}
and the $\ep$-dependent part is generally scheme-dependent. As we show, one may choose a scheme such that the scheme-dependent part depends only on the dimensions of the operators under consideration, and not on any other labels of the operators. However, renormalisation conditions may require different choices of scheme-dependent constants for different operators.

\end{itemize}

\subsection{Every amplitude is a correlator} \label{sec:every}

Here we consider the holographic theory that realizes each amplitude (Witten diagram) as a stand-alone correlation function. We consider five scalar fields $\Phi_j$ with $j=1,2,3,4,x$ dual to operators $\O_j$ of arbitrary dimensions $\Delta_j$. The (unregulated) bulk action, up to quartic order, is
\begin{align} \label{S4asym}
S^{\text{asym}} & = \frac{1}{2} \int \D^{d} x \sqrt{g} \sum_{j =1,2,3,4,x} \left[ \partial_\mu \Phi_j \partial^\mu \Phi_j + m^2_{\Delta_j} \Phi_j^2 \right] \nn\\
& \qquad + \int \D^{d} x \sqrt{g} \left[ \lambda_{12x} \Phi_1 \Phi_2 \Phi_x + \lambda_{34x} \Phi_x \Phi_3 \Phi_4 - \lambda_{1234} \Phi_1 \Phi_2 \Phi_3 \Phi_4 \right],
\end{align}
where
\begin{align}
m^2_{\Delta} = \Delta (\Delta - d)
\end{align}
and $\lambda_{12x}, \lambda_{34x}, \lambda_{1234}$ are arbitrary AdS couplings. We refer to this theory as the \emph{asymmetric theory}, since the resulting correlators have as few discrete symmetries as possible.  
Later on, in Section \ref{sec:symmetric}, we will  introduce the  {\it symmetric theory} where all correlators exhibit some form of  crossing symmetry.

\subsubsection{Regulated correlators}

Let us now derive the regulated correlators of the asymmetric theory \eqref{S4asym}, following \cite{Skenderis:2002wp}.

\begin{itemize}
\item Since the action \eqref{S4asym} is symmetric under the exchange of the fields $\Phi_1$ and $\Phi_2$, their dynamics is identical. The same conclusion holds for $\Phi_3$ and $\Phi_4$. The equations of motion for the bulk fields read
\begin{align}
& (-\Box_{AdS} + \reg{m}^2_{\Delta_{1}}) \Phi_{1} = - \lambda_{12x} \Phi_2 \Phi_x + \lambda_{1234} \Phi_2 \Phi_3 \Phi_4, \\
& (-\Box_{AdS} + \reg{m}^2_{\Delta_{3}}) \Phi_{3} = - \lambda_{34x} \Phi_4 \Phi_x + \lambda_{1234} \Phi_1 \Phi_2 \Phi_4, \\
& (-\Box_{AdS} + \reg{m}^2_{\Delta_{x}}) \Phi_{x} = - \lambda_{12x} \Phi_1 \Phi_2 - \lambda_{34x} \Phi_3 \Phi_4.
\end{align}

\item The equations of motion can now be solved perturbatively to obtain terms with up to and including three sources. Let us denote by $\Phi_{n\{j\}}$ the term in the solution depending on any $(j+1)$ sources. Thus, for example, $\Phi_{n\{0\}}$ is the solution to the free field equation. Concentrating on $\Phi_1$, in position space we find 
\begin{align}
\Phi_{1\{0\}} & = \K_{[\Delta_1]} \ast \phi_{1}, \label{neweq}\\
\Phi_{1\{1\}} & = - \lambda_{12x} \G_{[\Delta_1]} \ast ( \Phi_{2 \{0\}} \Phi_{x \{0\}} ), \label{asym_to_3pt} \\
\Phi_{1\{2\}} & = - \lambda_{12x} \G_{[\Delta_1]} \ast \left[ \Phi_{2 \{1\}} \Phi_{x \{0\}} + \Phi_{2 \{0\}} \Phi_{x \{1\}} \right] \nn\\
& \qquad\qquad + \lambda_{1234} \G_{[\Delta_1]} \ast ( \Phi_{2 \{0\}} \Phi_{3 \{0\}} \Phi_{4 \{0\}} ), \label{asym_to_4pt}
\end{align}
where $\ast$ indicates the convolution of the position-space variables,\footnote{In \eqref{neweq} the integral is taken only over the boundary directions. } {\it e.g.,}
\begin{align} \label{convolution}
\Phi_{1\{1\}}(x) =- \lambda_{12x}\int \D^{d+1} x' \, \sqrt{g(x')} \G_{[\Delta_1]}(x,x')  \Phi_{2 \{0\}}(x') \Phi_{x \{0\}}(x'). 
\end{align}

\item The holographic 2-point functions are diagonal and given by the standard expression
\begin{align}
\lla \O_j(\bs{k}) \O_j(-\bs{k}) \rra = (2 \Delta - d) \times \text{ coefficient of } z^{\Delta} \text{ in } \K_{\Delta}(z, k).
\end{align}
In Section \ref{sec:2pt}, we defined the 2-point amplitudes $\ino_{[\Delta \Delta]}$ in \eqref{amp2} in such a way that they are equal to the holographic 2-point function,
\begin{align} \label{asym2}
\lla \O_j(\bs{k}) \O_j(-\bs{k}) \rra = \ino_{[\Delta_j \Delta_j]}.
\end{align}

\item To go to the boundary for higher-point functions, we use the fact that
\begin{align}
\G_{[\Delta]}(z, k; \zeta) = \frac{z^{\Delta}}{2 \Delta - d} \K_{[\Delta]}(\zeta, k) + O(z^{\Delta + 2}).
\end{align}
Thus, from \eqref{1pt} we see that regulated 1-point functions with sources reads
\begin{align}
\< \O_{1} \>_{\{1\}s, \text{reg}} & = \lambda_{12x} \K_{[\Delta_1]} \ast ( \K_{[\Delta_2]} \ast \phi_2) ( \K_{[\Delta_x]} \ast \phi_x), \label{O11} \\
\< \O_{1} \>_{\{2\}s, \text{reg}} & = - \lambda_{12x} \lambda_{34x} \K_{[\Delta_1]} \ast \left[ (\K_{[\Delta_2]} \ast \phi_2) \G_{[\Delta_x]} \ast (\K_{[\Delta_3]} \ast \phi_3 ) (\K_{[\Delta_4]} \ast \phi_4 ) \right] \nn\\
& \quad- \lambda_{12x}^2 \K_{[\Delta_1]} \ast \left[ (\K_{[\Delta_x]} \ast \phi_x) \G_{[\Delta_2]} \ast (\K_{[\Delta_1]} \ast \phi_1 ) (\K_{[\Delta_x]} \ast \phi_x ) \right] \nn\\
& \quad- \lambda_{1234} \K_{[\Delta_1]} \ast ( \K_{[\Delta_2]} \ast \phi_2) ( \K_{[\Delta_3]} \ast \phi_3) ( \K_{[\Delta_4]} \ast \phi_4). \label{O12}
\end{align}
The $(n+1)$-point function is then obtained by taking $n$ functional derivatives with the factor $(-1)^n$.

These expressions represent the 3- and 4-point Witten diagrams, but with sources present. For example, \eqref{O11} written out explicitly is
\begin{align}
& \< \O_{1}(x_1) \>_{\{1\}s, \text{reg}} = \lambda_{12x} \int \D^{d+1} x \, \sqrt{g(x)} \K_{[\Delta_1]}(x_1, x) \times \nn\\
& \qquad \times \int \D^{d+1} x_2 \, \sqrt{g(x_2)} \K_{[\Delta_2]}(x, x_2) \phi_2 (x_2) \int \D^{d+1} x_3 \, \sqrt{g(x_3)} \K_{[\Delta_x]}(x, x_3) \phi_x(x_3)  \nn\\
& = \lambda_{12x} \int \D^{d+1} x_2 \sqrt{g(x_2)} \int \D^{d+1} x_3 \sqrt{g(x_3)} \, \ino_{[\Delta_1 \Delta_2 \Delta_3]}(x_1, x_2, x_3) \phi_2(x_2) \phi_x(x_3),
\end{align}
where here $\ino_{[\Delta_1 \Delta_2 \Delta_3]}(x_1, x_2, x_3)$ denotes the position space expression for the 3-point amplitude $\ino_{[\Delta_1 \Delta_2 \Delta_3]}$. We will not work in position space any further in this paper.

\begin{figure}[t]
\begin{tikzpicture}[scale=0.9]
\draw (0,0) circle [radius=3];
\draw [fill=black] (-3,0) circle [radius=0.1];
\draw [fill=black] (1.5,2.598) circle [radius=0.1];
\draw [fill=black] (1.5,-2.598) circle [radius=0.1];
\draw [fill=black] ( 0, 0) circle [radius=0.1];
\draw (-3,0) -- (0,0) -- (1.5,2.598);
\draw (0,0) -- (1.5,-2.598);
\node [left] at (-3,0) {$\O_1(x_1)$}; 
\node [right] at (1.5,2.7) {$\phi_3(x_3)$};
\node [right] at (1.5,-2.7) {$\phi_2(x_2)$};
\node [above] at (-1.5,0) {$\K_{[\Delta_1]}$};
\node [right] at (0.75,1.299) {$\K_{[\Delta_x]}$};
\node [right] at (0.75,-1.299) {$\K_{[\Delta_2]}$};
\node [right] at (0,0) {$\lambda_{12x}$};
\end{tikzpicture}
\qquad
\begin{tikzpicture}[scale=0.9]
\draw (0,0) circle [radius=3];
\draw [fill=black] (-2.121,-2.121) circle [radius=0.1];
\draw [fill=black] (-2.121, 2.121) circle [radius=0.1];
\draw [fill=black] ( 2.121,-2.121) circle [radius=0.1];
\draw [fill=black] ( 2.121, 2.121) circle [radius=0.1];
\draw [fill=black] (-1, 0) circle [radius=0.1];
\draw [fill=black] ( 1, 0) circle [radius=0.1];
\draw (-2.121,-2.121) -- (-1,0) -- (-2.121, 2.121);
\draw ( 2.121, 2.121) -- ( 1,0) -- ( 2.121,-2.121);
\draw (-1,0) -- (1,0);
\node [left] at (-2.121, 2.2) {$\O_1(x_1)$}; 
\node [left] at (-2.121,-2.2) {$\phi_2(x_2)$}; 	
\node [right] at ( 2.121,-2.2) {$\phi_3(x_3)$}; 
\node [right] at ( 2.121, 2.2) {$\phi_4(x_4)$}; 	
\node [right] at (-1.5, 1.2) {$\K_{[\Delta_1]}$};
\node [right] at (-1.5, -1.2) {$\K_{[\Delta_2]}$};
\node [left] at ( 1.5, -1.2) {$\K_{[\Delta_3]}$};
\node [left] at ( 1.5, 1.2) {$\K_{[\Delta_4]}$};
\node [above] at (0,0) {$\G_{[\Delta_x]}$};
\node [right] at (1,0) {$\lambda_{34x}$};
\node [left] at (-1,0) {$\lambda_{12x}$};
\end{tikzpicture}
\centering
\caption{3- and 4-point Witten diagrams corresponding to \eqref{O11} and the first line of \eqref{O12}.\label{fig:34pt}}
\end{figure}
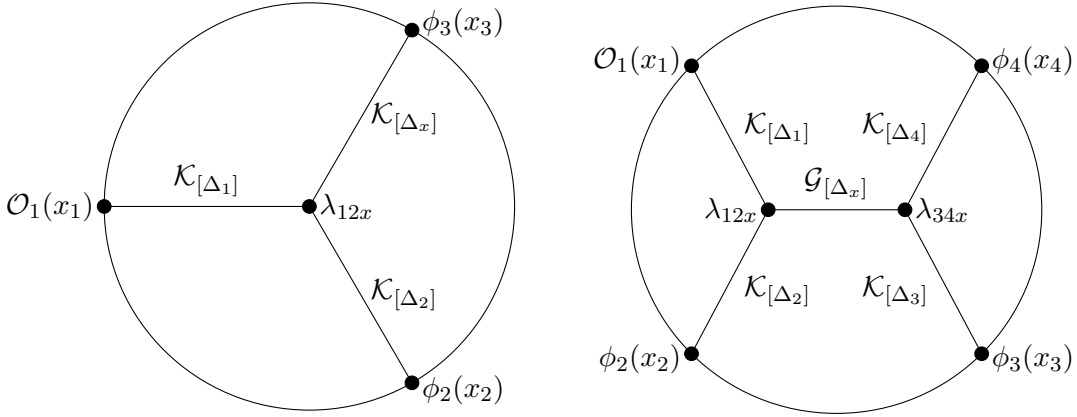

\item Directly from the action \eqref{S4asym}, we see that there are two non-vanishing 3-point functions, $\< \O_1 \O_2 \O_x \>$ and $\< \O_x \O_3 \O_4 \>$.  From \eqref{O11}, we see that
\begin{align} \label{asym3}
\lla \O_i(\bs{k}_1) \O_j(\bs{k}_2) \O_x(\bs{k}_3) \rra & = \lambda_{ijx} \, \ino_{[\Delta_i \Delta_j \Delta_x]}(k_1, k_2, k_3),
\end{align}
where $(ij) = (12)$ or $(34)$.

\item Among the non-vanishing 4-point functions, the most important is $\< \O_1 \O_2 \O_3 \O_4 \>$.  This contains two contributions: a single exchange diagram and  a single contact diagram. From \eqref{O12},
\begin{align} \label{asym4}
& \lla \O_1(\bs{k}_1) \O_2(\bs{k}_2) \O_3(\bs{k}_3) \O_4(\bs{k}_4) \rra 
 \nn\\& \qquad\qquad 
=
 \lambda_{12x} \lambda_{34x} \, \ino_{[\Delta_1 \Delta_2, \Delta_3 \Delta_4 x \Delta_x]} + \lambda_{1234} \, \ino_{[\Delta_1 \Delta_2 \Delta_3 \Delta_4]}.
\end{align}
If we want a bulk theory where the contact diagram $\ino_{[\Delta_1 \Delta_2 \Delta_3 \Delta_4]}$ is a correlator on its own, it therefore suffices to consider a bulk action with  $\lambda_{12x}=0$ or  $\lambda_{34x}=0$. On the other hand, if we want the exchange diagram to be a correlator on its own, then we need $\lambda_{1234}=0$. 

Note that $\ino_{[\Delta_1 \Delta_2 \Delta_3 \Delta_4]}$  and 
$\ino_{[\Delta_1 \Delta_2, \Delta_3 \Delta_4 x \Delta_x]}$ only depend on the dimensions on the operators involved -- not on any other labels that the operators may have, or whether the operators are identical or not -- and so we have just demonstrated that each  is on its own a correlator  in a specific bulk theory.

\item For completeness, we note that the asymmetric theory has additional non-vanishing 4-point functions:
\begin{align} \label{asym4_rest}
& \lla \O_i(\bs{k}_1) \O_j(\bs{k}_2) \O_j(\bs{k}_3) \O_i(\bs{k}_4) \rra  \nn\\
& \quad = \lambda_{ijk}^2 \, \left[ \ino_{[\Delta_i \Delta_j, \Delta_j \Delta_i x \Delta_k]}(\bs{k}_1, \bs{k}_2; \bs{k}_3, \bs{k}_4) + \ino_{[\Delta_i \Delta_j, \Delta_j \Delta_i x \Delta_k]}(\bs{k}_1, \bs{k}_3; \bs{k}_2, \bs{k}_4) \right],
\end{align}
where $(ijk) = (12x)$ or $(34x)$ or any permutation within each bracket (see Figure \ref{fig:asym_vertices} on page \pageref{fig:asym_vertices}). 
\end{itemize}

In conclusion, all  $3$- and $4$-point Witten diagrams are valid $3$- and $4$-point functions as they can be realized in the holographic theory given by \eqref{S4asym} for specific choices for couplings. Additional symmetry factors may appear if correlation functions of the same operator are considered. Similar conclusions can be reached for all higher-point diagrams. 

\subsection{Renormalized amplitudes} \label{sec:ren_amp}

\subsubsection{3-point functions}

Since every amplitude is a correlator, we can now list the counterterm actions that renormalize the $3$- and $4$-point functions in (\ref{asym3}) - (\ref{asym4_rest}). We set $\lambda_{12x} =\lambda_{34x}=\lambda_{1234}=1$, so the counterterm contributions derived from the actions cure the divergences of the amplitudes only. We will restore the $\lambda$ dependence when we discuss specific bulk actions. Let us start with the 3-point function to give some more details on the procedure.

\begin{itemize}
\item The form of the counterterms depends on the dimensions of the operators involved, and read
\begin{align} \label{S3asymCt222}
S_{[222]}^{\text{ct}\,(3)} & = \mathfrak{s}_{[222]} \int \D^{3 + 2\ep} \bs{x} \, \src_{1} \src_{2} \src_{3}, \\
S_{[322]}^{\text{ct}\,(3)} & = \mathfrak{s}_{[322]} \int \D^{3 + 2 \ep} \bs{x} \, \Src_{1} (\src_{2} \op_{3} + \src_{3} \op_{2}),  \label{S3asymCt322} \\
S_{[332]}^{\text{ct}\,(3)} & = \mathfrak{s}_{[332]} \int \D^{3 + 2 \ep} \bs{x} \, \src_{3} \partial_\mu \Src_{1} \partial^\mu \Src_{2}, \\
S_{[333]}^{\text{ct}\,(3)} & = \mathfrak{s}_{[333]}  \int \D^{3 + 2 \ep} \bs{x} \, ( \Src_{1} \Src_{2} \Op_{3} + \Src_{2} \Src_{3} \Op_{1} + \Src_{3} \Src_{1} \Op_{2} ), \label{S3asymCt333}
\end{align}
where the values of the constants are
\begin{empheq}[box=\nicebox]{align}
& \mathfrak{s}_{[222]} = - \Div_1(\mu) \act_{[222]}, && \mathfrak{s}_{[322]} = \Div_1(\mu) \act_{[322]}, \label{s222ands322} \\
& \mathfrak{s}_{[332]} = \Div_1(\mu) \act_{[332]}, && \mathfrak{s}_{[333]} = \frac{1}{3} \Div_1(\mu) \act_{[333]}. \label{s332}
\end{empheq}
The scheme-dependent part of the constants $\mathfrak{a}$ could in principle be different for different operators, but one may choose to work in a scheme where they depend only on the dimensions of the operators involved and are the same regardless of the permutation of the dimensions, \textit{e.g.}, $\mathfrak{a}_{[322]} = \mathfrak{a}_{[232]} = \mathfrak{a}_{[223]}$. In all cases, the constants have an $\ep$-expansion given by \eqref{a_exp}.

\item Given a counterterm action $S_{[\Delta_1 \Delta_2 \Delta_3]}^{\text{ct}\:(3)}$, its contribution $\ict_{[\Delta_1 \Delta_2 \Delta_3]}$ to the 3-point amplitude in \eqref{asym3} is obtained by taking three functional derivatives, with respect to the sources, of the generating functional incorporating the counterterm action:
\begin{align} \label{diff3example}
\frac{-\delta^3}{\delta \phi_1(\bs{x}_1) \delta \phi_2(\bs{x}_2) \delta \phi_3(\bs{x}_3)} \< \exp \left( -S_{[\Delta_1 \Delta_2 \Delta_3]}^{\text{ct}\:(3)} \right) \>_{s, \text{reg}} = \ireg_{[\Delta_1 \Delta_2 \Delta_3]} + \ict_{[\Delta_1 \Delta_2 \Delta_3]}.
\end{align}
We view $\ireg_{[\Delta_1 \Delta_2 \Delta_3]}$ as following from the differentiation of the regulated generating functional. The expression on the right-hand side should now possesses a finite limit as $\ep \rightarrow 0$.

\item From the counterterm actions above, the counterterm contributions to the 3-point amplitudes are 
\begin{align}
\ict_{[222]} & = \mathfrak{s}_{[222]}, \label{ict222} \\
\ict_{[322]} & = - \mathfrak{s}_{[322]} \times \left(\ireg_{[22]}(k_2) + \ireg_{[22]}(k_3) \right), \\
\ict_{[332]} & = - \mathfrak{s}_{[332]} \times (\bs{k}_1 \cdot \bs{k}_2), \\
\ict_{[333]} & = - \mathfrak{s}_{[333]} \times \left( \ireg_{[33]}(k_1) + \ireg_{[33]}(k_2) + \ireg_{[33]}(k_3) \right). \label{ict333}
\end{align}
It is easy to see they match perfectly the divergences in \eqref{ireg222} - \eqref{ireg333}.

\item A brief comment on notation is in order. 
The form of the counterterms \eqref{S3asymCt222}  -  \eqref{S3asymCt333} depends on the dimensions of the operators involved, as indicated by the subscript in square brackets. Here, we assume that all three operators and their sources featuring in the counterterms are distinct, as indicated by the indices. In particular, the counterterms functionally depend on the sources and operators, so more accurately, listing explicitly the functional dependence, we have
\begin{align} \label{S3full_arg}
S_{[\Delta_1 \Delta_2 \Delta_3]}^{\text{ct}\,(3)} & = S_{[\Delta_1 \Delta_2 \Delta_3]}^{\text{ct}\,(3)} (\phi_1^{[3 - \Delta_1]}, \phi_2^{[3 - \Delta_2]}, \phi_3^{[3 - \Delta_3]}; \O_1^{[\Delta_1]}, \O_2^{[\Delta_2]}, \O_3^{[\Delta_3]}) \nn\\
& = S_{[\Delta_1 \Delta_2 \Delta_3]}^{\text{ct}\,(3)} (\phi_1, \phi_2, \phi_3).
\end{align}
We will not list the arguments explicitly however unless some of them coincide. We will also drop the operators and dimensions, as shown in the second line, assuming that the pairing of the sources and operators is fixed. Note that the conformal dimensions in $S_{[\Delta_1 \Delta_2 \Delta_3]}^{\text{ct}\,(3)}$ cannot be dropped, as they indicate which form of the counterterm we consider.

If two or more fields coincide, additional symmetry factors may be required. If $n$ sources in a given counterterm coincide, its counterterm contribution equals $n! \, \ict$. For example, $S_{[222]}^{\text{ct}\,(3)} (\phi_1, \phi_2, \phi_3)$ produces $\ict_{[222]}$ when the three sources are different, $S_{[222]}^{\text{ct}\,(3)} (\phi_1, \phi_1, \phi_3)$ produces $2 \ict_{[222]}$, while with the identical sources $S_{[222]}^{\text{ct}\,(3)}(\phi, \phi, \phi)$ gives $6 \ict_{[222]}$. Note also that some counterterms contain several terms, which would coincide when some sources are equal.

\end{itemize}

\subsubsection{4-point functions}

In order to write the remaining counterterms for $4$-point functions, we notice two facts. First, the $3$-point counterterms \eqref{S3asymCt222} - \eqref{S3asymCt333} will generally contribute to 4-point amplitudes, and this contribution must be taken into account before the 4-point counterterm is added. Second, the form of the 4-point counterterms cannot depend on the operator in the exchange channel, because the counterterm action can only depend on the four external sources and/or operators, $\phi_j$ and $\O_j$ for $j=1,2,3,4$. 

\begin{itemize}
\item It is convenient  to list first all the types of counterterms contributing to the 4-point amplitudes of interest, as given in Table \ref{fig:sing_type}.

\begin{table}[t]
\begin{tabular}{|c|c|c|c|c|} \hline
Amplitude & Singularity type & Counterterm & Type & Contributes to \\ \hline
$2222$ & -- & -- & -- & -- \\ \hline
$3222$ & ultralocal & $\Src \src \src \src$ & anomaly & 4-pt \\
& 3-pt & $\Src \src \op$ & beta for $\src$ & 3, 4-pt \\ \hline
& 3-pt & $\Src \src \op$ & beta for $\src$ & 3, 4-pt \\
$3322$ & 3-pt & $\Src \Src \Op$ & beta for $\Src$ & 3, 4-pt \\
& 2-pt & $\Src \Src \src \op$ &  beta for $\src$ & 4-pt \\ \hline
& ultralocal & $\Src \Src \Src \src \partial^2$ & anomaly & 4-pt \\
$3332$ & 3-pt & $\Src \src \op$ & beta for $\src$ & 3, 4-pt \\
& 3-pt & $\Src \Src \Op$ & beta for $\Src$ & 3, 4-pt \\ \hline
$3333$ & 3-pt & $\Src \Src \Op$ & beta for $\Src$ & 3, 4-pt \\
& 2-pt & $\Src \Src \Src \Op$ & beta for $\Src$ & 4-pt \\ \hline
\end{tabular}
\centering
\caption{Singularity and counterterm types for various 4-point amplitudes. The first column specifies the type of amplitude: only the dimensions of the external operators as listed are relevant.  
The second column presents the general form of the singularities arising in the given amplitudes. 
The third column shows the general form of the counterterm curing the divergence. Each counterterm induces either an anomaly or a beta function for the couplings, as shown by the fourth column. Finally, the fifth column lists the $n$-point functions to which the counterterm contributes.\label{fig:sing_type}}
\end{table}

\item Starting with the counterterm actions \eqref{S3asymCt222} - \eqref{S3asymCt333}, one can take four functional derivatives similar to \eqref{diff3example} and obtain their contributions to the 4-point amplitudes. For $\Delta_x = 2$ these are as follows,
\begin{align}
\icts{3}_{[22,22x2]} & = 0, \label{icts(3)2222x2} \\
\icts{3}_{[32,22x2]} & = - \mathfrak{s}_{[322]} \ireg_{[222]}(s, k_3, k_4), \label{icts(3)3222x2} \\
\icts{3}_{[33,22x2]} & = 0, \\
\icts{3}_{[32,32x2]} & = - \mathfrak{s}_{[322]} \left[ \ireg_{[322]}(k_1, k_2, s) + \ireg_{[322]}(k_3, k_4, s) \right] + \mathfrak{s}_{[322]}^2 \ireg_{[22]}(s), \\
\icts{3}_{[32,33x2]} & = - \mathfrak{s}_{[322]} \ireg_{[332]}(k_3, k_4, s), \\
\icts{3}_{[33,33x2]} & = 0 \label{icts(3)3333x2}
\end{align}
and for $\Delta_x = 3$ one finds,
\begin{align}
\icts{3}_{[22,22x3]} & = 0, \label{icts(3)2222x3} \\
\icts{3}_{[32,22x3]} & = 0, \\
\icts{3}_{[33,22x3]} & = - \mathfrak{s}_{[333]} \ireg_{[322]}(s, k_3, k_4), \label{icts(3)3322x3}\\
\icts{3}_{[32,32x3]} & = 0, \\
\icts{3}_{[32,33x3]} & = - \mathfrak{s}_{[333]} \ireg_{[332]}(s, k_1, k_2), \\
\icts{3}_{[33,33x3]} & = - \mathfrak{s}_{[333]} \left[ \ireg_{[333]}(k_1, k_2, s) + \ireg_{[333]}(s, k_3, k_4) \right] + \mathfrak{s}_{[333]}^2 \ireg_{[33]}(s). \label{icts(3)3333x3}
\end{align}

\item We can now write down the most general local terms consistent with the symmetries that are not fixed by the renormalization of the 3-point functions. These are 
\begin{align}
S^{\text{ct} \, (4)}_{[22,22x\Delta_x]} & = 0, \label{S4Ct2222x} \\
S^{\text{ct} \, (4)}_{[32,22x\Delta_x]} & = \mathfrak{s}_{[32,22x\Delta_x]} \int \D^{3 + 2 \ep} \bs{x} \, \phi^{[0]}_{1} \phi^{[1]}_{2} \phi^{[1]}_{3} \phi^{[1]}_{4}, \label{S4Ct3222x} \\
S^{\text{ct} \, (4)}_{[33,22x\Delta_x]} & = \mathfrak{s}_{[33,22x\Delta_x]} \int \D^{3 + 2 \ep} \bs{x} \, \phi^{[0]}_{1} \phi^{[0]}_{2} \left( \phi^{[1]}_{3} \O^{[2]}_{4} + \phi^{[1]}_{4} \O^{[2]}_{3} \right), \\
S^{\text{ct} \, (4)}_{[32,32x\Delta_x]} & = \mathfrak{s}_{[32,32x\Delta_x]} \int \D^{3 + 2 \ep} \bs{x} \, \phi^{[0]}_{1} \phi^{[0]}_{3} \left( \phi^{[1]}_{2} \O^{[2]}_{4} + \phi^{[1]}_{4} \O^{[2]}_{2} \right), \\
S^{\text{ct} \, (4)}_{[32,33x\Delta_x]} & = \int \D^{3 + 2 \ep} \bs{x} \, \left[ \mathfrak{s}^{\{3+4\}}_{[33,23x\Delta_x]} \left(  \phi^{[0]}_{1} \phi^{[1]}_{2} (\partial^2 \phi^{[0]}_{3}) \, \phi^{[0]}_{4} + \phi^{[0]}_{1} \phi^{[1]}_{2} \phi^{[0]}_{3} (\partial^2 \phi^{[0]}_{4}) \right) \right. \label{S4Ct3233x} \\
& \qquad\qquad + \mathfrak{s}^{\{1\}}_{[32,33x\Delta_x]} (\partial^2 \phi^{[0]}_{1}) \, \phi^{[1]}_{2} \phi^{[0]}_{3} \phi^{[0]}_{4} + \mathfrak{s}^{\{2\}}_{[32,33x\Delta_x]} \phi^{[0]}_{1} (\partial^2 \phi^{[1]}_{2}) \, \phi^{[0]}_{3} \phi^{[0]}_{4} \nn\\
& \qquad\qquad \left. + \mathfrak{s}^{\{34\}}_{[32,33x\Delta_x]} \phi^{[0]}_{1} \phi^{[1]}_{2} \partial_\mu \phi^{[0]}_{3} \partial^\mu \phi^{[0]}_{4} + \mathfrak{s}^{\{12\}}_{[32,33x\Delta_x]} \partial_\mu ( \partial^\mu \phi^{[0]}_{1} \, \phi^{[1]}_{2} ) \phi^{[0]}_{3} \phi^{[0]}_{4} \right], \nn \\
S^{\text{ct} \, (4)}_{[33,33x\Delta_x]} & = \mathfrak{s}_{[33,33x\Delta_x]} \int \D^{3 + 2 \ep} \bs{x} \, \left( \phi^{[0]}_{1} \phi^{[0]}_{2} \phi^{[0]}_{3} \O^{[3]}_{4} + \phi^{[0]}_{2} \phi^{[0]}_{3} \phi^{[0]}_{4} \O^{[3]}_{1} \right.\nn\\
& \qquad\qquad\qquad \left. + \phi^{[0]}_{3} \phi^{[0]}_{4} \phi^{[0]}_{1} \O^{[3]}_{2} + \phi^{[0]}_{4} \phi^{[0]}_{1} \phi^{[0]}_{2} \O^{[3]}_{3} \right), \label{S4Ct3333x}
\end{align}
where the labels in curly brackets in $\mathfrak{s}^{\{\ldots\}}_{[32,33x\Delta_x]}$ indicate on which sources the $\partial^2$ acts. For example, $\{3+4\}$ indicates it acts on $\phi^{[0]}_{3}$ and $\phi^{[0]}_{4}$, $\{1\}$ that it acts on $\phi^{[0]}_{1}$,
$\{34\}$ in the product $\phi^{[0]}_{3} \phi^{[0]}_{4}$ and so on, with the precise structure as listed above.  
Note the representation \eqref{S4Ct3233x} is over-complete: only four of the terms listed are  independent, and the fifth can always be obtained through integration by parts. In particular, we can always set one of the four constants $\mathfrak{s}^{\{\ldots\}}_{[32,33x\Delta_x]}$ to zero. We will retain this over-complete representation, however, as it will prove convenient to use different non-vanishing structures for $\Delta_x = 2$ and $\Delta_x = 3$.

\item The contributions from the counterterm actions \eqref{S4Ct2222x} - \eqref{S4Ct3333x} are 
\begin{align}
\icts{4}_{[22,22x\Delta_x]} & = 0, \\
\icts{4}_{[32,22x\Delta_x]} & = - \mathfrak{s}_{[32,22x\Delta_x]}, \label{ict(4)32,22x} \\
\icts{4}_{[33,22x\Delta_x]} & = \mathfrak{s}_{[33,22x\Delta_x]} \left[ \ireg_{[22]}(k_3) + \ireg_{[22]}(k_4) \right], \\
\icts{4}_{[32,32x\Delta_x]} & = \mathfrak{s}_{[32,32x\Delta_x]} \left[ \ireg_{[22]}(k_2) + \ireg_{[22]}(k_4) \right], \\
\icts{4}_{[32,33x\Delta_x]} & = \mathfrak{s}^{\{1\}}_{[32,33x\Delta_x]} k_1^2 + \mathfrak{s}^{\{2\}}_{[32,33x\Delta_x]} k_2^2 + \mathfrak{s}^{\{3+4\}}_{[32,33x\Delta_x]} ( k_3^2 + k_4^2 ) \nn\\
& \qquad\qquad + \frac{1}{2} \mathfrak{s}^{\{12\}}_{[32,33x\Delta_x]} (s^2 + k_1^2 - k_2^2) + \frac{1}{2} \mathfrak{s}^{\{34\}}_{[32,33x\Delta_x]} (s^2 - k_3^2 - k_4^2), \\
\icts{4}_{[33,33x\Delta_x]} & = \mathfrak{s}_{[33,33x\Delta_x]} \left[ \ireg_{[33]}(k_1) + \ireg_{[33]}(k_2) + \ireg_{[33]}(k_3) + \ireg_{[33]}(k_4) \right],
\end{align}
where the values of the constants $\mathfrak{s}$ are listed in Table \ref{table:s}.
\begin{table}
\begin{center}
\begin{tabular}{|c|c|c|} \hline
 & $\Delta_x = 2$ & $\Delta_x = 3$ \\ \hline
$\mathfrak{s}_{[32,22x\Delta_x]}$ & $-\tfrac{1}{2} \Div_1^2(\mu) \mathfrak{a}_{[32,22x2]}$ & $\tfrac{1}{2} \Div_2(\mu) \mathfrak{a}_{[32,22x3]}$ \\ \hline
$\mathfrak{s}_{[33,22x\Delta_x]}$ & $\tfrac{1}{2} \Div_2(\mu) \mathfrak{a}_{[33,22x2]}$ & $\tfrac{1}{6} \Div_1^2(\mu) \mathfrak{a}_{[33,22x3]}$ \\ \hline
$\mathfrak{s}_{[32,32x\Delta_x]}$ & $\tfrac{1}{2} \Div_1^2(\mu) \mathfrak{a}_{[32,32x2]}$ & $-\tfrac{1}{2} \Div_2(\mu) \mathfrak{a}_{[32,32x3]}$ \\ \hline
$\mathfrak{s}^{\{12\}}_{[32,33x\Delta_x]}$ & $0$ & $- \tfrac{1}{6} \Div_1^2(\mu) \mathfrak{a}_{[32,33x3]}^{\{12\}}$ \\
$\mathfrak{s}^{\{34\}}_{[32,33x\Delta_x]}$ & $\tfrac{1}{2} \Div_1^2(\mu) \mathfrak{a}_{[32,33x2]}^{\{34\}}$ & $0$ \\
$\mathfrak{s}^{\{3+4\}}_{[32,33x\Delta_x]}$ & $\tfrac{3}{4} \Div_2(\mu) \mathfrak{a}_{[32,33x2]}^{\{3+4\}}$ & $\tfrac{1}{4} \Div_2(\mu) \mathfrak{a}_{[32,33x3]}^{\{3+4\}}$ \\
$\mathfrak{s}^{\{1\}}_{[32,33x\Delta_x]}$ & $-\tfrac{1}{4} \Div_2(\mu) \mathfrak{a}_{[32,33x2]}^{\{1\}}$ & $-\tfrac{1}{12} \Div_2(\mu) \mathfrak{a}_{[32,33x3]}^{\{1\}}$ \\
$\mathfrak{s}^{\{2\}}_{[32,33x\Delta_x]}$ & $\tfrac{1}{4} \Div_2(\mu) \mathfrak{a}_{[32,33x2]}^{\{2\}}$ & $\tfrac{1}{12} \Div_2(\mu) \mathfrak{a}_{[32,33x3]}^{\{2\}}$ \\ \hline
$\mathfrak{s}_{[33,33x\Delta_x]}$ & $\tfrac{1}{6} \Div_2(\mu) \mathfrak{a}_{[33,33x2]}$ & $\tfrac{1}{18} \Div_1^2(\mu) \mathfrak{a}_{[33,33x3]}$ \\ \hline
\end{tabular}
\end{center}
\caption{The values of the constants $\mathfrak{s}$ for exchange diagrams with exchange $\Delta_x=2$ and $\Delta_x=3$. \label{table:s}}
\end{table}

\item If the degree of divergence is higher than one, the subleading terms in the $\act$ constants are related to the 3-point $\act$ constants as explained in the following subsection.  

This leads to the following set of identities, which it is important to impose in all subsequent calculations:
\begin{empheq}[box=\nicebox]{align} \label{act2_fix_first}
\act_{[32,22x2]}^{(1)} & = 2 \act_{[322]}^{(1)} - 1, & \act_{[33,22x3]}^{(1)} & = 2 \act_{[333]}^{(1)} - \tfrac{1}{3}, \\
\act_{[32,32x2]}^{(1)} & = 2 \act_{[322]}^{(1)} - 1, & \act_{[32,33x3]}^{\{12\}(1)} & = 2 \act_{[333]}^{(1)} - \tfrac{5}{6}, \\
\act_{[32,33x2]}^{\{34\}(1)} & = 2 \act_{[322]}^{(1)} + \tfrac{1}{2}, & \act_{[33,33x3]}^{(1)} & = 2 \act_{[333]}^{(1)} - \tfrac{1}{3}. \label{act2_fix_last}
\end{empheq}

\end{itemize}

\subsubsection{Example}

To understand this last point better, and as an example  calculation, consider the 4-point amplitude $\ireg_{[32,22x2]}$. This amplitude exhibits a double pole at $\ep = 0$ and its divergent terms are given by \eqref{i3222x2div}. While one can work with the divergences expressed in terms of $\Div_1$ and $\Div_2$, it is easier to understand the situation when fully expanded in the regulator,
\begin{align} \label{ex3222x2reg}
\ireg_{[32,22x2]} & = \frac{1}{2 \ep^2} - \frac{1}{\ep} \left[ \log \p{34} + \gamma_E - \tfrac{1}{2} \right] + O(\ep^0).
\end{align}
The renormalization of the 3-point amplitude $\ireg_{[322]}$ introduces the counterterm \eqref{S3asymCt322}, which also contributes to the 4-point amplitude under consideration. Its contribution is given by \eqref{icts(3)3222x2}. With the value of $\mathfrak{s}_{[322]}$ given in \eqref{s222ands322}, we find
\begin{align} \label{ex3222x2ct3}
\icts{3}_{[32,22x2]} & = -\frac{1}{\ep^2} + \frac{1}{\ep} \left[ \log \p{34} + \log \mu + 2 \gamma_E - \act_{[322]}^{(1)} \right] + O(\ep^0).
\end{align}
Before moving to the renormalization of the 4-point functions, the renormalization procedure for 3-point functions must be carried out. This means that we treat $\act_{[322]}^{(1)}$ as fixed at this point. 

To fully renormalize the 4-point amplitude, we have to introduce the counterterm \eqref{S4Ct3222x}. Its contribution to the 4-point function is then given by taking four functional derivatives with respect to the sources, with the result  given by \eqref{ict(4)32,22x}. Thus, the sum of \eqref{ict(4)32,22x}, \eqref{ex3222x2reg} and \eqref{ex3222x2ct3} must be finite. With $\mathfrak{s}_{[32,22x2]}^{(n)}$ denoting terms of order $\ep^n$ in the expansion of $\mathfrak{s}_{[32,22x2]}$, we find
\begin{align}
& \ireg_{[32,22x2]} + \icts{3}_{[32,22x2]} + \icts{4}_{[32,22x2]} = \frac{1}{\ep^2} \left[ \mathfrak{s}_{[32,22x2]}^{(-2)} - \frac{1}{2} \right] \nn\\
& \qquad\qquad + \frac{1}{\ep} \left[ \mathfrak{s}_{[32,22x2]}^{(-1)}  + \gamma_E + \log \mu - \mathfrak{a}_{[322]}^{(1)} + \frac{1}{2} \right] + O(\ep^0).
\end{align}
Thus, we must choose
\begin{align}
\mathfrak{s}_{[32,22x2]}^{(-2)} & = \frac{1}{2}, & \mathfrak{s}_{[32,22x2]}^{(-1)} & = - \gamma_E - \log \mu + \mathfrak{a}_{[322]}^{(1)} - \frac{1}{2}.
\end{align}
As we can see, both the leading $\mathfrak{s}_{[32,22x2]}^{(-2)}$ as well as the subleading $\mathfrak{s}_{[32,22x2]}^{(-1)}$ term is completely fixed by the renormalization. Furthermore, the subleading term $\mathfrak{s}_{[32,22x2]}^{(-1)}$ depends on the subleading part of the renormalization constant $\mathfrak{a}_{[322]}^{(1)}$. Only the third term, $\mathfrak{s}_{[32,22x2]}^{(0)}$, is unrestricted and will contribute to the finite part of the renormalized correlator.

In the discussion above we presented our results  using the divergences $\Div_1$ and $\Div_2$. These objects also keep track of the scale-dependence, so that the $\log \mu$ term is automatically taken care of. For example, we can rewrite
\begin{align}
\frac{\mathfrak{s}_{[32,22x2]}^{(-2)}}{\ep^2} + \frac{\mathfrak{s}_{[32,22x2]}^{(-1)}}{\ep} + O(\ep^0) = \frac{1}{2} \left[ 1 + \ep (2 \act_{[322]}^{(1)} - 1) \right] \Div_1^2(\mu) + O(\ep^0).
\end{align}
Thus, if we identify the term in square brackets with the counterterm constant $\act_{[32,22x2]} = 1 + \ep \act_{[32,22x2]}^{(1)} + O(\ep^2)$, we must impose $\act_{[32,22x2]}^{(1)} = 2 \act_{[322]}^{(1)} - 1$. This is indeed the first relation in \eqref{act2_fix_first}.

Finally, let us stress here that the relations  \eqref{act2_fix_first} - \eqref{act2_fix_last} are essential for the theory to be renormalizable. Indeed, if $\mathfrak{s}_{[32,22x2]}^{(-1)}$ were any different in the example above, the 4-point amplitude $\ireg_{[32,22x2]}$ could not be renormalized.

\subsubsection{Contact diagrams}

\begin{itemize}
\item The form of the counterterms depends on the exchange operator only through the constants $\mathfrak{s}$. This also includes the contact diagrams, which we can identify with exchange diagrams with $\Delta_x = 0$, but for the sake of clarity, let us rewrite these expressions,
\begin{align}
S^{\text{ct} \, (4)}_{[2222]} & = 0, \\
S^{\text{ct} \, (4)}_{[3222]} & = \mathfrak{s}_{[3222]} \int \D^{3 + 2 \ep} \bs{x} \, \phi^{[0]}_{1} \phi^{[1]}_{2} \phi^{[1]}_{3} \phi^{[1]}_{4}, \\
S^{\text{ct} \, (4)}_{[3322]} & = \mathfrak{s}_{[3322]} \int \D^{3 + 2 \ep} \bs{x} \, \phi^{[0]}_{1} \phi^{[0]}_{2} \left( \phi^{[1]}_{3} \O^{[2]}_{4} + \phi^{[1]}_{4} \O^{[2]}_{3} \right), \\
S^{\text{ct} \, (4)}_{[3332]} & = \mathfrak{s}_{[3332]} \int \D^{3 + 2 \ep} \bs{x} \, \left(  \partial^2 \phi^{[0]}_{1} \, \phi^{[0]}_{2} \phi^{[0]}_{3} \phi^{[1]}_{4} + \phi^{[0]}_{1} \partial^2 \phi^{[0]}_{2} \, \phi^{[0]}_{3} \phi^{[1]}_{4} \right.\nn\\
& \qquad\qquad\qquad\qquad \left. + \phi^{[0]}_{1} \phi^{[0]}_{2} \partial^2 \phi^{[0]}_{3} \, \phi^{[1]}_{4} - \phi^{[0]}_{1} \phi^{[0]}_{2} \phi^{[0]}_{3} \partial^2 \phi^{[1]}_{4} \right), \\
S^{\text{ct} \, (4)}_{[3333]} & = \mathfrak{s}_{[3333]} \int \D^{3 + 2 \ep} \bs{x} \, \left( \phi^{[0]}_{1} \phi^{[0]}_{2} \phi^{[0]}_{3} \O^{[3]}_{4} + \phi^{[0]}_{2} \phi^{[0]}_{3} \phi^{[0]}_{4} \O^{[3]}_{1} \right.\nn\\
& \qquad\qquad\qquad\qquad \left. + \phi^{[0]}_{3} \phi^{[0]}_{4} \phi^{[0]}_{1} \O^{[3]}_{2} + \phi^{[0]}_{4} \phi^{[0]}_{1} \phi^{[0]}_{2} \O^{[3]}_{3} \right).
\end{align}
The values of the constants are
\begin{align}
\mathfrak{s}_{[3222]} & = \Div_2(\mu) \act_{[3222]}, \\
\mathfrak{s}_{[3322]} & = - \Div_2(\mu) \act_{[3322]}, \\
\mathfrak{s}_{[3332]} & = \frac{1}{2} \Div_2(\mu) \act_{[3332]}, \\
\mathfrak{s}_{[3333]} & = - \frac{1}{3} \Div_2(\mu) \act_{[3333]},
\end{align}
and the counterterm contributions are
\begin{align}
\icts{4}_{[2222]} & = 0, \\
\icts{4}_{[3222]} & = - \mathfrak{s}_{[3222]}, \\
\icts{4}_{[3322]} & = \mathfrak{s}_{[3322]} \left[ \ireg_{[22]}(k_3) + \ireg_{[22]}(k_4) \right], \\
\icts{4}_{[3332]} & = \mathfrak{s}_{[3332]} ( k_1^2 + k_2^2 + k_3^2 - k_4^2 ), \\
\icts{4}_{[3333]} & = \mathfrak{s}_{[3333]} \left[ \ireg_{[33]}(k_1) + \ireg_{[33]}(k_2) + \ireg_{[33]}(k_3) + \ireg_{[33]}(k_4) \right].
\end{align}

\item Similarly to the case of the 3-point counterterms in \eqref{S3full_arg}, the 4-point counterterms depend functionally on four sources and operators,
\begin{align} \label{S4full_arg}
S_{[\Delta_1 \Delta_2, \Delta_3 \Delta_4 x \Delta_x]}^{\text{ct}\,(4)} = S_{[\Delta_1 \Delta_2, \Delta_3 \Delta_4 x \Delta_x]}^{\text{ct}\,(4)}(\phi_1, \phi_2, \phi_3, \phi_4).
\end{align}
We omit the explicit arguments, if all the sources and operators are as listed.

\item In the remaining subsections, we will present two examples of how to use these results in practice.

\end{itemize}

\subsubsection{Renormalized amplitudes}

When the counterterm contributions $\ict$ are added to the regulated amplitudes $\ireg$, the   limit $\ep \rightarrow 0$ exists. Here we list the \emph{renormalized amplitudes} $\iren$ defined as
\begin{align}
\iren = \lim_{\epsilon \rightarrow 0} \left[ \ireg + \ict \right].
\end{align}

\begin{itemize}
\item 2-point amplitudes are finite and do not require renormalization,
\begin{align}
\iren_{[22]} & = - k, \label{iren22} \\
\iren_{[33]} & = k^3, \label{iren33} \\
\iren_{[23]} & = 0.
\end{align}
We do not allow for a counterterm of the form $\int \Src \Box \src$, which would result in a local contribution to $\< \O_{[2]} \O_{[3]} \>$. We keep the 2-point functions diagonal.

\item The renormalized 3-point amplitudes are
\begin{align}
\iren_{[222]} & = - \log \left( \frac{k_t}{\mu} \right) - \act_{[222]}^{(1)}, \\
\iren_{[322]} & = (k_2 + k_3) \left[ \log \left( \frac{k_t}{\mu} \right) + \act_{[322]}^{(1)} - 1 \right] - k_1, \\[1ex]
\iren_{[332]} & = \frac{k_1^2 + k_2^2 - k_3^2}{2} \left[ \log \left( \frac{k_t}{\mu} \right) + \act_{[332]}^{(1)} - \tfrac{3}{2} \right] + \frac{1}{2} \left[ k_1^2 + k_2^2 - k_1 k_2 + k_3 (k_1 + k_2) \right], \\
\iren_{[333]} & = -\frac{k_1^3 + k_2^3 + k_3^3}{3} \left[ \log \left( \frac{k_t}{\mu} \right) + \act_{[333]}^{(1)} - \tfrac{4}{3} \right] + \frac{1}{3} \left[ k_1^2 k_2 + 5 \text{ perms.} - k_1 k_2 k_3 \right].
\end{align}

\item The renormalized contact diagrams are
\begin{align}
\iren_{[2222]} & = \frac{1}{k_T}, \\
\iren_{[3222]} & = - \left[ \log \left( \frac{k_T}{\mu} \right) + \frac{1}{2} \act_{[3222]}^{(1)} \right] + \frac{k_1}{k_T}, \\[1ex]
\iren_{[3322]} & =  (k_3 + k_4) \left[ \log \left( \frac{k_T}{\mu} \right) + \frac{1}{2} \act_{[3322]}^{(1)} \right] + \frac{k_1 k_2}{k_T} - k_T, \\[1ex]
\iren_{[3332]} & = \frac{k_1^2 + k_2^2 + k_3^2 - k_4^2}{2} \left[ \log \left( \frac{k_T}{\mu} \right) + \frac{1}{2} \act_{[3332]}^{(1)} \right] + \frac{k_1 k_2 k_3}{k_T} + k_T  \left( k_4 - \frac{k_T}{4} \right), \\[1ex]
\iren_{[3333]} & = - \frac{k_1^3 + k_2^3 + k_3^3 + k_4^3}{3} \left[ \log \left( \frac{k_T}{\mu} \right) + \frac{1}{2} \act_{[3333]}^{(1)} \right] \nn\\
& \qquad\qquad + \frac{k_1 k_2 k_3 k_4}{k_T} + k_T \left( - \s{2}{1234} + \frac{4}{9} k_T^2 \right).
\end{align}

\item For exchange diagrams we only list the \emph{logarithmic part} here, $\ilog$, defined as
\begin{align}
\ilog_{[\Delta_1 \Delta_2, \Delta_3 \Delta_4 x \Delta_x]} = \lim_{\epsilon \rightarrow 0} \left[ \idiv_{[\Delta_1 \Delta_2, \Delta_3, \Delta_4 x \Delta_x]} + \ict_{[\Delta_1 \Delta_2, \Delta_3 \Delta_4 x \Delta_x]} \right].
\end{align}
The full renormalized amplitude then reads
\begin{align}
\iren_{[\Delta_1 \Delta_2, \Delta_3 \Delta_4 x \Delta_x]} = \ilog_{[\Delta_1 \Delta_2, \Delta_3 \Delta_4 x \Delta_x]} + \ifin_{[\Delta_1 \Delta_2, \Delta_3 \Delta_4 x \Delta_x]},
\end{align}
where the finite parts $\ifin$ are listed in Section \ref{sec:reg_amp}.

\item All in all, the logarithmic parts of the renormalized amplitudes with $\Delta_x = 2$ are
\begin{align}
\ilog_{[22,22x2]} & = 0,\\[1ex]
\ilog_{[32,22x2]} & = \frac{1}{2} \log^2 \left( \frac{\p{34}}{\mu} \right) + \log \left( \frac{\p{34}}{\mu} \right) \left[ -1 + \act_{[322]}^{(1)} \right] + \left[ -\act_{[322]}^{(2)} + \frac{1}{2} \act_{[32,22x2]}^{(2)} \right],\\
\ilog_{[33,22x2]} & = - \frac{k_3 + k_4}{2} \left[ \log \left( \frac{k_T}{\mu} \right) + \frac{1}{2} \act_{[33,22x2]}^{(1)} \right],\\[1ex]
\ilog_{[32,32x2]} & = - \frac{k_2 + s}{2} \log^2 \left( \frac{\p{12}}{\mu} \right) - \frac{k_4 + s}{2} \log^2 \left( \frac{\p{34}}{\mu} \right) \nn\\
& \qquad\qquad + \left[ (\p{12} + k_2) - (k_2 + s) \act_{[322]}^{(1)} \right] \log \left( \frac{\p{12}}{\mu} \right) \nn\\
& \qquad\qquad + \left[ (\p{34} + k_4) - (k_4 + s) \act_{[322]}^{(1)} \right] \log \left( \frac{\p{34}}{\mu} \right) \nn\\
& \qquad\qquad + (2 s + k_T) \left[1 + \act_{[322]}^{(1)} \right] + (k_2 + k_4) \left[ \act_{[322]}^{(2)} - \frac{1}{2} \act_{[32,32x2]}^{(2)} \right]\nn\\&\qquad\qquad  - s (\act_{[322]}^{(1)})^2,\\[1ex]
\ilog_{[32,33x2]} & = \frac{s^2 - k_3^2 - k_4^2}{4} \log^2 \left( \frac{\p{34}}{\mu} \right) \nn\\
& \qquad\qquad + \frac{1}{4} \left[ -k_1^2 + k_2^2 + (3 - 2 \act_{[322]}^{(1)})(k_3^2 + k_4^2) + 2 k_3 k_4 - 2 s (k_3 + k_4) \right.\nn\\
& \qquad\qquad\qquad\qquad \left. + 2 s^2 ( -1 + \act_{[322]}^{(1)} ) \right] \log \left( \frac{\p{34}}{\mu} \right) \nn\\
& \qquad\qquad + \frac{(k_3 + k_4)^2}{8} \left[ 1 + 2 \act_{[322]}^{(1)} \right] - \frac{s(k_3 + k_4)}{4} \left[ 3 + 2 \act_{[322]}^{(1)} \right] \nn\\[0.5ex]
& \qquad\qquad - \frac{s^2}{4} \left[ \frac{7}{2} + 3 \act_{[322]}^{(1)} \right] + \frac{s^2 - k_3^2 - k_4^2}{4} \left[ - 2 \act_{[322]}^{(2)} + \act_{[32,33x2]}^{\{34\}(2)} \right] \nn\\[0.5ex]
& \qquad\qquad  + \frac{1}{8} \left[ - k_1^2 \act_{[32,33x2]}^{\{1\}(1)} + k_2^2 \act_{[32,33x2]}^{\{2\}(1)} + 3 (k_3^2 + k_4^2) \act_{[32,33x2]}^{\{3+4\}(1)} \right],\\[1ex]
\ilog_{[33,33x2]} & = \frac{k_1^3 + k_2^3 + k_3^3 + k_4^3}{6} \left[ \log \left( \frac{k_T}{\mu} \right) + \frac{1}{2} \act_{[33,33x2]}^{(1)} \right].
\end{align}
\item The logarithmic parts of the renormalized amplitudes with $\Delta_x = 3$ are
\end{itemize}
\begin{align}
\ilog_{[22,22x3]} & = 0, \\
\ilog_{[32,22x3]} & = - \frac{1}{2} \left[ \log \left( \frac{\p{12}}{\mu} \right) + \frac{1}{2} \act_{[32,22x3]}^{(1)} \right],\\[1ex]
\ilog_{[33,22x3]} & = - \frac{k_3 + k_4}{6} \log^2 \left( \frac{\p{34}}{\mu} \right) + \frac{1}{3} \left[ s + \left( \frac{4}{3} - \act_{[333]}^{(1)} \right) (k_3 + k_4) \right] \log \left( \frac{\p{34}}{\mu} \right) \nn\\[0.5ex]
& \qquad\qquad + \frac{k_3 + k_4}{3} \left[ 1 + \act_{[333]}^{(1)} + \act_{[333]}^{(2)} - \frac{1}{2} \act_{[33,22x3]}^{(2)} \right] + \frac{s}{3} \left[1 + \act_{[333]}^{(1)} \right],\\[2ex]
\ilog_{[32,32x3]} & = \frac{k_2 + k_4}{2} \left[ \log \left( \frac{k_T}{\mu} \right) + \frac{1}{2} \act_{[32,32x3]}^{(1)} \right],\\[1ex]
\ilog_{[32,33x3]} & = \frac{k_2^2 - k_1^2 - s^2}{12} \log^2 \left( \frac{\p{12}}{\mu} \right) + \frac{1}{6} \left[ k_1^2 \left( \frac{5}{6} - \act_{[333]}^{(1)} \right) + k_2^2 \left( - \frac{11}{6} + \act_{[333]}^{(1)} \right) \right.\nn\\[0.5ex]
& \qquad\qquad \left. - k_1 k_2 + s(k_1 - k_2) + \frac{3}{2} ( k_3^2 + k_4^2) + s^2 \left( \frac{4}{3} - \act_{[333]}^{(1)} \right) \right] \log \left( \frac{\p{12}}{\mu} \right) \nn\\
& \qquad\qquad - \left( \frac{1}{4} + \frac{1}{6} \act_{[333]}^{(1)} \right) k_2 (k_1 + s) + \frac{1}{12} (1 + 2 \act_{[333]}^{(1)}) s k_1 \nn\\[0.5ex]
& \qquad\qquad + \frac{k_3^2 + k_4^2}{8} \act_{[32,33x3]}^{\{3+4\}(1)} + \frac{s^2 + k_1^2 - k_2^2}{24} \left[ 1 + 2 \act_{[333]}^{(1)} + 4 \act_{[333]}^{(2)} - 2 \act_{[32,33x3]}^{\{12\}(2)} \right] \nn\\[0.5ex]
& \qquad\qquad - \frac{k_1^2}{24} \act_{[32,33x3]}^{\{1\}(1)} - \frac{k_2^2}{24} \left[ 6 + 4 \act_{[333]}^{(1)} - \act_{[32,33x3]}^{\{2\}(1)} \right], 
\end{align}
\pagebreak
\begin{align}
\ilog_{[33,33x3]} & = \frac{s^3 + k_1^3 + k_2^3}{18} \log^2 \left( \frac{\p{12}}{\mu} \right) + \frac{s^3 + k_3^3 + k_4^3}{18} \log^2 \left( \frac{\p{34}}{\mu} \right) \nn\\
& \qquad\quad - \frac{1}{9} \left[ \left( \frac{5}{3} - \act_{[333]}^{(1)} \right) \s{1}{12s}^3 + (-4 + 3 \act_{[333]}^{(1)}) \s{1}{12s} \s{2}{12s} \right.\nn\\
& \qquad\quad\qquad\qquad \left. + (1 - 3 \act_{[333]}^{(1)}) \s{3}{12s} - \frac{s^3}{3} \right] \log \left( \frac{\p{12}}{\mu} \right) \nn\\
& \qquad\quad - \frac{1}{9} \left[ \left( \frac{5}{3} - \act_{[333]}^{(1)} \right) \s{1}{34s}^3 + (-4 + 3 \act_{[333]}^{(1)}) \s{1}{34s} \s{2}{34s} \right.\nn\\
& \qquad\quad\qquad\qquad \left. + (1 - 3 \act_{[333]}^{(1)}) \s{3}{34s} - \frac{s^3}{3} \right] \log \left( \frac{\p{34}}{\mu} \right) \nn\\
& \qquad\quad - \frac{4 + 3 \act_{[333]}^{(1)}}{27} \left[ k_1 k_2 (k_1 + k_2) + k_3 k_4 (k_3 + k_4) + s (k_1^2 + k_2^2 + k_3^2 + k_4^2) + s^2 k_T \right] \nn\\[0.5ex]
& \qquad\quad + \frac{s}{27} \left[ 1 + 3 \act_{[333]}^{(1)} \right] (k_1 k_2 + k_3 k_4) - \frac{s^3}{81} \left[26 + 24 \act_{[333]}^{(1)} - 9 (\act_{[333]}^{(1)})^2 \right] \nn\\[0.5ex]
& \qquad\quad - \frac{26 + 24 \act_{[333]}^{(1)} + 18 \act_{[333]}^{(2)} - 9 \act_{[33,33x3]}^{(2)}}{162} (k_1^3 + k_2^3 + k_3^3 + k_4^3).
\end{align}

\subsection{Asymmetric theory} \label{sec:asym_ren}

In the previous sections we listed: (\textit{i}) the counterterm contributions to the amplitudes that would render them finite, (\textit{ii}) the renormalized amplitudes, and (\textit{iii}) the counterterm actions that produce the counterterm contributions. In this section, we apply these results to completely renormalize 2-, 3-, and 4-point functions in the asymmetric theory \eqref{S4asym}.

The 2-point functions 
are finite and do not require renormalization. Thus, they already represent the renormalized correlators,
\begin{align} \label{2pt}
\lla \O_j^{[\Delta_j]}(\bs{k}) \O_j^{[\Delta_j]}(-\bs{k}) \rra_{\text{ren}} & = \iren_{[\Delta_j \Delta_j]},
\end{align}
with $\iren_{[\Delta_j \Delta_j]}$ given by \eqref{iren22} and \eqref{iren33}.

We will now show that the following CFT counterterm action renormalizes all 3- and 4-point functions in the asymmetric theory,
\begin{align} \label{SasymCt}
S^{\text{asym, ct}} = S^{\text{asym, ct} \, (3)} + S_{1234}^{\text{asym, ct} \, (4)} + S_{\text{cross}}^{\text{asym, ct} \, (4)},
\end{align}
where
\begin{align} \label{S3asymCt}
S^{\text{asym, ct} \, (3)} & = \lambda_{12x} S^{\text{ct} \, (3)}_{[\Delta_1 \Delta_2 \Delta_x]} (\phi_1, \phi_2, \phi_x) + \lambda_{34x} S^{\text{ct} \, (3)}_{[\Delta_3 \Delta_4 \Delta_x]} (\phi_3, \phi_4, \phi_x), \\[1ex]
S^{\text{asym, ct} \, (4)}_{1234} & = \lambda_{12x} \lambda_{34x} S^{\text{ct} \, (4)}_{[\Delta_1 \Delta_2, \Delta_3 \Delta_4 x \Delta_x]} (\phi_1, \phi_2, \phi_3, \phi_4)\nn\\[0.5ex]
& \qquad\qquad + \lambda_{1234} S^{\text{ct} \, (4)}_{[\Delta_1 \Delta_2 \Delta_3 \Delta_4]} (\phi_1, \phi_2, \phi_3, \phi_4), 
\end{align}
\begin{align}
S^{\text{asym, ct} \, (4)}_{\text{cross}} & = \frac{1}{2} \lambda_{12x}^2 \left[ S^{\text{ct} \, (4)}_{[\Delta_1 \Delta_2, \Delta_2 \Delta_1 x \Delta_x]} (\phi_1, \phi_2, \phi_2, \phi_1) + S^{\text{ct} \, (4)}_{[\Delta_1 \Delta_x, \Delta_x \Delta_1 x \Delta_2]} (\phi_1, \phi_x, \phi_x, \phi_1)\right.\nn\\[0.5ex]
& \qquad\qquad\qquad\qquad \left. + S^{\text{ct} \, (4)}_{[\Delta_2 \Delta_x, \Delta_x \Delta_2 x \Delta_1]} (\phi_2, \phi_x, \phi_x, \phi_2) \right] \nn\\
& + \frac{1}{2} \lambda_{34x}^2 \left[ S^{\text{ct} \, (4)}_{[\Delta_3 \Delta_4, \Delta_4 \Delta_3 x \Delta_x]} (\phi_3, \phi_4, \phi_4, \phi_3) + S^{\text{ct} \, (4)}_{[\Delta_3 \Delta_x, \Delta_x \Delta_3 x \Delta_4]} (\phi_3, \phi_x, \phi_x, \phi_3) \right.\nn\\[0.5ex]
& \qquad\qquad\qquad\qquad \left. + S^{\text{ct} \, (4)}_{[\Delta_4 \Delta_x, \Delta_x \Delta_4 x \Delta_3]} (\phi_4, \phi_x, \phi_x, \phi_4) \right] . \label{S4restCt}
\end{align}
The counterterm actions in these expressions are listed in Section \ref{sec:ren_amp}.

The first term in \eqref{SasymCt}, $S^{\text{asym, ct} \, (3)}$, renormalizes the 3-point functions \eqref{asym3}. Indeed, it contributes $\lambda_{12x} \ict_{[\Delta_1 \Delta_2 \Delta_x]}$ to the correlator $\< \O_1 \O_2 \O_x \>$ and $\lambda_{34x} \ict_{[\Delta_3 \Delta_4 \Delta_x]}$ to the correlator $\< \O_3 \O_4 \O_x \>$. It also contributes to the 4-point functions, contributing $\lambda_{12x} \lambda_{34x} \icts{3}_{[\Delta_1 \Delta_2, \Delta_3 \Delta_4 x \Delta_x]}$ to $\< \O_1 \O_2 \O_3 \O_4 \>$. To see this more explicitly, consider an example with $\Delta_1 = \Delta_2 = \Delta_x = 3$ and $\Delta_3 = \Delta_4 = 2$. In such a case $S^{\text{asym, ct} \, (3)}$ reads
\begin{align}
S^{\text{asym, ct} \, (3)} & = \Div_1(\mu) \int \D^{3 + 2 \ep} \bs{x} \, \left[ \frac{1}{3} \lambda_{12x} \mathfrak{a}_{[333]} \left( \phi_1 \phi_2 \O_x + \phi_2 \phi_x \O_1 + \phi_x \phi_1 \O_2 \right) \right.\nn\\
& \qquad\qquad\qquad \left. + \lambda_{34x} \mathfrak{a}_{[322]} \phi_x \left( \phi_3 \O_4 + \phi_4 \O_3 \right) \vphantom{\Big[}\right].
\end{align}
If we are interested in its contribution to the 4-point function $\< \O_1 \O_2 \O_3 \O_4 \>$, we can drop all sources $\phi_x$ of the exchange scalar leaving only the first term. Then, we differentiate
\begin{align}
& \left. \frac{\delta^4}{\delta \phi_1(\bs{x}_1) \delta \phi_2(\bs{x}_2) \delta \phi_3(\bs{x}_3) \delta \phi_4(\bs{x}_4)} \< \exp \left(- \frac{1}{3} \lambda_{12x} \Div_1(\mu) \mathfrak{a}_{[333]} \int \D^{3 + 2 \ep} \bs{x} \, \phi_1 \phi_2 \O_x \right) \> \right|_{\phi_j = 0}\nn\\
& = \left. \frac{\delta^2}{\delta \phi_1(\bs{x}_1) \delta \phi_2(\bs{x}_2)} \< \O_3(\bs{x}_3) \O_4(\bs{x}_4) \, \exp \left(- \frac{1}{3} \lambda_{12x} \Div_1(\mu) \mathfrak{a}_{[333]} \int \D^{3 + 2 \ep} \bs{x} \, \phi_1 \phi_2 \O_x \right) \> \right|_{\phi_j = 0} \nn\\
& = \left. \frac{\delta}{\delta \phi_1(\bs{x}_1)} \< \O_3(\bs{x}_3) \O_4(\bs{x}_4) \left( - \O_2(\bs{x}_2) - \frac{1}{3} \lambda_{12x} \Div_1(\mu) \mathfrak{a}_{[333]} \phi_1(\bs{x}_2) \O_x(\bs{x}_2) \right) \> \right|_{\phi_j = 0} \nn\\
& = \< \O_1(\bs{x}_1) \O_2(\bs{x}_2) \O_3(\bs{x}_3) \O_4(\bs{x}_4) \> - \frac{1}{3} \lambda_{12x} \Div_1(\mu) \mathfrak{a}_{[333]} \delta(\bs{x}_1 - \bs{x}_2) \< \O_x(\bs{x}_1) \O_3(\bs{x}_3) \O_4(\bs{x}_4) \>.
\end{align}
The Fourier transform of the second term yields $\lambda_{12x} \icts{3}_{[33,22x3]}$  (use (\ref{icts(3)3322x3}) and (\ref{s332})) as expected. Thus, the 4-point function $\< \O_1 \O_2 \O_3 \O_4 \>$ will be renormalized by the addition of $\icts{4}_{[33,22x3]}$. By design, this is what the counterterm action $\lambda_{12x} \lambda_{34x} S^{\text{ct} \, (4)}_{[\Delta_1 \Delta_2, \Delta_3 \Delta_4 x \Delta_x]}$ contributes.

We argued that the terms $S^{\text{asym, ct} \, (3)}$ and $S_{1234}^{\text{asym, ct} \, (4)}$ renormalize all 3-point functions as well as the 4-point function $\< \O_1 \O_2 \O_3 \O_4 \>$. The renormalized correlators are given by \eqref{asym3} and \eqref{asym4} with the amplitudes on the right hand side replaced by the renormalized amplitudes,
\begin{align} \label{asym3ren}
 \lla \O_i(\bs{k}_1) \O_j(\bs{k}_2) \O_x(\bs{k}_3) \rra_{\text{ren}} &= \lambda_{ijx} \, \iren_{[\Delta_i \Delta_j \Delta_x]}(k_1, k_2, k_3), \\
 \lla \O_1(\bs{k}_1) \O_2(\bs{k}_2) \O_3(\bs{k}_3) \O_4(\bs{k}_4) \rra_{\text{ren}} 
&= \lambda_{12x} \lambda_{34x} \, \iren_{[\Delta_1 \Delta_2, \Delta_3 \Delta_4 x \Delta_x]} + \lambda_{1234} \, \iren_{[\Delta_1 \Delta_2 \Delta_3 \Delta_4]}.
\end{align}
As we can see, we obtain the renormalized correlators by simply replacing the amplitudes by the renormalized amplitudes listed in Section \ref{sec:ren_amp}.

\begin{figure}[t]
\begin{tikzpicture}[scale=0.42]
\triL{-10}{0}{$1$}{$2$}{$x$}
\node at (-6.5, 0) {$\oplus$};
\triR{-3}{0}{$3$}{$4$}{$x$}
\node at (-9,-4) {$\lambda_{12x}$};
\node at (-3.5,-4) {$\lambda_{34x}$};
\draw[->] (0,0.5) -- (3,6);
\draw[->] (0,0) -- (3,0);
\draw[->] (0,-0.5) -- (3,-6);
\node at (6,6) {$\lambda_{12x}^2 \big($};
\triR{10}{6}{$1$}{$2$}{$x$}
\triL{10+0.121}{6}{$2$}{$1$}{}
\node at (13,6) {$+$};
\triR{16}{6}{$2$}{$x$}{$1$}
\triL{16+0.121}{6}{$x$}{$2$}{}
\node at (19,6) {$+$};
\triR{22}{6}{$x$}{$1$}{$2$}
\triL{22+0.121}{6}{$1$}{$x$}{}
\node at (25,6) {$\big)$};
\node at (11,0) {$\lambda_{12x} \lambda_{34x}$};
\triR{16}{0}{$1$}{$2$}{$x$}
\triL{16+0.121}{0}{$3$}{$4$}{}
\node at (6,-6) {$\lambda_{34x}^2 \big($};
\triR{10}{-6}{$3$}{$4$}{$x$}
\triL{10+0.121}{-6}{$4$}{$3$}{}
\node at (13,-6) {$+$};
\triR{16}{-6}{$4$}{$x$}{$3$}
\triL{16+0.121}{-6}{$x$}{$4$}{}
\node at (19,-6) {$+$};
\triR{22}{-6}{$x$}{$3$}{$4$}
\triL{22+0.121}{-6}{$3$}{$x$}{}
\node at (25,-6) {$\big)$};
\end{tikzpicture}
\centering
\caption{All non-vanishing exchange 4-point functions in the asymmetric theory. On the left, we have two 3-vertices present in the action \eqref{S4asym}. The diagrams at the top on the right-hand side represent schematically (dropping all symmetry factors) the 4-point functions $\< \O_1 \O_2 \O_2 \O_1 \>$, $\< \O_2 \O_x \O_x \O_2 \>$ and $\< \O_x \O_1 \O_1 \O_x \>$, while those at the bottom represent $\< \O_3 \O_4 \O_4 \O_3 \>$, $\< \O_4 \O_x \O_x \O_4 \>$ and $\< \O_x \O_3 \O_3 \O_x \>$. Together, these are the six correlators in \eqref{ijji_correlators}. The middle line corresponds to the 4-point function $\< \O_1 \O_2 \O_3 \O_4 \>$ obtained by combining the two vertices.\label{fig:asym_vertices}}
\end{figure}
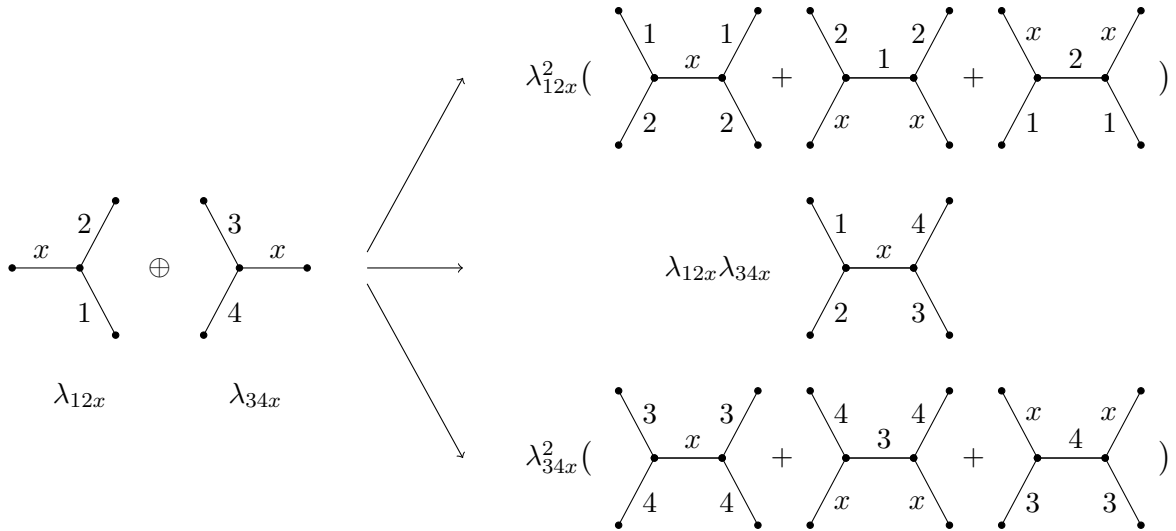

The asymmetric theory contains additional non-vanishing 4-point functions, namely those in \eqref{asym4_rest}. To complete our analysis, we should renormalize these as well. This is achieved by $S_{\text{cross}}^{\text{asym, ct} \, (4)}$, the last term in \eqref{SasymCt}. To see that this indeed works, first we have to consider the contributions to the 4-point function from the 3-point counterterms in $S^{\text{asym, ct} \, (3)}$. To do this, first notice that we only have three correlators to consider, depending on the dimensions of the operators involved: 
\[ \label{ijji_correlators}
\< \O_i^{[2]} \O_j^{[2]} \O_j^{[2]} \O_i^{[2]} \>, \qquad \< \O_i^{[3]} \O_j^{[2]} \O_j^{[2]} \O_i^{[3]} \>, \qquad \< \O_i^{[3]} \O_j^{[3]} \O_j^{[3]} \O_i^{[3]} \>.\] 
There are six possible correlators in each case, by choosing $(ij)$ from $(12)$, $(1x)$, $(2x)$, $(34)$, $(3x)$, $(4x)$ as shown in Figure \ref{fig:asym_vertices}.  The analysis is similar in all cases, however, so we will concentrate on, say, $(ij) = (12)$ and the counterterm $\lambda_{12x} S^{\text{ct} \, (3)}_{[\Delta_1 \Delta_2 \Delta_x]}$. Finally, for the counterterm to contribute, it must be non-vanishing when we set $\phi_x=0$, since we will not be differentiating with respect to this source. An inspection of (\ref{S3asymCt222})-(\ref{S3asymCt333}) shows that there are only two cases where the contribution to the 4-point function from the 3-point counterterm action is non-vanishing: (\textit{i}) $\Delta_1 = 3, \Delta_2 = \Delta_x = 2$ and (\textit{ii}) $\Delta_1 = \Delta_2 = \Delta_x = 3$. The relevant form of the counterterm actions, $S^{\text{ct}(3)}_{[322]}$ and $S^{\text{ct}(3)}_{[333]}$ with $\phi_x$ set to zero, is identical for both cases up to the overall normalization,
\begin{align}
S^{\text{ct}(3)}_I = c_I \Div_1(\mu) \mathfrak{a}_I \int \D^{3 + 2 \ep} \bs{x} \, \phi_1 \phi_2 \O_x,
\end{align}
where $c_I = 2$ for $I = [322]$ and $c_I = 1/3$ for $I = [333]$. By taking four functional derivatives, we find
\begin{align}
& \left. \frac{\delta^4}{\delta \phi_1(\bs{x}_1) \delta \phi_2(\bs{x}_2) \delta \phi_2(\bs{x}_3) \delta \phi_1(\bs{x}_4)} \< \exp \left(- c_I \Div_1(\mu) \mathfrak{a}_I \int \D^{3 + 2 \ep} \bs{x} \, \phi_1 \phi_2 \O_x \right) \> \right|_{\phi_j = 0}\nn\\
& = \< \O_1(\bs{x}_1) \O_2(\bs{x}_2) \O_2(\bs{x}_3) \O_1(\bs{x}_4) \> \nn\\
& \qquad\qquad - c_I \Div_1(\mu) \mathfrak{a}_I \left[ \, \delta(\bs{x}_3 - \bs{x}_4) \< \O_1(\bs{x}_1) \O_2(\bs{x}_2) \O_x(\bs{x}_3) \> \right.\nn\\
& \qquad\qquad\qquad\qquad + \delta(\bs{x}_1 - \bs{x}_2) \< \O_x(\bs{x}_2) \O_2(\bs{x}_3) \O_1(\bs{x}_4) \> \nn\\
& \qquad\qquad\qquad\qquad + \delta(\bs{x}_2 - \bs{x}_4) \< \O_1(\bs{x}_1) \O_2(\bs{x}_3) \O_x(\bs{x}_4) \> \nn\\
& \qquad\qquad\qquad\qquad \left. + \delta(\bs{x}_1 - \bs{x}_3) \< \O_x(\bs{x}_1) \O_2(\bs{x}_2) \O_1(\bs{x}_4) \> \right] \nn\\
& \qquad\qquad + c_I^2 \Div_1^2(\mu) \mathfrak{a}^2_I \left[ \delta(\bs{x}_1 - \bs{x}_2) \delta(\bs{x}_3 - \bs{x}_4) \< \O_x(\bs{x}_1) \O_x(\bs{x}_3) \> \right.\nn\\
& \qquad\qquad\qquad\qquad \left. + \delta(\bs{x}_1 - \bs{x}_3) \delta(\bs{x}_2 - \bs{x}_4) \< \O_x(\bs{x}_1) \O_x(\bs{x}_2) \> \right].
\end{align}
Fourier transforming this expression, we see that the counterterm contribution from the 3-point counterterm action $S^{\text{asym, ct}\,(3)}$ to the 4-point functions $\< \O_i \O_j \O_j \O_i \>$ reads
\begin{align}
\icts{3}_{[\Delta_i \Delta_j, \Delta_j \Delta_i x \Delta_x]}(\bs{k}_1, \bs{k}_2; \bs{k}_3, \bs{k}_4) + \icts{3}_{[\Delta_i \Delta_j, \Delta_j \Delta_i x \Delta_x]}(\bs{k}_1, \bs{k}_3; \bs{k}_2, \bs{k}_4).
\end{align}
These two terms correspond to the two terms in the regulated correlator itself in \eqref{asym4_rest}. Thus, what remains to be shown is that the counterterms $S^{\text{asym, ct} \, (4)}_{\text{cross}}$ in \eqref{S4restCt} correctly produce
\begin{align} \label{what_we_want1}
\icts{4}_{[\Delta_i \Delta_j, \Delta_j \Delta_i x \Delta_x]}(\bs{k}_1, \bs{k}_2; \bs{k}_3, \bs{k}_4) + \icts{4}_{[\Delta_i \Delta_j, \Delta_j \Delta_i x \Delta_x]}(\bs{k}_1, \bs{k}_3; \bs{k}_2, \bs{k}_4).
\end{align}
First, let us consider case (\textit{i}) with $\Delta_1 = 3$ and $\Delta_2 = \Delta_x = 2$. The counterterm action (\ref{S4Ct3233x}) becomes 
\begin{align}
S^{\text{ct} \, (4)}_{[32,23x2]}(\phi_1, \phi_2, \phi_2, \phi_1) = 2 \mathfrak{s}_{[32,23x2]} \int \D^{3 + 2 \ep} \bs{x} \, \phi_1^2 \phi_2 \O_2.
\end{align}
By taking four functional derivatives as above, we find that the contribution from this counterterm to the correlation function in momentum space equals twice the expression in \eqref{what_we_want1}. Thus, we have to introduce the symmetry factor of $1/2$ in \eqref{S4restCt}. Similarly, for  case (\textit{ii}) where $\Delta_1 = \Delta_2 = \Delta_x = 3$, the counterterm action (\ref{S4Ct3333x}) reads
\begin{align}
S^{\text{ct} \, (4)}_{[33,33x3]}(\phi_1, \phi_2, \phi_2, \phi_1) = 2 \mathfrak{s}_{[33,33x3]} \int \D^{3 + 2 \ep} \bs{x} \, \left[ \phi_1^2 \phi_2 \O_2 + \phi_1 \phi_2^2 \O_1 \right].
\end{align}
By taking  functional derivatives we again find that the counterterm contribution to the 4-point function is twice that of \eqref{what_we_want1}. Thus, the multiplicative factor of $1/2$ in \eqref{S4restCt} is correct in all cases.

\subsection{Symmetric theory} \label{sec:symmetric}

In this section we consider the most symmetric case: a bulk theory containing only two fields, $\Phi_{[2]}$ and $\Phi_{[3]}$, dual to two operators, $\O_{[2]}$ of dimension two and $\O_{[3]}$ of dimension three. The AdS action contains all possible couplings between the fields,
\begin{align} \label{Ssym}
S^{\text{sym}} & = \frac{1}{2} \int \D^{4 + 2 \ep} x \sqrt{g} \sum_{\Delta=2,3} \left[ \partial_\mu \Phi_{[\Delta]} \partial^\mu \Phi_{[\Delta]} + \reg{m}^2_{\Delta} \Phi_{[\Delta]}^2 \right] \nn\\
& \qquad + \int \D^{4 + 2 \ep} x \sqrt{g} \left[ \frac{1}{6} \lambda_{[222]} \Phi_{[2]}^3 + \frac{1}{2} \lambda_{[322]} \Phi_{[3]} \Phi_{[2]}^2 + \frac{1}{2}\lambda_{[332]} \Phi_{[3]}^2 \Phi_{[2]} + \frac{1}{6} \lambda_{[333]} \Phi_{[3]}^3 \right] \nn\\
& \qquad - \int \D^{4 + 2 \ep} x \sqrt{g} \left[ \frac{1}{24} \lambda_{[2222]} \Phi_{[2]}^4 + \frac{1}{6} \lambda_{[3222]} \Phi_{[3]} \Phi_{[2]}^3 + \frac{1}{4} \lambda_{[3322]} \Phi_{[3]}^2 \Phi_{[2]}^2 \right.\nn\\
& \qquad\qquad\qquad\qquad\qquad \left. + \frac{1}{6} \lambda_{[3332]} \Phi_{[3]}^3 \Phi_{[2]} + \frac{1}{24} \lambda_{[3333]} \Phi_{[3]}^4 \right],
\end{align}
where we inserted standard symmetry factors for convenience. The coupling constants $\lambda_{[\Delta_1 \Delta_2 \Delta_3]}$ and $\lambda_{[\Delta_1 \Delta_2 \Delta_3 \Delta_4]}$ depend only on the dimensions $\Delta_j$ of the operators interacting, and not on the order in which they appear.  When convenient, we can thus permute the dimensions determining the coupling, \textit{e.g.}, $\lambda_{[322]} = \lambda_{[232]} = \lambda_{[223]}$, and so on for other couplings.

\subsubsection{Regulated correlators}

Let us derive the regulated correlators in the symmetric theory \eqref{Ssym}.

\begin{itemize}
\item With the symmetry factors included in the action, the equations of motion read
\begin{align}
& (-\Box_{AdS} + \reg{m}^2_{2}) \Phi_{[2]} = - \frac{1}{2} \lambda_{[222]} \Phi_{[2]}^2 - \lambda_{[322]} \Phi_{[3]} \Phi_{[2]} - \frac{1}{2} \lambda_{[332]} \Phi_{[3]}^2 \nn\\
& \qquad + \frac{1}{6} \lambda_{[2222]} \Phi_{[2]}^3 + \frac{1}{2} \lambda_{[3222]} \Phi_{[3]} \Phi_{[2]}^2 + \frac{1}{2} \lambda_{[3322]} \Phi_{[3]}^2 \Phi_{[2]} + \frac{1}{6} \lambda_{[3332]} \Phi_{[3]}^3, \\
& (-\Box_{AdS} + \reg{m}^2_{3}) \Phi_{[3]} = - \frac{1}{2} \lambda_{[322]} \Phi_{[2]}^2 - \lambda_{[332]} \Phi_{[3]} \Phi_{[2]} - \frac{1}{2} \lambda_{[333]} \Phi_{[3]}^2 \nn\\
& \qquad + \frac{1}{6} \lambda_{[3222]} \Phi_{[2]}^3 + \frac{1}{2} \lambda_{[3322]} \Phi_{[3]} \Phi_{[2]}^2 + \frac{1}{2} \lambda_{[3332]} \Phi_{[2]} \Phi_{[3]}^2 + \frac{1}{6} \lambda_{[3333]} \Phi_{[3]}^3.
\end{align}

\item We now solve these equations of motion perturbatively, up to and including terms with three sources. Let us denote by $\Phi_{[\Delta]\{j\}}$ the term in the solution which depends on any $(j+1)$ sources. This means that $\Phi_{[\Delta]\{0\}}$ is the solution to the free field equation. We have
\begin{align}
\Phi_{[3]\{0\}} & = \K_{[3]} \ast \phi_{[3]}, \\[1ex]
\Phi_{[3]\{1\}} & = - \G_{[3]} \ast \left[ \frac{1}{2} \lambda_{[333]} \Phi_{[3]\{0\}} \Phi_{[3]\{0\}} + \lambda_{[332]} \Phi_{[3]\{0\}} \Phi_{[2]\{0\}} \right.\nn\\
& \qquad\qquad \left. + \frac{1}{2} \lambda_{[322]} \Phi_{[2]\{0\}} \Phi_{[2]\{0\}} \right], 
\end{align}
\begin{align}
\Phi_{[3]\{2\}} & = - \G_{[3]} \ast \left[ \lambda_{[333]} \Phi_{[3]\{1\}} \Phi_{[3]\{0\}} + \lambda_{[322]} \Phi_{[2]\{1\}} \Phi_{[2]\{0\}} \right.\nn\\[0.5ex]
& \qquad\qquad \left. + \lambda_{[332]} \left( \Phi_{[3]\{1\}} \Phi_{[2]\{0\}} + \Phi_{[3]\{0\}} \Phi_{[2]\{1\}} \right) \right] \nn\\
& \qquad + \G_{[3]} \ast \left[ \frac{1}{6} \lambda_{[3333]} \Phi_{[3]\{0\}} \Phi_{[3]\{0\}} \Phi_{[3]\{0\}} + \frac{1}{2} \lambda_{[3332]} \Phi_{[3]\{0\}} \Phi_{[3]\{0\}} \Phi_{[2]\{0\}} \right.\nn\\
& \qquad\qquad \left. + \frac{1}{2} \lambda_{[3322]} \Phi_{[3]\{0\}} \Phi_{[2]\{0\}} \Phi_{[2]\{0\}} + \frac{1}{6} \lambda_{[3222]} \Phi_{[2]\{0\}} \Phi_{[2]\{0\}} \Phi_{[2]\{0\}} \right],
\end{align}
where we recall that $\ast$ indicates the convolution of the position-space variables (see \eqref{convolution}).
Analogous expressions with $2 \leftrightarrow 3$ follow for $\Phi_{[2]}$. Setting $\psi_{[\Delta]} = \Phi_{[\Delta]\{0\}} = \K_{[\Delta]} \ast \phi_{[\Delta]}$,  we can write the expansion of $\Phi_{[3]\{2\}}$ in the form
\begin{align} \label{Phi2Sym}
\Phi_{[3]\{2\}} & = \G_{[3]} \ast \left[ \left( \lambda_{[333]} \psi_{[3]} + \lambda_{[332]} \psi_{[2]}   \right) \G_{[3]} \ast \left( \frac{1}{2} \lambda_{[333]} \psi_{[3]} \psi_{[3]} + \lambda_{[332]} \psi_{[3]} \psi_{[2]} + \frac{1}{2} \lambda_{[322]} \psi_{[2]} \psi_{[2]} \right) \right.\nn\\
& \qquad + \left( \lambda_{[332]} \psi_{[3]} + \lambda_{[322]} \psi_{[2]}   \right) \G_{[3]} \ast \left( \frac{1}{2} \lambda_{[332]} \psi_{[3]} \psi_{[3]} + \lambda_{[322]} \psi_{[3]} \psi_{[2]} + \frac{1}{2} \lambda_{[222]} \psi_{[2]} \psi_{[2]} \right) \nn\\
& \qquad \left. + \frac{1}{6} \lambda_{[3333]} \psi_{[3]}^3 + \frac{1}{2} \lambda_{[3332]} \psi_{[3]}^2 \psi_{[2]} + \frac{1}{2} \lambda_{[3322]} \psi_{[3]} \psi_{[2]}^2 + \frac{1}{6} \lambda_{[3222]} \psi_{[2]}^3 \right].
\end{align}

\item All 3-point functions then follow from $\Phi_{[2]\{1\}}$ and $\Phi_{[3]\{1\}}$. They have no additional symmetry factors and read
\begin{align} \label{3pt_sym}
\lla \O_{[\Delta_1]}(\bs{k}_1) \O_{[\Delta_2]}(\bs{k}_2) \O_{[\Delta_3]}(\bs{k}_3) \rra_{\text{reg}} = \lambda_{[\Delta_1 \Delta_2 \Delta_3]} \, \ireg_{[\Delta_1 \Delta_2 \Delta_3]}(k_1, k_2, k_3)
\end{align}
for $\Delta_1, \Delta_2, \Delta_3 = 2,3$.

\item The 4-point functions follow from $\Phi_{[2]\{2\}}$ and $\Phi_{[3]\{2\}}$. For example, starting with \eqref{Phi2Sym}, we replace the external  propagator $\G_{[3]}$ with $-\K_{[3]}$, multiply by $(-1)^3$ and take  three functional derivatives with respect to the sources. All in all, the symmetry factors are selected in such a way that, after all symmetries of the amplitudes are used, every correlator takes the form
\begin{align}
& \lla \O_{[\Delta_1]}(\bs{k}_1) \O_{[\Delta_2]}(\bs{k}_2) \O_{[\Delta_3]}(\bs{k}_3) \O_{[\Delta_4]}(\bs{k}_4) \rra_{\text{reg}}  \nn\\
& \qquad = \sum_{\Delta_x = 2,3} \left[ \lambda_{[\Delta_1 \Delta_2 \Delta_x]} \lambda_{[\Delta_x \Delta_3 \Delta_4]} \ireg_{[\Delta_1 \Delta_2, \Delta_3 \Delta_4 x \Delta_x]}(\bs{k}_1, \bs{k}_2; \bs{k}_3, \bs{k}_4) \right.\nn\\
& \qquad\qquad\qquad\qquad + \lambda_{[\Delta_1 \Delta_3 \Delta_x]} \lambda_{[\Delta_x \Delta_2 \Delta_4]} \ireg_{[\Delta_1 \Delta_3, \Delta_2 \Delta_4 x \Delta_x]}(\bs{k}_1, \bs{k}_3; \bs{k}_2, \bs{k}_4) \nn\\
& \qquad\qquad\qquad\qquad \left. + \lambda_{[\Delta_1 \Delta_4 \Delta_x]} \lambda_{[\Delta_x \Delta_2 \Delta_3]} \ireg_{[\Delta_1 \Delta_4, \Delta_2 \Delta_3 x \Delta_x]}(\bs{k}_1, \bs{k}_4; \bs{k}_2, \bs{k}_3) \right] \nn\\
& \qquad\qquad + \lambda_{[\Delta_1 \Delta_2 \Delta_3 \Delta_4]} \, \ireg_{[\Delta_1 \Delta_2 \Delta_3 \Delta_4]}(k_1, k_2, k_3, k_4). \label{4pt_sym}
\end{align}
Note that crossing symmetric terms appear with unit coefficients.

\end{itemize}

\subsubsection{Renormalization}

Let us now renormalize the symmetric theory with only two 
operators $\O_{[2]}$ and $\O_{[3]}$, governed by the bulk action \eqref{Ssym}. The regulated 3-point functions are given by \eqref{3pt_sym}, and 4-point functions by \eqref{4pt_sym}. As discussed, the renormalized 3- and 4-point functions are found by replacing the regulated amplitudes by the renormalized ones, as listed in Section \ref{sec:ren_amp}.  This gives 
\begin{align} \label{3pt_sym_ren}
\lla \O_{[\Delta_1]}(\bs{k}_1) \O_{[\Delta_2]}(\bs{k}_2) \O_{[\Delta_3]}(\bs{k}_3) \rra_{\text{ren}} = \lambda_{[\Delta_1 \Delta_2 \Delta_3]} \, \iren_{[\Delta_1 \Delta_2 \Delta_3]}(k_1, k_2, k_3),
\end{align}
and
\begin{align}
& \lla \O_{[\Delta_1]}(\bs{k}_1) \O_{[\Delta_2]}(\bs{k}_2) \O_{[\Delta_3]}(\bs{k}_3) \O_{[\Delta_4]}(\bs{k}_4) \rra_{\text{ren}}  \nn\\
& \qquad = \sum_{\Delta_x = 2,3} \left[ \lambda_{[\Delta_1 \Delta_2 \Delta_x]} \lambda_{[\Delta_x \Delta_3 \Delta_4]} \iren_{[\Delta_1 \Delta_2, \Delta_3 \Delta_4 x \Delta_x]}(\bs{k}_1, \bs{k}_2; \bs{k}_3, \bs{k}_4) \right.\nn\\
& \qquad\qquad\qquad\qquad + \lambda_{[\Delta_1 \Delta_3 \Delta_x]} \lambda_{[\Delta_x \Delta_2 \Delta_4]} \iren_{[\Delta_1 \Delta_3, \Delta_2 \Delta_4 x \Delta_x]}(\bs{k}_1, \bs{k}_3; \bs{k}_2, \bs{k}_4) \nn\\
& \qquad\qquad\qquad\qquad \left. + \lambda_{[\Delta_1 \Delta_4 \Delta_x]} \lambda_{[\Delta_x \Delta_2 \Delta_3]} \iren_{[\Delta_1 \Delta_4, \Delta_2 \Delta_3 x \Delta_x]}(\bs{k}_1, \bs{k}_4; \bs{k}_2, \bs{k}_3) \right] \nn\\
& \qquad\qquad + \lambda_{[\Delta_1 \Delta_2 \Delta_3 \Delta_4]} \, \iren_{[\Delta_1 \Delta_2 \Delta_3 \Delta_4]}(k_1, k_2, k_3, k_4), \label{4pt_sym_ren}
\end{align}
thus providing explicit expressions for all 3- and 4-point functions in the symmetric theory. 

Just as we did for the asymmetric theory in Section \ref{sec:asym_ren},  let us now specify the counterterm action  that renormalizes the theory and analyze its properties. Since we have only two fields that are distinguishable by their dimensions, we may drop all indices in the counterterm actions and define
\begin{align}
S^{\text{sym, ct} \, (3)}_{[\Delta_1 \Delta_2 \Delta_3]} & = S^{\text{ct} \, (3)}_{[\Delta_1 \Delta_2 \Delta_3]} (\phi_{[d - \Delta_1]}, \phi_{[d - \Delta_2]}, \phi_{[d - \Delta_3]}), \\[1ex]
S^{\text{sym, ct} \, (4)}_{[\Delta_1 \Delta_2, \Delta_3 \Delta_4 x \Delta_x]} & = S^{\text{ct} \, (4)}_{[\Delta_1 \Delta_2, \Delta_3 \Delta_4 x \Delta_x]} (\phi_{[d - \Delta_1]}, \phi_{[d - \Delta_2]}, \phi_{[d - \Delta_3]}, \phi_{[d - \Delta_4]}).
\end{align}
Note that some symmetry factors may appear, for example,
\begin{align}
S^{\text{sym, ct} \, (3)}_{[322]} = 2 \mathfrak{s}_{[322]} \int \D^{3 + 2 \ep} \bs{x} \, \phi_{[0]} \phi_{[1]} \O_{[2]}.
\end{align}
The complete counterterm action is given by 
\begin{align} \label{SCtSym}
S^{\text{sym, ct}} & = \frac{1}{6} \lambda_{[222]} S^{\text{sym, ct} \, (3)}_{[222]} + \frac{1}{2} \lambda_{[322]} S^{\text{sym, ct} \, (3)}_{[322]} + \frac{1}{2} \lambda_{[332]} S^{\text{sym, ct} \, (3)}_{[332]} + \frac{1}{6} \lambda_{[333]} S^{\text{sym, ct} \, (3)}_{[333]} \nn\\
& \quad + \frac{1}{6} \lambda_{[3222]} S^{\text{sym, ct} \, (4)}_{[3222]} + \frac{1}{4} \lambda_{[3322]} S^{\text{sym, ct} \, (4)}_{[3322]} + \frac{1}{6} \lambda_{[3332]} S^{\text{sym, ct} \, (4)}_{[3332]} + \frac{1}{24} \lambda_{[3333]} S^{\text{sym, ct} \, (4)}_{[3333]} \nn\\
& \quad + \sum_{\Delta_x = 2,3} \left[ \frac{1}{2} \lambda_{[32 \Delta_x]} \lambda_{[22 \Delta_x]} S^{\text{sym, ct} \, (4)}_{[32,22x\Delta_x]} \right.\nn\\
& \quad\qquad\qquad + \frac{1}{4} \left( \lambda_{[33 \Delta_x]} \lambda_{[22 \Delta_x]} S^{\text{sym, ct} \, (4)}_{[33,22x\Delta_x]} + 2 \lambda_{[32\Delta_x]}^2 S^{\text{sym, ct} \, (4)}_{[32,32x\Delta_x]} \right) \nn\\
& \quad\qquad\qquad + \frac{1}{6} \lambda_{[33 \Delta_x]} \lambda_{[32 \Delta_x]} \left( S^{\text{sym, ct} \, (4)}_{[32,33x\Delta_x]} + S^{\text{sym, ct} \, (4)}_{[33,23x\Delta_x]} + S^{\text{sym, ct} \, (4)}_{[33,32x\Delta_x]} \right) \nn\\
& \quad\qquad\qquad \left. + \frac{1}{8} \lambda_{[33 \Delta_x]}^2 S^{\text{sym, ct} \, (4)}_{[33,33x\Delta_x]} \right].
\end{align}
The symmetry factors here are quite straightforward for the 3-point counterterm actions, but more complicated for the 4-point amplitudes. Since each 3-point function in \eqref{3pt_sym} is given by a single amplitude, the corresponding symmetry factors are equal to one over the symmetry factor of the Witten diagram. The same argument holds for contact 4-point functions, which leads to the values listed.

For the exchange 4-point counterterms, we have to remember that the expressions in \eqref{4pt_sym_ren} always contain three crossing-symmetric terms. 
Consider, for example,  the contribution to the 4-point function $\< \O_{[3]} \O_{[2]} \O_{[2]} \O_{[2]} \>$. We need the counterterm contribution to each of the three crossing symmetric terms in \eqref{4pt_sym_ren}. Notice, however, that $\icts{4}_{[32,22x\Delta_x]}$ in \eqref{ict(4)32,22x} is a constant and thus is trivially crossing symmetric. Hence, we need the contribution of $3 \icts{4}_{[32,22x\Delta_x]}$ from the action $S^{\text{sym, ct}\,(4)}_{[32,22x\Delta_x]}$. On the other hand, $S^{\text{sym, ct}\,(4)}_{[32,22x\Delta_x]} = \mathfrak{s}_{[32,22x\Delta_x]} \int \phi_{[0]} \phi_{[1]}^3$ and thus after taking the functional derivatives, its contribution to the 4-point function equals $6 \icts{4}_{[3222]}$. This leads to the factor of $1/2$ in \eqref{SCtSym} multiplying $S^{\text{sym, ct}\,(4)}_{[32,22x\Delta_x]}$.
Similarly, let us look at the counterterm contribution rendering the 4-point function $\< \O_{[3]} \O_{[3]} \O_{[2]} \O_{[2]} \>$ finite. The three terms in \eqref{4pt_sym_ren} become
\begin{align}
& \left( \lambda_{[33 \Delta_x]} \lambda_{[22 \Delta_x]} \mathfrak{s}_{[33,22x\Delta_x]} + 2 \lambda_{[32\Delta_x]}^2 \mathfrak{s}_{[32,32x\Delta_x]} \right) \times \left[ \ireg_{[22]}(k_3) + \ireg_{[22]}(k_4)  \right].
\end{align}
On the other hand, the contributions to the 4-point function from the counterterm actions $S^{\text{sym, ct} \, (4)}_{[33,22x\Delta_x]}$ and $S^{\text{sym, ct} \, (4)}_{[32,32x\Delta_x]}$ are  $4 \mathfrak{s}_{[33,22x\Delta_x]} ( \ireg_{[22]}(k_3) + \ireg_{[22]}(k_4) )$ and $4 \mathfrak{s}_{[32,32x\Delta_x]} ( \ireg_{[22]}(k_3) + \ireg_{[22]}(k_4) )$ respectively. This leads to the factor of $1/4$ in \eqref{SCtSym}.

\subsubsection{Beta functions} \label{sec:beta}

From a QFT point of view, it is more convenient to organize the counterterm action as 
\begin{align}\label{SCtSymZ}
S^{\text{sym}}_{\text{ct}} & = \int \D^{3 + 2 \ep} \bs{x} \left[ Z_{[0]}(\phi_{[0]} \mu^{-\ep}) \phi_{[0]} \O_{[3]} + Z_{[1]}(\phi_{[0]} \mu^{-\ep}) \phi_{[1]} \O_{[2]} \right.\nn\\
& \qquad\qquad\qquad\qquad \left. + \mu^{2 \ep} Z_{[3]}(\phi_{[0]} \mu^{-\ep}, \phi_{[1]} \mu^{-\ep}) \bs{1} \ \right].
\end{align}
In the language of textbook QFT,  
$Z_{[0]}$ and $Z_{[1]}$ are  multiplicative renormalization factors inducing beta functions for the sources $\phi_{[0]}$ and $\phi_{[1]}$ respectively, while $Z_{[3]}$ is the anomaly-inducing additive renormalization factor. This means that we identify $\phi_{[0]}$ and $\phi_{[1]}$ as \emph{renormalized sources}, which depend implicitly  on the renormalization scale $\mu$. The \emph{bare sources} are the combinations that couple directly to the operators in \eqref{SCtSymZ}, \textit{i.e.},
\begin{align} \label{bare_sources}
\phi^{\text{bare}}_{[0]} & = Z_{[0]}(\phi_{[0]} \mu^{-\ep}) \phi_{[0]}, & \phi^{\text{bare}}_{[1]} & = Z_{[1]}(\phi_{[0]} \mu^{-\ep}) \phi_{[1]}.
\end{align}

By re-organizing the action \eqref{SCtSym}, we find the  multiplicative renormalization factors
\begin{align}
Z_{[0]} & = 1 + \frac{1}{2} \lambda_{[333]} \mathfrak{s}_{[333]} \phi_{[0]} \nn\\
& \qquad + \left[ \frac{1}{6} \lambda_{[3333]} \mathfrak{s}_{[3333]} + \frac{1}{2} \sum_{\Delta_x = 2,3} \lambda_{[33 \Delta_x]}^2 \mathfrak{s}_{[33,33x\Delta_x]} \right] \phi_{[0]}^2 + O(\phi_{[0]}^3), \\
Z_{[1]} & = 1 + \lambda_{[322]} \mathfrak{s}_{[322]} \phi_{[0]} + \left[ \frac{1}{2} \lambda_{[3322]} \mathfrak{s}_{[3322]} \right. \nn\\
& \qquad \left. + \frac{1}{2} \sum_{\Delta_x = 2,3} \left( 2 \lambda_{[32 \Delta_x]}^2 \mathfrak{s}_{[32,32x\Delta_x]} + \lambda_{[33\Delta_x]} \lambda_{[22\Delta_x]} \mathfrak{s}_{[33,22x\Delta_x]} \right) \right] \phi_{[0]}^2 + O(\phi_{[0]}^3).
\end{align}
It is now straightforward to compute the beta functions for the two couplings,
\begin{align}
\beta_{\phi_{[n]}} = \mu \frac{\D}{\D \mu} \phi_{[n]}.
\end{align}
To do this, we invert the relations \eqref{bare_sources} in order to express the renormalized sources in terms of the bare, scale-independent sources. In this way, we find
\begin{align} \label{beta0}
\beta_{\phi_{[0]}} & = \frac{1}{6} \lambda_{[333]} \phi_{[0]}^2 + \left[ - \frac{1}{18} \lambda_{[3333]} + \frac{1}{12} \lambda_{[332]}^2 - \frac{1}{54} \lambda_{[333]}^2 \right] \phi_{[0]}^3 + O(\phi_{[0]}^4), \\
\beta_{\phi_{[1]}} & = \lambda_{[322]} \phi_{[0]} \phi_{[1]} + \left[ - \frac{1}{2} \lambda_{[3322]} - \lambda_{[322]}^2 + \frac{1}{4} \lambda_{[222]} \lambda_{[332]} - \frac{1}{2} \lambda_{[332]}^2 \right.\nn\\
& \qquad\qquad \left. - \frac{1}{18} (1 + 3 \mathfrak{a}^{(1)}_{[322]} - 3 \mathfrak{a}_{[333]}^{(1)} ) \lambda_{[322]} \lambda_{[333]} \right] \phi_{[0]}^2 \phi_{[1]} + O(\phi_{[0]}^3). \label{beta1}
\end{align}
Notice that the second term in $\beta_{\phi_{[1]}}$ is scheme-dependent. However, this is not in contradiction with the well-known statement that the first two terms in the beta function are scheme-independent.  
Rather, this happens because the beta functions are calculated for the dimensionful regulated sources $\phi_{[0]}$ and $\phi_{[1]}$. If  instead we were to look at the dimensionless couplings $g_{[0]}$ and $g_{[1]}$, defined as
\begin{align}
& g_{[0]} = \phi_{[0]} \mu^{-\ep}, \qquad g_{[1]} = \phi_{[1]} \mu^{-1-\ep},
\end{align}
we would obtain beta functions
\begin{align}
\beta_{g_{[0]}} & = - \ep g_{[0]} + \frac{1}{6} \lambda_{[333]} g_{[0]}^2 + \left[ - \frac{1}{18} \lambda_{[3333]} + \frac{1}{12} \lambda_{[332]}^2 - \frac{1}{54} \lambda_{[333]}^2 \right] g_{[0]}^3 + O(g_{[0]}^4), \\
\beta_{g_{[1]}} & = -(1 + \ep) g_{[1]} + \lambda_{[322]} g_{[0]} g_{[1]} + \left[ - \frac{1}{2} \lambda_{[3322]} - \lambda_{[322]}^2 + \frac{1}{4} \lambda_{[222]} \lambda_{[332]} - \frac{1}{2} \lambda_{[332]}^2 \right.\nn\\
& \qquad\qquad \left. - \frac{1}{18} (1 + 3 \mathfrak{a}^{(1)}_{[322]} - 3 \mathfrak{a}_{[333]}^{(1)} ) \lambda_{[322]} \lambda_{[333]} \right] g_{[0]}^2 g_{[1]} + O(g_{[0]}^3),
\end{align}
where, as customary, we kept the terms of order $\ep$ in the classical scaling term.  The beta function for $g_{[1]}$ now contains a non-vanishing linear term so that the first two terms are indeed scheme-independent.

\subsubsection{Anomalies}

The dependence of the correlation functions on the renormalization scale $\mu$ is governed by the Callan-Symanzik equation. Let $W[\phi_{[0]}, \phi_{[1]}; \mu]$ be the renormalized generating functional and $\phi_{[0]}$ and $\phi_{[1]}$ denote the renormalized sources. The chain rule then leads to the Callan-Symanzik equation,
\begin{align} \label{Callan-Symanzik}
\mu \frac{\D}{\D \mu} W = \left[ \mu \frac{\partial}{\partial \mu} + \sum_{j=0,1} \int \D^d \bs{x} \, \beta_{\phi_{[j]}} \frac{\delta}{\delta \phi_{[j]}(\bs{x})}  \right] W = \int \D^d \bs{x} \, \mathcal{A},
\end{align}
where $\mathcal{A}$ is the anomaly. The anomaly reads
\begin{align}
\mathcal{A} = \mathcal{A}^{(3)} + \mathcal{A}^{(4)},
\end{align}
where
\begin{align} \label{An3}
\mathcal{A}^{(3)} & = -\frac{1}{6} \lambda_{[222]} \phi_{[1]}^3 + \frac{1}{2} \lambda_{[332]} \phi_{[1]} \partial_\mu \phi_{[0]} \partial^\mu \phi_{[0]}, \\
\mathcal{A}^{(4)} & = \phi_{[0]} \phi_{[1]}^3 \, \left[ \frac{1}{2} \lambda_{[322]} \lambda_{[222]} (1 + \act_{[222]}^{(1)} - \act_{[322]}^{(1)} ) + \frac{1}{4} \lambda_{[332]} \lambda_{[322]} + \frac{1}{6} \lambda_{[3222]} \right] \nn\\
& \qquad + \left( \frac{1}{6} \phi_{[0]}^3 \partial^2 \phi_{[1]} - \frac{1}{2} \phi_{[0]}^2 \phi_{[1]} \partial^2 \phi_{[0]} \right) \times \left[ - \frac{1}{2} \lambda_{[3332]} \right.\nn\\
& \qquad\qquad \left. - \frac{1}{2} \lambda_{[332]} \lambda_{[322]} (2 + \act_{[332]}^{(1)} - \act_{[322]}^{(1)} ) - \frac{1}{3} \lambda_{[333]} \lambda_{[332]} ( \tfrac{25}{12} + \act_{[332]}^{(1)} - \act_{[333]}^{(1)} ) \right] \nn\\
& \qquad + \phi_{[0]}^3 \partial^2 \phi_{[1]} \, \left[ \frac{1}{3} \lambda_{[332]} \lambda_{[322]} + \frac{1}{9} \lambda_{[333]} \lambda_{[332]} \right] .
\end{align}
As we can see the anomaly is scheme-dependent, which means that some terms can be eliminated by the suitable choice of the counterterm constants.

The Callan-Symanzik equation \eqref{Callan-Symanzik} serves as a check on our results. First, we act with suitable functional derivatives on both sides in order to translate it into a statement on the correlators. As an example, consider the renormalized 4-point function $\< \O_{[3]} \O_{[2]} \O_{[2]} \O_{[2]} \>$. The explicit derivative of \eqref{4pt_sym} with respect to the scale $\mu$ reads
\begin{align}
& \mu \frac{\partial}{\partial \mu} \lla \O_{[3]}(\bs{k}_1) \O_{[2]}(\bs{k}_2) \O_{[2]}(\bs{k}_3) \O_{[2]}(\bs{k}_4) \rra = \frac{3}{2} \lambda_{[332]} \lambda_{[322]} + \lambda_{[3222]} \nn\\[1ex]
& \qquad\qquad + \lambda_{[322]} \lambda_{[222]} \left[ - \log \left( \frac{l_{34+} l_{24+} l_{23+}}{\mu^3} \right) + 3 - 3 \act_{[322]}^{(1)} \right].
\end{align}
It is easy to see that the only relevant term in the beta functions is $\lambda_{[322]} \phi_{[0]} \phi_{[1]}$ present in $\beta_{[1]}$. The corresponding term in the Callan-Symanzik equation, Fourier-transformed into momentum space, reads
\begin{align}
& \left. \frac{\delta^4}{\delta \phi_{[0]}(\bs{x}_1) \delta \phi_{[1]}(\bs{x}_2) \delta \phi_{[1]}(\bs{x}_3) \delta \phi_{[1]}(\bs{x}_4) } \int \D^3 \bs{x} \, \lambda_{[322]} \phi_{[0]} \phi_{[1]} \frac{\delta W}{\delta \phi_{[1]}} \right|_{\text{Fourier}}  \nn\\[1ex]
& \qquad\qquad = \lambda_{[322]} \lambda_{[222]} \left[ \log \left( \frac{l_{34+} l_{24+} l_{23+}}{\mu^3} \right) + 3 \act_{[222]}^{(1)} \right].
\end{align}
When the two terms are added together, the logarithmic terms cancel, leading to an ultralocal expression. This is an important check on our results as the anomaly in a local theory must be ultralocal. It is now straightforward to check that the remaining sum is obtained by acting with the functional derivatives on the anomaly $\mathcal{A}^{(3)}$ in \eqref{An3}.

\section{Change of scheme}\label{sec:schemechange}

The aim of this section is to present a comprehensive analysis allowing for a change of the regularization scheme. The amplitudes presented in Section \ref{sec:reg_amp} are regulated in the half-integer scheme \eqref{special}. In this section, we show how to obtain the amplitudes regulated in the general scheme \eqref{genreg}, \textit{i.e.},
\begin{align} \label{general_rep}
& d \longmapsto \dreg = d + 2 u \ep, && \Delta_j \longmapsto \Dreg_j = \Delta_j + (u + v_j) \ep,
\end{align}
where $u$ and $v_j$ for $j=1,2,3,4,x$ are arbitrary regularization parameters.  There are several motivations for this:

\begin{itemize}

\item The half-integer scheme we used so far is very convenient, but different regularization schemes may be needed for other 
computations. In general, one needs to shift all parameters by different amounts, as in \eqref{general_rep}, in order to regulate all correlators.
For example, in the computation of tensorial correlators involving conserved currents and/or stress tensors, the general scheme is  typically required as discussed in \cite{Bzowski:2015pba, Bzowski:2017poo, Bzowski:2018fql}.

\item Another motivation arises from applications to inflationary cosmology. 
As discussed in \cite{Bzowski:2012ih, McFadden:2013ria}, a certain class of slow-roll models can be described holographically in terms of a three-dimensional CFT deformed by a marginally relevant scalar of dimension $\Delta = 3 - \lambda$.  The small parameter $\lambda\ll 1$ is related to the slow-roll parameters and the spectral tilt, and allows one to  compute cosmological correlators using conformal perturbation theory.  For such purposes, we want to  know amplitudes in the regularization scheme $u = 0$ and $v_j = -1$, \textit{i.e.}, for $d = 3$ and $\Delta = 3 - \ep$, where the regulator $\ep$  now plays the role of $\lambda$.

\end{itemize}

In the following, since our focus will now be on the $u$ and $v_j$ dependence of the regulated amplitudes, we will drop all other variables and denote $\ireg = \ireg(u, v_j)$.

\subsection{2-point amplitudes}

In the general regularization scheme \eqref{general_rep}, 2-point amplitudes follow from \eqref{1pt} and the expansion of the bulk-to-boundary propagator \eqref{KPropagator}. With $\dreg = 3 + 2 u \ep$ and $\Dreg = \Delta + (u + v) \ep$, one finds
\begin{align}
\ireg_{[22]} & = - \frac{\Gamma \left( \frac{1}{2} - v \ep \right)}{4^{v \ep} \Gamma \left( \frac{1}{2} + v \ep \right)} k^{1 + 2 v \ep}, \qquad\quad
\ireg_{[33]}  = - \frac{\Gamma \left( -\frac{1}{2} - v \ep \right)}{4^{1 + v \ep} \Gamma \left( \frac{3}{2} + v \ep \right)} k^{3 + 2 v \ep}.
\end{align}

\subsection{3-point amplitudes}

In this subsection we address the scheme-change procedure for 3-point amplitudes, following the discussion in \cite{Bzowski:2015pba} and \cite{Bzowski:2015yxv}. In general, it is difficult to calculate directly the amplitudes in the general regularization scheme \eqref{general_rep} with arbitrary $u, v_j$. However, if we know the regulated amplitude for some specific choice $\bar{u}, \bar{v}_j$, $\ireg(\bar{u}, \bar{v}_j)$, then it is possible to compute the additional terms needed to obtain  
the regulated amplitude $\ireg(u, v_j)$ for general $u, v_j$.  As we already know the results in the half-integer scheme with $\bar{u} = 1$ and $\bar{v}_j = 0$ from Section \ref{sec:reg_amp}, this will be our starting point.

As discussed in \cite{Bzowski:2015yxv}, the scheme-dependent terms are contained in a specific part of the regulated amplitude, which also contains all divergent terms. The idea is then to isolate this part, which we will call the \emph{parameter-dependent} or \emph{scheme-dependent} part. The remaining part is finite and scheme-independent, and can be obtained from the half-integer scheme.
We computed in full generality the finite scheme-dependent part  of 3-point functions in  \cite{Bzowski:2015yxv}, but for our purposes here we also need the order $\ep$ scheme-dependent part. The reason is that this part will be needed in the computation of finite scheme-dependent terms in 4-point exchange diagrams.

We begin this section by reviewing the procedure for scheme change discussed in \cite{Bzowski:2015yxv}, and showing how the procedure can be simplified when the starting point is the half-integer scheme \eqref{special}. Our main goal is to derive the 3-point amplitudes, including parameter-dependent terms up to and including terms of order $\ep$. This will be essential for the derivation of the exchange 4-point functions in the general regularization scheme \eqref{general_rep}  in Section \ref{scheme:exchange}. Furthermore, for this same reason, we also must calculate 3-point-like amplitudes with one Bessel-$K$ function in \eqref{amp3} replaced by a Bessel-$I$. 

\subsubsection{Definitions}

Let $\mathcal{I}$ denote the integrand in \eqref{amp3}. Its power expansion around $z = 0$ reads
\begin{align} \label{ireg_exp}
\mathcal{I} & = \sum_{\sigma_1, \sigma_2, \sigma_3 = \pm 1} \: \sum_{n_1, n_2, n_3=0}^{\infty} b_{n_1}(\sigma_1 \breg_1) b_{n_2}(\sigma_2 \breg_2) b_{n_3}(\sigma_3 \breg_3) z^{\frac{\dreg}{2} - 1 + \sum_{j=1}^3 ( \sigma_j \breg_j + 2 n_j )},
\end{align}
where the coefficients $b_n$ are defined in \eqref{bcf} and the regulated $\beta_j$ parameters are given as usual by
\begin{align} \label{alphabeta_reg}
& \breg_j = \Dreg_j - \frac{\dreg}{2} = \beta_j + v_j \ep, && \beta_j = \Delta_j - \frac{d}{2}.
\end{align}
Let us rearrange the sum as 
\begin{align} \label{curlyIdef}
\mathcal{I} & 
= \sum_{\sigma_1, \sigma_2, \sigma_3 = \pm 1} \sum_{n=0}^{\infty} \mathcal{I}_{(\sigma_j, n)} z^{L(\sigma_j, n) - 1},
\end{align}
where the non-vanishing coefficients and the corresponding $L$s are parametrized by three signs, $\sigma_1, \sigma_2, \sigma_3 = \pm 1$ and a non-negative integer $n$,
\begin{align} \label{IL}
L(\sigma_j, n) & = \frac{\dreg}{2} + \sum_{j=1}^3 \sigma_j \breg_j + 2 n, \\
\mathcal{I}_{(\sigma_j, n)} & = \sum_{\substack{n_1, n_2, n_3 \geq 0\\n_1 + n_2 + n_3=n}} b_{n_1}(\sigma_1 \breg_1) b_{n_2}(\sigma_2 \breg_2) b_{n_3}(\sigma_3 \breg_3). \label{I_sigman}
\end{align}
Next we analyze the $\ep$-dependence. From \eqref{IL}, we see that $L$ is linear in the regulator, 
\begin{align} \label{L_split}
L(\sigma_j, n) = L_0(\sigma_j, n) + \ep \, L_{\sigma_1 \sigma_2 \sigma_3}(u,v_j) ,
\end{align}
where
\begin{align}
L_0(\sigma_j, n) & = \frac{d}{2} + \sum_{j=1}^3 \sigma_j \beta_j + 2 n, &
L_{\sigma_1 \sigma_2 \sigma_3}(u,v_j)  = u +\sum_{j=1}^3 \sigma_j v_j \, , \label{L1}
\end{align}
and all scheme-dependence in $L$ enters via $L_{\sigma_1 \sigma_2 \sigma_3}(u,v_j) $.

The coefficients $\mathcal{I}_{(\sigma_j, n)}$ depend on the regulator $\ep$ and $v_j$ but not on $u$. This follows from the fact that all dependence on the regulators enters through $\hat{\beta}_j$, and  $\hat{\beta}_j$ depends only on $\ep$ and $v_j$. Actually, 
in the cases we analyze in this paper ($\beta_j$ non-integral and positive) the coefficients are smooth as $\ep \to 0$. It follows that 
\begin{equation} \label{I_sch}
\mathcal{I}_{(\sigma_j, n)} = \mathcal{I}^0_{(\sigma_j, n)} + \ep \sum_{k=1}^3 \mathcal{I}^1_{(\sigma_j, n) k} v_k + O(\ep^2),
\end{equation}
where both $\mathcal{I}^0_{(\sigma_j, n)}$ and $\mathcal{I}^1_{(\sigma_j, n) k}$ are independent of   $u$ and $v_j$.

As shown in  \cite{Bzowski:2015yxv}, all divergences and finite $u$ and $v_j$-dependent terms for 3-point amplitudes \eqref{amp3} are contained in $\idiv$ defined by
\begin{align} \label{idiv_def}
\idiv(u, v_j; \mu) & = \sum_{\{L_0=0\}} \int_0^{\mu^{-1}} \D z \, \mathcal{I}_{(\sigma_j,n)} z^{-1 + \ep L_{\sigma_1 \sigma_2 \sigma_3}} 
 = \sum_{\{L_0=0\}} \frac{\mathcal{I}_{(\sigma_j,n)} \mu^{- \ep L_{\sigma_1 \sigma_2 \sigma_3}}}{\ep L_{\sigma_1 \sigma_2 \sigma_3}}, 
\end{align}
where  the sum is over all signs $\sigma_j$ and $n$ such that $L_0=0$ and $\mu > 0$ is an arbitrary number.
Using \eqref{L1}, the condition $L_0=0$ may be solved to determine $n$ in terms of the signs $\sigma_j$, $n=n(\sigma_j)$, 
and we define
\begin{equation} \label{def_isigns}
 \mathcal{I}_{\sigma_1\sigma_2 \sigma_3} = 
 \left\{
 \begin{array}{cc}
 \mathcal{I}_{(\sigma_j,n(\sigma_j))} & \qquad n(\sigma_j) =0,1, 2, \ldots ,\\
 0 & \qquad {\rm otherwise}\, .
 \end{array}
 \right.
 \end{equation}
Then 
 \begin{equation}
 \idiv(u, v_j; \mu) \label{Idiv}
 = \sum_{\sigma_1, \sigma_2, \sigma_3 = \pm 1} \frac{\mathcal{I}_{\sigma_1\sigma_2 \sigma_3} \mu^{- \ep L_{\sigma_1 \sigma_2 \sigma_3}}}{\ep L_{\sigma_1 \sigma_2 \sigma_3}},
 \end{equation}
where the sum is now  over all signs $\sigma_j$. Due to the fact that all amplitudes discussed in this paper are at most linearly divergent and renormalizable by counterterms, there are only two possibilities to yield a non-zero $\mathcal{I}_{\sigma_1\sigma_2 \sigma_3}$. Either $\mathcal{I}_{---} \neq 0$ with all other choices of signs vanishing, or some of $\mathcal{I}_{+--}, \mathcal{I}_{-+-}, \mathcal{I}_{--+}$ are non-zero (possibly all three), while all other combinations vanish. This can be seen from Table \ref{fig:sing_type}: as discussed at length in \cite{Bzowski:2015pba}, anomalies are associated with $(---)$ singularities and beta functions with $(--+)$ ones.

In \cite{Bzowski:2015yxv} we have shown the following procedure for scheme change:
\begin{fact} \label{fact:diff}
If we already know the amplitude $\ireg(\bar{u}, \bar{v}_j)$ regulated in \emph{some} scheme with parameters $\bar{u}$ and $\bar{v}_j$, we can use \eqref{Idiv} to obtain the amplitude in \emph{any} regularization scheme,
\end{fact}
\vspace{-0.3cm}
\begin{align} \label{Isch}
\ireg(u, v_j) = \ireg(\bar{u}, \bar{v}_j) + \left[ \idiv(u, v_j) - \idiv(\bar{u}, \bar{v}_j)  \right] + O(\ep).
\end{align}
From now on, we will concentrate on the amplitudes we are interested in for this paper, \textit{i.e.}, $d = 3$ and all conformal dimensions $\Delta_j$ equal to $2$ or $3$. For these $d$ and $\Delta_j$ the 3-point functions are at most linearly divergent. The divergence thus can only emerge from the explicit divergence in \eqref{Idiv}. In particular, the coefficients $\mathcal{I}_{\sigma_1\sigma_2\sigma_3} $ are finite and continuous in the $\ep \rightarrow 0$ limit.

\subsubsection{Scheme change}

Equation \eqref{Isch} can be greatly simplified and generalized when we start from the half-integer scheme with $\bar{v}_j = 0$. First note that, given an amplitude regulated in the half-integer scheme with $u = 1$ and $v_j = 0$, we can obtain for free the amplitude regulated with general $u$ simply by rescaling $\ep \mapsto u \ep$. Thus, knowing $\ireg(1, 0)$ to all orders in $\ep$, we also know $\ireg(u, 0)$ to all orders in $\ep$.

Let us also define $\ireg|_{\ep^n}$ to represent the terms of order $\ep^n$ in a regulated amplitude $\ireg$, so that up to linear order a linearly divergent amplitude is given by
\begin{align}
\ireg(u, v_j) = \frac{\ireg(u, v_j)|_{\ep^{-1}}}{\ep} + \ireg(u, v_j)|_{\ep^0} + \ep \, \ireg(u, v_j)|_{\ep^1} + O(\ep^2).
\end{align}
With this notation in place, one can show the following:
\begin{fact} \label{fact:sch}
Consider a linearly divergent amplitude $\ireg_{[\Delta_1 \Delta_2 \Delta_3]}$ regulated in the general $(u, v_j)$-scheme. The expansion coefficients are given by
\end{fact}
\vspace{-0.3cm}
\begin{align}
\ireg(u, v_j)|_{\ep^{-1}} & = \idiv \left(u,v_j; 1 \right) |_{\ep^{-1}}, \label{ischM1} \\
\ireg(u, v_j)|_{\ep^{0}} & = \ireg(1, 0)|_{\ep^0} + \idiv \left(u,v_j; 1 \right)|_{\ep^0}, \label{isch0} \\
\ireg(u, v_j)|_{\ep^{1}} & = \ireg(u, 0)|_{\ep^1} + \sum_{j=1}^3 \ep v_j \frac{\partial \ireg}{\partial \beta_j}(u, 0)|_{\ep^1} 
+  \idiv \left(u, v_j; 1 \right)|_{\ep^1}. \label{ischP1}
\end{align}
Not only can we evaluate the finite part for general $(u, v_j)$,  we can also evaluate the subleading terms of order $\ep$ provided we know the derivatives of the regulated amplitude with respect to the Bessel indices $\beta_j$. At half-integral Bessel indices the derivatives with respect to the order are known explicitly, see equations \eqref{dKdb12} -- \eqref{dIdb32}. Furthermore, we will need the expansion of the amplitude to order $\ep^1$ when we discuss the scheme change for exchange 4-point functions.

\subsubsection{Example}

To illustrate our discussion  consider the amplitude $\ireg_{[222]}$ regulated in a scheme with general $u$ and $v_{[2]}$ for all dimensions, $\Dreg_j = 2 + (u + v_{[2]}) \ep$ for $j=1,2,3$.  First, let us recall the expression we have already computed in the 
$(1,0)$ scheme,
\begin{align} \label{i2_01}
\ireg_{[222]}(1,0)  &= \Div_1(k_t) = \Gamma( \ep) k_t^{-\ep} \nn\\
&=\frac{1}{ \ep} - ( \log k_t + \gamma_E) + \frac{ \ep}{2} \left( (\log k_t + \gamma_E)^2  + \frac{\pi^2}{6} \right) + O(\ep^2),
\end{align}
where $k_t = k_1+k_2+k_3$.

To work out the change of scheme we need $\idiv_{[222]}$. In this case, there is only one choice of signs for which 
$L_0=0$ yields a non-negative integer $n$, namely all minus, which yields $n=0$ and $L_{---}=u - 3 v_{[2]}$ so that
\begin{align} \label{isch222}
\idiv_{[222]}(u,v_{[2]}; \mu) & = \int_0^{\mu^{-1}} \D z \, z^{-1 + \ep (u - 3 v_{[2]})}  = \frac{\mu^{- \ep (u - 3 v_{[2]})}}{\ep \, (u - 3 v_{[2]})}  \nn\\
&= \frac{1}{\ep \, (u - 3 v_{[2]})} + \log \mu + \frac{1}{2} (u - 3 v_{[2]})\log^2 \mu \, \ep +  O(\ep^2).
\end{align}
To use Fact \ref{fact:sch},  we evaluate at $\mu = 1$ to obtain
\begin{align}
\idiv_{[222]}(u, v_{[2]}; 1)|_{\ep^{-1}} & = \frac{1}{\ep \, (u - 3 v_{[2]})}, \qquad
\idiv_{[222]}(u, v_{[2]}; 1)|_{\ep^n}  = 0, \ n \geq 0.
\end{align}
We also need  $\ireg_{[222]}(u,0)$ which may be obtained from 
\eqref{i2_01} by $\ep \mapsto u \ep$,
\begin{align}
\ireg_{[222]}(u, 0) & = \frac{1}{u \ep} - \left( \log k_t + \gamma_E  \right) + \frac{u \ep}{2} \left[ ( \log k_t + \gamma_E)^2 + \frac{\pi^2}{6} \right] + O(\ep^2).
\end{align}
The final and the most difficult step is to calculate the terms involving the derivatives with respect to the Bessel order. Using \eqref{dKdb12} and \eqref{intE1} we find
\begin{align}
& \left. \ep v_{[2]} \frac{\partial \ireg_{[222]}}{\partial \beta_1} \right|_{\beta_1 = \frac{1}{2}}  \nn\\
& \qquad = \ep v_{[2]} \int_0^{\infty} \D z \, e^{-z k_t} z^{-1 + u \ep} \left( e^{2 z k_1} E_1(2 z k_1) + \log k_1 + \log 2 + \gamma_E \right) \nn\\
& \qquad = \frac{v_{[2]}}{u^2 \ep} + \left[ - \Li_2 \left( \frac{-k_1 + k_2 + k_3}{k_t} \right) + \frac{1}{2} \left( \log k_t + \gamma_E \right)^2 + \frac{\pi^2}{12} \right] v_{[2]} \ep + O(\ep^2).
\end{align}
We can put everything together to find the amplitude $\ireg_{[222]}$ to linear order in $\ep$ and in an arbitrary regularization scheme. According to Fact \ref{fact:sch},
\begin{align}
\ireg_{[222]}(u, v_j)|_{\ep^{-1}} & = \frac{1}{u - 3 v_{[2]}}, \\[0.5ex]
\ireg_{[222]}(u, v_j)|_{\ep^{0}} & = - \log k_t - \gamma_E, \\[0.5ex]
\ireg_{[222]}(u, v_j)|_{\ep^{1}} & = \frac{u + 3 v_{[2]}}{2} \left[ ( \log k_t + \gamma_E)^2 + \frac{\pi^2}{6} \right] \nn\\[0.5ex]
& \qquad - v_{[2]} \left[ \Li_2 \left( \frac{\m{12}}{k_t} \right) + \Li_2 \left( \frac{\m{13}}{k_t} \right) + \Li_2 \left( \frac{\m{23}}{k_t} \right) \right],
\end{align}
where $\m{12} = k_1 + k_2 - | \bs{k}_1 + \bs{k}_2 | = k_1 + k_2 - k_3$ and similarly for $\m{13}$ and $\m{23}$.

This example is slightly degenerate as it does not contain any parameter-dependent finite piece. To illustrate this point consider the case $\ireg_{[332]}$, and to simplify the discussion, we consider the two operators of dimension three as identical so that $\Dreg_j = 3 + (u + v_{[3]}) \ep$ for $j=1, 2$ and $\Dreg_2 = 2 + (u + v_{[2]}) \ep$. The $L_0=0$ condition is satisfied with all signs negative and $n=1$ 
so that $L_{---}= u - 2 v_{[3]} - v_{[2]}$. Following the same steps, we find
\begin{align} \label{isch332}
\idiv_{[332]}(u, v_{[2]}, v_{[3]}; \mu) & = \int_0^{\mu^{-1}} \D z \, z^{-1 + \ep \, L_{---}} 
\frac{1}{2} \left[ \frac{k_3^2}{1 - 2 v_{[2]} \ep} - \frac{k_1^2 + k_2^2}{1 + 2 v_{[3]} \ep} \right] \nn\\[1ex]
& = \frac{k_3^2 - k_1^2 - k_2^2}{2 \ep \, L_{---}} + \frac{(k_1^2 + k_2^2) v_{[3]} + k_3^2 v_{[2]}}{L_{---}} \nn\\
& \qquad\qquad\qquad - \frac{1}{2} (k_3^2 - k_1^2 - k_2^2) \log \mu + O(\ep).
\end{align}
The first term is an unambiguous divergence, which for $u = 1$ and $v_{[j]} = 0$ matches $\ireg_{[322]}$. The second term is a finite, scheme-dependent piece and depends only on $v_j$ but not $u$. The last term contains the cut-off scale, but otherwise is scheme-independent. This is the general structure, which holds in the generic case. Using Fact \ref{fact:sch}, we find
\begin{align} \label{ex332a}
\ireg_{[332]}(u, v_j) = \frac{k_3^2 - k_1^2 - k_2^2}{2 (u - v_{[2]} - 2 v_{[3]}) \, \ep} + \ireg_{[332]}(1, 0)|_{\ep^0} + \frac{(k_1^2 + k_2^2) v_{[3]} + k_3^2 v_{[2]}}{u - v_{[2]} - 2 v_{[3]}} + O(\ep).
\end{align}
The finite part here requires the expansion of the regulated amplitude \eqref{ireg332} to order $\ep^0$, yielding
\begin{align}
\ireg_{[332]}(1, 0)|_{\ep^0} & = \frac{1}{2} (k_1^2 + k_2^2 - k_3^2) \left( \log k_t + \gamma_E \right)  - \frac{1}{4} ( k_1 + k_2 - 3 k_3)(k_1 + k_2 + k_3).
\end{align}
This provides the expression for $\ireg_{[332]}$ regulated in an arbitrary scheme through order $\ep^0$. The $\ep^1$ terms may be computed as in the previous example. The final result through order $\ep^1$  is given in \eqref{i332-f}-\eqref{i332-l}.

\subsection{4-point contact amplitudes}

The scheme change procedure for 4-point (and in fact, any $n$-point) contact amplitudes is completely analogous to that of 3-point amplitudes. We define the analogue of \eqref{L1} with four signs,
\begin{align} \label{Lpm4}
L_0(\sigma_j, n) & = d + \sum_{j=1}^4 \sigma_j \beta_j + 2 n, \qquad L_{\sigma_1 \sigma_2 \sigma_3 \sigma_4}(u,v_j) = u + \sum_{j=1}^4 \sigma_j v_j
\end{align}
and then \eqref{Idiv} becomes
\begin{align} \label{Isch4}
\idiv(u,v_j; \mu) & = \sum_{\sigma_1, \sigma_2, \sigma_3, \sigma_4 = \pm 1} \frac{\mathcal{I}_{{\sigma_1 \sigma_2 \sigma_3 \sigma_4}}(v_j) \mu^{- \ep L_{\sigma_1 \sigma_2 \sigma_3 \sigma_4}}}{\ep L_{\sigma_1 \sigma_2 \sigma_3 \sigma_4}},
\end{align}
where $\mathcal{I}_{{\sigma_1 \sigma_2 \sigma_3 \sigma_4}}(v_j) $ is defined by the analogous equation to \eqref{def_isigns}.
As in the case of 3-point functions, in all the cases we discuss in this paper only one combination signs gives a non-zero $\mathcal{I}_{{\sigma_1 \sigma_2 \sigma_3 \sigma_4}} $, and in all cases these signs are either all minus (corresponding to anomalies), or one plus and three minus (corresponding to beta functions). 
The analogue of Fact \ref{fact:sch} then holds:
\begin{fact} \label{fact:sch4}
Consider a linearly divergent amplitude $\ireg_{[\Delta_1 \Delta_2 \Delta_3 \Delta_4]}$ regulated in an arbitrary $u, v_j$-scheme. The expansion coefficients can be extracted as follows,
\end{fact}
\vspace{-0.3cm}
\begin{align}
\ireg(u, v_j)|_{\ep^{-1}} & = \idiv \left(u,v_j; 1 \right) |_{\ep^{-1}}, \\[1ex]
\ireg(u, v_j)|_{\ep^{0}} & = \ireg(1, 0)|_{\ep^0} + \idiv \left(u,v_j; 1 \right)|_{\ep^0}, \\
\ireg(u, v_j)|_{\ep^{1}} & = \ireg(u, 0)|_{\ep^1} + \sum_{j=1}^4 \ep v_j \frac{\partial \ireg}{\partial \beta_j}(u, 0)|_{\ep^1} 
+  \idiv \left(u, v_j; 1 \right)|_{\ep^1}.
\end{align}
The generalization to $n$-point contact amplitudes is straightforward.

\subsection{4-point exchange diagrams} \label{scheme:exchange}

In this subsection, we present a procedure that allows one to evaluate the 4-point exchange amplitude in the general regularization scheme \eqref{genreg}. The idea is the same as in our discussion of 3-point functions: we would like  to split  the regulated exchange amplitude $\ireg$ into two parts, with one of them, $\idiv$, containing all scheme-dependent terms, and then apply the analogue of Fact \ref{fact:diff}. We will present such a split, but our construction is not optimal. There is a natural ambiguity in that one may always add  to a given $\idiv$ additional scheme-independent terms, and such terms would not contribute to $\ireg(u, v_j)$. Ideally,  $\idiv$ should not contain any scheme-independent terms, but our procedure will produce  $\idiv$ containing  (a large number of) such scheme-independent terms. 

\subsubsection{Definitions} \label{def:scheme:exchange}

We begin with a number of definitions. Since the bulk-to-bulk propagator \eqref{GPropagator} contains a Bessel-$I$ function, we will need  3-point-like amplitudes with one Bessel-$K$ replaced by a Bessel-$I$. To do this, we first define
\begin{align} \label{IPropagator}
\II_{d, \Delta}(z, k) = 2^{\Delta - \frac{d}{2} - 1} \Gamma \left( \Delta - \frac{d}{2} \right) k^{\Delta - \frac{d}{2}} z^{\frac{d}{2}} I_{\Delta - \frac{d}{2}}(k z)
\end{align}
so that we can rewrite the bulk-to-bulk propagator as
\begin{align} \label{BB}
\G_{d, \Delta}(z, k; \z) = k^{-(2 \Delta - d)} \times \left\{ \begin{array}{ll}
\II_{\Delta - \frac{d}{2}}(k z) \K_{\Delta - \frac{d}{2}}(k \z) & \text{ for } z < \z, \\
\K_{\Delta - \frac{d}{2}}(k z) \II_{\Delta - \frac{d}{2}}(k \z) & \text{ for } z > \z.
\end{array} \right.	
\end{align}
Next, we define the 3-point amplitude \eqref{amp3} with $\Kreg_{[\Delta_3]}$ replaced by $\Ireg_{[\Delta_3]}$ (\textit{i.e.}, the regulated amplitude $\II_{d, \Delta}$) as
\begin{align} \label{amp3i}
\jreg_{[\Delta_1 \Delta_2; \Delta_3]} = \int_0^{\infty} \D z \, z^{-\dreg-1} \, \Kreg_{[\Delta_1]}(z, k_1) \Kreg_{[\Delta_2]}(z, k_2) \Ireg_{[\Delta_3]}(z, k_3).
\end{align}
The properties of these amplitudes are very similar to those of the actual 3-point amplitudes \eqref{amp3}. In particular $\jdiv_{[\Delta_1 \Delta_2; \Delta_3]}$ is  defined analogously to \eqref{Idiv} and Fact \ref{fact:sch} holds as well.

Furthermore, we generalize the amplitudes $\ireg$ and $\jreg$ by leaving the upper limits of integration open,
\begin{align} \label{iregz}
\ireg_{[\Delta_1 \Delta_2 \Delta_3]}(z; k_1, k_2, k_3) & = \int_0^{z} \D \z \, \z^{-\dreg-1} \, \Kreg_{[\Delta_1]}(\z, k_1) \Kreg_{[\Delta_2]}(\z, k_2) \Kreg_{[\Delta_3]}(\z, k_3), \\
\jreg_{[\Delta_1 \Delta_2; \Delta_3]}(z; k_1, k_2, k_3) & = \int_0^{z} \D \z \, \z^{-\dreg-1} \, \Kreg_{[\Delta_1]}(\z, k_1) \Kreg_{[\Delta_2]}(\z, k_2) \Ireg_{[\Delta_3]}(\z, k_3), \label{jregz}
\end{align}
so that $\ireg_{[\Delta_1 \Delta_2 \Delta_3]} = \ireg_{[\Delta_1 \Delta_2 \Delta_3]}(\infty)$ and $\jreg_{[\Delta_1 \Delta_2; \Delta_3]} = \jreg_{[\Delta_1 \Delta_2; \Delta_3]}(\infty)$. Finally, we promote the explicit power of the integration variable to another parameter,
\begin{align}
\ireg_{\alpha [\Delta_1 \Delta_2 \Delta_3]} & = \int_0^{\infty} \D \z \, \z^{\alpha} \, \Kreg_{[\Delta_1]}(\z, k_1) \Kreg_{[\Delta_2]}(\z, k_2) \Kreg_{[\Delta_3]}(\z, k_3), \\
\jreg_{\alpha [\Delta_1 \Delta_2; \Delta_3]} & = \int_0^{\infty} \D \z \, \z^{\alpha} \, \Kreg_{[\Delta_1]}(\z, k_1) \Kreg_{[\Delta_2]}(\z, k_2) \Ireg_{[\Delta_3]}(\z, k_3), \label{jreg_a}
\end{align}
so that $\ireg_{[\Delta_1 \Delta_2 \Delta_3]} = \ireg_{-\dreg - 1, [\Delta_1 \Delta_2 \Delta_3]}$ and $\jreg_{[\Delta_1 \Delta_2; \Delta_3]} = \jreg_{-\dreg - 1, [\Delta_1 \Delta_2; \Delta_3]}$.

\subsubsection{Change of scheme}

Using the bulk-to-bulk propagator \eqref{BB} and 
$\int_z^{\infty} = \int_0^{\infty} - \int_0^{z}$ 
we can rewrite the 4-point exchange amplitude \eqref{amp4x} as 
\begin{align} \label{4exp}
& \ireg_{[\Delta_1 \Delta_2, \Delta_3 \Delta_4 x \Delta_x]} = s^{-2 \breg_x} \jreg_{[\Delta_1 \Delta_2; \Delta_x]}(k_1, k_2, s) \, \ireg_{[\Delta_3 \Delta_4 \Delta_x]}(k_3, k_4, s) \nn\\
& \qquad - s^{-2 \breg_x} \int_0^\infty \D z \, z^{-\dreg-1} \Kreg_{[\Delta_1]}(z, k_1) \Kreg_{[\Delta_2]}(z, k_2) \reg{\II}_{[\Delta_x]}(z, s) 
\ireg_{[\Delta_3 \Delta_4 \Delta_x]}(z; k_3, k_4, s) \nn\\
& \qquad + s^{-2 \breg_x} \int_0^\infty \D z \, z^{-\dreg-1} \Kreg_{[\Delta_1]}(z, k_1) \Kreg_{[\Delta_2]}(z, k_2) \Kreg_{[\Delta_x]}(z, s) 
\jreg_{[\Delta_3 \Delta_4; \Delta_x]}(z; k_3, k_4, s).
\end{align}
The $z$-dependent amplitudes $\ireg_{[\Delta_3 \Delta_4 \Delta_x]}(z; k_3, k_4, s)$ and $\jreg_{[\Delta_3 \Delta_4; \Delta_x]}(z; k_3, k_4, s)$ can now be computed after 
power expanding the integrands using \eqref{I_Bessel} and \eqref{Iexp}. Analogously to \eqref{curlyIdef}, let $\mathcal{I}$ and $\mathcal{J}$ denote the integrands in \eqref{iregz} and \eqref{jregz} respectively. Furthermore, let $\mathcal{I}_{(\sigma_j,n)}$ denote the coefficient of $\zeta^{L-1}$ in the power expansion of $\mathcal{I}$, and $\mathcal{J}_{(\sigma_1, \sigma_2, n)}$ the coefficient of $\zeta^{L'-1}$ in the power expansion of $\mathcal{J}$. Then
\begin{align}
\ireg_{[\Delta_3 \Delta_4 \Delta_x]}(z; k_3, k_4, s) & = \sum_{\sigma_1, \sigma_2 , \sigma_3= \pm 1} \sum_{n=0}^\infty \frac{\mathcal{I}_{(\sigma_j,n)}(k_3, k_4, s) z^{L(\sigma_j,n)}}{L(\sigma_j,n)}, \\
\jreg_{[\Delta_3 \Delta_4; \Delta_x]}(z; k_3, k_4, s) & = \sum_{\sigma_1, \sigma_2 = \pm 1} \sum_{n=0}^\infty \frac{\mathcal{J}_{(\sigma_1, \sigma_2, n)}(k_3, k_4, s) z^{L'(\sigma_1, \sigma_2, n)}}{L'(\sigma_1, \sigma_2, n)}.
\end{align}
The definition of $\mathcal{I}_{(\sigma_j,n)}(k_3, k_4, s)$ and the value of $L$ are given in \eqref{I_sigman}, and 
$\mathcal{J}_{(\sigma_1, \sigma_2, n)}$ is defined analogously.  
For the values of $L'$, notice that the series expansion of the Bessel-$I$ function in \eqref{I_Bessel} is similar  to that of the
Bessel-$K$ function \eqref{Iexp} but there are only  $z^{\beta}$ terms, so relative to Bessel-$K$ we have $\sigma=1$.  
Thus, the values of $L'$ are identical to $L$, but with $\sigma_3 = 1$ fixed,
\begin{align} \label{fullLp}
L'(\sigma_1, \sigma_2, n) = L(\sigma_1, \sigma_2, +1, n) = \frac{\dreg}{2} + \sum_{j=1}^2 \sigma_j \breg_j + \breg_3 + 2 n,
\end{align}
and we define
\begin{equation}
L' = L'_0+\ep L'_{\sigma_1 \sigma_2}, \quad 
L'_0= \frac{d}{2} + \sum_{j=1}^2 \sigma_j \beta_j + \beta_3 + 2 n, \quad
L'_{\sigma_1 \sigma_2} =  u + \sum_{j=1}^2 \sigma_j v_j  + v_3 \, . \label{L'1}
\end{equation}

With this notation in place, we rewrite \eqref{4exp} as
\begin{align} \label{4expa}
& \ireg_{[\Delta_1 \Delta_2, \Delta_3 \Delta_4 x \Delta_x]} = s^{-2 \breg_x} 
\jreg_{[\Delta_1 \Delta_2; \Delta_x]}(k_1, k_2, s) \, \ireg_{[\Delta_3 \Delta_4 \Delta_x]}(k_3, k_4, s) \nn\\
& \qquad - s^{-2 \breg_x} \sum_{\sigma_1, \sigma_2 , \sigma_3= \pm 1} \sum_{n=0}^\infty \frac{\mathcal{I}_{(\sigma_j,n)}(k_3, k_4, s) }{L(\sigma_j,n)}\, \jreg_{-\dreg - 1 + L, [\Delta_1 \Delta_2; \Delta_x]}(k_1, k_2, s) \nn\\
& \qquad + s^{-2 \breg_x} \sum_{\sigma_1, \sigma_2 = \pm 1} \sum_{n=0}^\infty \frac{\mathcal{J}_{(\sigma_1, \sigma_2, n)}(k_3, k_4, s)}{L'(\sigma_1, \sigma_2, n)} \, \ireg_{-\dreg - 1 + L', [\Delta_1 \Delta_2 \Delta_x]}(k_1, k_2, s).
\end{align}
Thus, the divergent terms $\idiv$ for the exchange 4-point function can be extracted from our knowledge of the scheme-dependent terms for all relevant 3-point functions. We arrive at:
\begin{fact} \label{fact4}
The piece $i^{\rm div}_{[\Delta_1 \Delta_2, \Delta_3 \Delta_4 x \Delta_x]}$ containing all divergences and all finite scheme-dependent terms  for the exchange 4-point amplitudes with $d = 3$ and $\Delta_j = 2$ or $3$ can be computed as
\begin{align} \label{isch4}
& s^{2 \breg_x} i^{\rm div}_{[\Delta_1 \Delta_2, \Delta_3 \Delta_4 x \Delta_x]} = \jreg_{[\Delta_1 \Delta_2; \Delta_x]}(k_1, k_2, s) \, \ireg_{[\Delta_3 \Delta_4 \Delta_x]}(k_3, k_4, s) \nn\\[1ex]
& \qquad - \sum_{\sigma_1, \sigma_2 , \sigma_3= \pm 1} 
\sum_{\substack{n=0 \\L_0 \neq 0}}^\infty 
\frac{\mathcal{I}_{(\sigma_j,n)}(k_3, k_4, s) }{L(\sigma_j,n)} \, j^{\rm div}_{-\dreg - 1 + L, [\Delta_1 \Delta_2; \Delta_x]}(k_1, k_2, s) \nn\\
& \qquad - \sum_{\substack{\sigma_1, \sigma_2, \sigma_3 = \pm 1 \\ \{L_0=0\}}} \frac{\mathcal{I}_{{\sigma_1 \sigma_2 \sigma_3}}(k_3, k_4, s)}{\ep \, L_{\sigma_1 \sigma_2 \sigma_3}} \, \jreg_{-\dreg - 1 + \ep \, L_{\sigma_1 \sigma_2 \sigma_3}, [\Delta_1 \Delta_2; \Delta_x]}(k_1, k_2, s) \nn\\
& \qquad + \sum_{\sigma_1, \sigma_2 = \pm 1} 
 \sum_{\substack{n=0 \\L'_0 \neq 0}}^\infty 
\frac{\mathcal{J}_{(\sigma_1, \sigma_2, n)}(k_3, k_4, s)}{L'(\sigma_1, \sigma_2, n)} \, i^{\rm div}_{-\dreg - 1 + L', [\Delta_1 \Delta_2 \Delta_x]}(k_1, k_2, s) \nn\\
& \qquad + \sum_{\substack{\sigma_1, \sigma_2 = \pm 1 \\ \{L'_0=0\}}} \frac{\mathcal{J}_{\sigma_1 \sigma_2}(k_3, k_4, s)}{\ep \, L'_{\sigma_1 \sigma_2}} \, \ireg_{-\dreg - 1 + \ep \, L'_{\sigma_1 \sigma_2}, [\Delta_1 \Delta_2 \Delta_x]}(k_1, k_2, s),
\end{align}
where $\mathcal{I}_{{\sigma_1 \sigma_2 \sigma_3}}(k_3, k_4, s)$ is defined in \eqref{def_isigns} and $\mathcal{J}_{(\sigma_1, \sigma_2, n)}(k_3, k_4, s)$ has an analogous definition. The summation in the second and fourth terms is over all signs and all integers such that $L_0$ and $L_0'$ are not equal to zero, respectively. These summations are finite, since we only need to consider the amplitudes with
\begin{align} \label{L_bound}
& L \leq \breg_1 + \breg_2 - \breg_x - \frac{\dreg}{2}, && L' \leq \breg_1 + \breg_2 + \breg_x - \frac{\dreg}{2}.
\end{align}
The summations in the third and fifth terms are over the locus of $L_0=0$ and $L_0'=0$, respectively.

Thus, if we already know the amplitude $\ireg(\bar{u}, \bar{v}_j)$ regulated in \emph{some} scheme with parameters $\bar{u}$ and $\bar{v}_j$, we can use it to obtain the amplitude in \emph{any} regularization scheme,
\begin{align} \label{change_scheme_4}
\ireg(u, v_j) = \ireg(\bar{u}, \bar{v}_j) + \left[ i^{\rm div}(u, v_j) - i^{\rm div}(\bar{u}, \bar{v}_j)  \right] + O(\ep).
\end{align}
\end{fact}

The bound \eqref{L_bound} reflects the fact that for $L$ and $L'$ sufficiently large, the amplitudes in \eqref{4expa} become finite and thus scheme-independent. To figure out how large $L$ and $L'$ need to be for the amplitudes to be finite, we power expand the integrands of $\jno_{-d - 1 + L, [\Delta_1 \Delta_2; \Delta_x]}$ and $\ino_{-d - 1 + L', [\Delta_1 \Delta_2 \Delta_x]}$. The lowest powers of the integration variables in the two cases are $\tfrac{1}{2} d - 1 + L - \beta_1 - \beta_2 + \beta_3$ for the $\jno$ amplitude and $\tfrac{1}{2} d - 1 + L' - \beta_1 - \beta_2 - \beta_3$ for $\ino$. The amplitudes are finite if these combinations are strictly greater than $-1$. Equivalently, divergent amplitudes can only occur for $L$ and $L'$ satisfying \eqref{L_bound}.

In order to use this fact we must be careful to evaluate \eqref{isch4} including all finite terms of order $\ep^0$. This means that one generally needs to evaluate all 3-point amplitudes up to and including terms of order $\ep^1$. For example, in the first line both 3-point amplitudes may be linearly divergent. Thus, in order to extract correctly the finite $u$ and $v_j$-dependent terms we have to multiply the divergence of one amplitude by terms of order $\ep^1$ in the other. This is the reason why we computed the order $\ep$ parameter-dependent terms in 3-point amplitudes in Fact \ref{fact:sch}.

\subsubsection{Example}

Let us discuss an example of the scheme change for the amplitude $\ireg_{[33,22x2]}$. To make the expressions slightly shorter, let us use only two parameters $v_{[2]}$ and $v_{[3]}$ to regulate the corresponding dimensions, \textit{i.e.}, for $\Delta = 2$ the regulated dimension is $\Dreg = 2 + (u + v_{[2]}) \ep$, while for $\Delta = 3$ we take $\Dreg = 3 + (u + v_{[3]}) \ep$.

First, we find out which 3-point amplitudes are needed in \eqref{isch4}. The first term is the product of $\ireg_{[222]}$ and $\jreg_{[33;2]}$. Since the amplitude $\jreg_{[33;2]}$ is finite (see Table \ref{fig:sing3}), it is enough to know $\ireg_{[222]}$ only up to finite order. On the other hand, we need all $u$ and $v_j$-dependent terms, including terms of order $\ep$ in $\jreg_{[33;2]}$ as they get multiplied by the divergence of $\ireg_{[222]}$.

For the next two terms we must power expand the integrands of $\ireg_{[222]}$ and $\jreg_{[22;2]}$. We must include all terms of order $z^{L-1}$ with $L$ and $L'$ bounded by \eqref{L_bound}, which in this case evaluates to $L \leq 1$ and $L' \leq 2$. Power expanding the integrands of $\jreg_{-4 - 2 u \ep + L, [22;2]}$ and $\ireg_{-4 - 2 u \ep + L', [222]}$ we find the divergent pieces $\jdiv_{-4 - 2 u \ep + L, [22;2]}$ and $\idiv_{-4 - 2 u \ep + L', [222]}$ in \eqref{isch4},
\begin{align} \label{expI}
\mathcal{I} & = \z^{-1 + \ep(u - 3 v_{[2]})} \nn\\
& \qquad + \z^{\ep(u - v_{[2]})} \frac{\Gamma (-\tfrac{1}{2} - \ep v_{[2]})}{2^{1 + 2 \ep v_{[2]}} \Gamma( \tfrac{1}{2} + \ep v_{[2]})} \left( k_3^{1 + 2 \ep v_{[2]}} + k_4^{1 + 2 \ep v_{[2]}} + s^{1 + 2 \ep v_{[2]}} \right) \nn\\
& \qquad + O(\z),\\[1ex]
\mathcal{J} & = \z^{\ep (u - v_{[2]})} \frac{\Gamma ( \tfrac{1}{2} + \ep v_{[2]})}{2 \Gamma( \tfrac{3}{2} + \ep v_{[2]})} s^{1 + 2 \ep v_{[2]}} \nn\\
& \qquad + \z^{1 + \ep (u + v_{[2]})} \frac{\Gamma( -\tfrac{1}{2} - \ep v_{[2]} )}{2^{2 + 2 \ep v_{[2]}} \Gamma( \tfrac{3}{2} + \ep v_{[2]})} \left[ (k_3 s)^{1 + 2 \ep v_{[2]}} + (k_4 s)^{1 + 2 \ep v_{[2]}} \right] \nn\\
& \qquad + O(\z^2). \label{expJ}
\end{align}
When integrated term-by-term, the first term in $\mathcal{I}$ produces a singularity at $\ep = 0$. This term corresponds to the third line of \eqref{isch4}. Indeed, the condition $L_0 = 0$, with $L_0$ defined in \eqref{L1} has a unique solution corresponding to all $\sigma_j = -1$ and $n(\sigma_j) = 0$ as $\beta_1 = \beta_2 = \beta_x = 1/2$. Similarly, the fact that the expansion \eqref{expJ} does not contain terms of order $-1 + O(\ep)$ indicates that the last line of \eqref{isch4} is absent. Equivalently, $L'_0$ defined in \eqref{L'1} does not vanish for any combination of signs. This means that all remaining terms in \eqref{expI} and \eqref{expJ} give rise to the scheme-dependent terms in the second and fourth line of \eqref{isch4}. In total, we find
\begin{align} \label{ex_scheme4ptX}
s^{1 + 2 v_{[2]} \ep} \idiv_{[33,22x2]} & = \jreg_{[33;2]}(k_1, k_2, s) \ireg_{[222]}(k_3, k_4, s) \nn\\
& \qquad - \frac{1}{\ep(u - 3 v_{[2]})} \times \jreg_{-4 - \ep (u + 3 v_{[2]}), [33;2]}(k_1, k_2, s) \nn\\
& \qquad - \frac{\Gamma (-\tfrac{1}{2} - \ep v_{[2]})}{2^{1 + 2 \ep v_{[2]}} (1 + \ep(u - v_{[2]})) \Gamma( \tfrac{1}{2} + \ep v_{[2]})} \left( k_3^{1 + 2 \ep v_{[2]}} + k_4^{1 + 2 \ep v_{[2]}} + s^{1 + 2 \ep v_{[2]}} \right)\times  \nn\\
& \qquad\qquad\qquad\qquad \times \jdiv_{-3 - \ep(u + v_{[2]}), [33;2]}(k_1, k_2, s) \nn\\[1ex]
& \qquad + \frac{\Gamma ( \tfrac{1}{2} + \ep v_{[2]})}{2 (1 + \ep (u - v_{[2]})) \Gamma( \tfrac{3}{2} + \ep v_{[2]})} s^{1 + 2 \ep v_{[2]}} \idiv_{-3-\ep (u + v_{[2]}), [332]}(k_1, k_2, s) \nn\\[1ex]
& \qquad + \frac{\Gamma( -\tfrac{1}{2} - \ep v_{[2]} )}{2^{2 + 2 \ep v_{[2]}} (2 + \ep (u + v_{[2]})) \Gamma( \tfrac{3}{2} + \ep v_{[2]})} \left[ (k_3 s)^{1 + 2 \ep v_{[2]}} + (k_4 s)^{1 + 2 \ep v_{[2]}} \right]\times \nn\\
& \qquad\qquad\qquad\qquad \times \idiv_{-2 - \ep(u - v_{[2]}), [332]}(k_1, k_2, s).
\end{align}
To complete the calculation, we must be able to evaluate the amplitudes on the right-hand side. The divergent terms can always be evaluated by means of the definition \eqref{Idiv}. For example, to calculate $\jdiv_{-3 - \ep(u + v_{[2]}), [33;2]}(k_1, k_2, s)$ we power expand the integrand in \eqref{jreg_a} and pick the term of order $\zeta^{-1 + O(\ep)}$. To be precise, the power expansion reads
\begin{align}
\zeta^{-3 - \ep(u + v_{[2]})} \Kreg_{[3]}(\z, k_1) \Kreg_{[3]}(\z, k_2) \Ireg_{[2]}(\z, s) = \frac{s^{1 + 2 \ep v_{[2]}} \Gamma ( \tfrac{1}{2} + \ep v_{[2]})}{2 \Gamma ( \tfrac{3}{2} + \ep v_{[2]} )} \zeta^{-1 + 2 \ep (u - v_{[3]})} + O(\zeta^0).
\end{align}
Thus,
\begin{align}
\jdiv_{-3 - \ep(u + v_{[2]}), [33;2]}(k_1, k_2, s) & = \frac{s^{1 + 2 \ep v_{[2]}} \Gamma ( \tfrac{1}{2} + \ep v_{[2]})}{2 \Gamma ( \tfrac{3}{2} + \ep v_{[2]} )} \int_0^{\mu^{-1}} \D \zeta \, \zeta^{-1 + 2 \ep (u - v_{[3]})} \nn\\
& = \frac{s^{1 + 2 \ep v_{[2]}} \Gamma ( \tfrac{1}{2} + \ep v_{[2]})}{2 \Gamma ( \tfrac{3}{2} + \ep v_{[2]} )} \times \frac{\mu^{-2 \ep (u - v_{[3]})}}{2 \ep \, (u - v_{[3]})}.
\end{align}
In the same fashion we find
\begin{align}
\idiv_{-3 - \ep(u + v_{[2]}), [332]}(k_1, k_2, s) & = \frac{s^{1 + 2 \ep v_{[2]}} \mu^{-2 \ep(u - v_{[3]})} \Gamma ( -\tfrac{1}{2} - \ep v_{[2]})}{2^{2+2 \ep v_{[2]}} \ep \, (u - v_{[3]}) \Gamma ( \tfrac{1}{2} + \ep v_{[2]} )}, \\
\idiv_{-2 - \ep(u - v_{[2]}), [332]}(k_1, k_2, s) & = \frac{\mu^{-2 \ep (u - v_{[3]})}}{2 \ep \, (u - v_{[3]})}.
\end{align}

The remaining term can be evaluated including terms of order $\ep^1$, since $\jreg_{-4 - \ep (u + 3 v_{[2]}), [33;2]}$ equals  $\jreg_{[33;2]}$ evaluated in the scheme with the $u$-parameter equal to $2u - 3v_{[2]}$, \textit{i.e.},
\begin{align}
\jreg_{-4 - \ep (u + 3 v_{[2]}), [33;2]}(u; v_{[2]}, v_{[3]}) = \jreg_{[33;2]}(2u - 3v_{[2]}; v_{[2]}, v_{[3]}).
\end{align}
When these results are substituted back to \eqref{ex_scheme4ptX} and power expanded we arrive at
\begin{align}
\idiv_{[33,22x2]} & = s^{-1 - 2 \ep v_{[2]}} \left[ \frac{\jreg^{2 u - 3 v_{[2]} \{v_{[3]} v_{[3]} v_{[2]}\}}_{[33;2]}(k_1, k_2, s)}{\ep \, (-u + 3 v_{[2]}) } + \jreg_{[33;2]}(k_1, k_2, s) \ireg_{[222]}(s, k_3, k_4) \right] \nn\\
& \qquad + \frac{k_3}{2 (u - v_{[3]})} \left[ \frac{1}{2 \ep} + v_{[2]} ( \log k_3 + \gamma_E + \log 2 - \tfrac{7}{4} ) + \frac{1}{4} ( -3 u + 4 v_{[2]} ) \right] \nn\\
& \qquad + \frac{k_4}{2 (u - v_{[3]})} \left[ \frac{1}{2 \ep} + v_{[2]} ( \log k_4 + \gamma_E + \log 2 - \tfrac{7}{4} ) + \frac{1}{4} ( -3 u + 4 v_{[2]} ) \right] + O(\ep).
\end{align}
In the first term we used the convention where, in the superscript, we indicate the value of the $u$ and $v_j$ parameters at which the 3-point amplitude should be evaluated,
\begin{align}
\ireg^{u \{v_1 v_2 v_3\}}_{[\Delta_1 \Delta_2 \Delta_3]}(k_1, k_2, k_3) & = \ireg_{ [\Delta_1 \Delta_2 \Delta_3]}(k_1, k_2, k_3; \dreg = 3 + 2 u \ep, \Dreg_j = \Delta_j + (u + v_j) \ep), \\[1ex]
\jreg^{u \{v_1 v_2 v_3\}}_{[\Delta_1 \Delta_2; \Delta_3]}(k_1, k_2, k_3) & = \ireg_{ [\Delta_1 \Delta_2; \Delta_3]}(k_1, k_2, k_3; \dreg = 3 + 2 u \ep, \Dreg_j = \Delta_j + (u + v_j) \ep).
\end{align}

All results are listed in Section \ref{sec:scheme_change_results}. There, we  list first the 3-point amplitudes evaluated to order $\ep^1$, then the regulated 4-point amplitudes evaluated in an arbitrary scheme.

\subsubsection{Proofs}

Our goal in this subsection is to prove Fact \ref{fact:sch} and Fact \ref{fact4}. The starting point is the analysis of singularities of the regulated 3-point amplitude \eqref{amp3}. The full analysis is carried out in \cite{Bzowski:2015yxv}, but here we will limit our attention to the cases at hand, \textit{i.e.}, we assume unregulated $\beta_j$ parameters that are non-integral. This implies that all expansion coefficients in \eqref{bcf} are continuous at $\ep = 0$, meaning  one can split the integral in \eqref{amp3} into two regions. Choosing $\mu > 0$, we write
\begin{align} \label{ireg_split}
\ireg & = \int_{\mu^{-1}}^{\infty} \D z \, \mathcal{I} + \int_0^{\mu^{-1}} \D z \sum_{L} \mathcal{I}_{L} z^{L - 1}.
\end{align}
Due to the exponential fall-off of the Bessel functions, the first integral converges for all $\alpha$ and $\beta_j$. By the dominated convergence theorem, one can pass the $\epsilon \rightarrow 0$ limit through the integral. For the second integral, recall that the regulated amplitude $\ireg$ is defined by the analytic continuation of the parameters $\alpha$ and $\beta_j$ to the region of convergence. In other words, we can assume that the powers of $z$ are all greater than $-1$ and at least the integral of each term separately exists. Thus, to commute the sum and the integral it is enough to show that the series converges to an integrable function when the $\mathcal{I}_{L}$ are replaced by their absolute values, $|\mathcal{I}_{L}|$. This is indeed the case, since almost all coefficients $a_n(\nu)$ are positive, \textit{i.e.}, $a_n(\nu) > 0$ for all $n \geq \nu$. Since the radius of convergence of \eqref{Iexp} is infinite, we can reverse the order of summation and integration in \eqref{ireg_split} by the Fubini's theorem. This shows the following:
\begin{fact}
The amplitude $\ireg_{[\Delta_1 \Delta_2 \Delta_3]}$ is divergent at $\ep \rightarrow 0$ if and only if there exist independent choices of signs $\sigma_1, \sigma_2, \sigma_3 = \pm 1$ and a non-negative integer $n$ such that $L_0(\sigma_j, n) = 0$, \textit{i.e.},
\begin{align} \label{cond}
\frac{d}{2} + \sigma_1 \beta_1 + \sigma_2 \beta_2 + \sigma_3 \beta_3 = - 2 n.
\end{align}
\end{fact}
The same argument holds for the amplitudes $\jreg_{[\Delta_1 \Delta_2; \Delta_3]}$, except that the corresponding condition is
\begin{align}
L'_0(\sigma_1, \sigma_2, n) = L_0(\sigma_1, \sigma_2, +1, n) = 0.
\end{align}
Since the sign of $\sigma_3$ is fixed to $+1$, the amplitudes $\jreg_{[\Delta_1 \Delta_2 \Delta_3]}$ cannot be more singular than $\ireg_{[\Delta_1 \Delta_2 \Delta_3]}$. The following table presents the rank of the divergence at $\ep = 0$ of the 3-point amplitudes:
\begin{table}[ht]
\begin{tabular}{|c|c|c|} \hline
Dimensions & $\ireg_{[\Delta_1 \Delta_2 \Delta_3]}$ & $\jreg_{[\Delta_1 \Delta_2; \Delta_3]}$ \\ \hline
$[22;2]$ & $1$ & $0$ \\ \hline
$[22;3]$ & $1$ & $0$ \\ \hline
$[32;2]$ & $1$ & $1$ \\ \hline
$[32;3]$ & $1$ & $0$ \\ \hline
$[33;2]$ & $1$ & $0$ \\ \hline
$[33;3]$ & $1$ & $1$ \\ \hline
\end{tabular}
\centering
\caption{Degrees of singularities of 3-point amplitudes.\label{fig:sing3}}
\end{table}

The reasoning above is sufficient to prove Fact \ref{fact:diff}. Indeed, as we can see, all terms in \eqref{ireg_split} have a finite $\ep = 0$ limit except the ones connected with $L_0 = 0$. Thus, all parameter-dependent terms are contained in $\idiv$. This is sufficient to prove the first two equations in Fact \ref{fact:sch}. However, since we also want to derive the scheme-dependent terms in terms of order $\ep^1$, we must first prove the following:

\begin{fact} \label{fact:struct}
Consider a regulated amplitude  $\ireg(u, v_j)$ exhibiting a first order pole as $\ep \rightarrow 0$. As far as the $u$ and $v_j$-dependence of the amplitude is concerned, one has the decomposition

\begin{align} \label{Istruct}
i^{\rm div}(u, v_j) & = \frac{1}{\ep} \, \sum_{\sigma_1, \sigma_2, \sigma_3 = \pm 1} \frac{A_{\sigma_1 \sigma_2 \sigma_3}}{L_{\sigma_1 \sigma_2 \sigma_3}} \nn\\
& \qquad + B + \sum_{\sigma_1, \sigma_2, \sigma_3 = \pm 1} \sum_{j=1}^3 \frac{B_{\sigma_1 \sigma_2 \sigma_3}^{(j)} v_j}{L_{\sigma_1 \sigma_2 \sigma_3}} \nn\\
& \qquad + \ep \left[ C u + \sum_{\sigma_1, \sigma_2, \sigma_3 = \pm 1} \sum_{j=1}^3 C_{\sigma_1 \sigma_2 \sigma_3}^{(j)} v_j + \sum_{\sigma_1, \sigma_2, \sigma_3 = \pm 1} \sum_{i,j=1}^3 \frac{C^{(ij)}_{\sigma_1 \sigma_2 \sigma_3} v_i v_j}{L_{\sigma_1 \sigma_2 \sigma_3}} \right] + O(\ep^2),
\end{align}
where $L_{\sigma_1 \sigma_2 \sigma_3}$ is given by \eqref{L1} and $A_{\sigma_1 \sigma_2 \sigma_3}, B, B_{\sigma_1 \sigma_2 \sigma_3}^{(j)}, C, C_{\sigma_1 \sigma_2 \sigma_3}^{(j)}, C^{(ij)}_{\sigma_1 \sigma_2 \sigma_3}$ are all $u$- and $v_j$-independent. The same structure holds for $\ireg(u, v_j)$.
\end{fact}
This fact follows from the structure of terms in  \eqref{ireg_split}. All terms in \eqref{ireg_split}, except \eqref{Idiv}, are finite as $\ep \to 0$.
The upper integral in \eqref{ireg_split} is finite and scheme-independent. The finite terms of the lower integral are given by
\begin{equation}
i^{\rm finite} = \sum_{\sigma_1, \sigma_2 , \sigma_3= \pm 1} 
\sum_{\substack{n=0 \\L_0 \neq 0}}^\infty \frac{\mathcal{I}_{(\sigma_j,n)} \mu^{-L(\sigma_j,n)}}{L(\sigma_j,n)}. 
\end{equation}
As shown in \eqref{I_sch},  
$\mathcal{I}_{(\sigma_j,n)}$ is finite and scheme-independent as $\ep \to 0$, and the order $\ep$ term is linear in $v_j$ but independent of $u$ (more generally, the $n$th order term in their Taylor expansion around $\ep=0$
is a polynomial of order $n$ in $v_j$ and has no $u$ dependence). Similarly, the $\ep, u, v_j$ dependence of terms depending on $L(\sigma_j,n)$ follow from \eqref{L_split}-\eqref{L1}. When $L_0 \neq 0$, such terms are finite and scheme-independent as $\ep \to 0$, and the $n$th order term in their Taylor expansion around $\ep=0$
is a polynomial of order $n$ in $u, v_j$. It follows that in the absence of divergences the $\ep$-expansion of any amplitude takes the form \eqref{Istruct} with  $A_{\sigma_1 \sigma_2 \sigma_3} = B_{\sigma_1 \sigma_2 \sigma_3}^{(j)} = C^{(ij)}_{\sigma_1 \sigma_2 \sigma_3} = 0$.  

Divergent terms appear only from \eqref{Idiv} and have the form presented. Expanding these in $\ep$, we see that we can obtain a finite scheme-dependent term from the order $\ep$ part  of $\mathcal{I}_{(\sigma_j,n)}$, which thus depends only on $v_j$, and their expansion to order $\ep$ yields terms that linear in $u$ and quadratic in $v_j$. Thus \eqref{Istruct} contains all divergent and all scheme-dependent terms through order $\ep$.

Using the above fact we can derive Fact \ref{fact:sch}. Consider the finite terms first. Notice that $B$ is scheme-independent and the remaining term in the second line vanishes at $v_j = 0$. Thus $B$ is the finite part of $\ireg(1, 0)$ evaluated in the half-integer scheme. To recover $B^{(j)}_{\sigma_1 \sigma_2 \sigma_3}$ we use \eqref{Isch} with $\bar{u} = 1$ and $\bar{v}_j = 0$, which tells us that
\begin{align}
\left[ \ireg(u, v_j) - \ireg(1, 0) \right]_{\ep^0} = \left[ \idiv(u, v_j) - \idiv(1, 0) \right]_{\ep^0}.
\end{align}
Since only $\idiv(u, v_j)$ is $v_j$-dependent on the right hand side, we can find $B^{(j)}_{\sigma_1 \sigma_2 \sigma_3}$ by retrieving the coefficient of $v_j/L_{\sigma_1 \sigma_2 \sigma_3}$ in $\idiv(u, v_j)$. Indeed, using \eqref{I_sch} we see that $\idiv$ contains precise a terms of this form. In conclusion,
\begin{align} \label{ireg_fin}
\ireg(u, v_j) |_{\ep^0} = \ireg(1, 0)|_{\ep^0} + \sum_{\sigma_1, \sigma_2, \sigma_3 = \pm 1} \sum_{j=1}^3 \idiv(u, v_j; 1) |_{\ep^0}.
\end{align}

An analogous argument holds for terms of order $\ep^1$. Since for $v_j =0$ only the first term $C$ survives, its value is equal to the term of order $\epsilon$ in $\ireg(1, 0)$. The term linear in $v_j$ is similarly determined by the derivative $\partial \ireg/\partial \beta_j$, and  the last term, with $L_{\sigma_1 \sigma_2 \sigma_3}$ in the denominator, is accounted for by $\idiv$.
We simply power expand \eqref{Idiv} to order $\epsilon$ and keep terms quadratic in the $v_j$. This determines $C^{(ij)}_{\sigma_1 \sigma_2 \sigma_3}$. All in all, we have found a simple prescription to calculate any 3-point amplitude $\ireg(u, v_j)$ to order $\ep^1$ in any regularization scheme.

Let us now discuss Fact \ref{fact4}. To prove Fact \ref{fact4}, it suffices to show that $\ireg_{[\Delta_1 \Delta_2, \Delta_3 \Delta_4 x \Delta_x]} -\idiv_{[\Delta_1 \Delta_2, \Delta_3 \Delta_4 x \Delta_x]}$ is finite and scheme-independent. The difference is equal to
\begin{align} \label{proof_fact4}
& s^{2 \breg_x} \left(\ireg_{[\Delta_1 \Delta_2, \Delta_3 \Delta_4 x \Delta_x]} -i^{\rm div}_{[\Delta_1 \Delta_2, \Delta_3 \Delta_4 x \Delta_x]}\right) = \\
& \qquad - \sum_{\sigma_1, \sigma_2 , \sigma_3= \pm 1} 
\sum_{\substack{n=0 \\L_0 \neq 0}}^\infty 
\frac{\mathcal{I}_{(\sigma_j,n)}(k_3, k_4, s) }{L(\sigma_j,n)} \,\left( \hat{j}_{-\dreg - 1 + L, [\Delta_1 \Delta_2; \Delta_x]}
 -j^{\rm div}_{-\dreg - 1 + L, [\Delta_1 \Delta_2; \Delta_x]} \right)\nn\\
&\qquad + \sum_{\sigma_1, \sigma_2 = \pm 1} 
 \sum_{\substack{n=0 \\L'_0 \neq 0}}^\infty 
\frac{\mathcal{J}_{(\sigma_1, \sigma_2, n)}(k_3, k_4, s)}{L'(\sigma_1, \sigma_2, n)} \, \left(\ireg_{-\dreg - 1 + L', [\Delta_1 \Delta_2; \Delta_x]} - i^{\rm div}_{-\dreg - 1 + L', [\Delta_1 \Delta_2; \Delta_x]} \right) 
\end{align}
By construction, $(\hat{j}-j^{\rm div})$ and $(\ireg-\idiv)$ are finite and scheme-independent, and we also showed earlier that 
$\mathcal{I}_{(\sigma_j,n)}, L(\sigma_j,n), \mathcal{J}_{(\sigma_1, \sigma_2, n)}, L'(\sigma_1, \sigma_2, n)$ are finite and scheme-independent as $\ep \to 0$. It follows that the right hand side of \eqref{proof_fact4} is indeed finite and scheme-independent.

\subsection{List of results} \label{sec:scheme_change_results}

In this section we list:
\begin{itemize}
\item All 3-point amplitudes $\ireg_{[\Delta_1 \Delta_2 \Delta_3]}(u, v_j)$ and $\jreg_{[\Delta_1 \Delta_2; \Delta_3]}(u, v_j)$ for $\Delta_j = 2$ or $3$ in an arbitrary regularization scheme, up to and including terms of order $\ep^1$.
\item All 4-point amplitudes $\ireg_{[\Delta_1 \Delta_2 \Delta_3 \Delta_4]}(u, v_j)$ and $\ireg_{[\Delta_1 \Delta_2, \Delta_3 \Delta_4 x \Delta_x]}(u, v_j)$ regulated in an arbitrary regularization scheme.
\end{itemize}
To do so, we introduce the following definitions.

\subsubsection{Definitions}

\begin{itemize}
\item The amplitudes $\ireg_{[\Delta_1 \Delta_2 \Delta_3]}$ and $\jreg_{[\Delta_1 \Delta_2; \Delta_3]}$ are defined in \eqref{amp3} and \eqref{amp3i}. We consider the general regularization scheme \eqref{genreg}.
\item 3-point amplitudes evaluated up to and including order $\ep^1$ are decomposed as follows:
\begin{align}
\ireg_{[\Delta_1 \Delta_2 \Delta_3]} & = \idiv_{[\Delta_1 \Delta_2 \Delta_3]} + \ifin_{[\Delta_1 \Delta_2 \Delta_3]} + \ep \left[ u \, \ireg^{u \ep}_{[\Delta_1 \Delta_2 \Delta_3]} + \sum_{j=1}^3 v_{j} \ireg^{v_j \ep}_{[\Delta_1 \Delta_2 \Delta_3]} \right] + O(\ep^2), \\
\jreg_{[\Delta_1 \Delta_2; \Delta_3]} & = \jdiv_{[\Delta_1 \Delta_2; \Delta_3]} + \jfin_{[\Delta_1 \Delta_2; \Delta_3]} + \ep \left[ u \, \jreg^{u \ep}_{[\Delta_1 \Delta_2; \Delta_3]} +\sum_{j=1}^3 v_{j} \jreg^{v_j \ep}_{[\Delta_1 \Delta_2; \Delta_3]} \right] + O(\ep^2),
\end{align}
where
\begin{align}
\idiv_{[\Delta_1 \Delta_2 \Delta_3]} & = \sum_{\sigma_1, \sigma_2, \sigma_3 = \pm 1} \ino^{\sigma_1 \sigma_2 \sigma_3}_{[\Delta_1 \Delta_2 \Delta_3]}, \\[1ex]
\jdiv_{[\Delta_1 \Delta_2; \Delta_3]} & = \sum_{\sigma_1, \sigma_2 = \pm 1} \jno^{\sigma_1 \sigma_2 +}_{[\Delta_1 \Delta_2; \Delta_3]}.
\end{align}
Only the non-vanishing terms $\ino^{\sigma_1 \sigma_2 \sigma_3}_{[\Delta_1 \Delta_2 \Delta_3]}$ and $\jno^{\sigma_1 \sigma_2 +}_{[\Delta_1 \Delta_2 \Delta_3]}$ are listed.
\item The divergences of the amplitude are contained entirely in the divergent parts $\idiv$ and $\jdiv$. All remaining terms are parameter-independent.
\item 3-point amplitudes are treated as functions of the momentum magnitudes $k_1, k_2, k_3$. We also use the notation
\begin{align}
k_t & = k_1 + k_2 + k_3, & v_t & = v_1 + v_2 + v_3, &
\m{12} & = k_1 + k_2 - k_3,
\end{align}
and analogously for $\m{13}$ and $\m{23}$. This follows from  \eqref{mij} as $| \bs{k}_1 + \bs{k}_2 | = k_3$.
\item For 4-point functions, we use the notations
\begin{align}
k_T & = k_1 + k_2 + k_3 + k_4, & v_T & = v_1 + v_2 + v_3 + v_4.
\end{align}
\item An exchange 4-point function $\ireg_{[\Delta_1 \Delta_2, \Delta_3 \Delta_4 x \Delta_x]}(u, v_1, v_2, v_3, v_4, v_x)$ can be evaluated in the regularization scheme \eqref{genreg} by means of the formula
\begin{align} \label{change_scheme_4ptX}
\ireg_{[\Delta_1 \Delta_2, \Delta_3 \Delta_4 x \Delta_x]}(u, v_j) &= \ireg_{[\Delta_1 \Delta_2, \Delta_3 \Delta_4 x \Delta_x]}(1, 0) \nn\\&\quad + \left[ \idiv_{[\Delta_1 \Delta_2, \Delta_3 \Delta_4 x \Delta_x]}(u, v_j) - \idiv_{[\Delta_1 \Delta_2, \Delta_3 \Delta_4 x \Delta_x]}(1, 0) \right].
\end{align}
The amplitudes $\ireg_{[\Delta_1 \Delta_2, \Delta_3 \Delta_4 x \Delta_x]}(1, 0)$ are evaluated in the half-integer scheme and they are listed in Section \ref{sec:reg_amp}. The divergent terms are then listed in Section \ref{sec:div4ptX}.
\item The divergent terms for the 4-point amplitudes contain 3-point amplitudes evaluated in some specific schemes. To avoid clutter, we use the following notation,
\begin{align}
\ireg^{u \{v_1 v_2 v_3\}}_{[\Delta_1 \Delta_2 \Delta_3]}(k_1, k_2, k_3) & = \ireg_{ [\Delta_1 \Delta_2 \Delta_3]}(k_1, k_2, k_3; \dreg = 3 + 2 u \ep, \Dreg_j = \Delta_j + (u + v_j) \ep), \\
\jreg^{u \{v_1 v_2 v_3\}}_{[\Delta_1 \Delta_2; \Delta_3]}(k_1, k_2, k_3) & = \ireg_{ [\Delta_1 \Delta_2; \Delta_3]}(k_1, k_2, k_3; \dreg = 3 + 2 u \ep, \Dreg_j = \Delta_j + (u + v_j) \ep),
\end{align}
where the subscripts indicate the values of the $u$ and $v_j$-parameters.
\item Certain expressions are  long and unwieldy -- despite our best efforts to avoid  typos, when in doubt, we recommend using the results stored in the Mathematica notebooks.
\end{itemize}

\subsubsection{3-point amplitudes $\ireg_{[\Delta_1 \Delta_2 \Delta_3]}$}

The expansions of the amplitudes $\ireg_{[\Delta_1 \Delta_2 \Delta_3]}$ to order $\ep^1$ are as follows:
\begin{itemize}
\item For $\ireg_{[222]}$:
\begin{align}
\ino^{---}_{[222]} & = \frac{1}{\ep \, (u - v_t)}, \\[0.5ex]
\ifin_{[222]} & = -\log k_t - \gamma_E, \\
\ireg^{u \ep}_{[222]} & = \frac{1}{2} ( \log k_t + \gamma_E )^2 + \frac{\pi^2}{12}, \\[0.5ex]
\ireg^{v_1 \ep}_{[222]} & = - \Li_2 \left( \frac{\m{23}}{k_t} \right) - \frac{1}{2} ( \log k_t + \gamma_E )^2 + \frac{\pi^2}{12}, \\[0.5ex]
\ireg^{v_2 \ep}_{[222]} & = \ireg^{v_1 \ep}_{[222]}(k_1 \leftrightarrow k_2), \\[1ex]
\ireg^{v_3 \ep}_{[222]} & = \ireg^{v_1 \ep}_{[222]}(k_1 \leftrightarrow k_3).
\end{align}
\item For $\ireg_{[322]}$:
\begin{align}
\ino^{-+-}_{[322]} & = \frac{k_2}{u - v_1 + v_2 - v_3} \left[ - \frac{1}{\ep} - 2 v_2 ( \log k_2 + \gamma_E + \log 2 - 1 ) \right.\nn\\
& \qquad\qquad\qquad \left. - 2 v_2^2 \ep \left( ( \log k_2 + \gamma_E + \log 2 - 1 )^2 + 1 \right) \right], \\[0.5ex]
\ino^{--+}_{[322]} & = \ino^{-+-}_{[322]}( k_2 \leftrightarrow k_3, v_2 \leftrightarrow v_3 ), \\
\ifin_{[322]} & = (k_2 + k_3) ( \log k_t + \gamma_E - 1) - k_1, \\
\ireg^{u \ep}_{[322]} & = - \frac{k_2 + k_3}{2} \left[  ( \log k_t + \gamma_E - 1)^2 + 1 + \frac{\pi^2}{6} \right] + k_1 \left( \log k_t + \gamma_E - 1 \right), \\[0.5ex]
\ireg^{v_1 \ep}_{[322]} & = \frac{k_2 + k_3}{2} \left[ 2 \Li_2 \left( \frac{\m{23}}{k_t} \right) + ( \log k_t + \gamma_E - 1 )^2 + 1 - \frac{\pi^2}{6} \right] \nn\\
& \qquad + k_1 \left( \log k_t - 2 \log k_1 - \gamma_E - 2 \log 2 + 3 \right), \\[0.5ex]
\ireg^{v_2 \ep}_{[322]} & = \frac{k_3 - k_2}{2} \left[ 2 \Li_2 \left( \frac{\m{13}}{k_t} \right) + ( \log k_t + \gamma_E - 1)^2 + 1 - \frac{\pi^2}{6} \right] \nn\\
& \qquad + 2 k_2 \left[ ( \log k_t + \gamma_E - 1 ) ( \log k_2 + \gamma_E + \log 2 - 1) + 1 \right] 
\nn\\& \qquad 
- k_1 ( \log k_t + \gamma_E - 1), \\[0.5ex]
\ireg^{v_3 \ep}_{[322]} & = \ireg^{v_2 \ep}_{[322]}(k_2 \leftrightarrow k_3).
\end{align}
\item For $\ireg_{[332]}$:
\begin{align}
\ino^{---}_{[332]} & = \frac{1}{u - v_t} \left[ \frac{k_3^2 - k_1^2 - k_2^2}{2 \ep} + (v_1 k_1^2 + v_2 k_2^2 + v_3 k_3^2) \right.\nn\\
& \qquad\qquad\qquad \left. - 2 \ep ( v_1^2 k_1^2 + v_2^2 k_2^2 - v_3^2 k_3^2 ) \right],  \label{i332-f}\\[0.5ex]
\ifin_{[332]} & = \frac{1}{2} (k_1^2 + k_2^2 - k_3^2) \left( \log k_t + \gamma_E \right)  - \frac{1}{4} k_t ( k_1 + k_2 - 3 k_3), \\
\ireg^{u \ep}_{[332]} & = - \frac{k_1^2 + k_2^2 - k_3^2}{4} \left[ ( \log k_t + \gamma_E - \tfrac{1}{2})^2 + \frac{\pi^2}{6} \right] \nn\\
& \qquad + \frac{1}{2} ( \log k_t + \gamma_E - \tfrac{1}{2} ) ( k_1 k_2 - k_1 k_3 - k_2 k_3 - k_3^2 ) \nn\\
& \qquad + \frac{k_3 (k_1 + k_2)}{2} - \frac{k_1^2 + k_2^2 - 9 k_3^2}{16}, \\[0.5ex]
\ireg^{v_1 \ep}_{[332]} & = \frac{k_1^2 + k_2^2 - k_3^2}{4} \left[ 2 \Li_2 \left( \frac{\m{23}}{k_t} \right) + ( \log k_t + \gamma_E - \tfrac{1}{2} )^2 + \frac{1}{4} - \frac{\pi^2}{6} \right] \nn\\
& \qquad + k_1 (k_3 - k_2) ( \log k_1 + \gamma_E + \log 2 ) \nn\\
& \qquad - \frac{1}{2} (k_1 + k_3)(2 k_1 - k_2 - k_3) ( \log k_t + \gamma_E - \tfrac{1}{2} ) \nn\\
& \qquad + \frac{1}{2} ( 3 k_1 k_2 - 4 k_1 k_3 - k_2 k_3 - k_3^2 ), \\
\ireg^{v_2 \ep}_{[332]} & = \ireg^{v_1 \ep}_{[332]}( k_1 \leftrightarrow k_2 ), 
\end{align}
\begin{align}
\ireg^{v_3 \ep}_{[332]} & = \frac{k_1^2 + k_2^2 - k_3^2}{4} \left[ 2 \Li_2 \left( \frac{\m{12}}{k_t} \right) + ( \log k_t + \gamma_E - \tfrac{1}{2} )^2 + \frac{1}{4} - \frac{\pi^2}{6}  \right] \nn\\
& \qquad - \frac{1}{2} (k_1 + k_3)(k_2 + k_3) ( \log k_t + \gamma_E - \tfrac{1}{2} ) \nn\\
& \qquad + (k_1 + k_2) k_3 ( \log k_3 + \gamma_E + \log 2 ) + \frac{3 k_3^2}{2}. \label{i332-l}
\end{align}
\item For $\ireg_{[333]}$:
\begin{align}
\ino^{+--}_{[333]} & = \frac{k_1^3}{u + v_1 - v_2 - v_3} \left[ \frac{1}{3 \ep} + \frac{2 v_1}{3} ( \log k_1 + \gamma_E + \log 2 - \tfrac{7}{3} ) \right. \nn\\
& \qquad\qquad\qquad \left. \frac{2 v_1^2 \ep}{3} \left( ( \log k_1 + \gamma_E + \log 2 - \tfrac{7}{3} )^2 + \frac{19}{9} \right) \right], \\[0.5ex]
\ino^{-+-}_{[333]} & = \ino^{+--}_{[333]}( k_1 \leftrightarrow k_2, v_1 \leftrightarrow v_2 ), \\[0.5ex]
\ino^{--+}_{[333]} & = \ino^{+--}_{[333]}( k_1 \leftrightarrow k_3, v_1 \leftrightarrow v_3 ), \\[0.5ex]
\ifin_{[333]} & = -\frac{k_1^3+k_2^3+k_3^3}{3} ( \log k_t + \gamma_E - \tfrac{4}{3} ) + \frac{1}{3} ( \s{1}{123} \s{2}{123} - 4 \s{3}{123} ), 
\end{align}
\begin{align}
\ireg^{u \ep}_{[333]} & = \frac{k_1^3 + k_2^3 + k_3^3}{6} \left[ ( \log k_t + \gamma_E - \tfrac{4}{3} )^2 + \frac{\pi^2}{6} \right] \nn\\
& \qquad - \frac{1}{3} ( \s{1}{123} \s{2}{123} - 4 \s{3}{123} ) ( \log k_t + \gamma_E - \tfrac{4}{3} ) \nn\\
& \qquad + \frac{5(k_1^3 + k_2^3 + k_3^3) + 9 k_1 k_2 k_3}{27}, \\[0.5ex]
\ireg^{v_1 \ep}_{[333]} & = \frac{k_1^3 - k_2^3 - k_3^3}{6} \left[ 2 \Li_2 \left( \frac{\m{23}}{k_t} \right) + (\log k_t + \gamma_E - \tfrac{4}{3})^2 + \frac{10}{9} - \frac{\pi^2}{6} \right] \nn\\
& \qquad - \frac{2}{3} k_1^3 ( \log k_1 + \gamma_E + \log 2 - \tfrac{7}{3} ) ( \log k_t + \gamma_E - \tfrac{4}{3} ) \nn\\
& \qquad + \frac{2}{3} k_1 (k_2^2 + k_3^2 - k_2 k_3) ( \log k_1 + \gamma_E + \log 2 - \tfrac{7}{3} ) \nn\\
& \qquad + \frac{1}{3} \left[ \s{1}{123} \s{2}{123} - 2 \s{3}{123} - 2 k_1 (k_2^2 + k_3^2) \right] ( \log k_t + \gamma_E - \tfrac{4}{3} ) \nn\\
& \qquad - \frac{k_1}{27} \left[ 20 k_1^2 + 9 k_2 k_3 + 18 k_1 (k_2 + k_3) \right], \\[0.5ex]
\ireg^{v_2 \ep}_{[333]} & = \ireg^{v_1 \ep}_{[333]}(k_1 \leftrightarrow k_2), \\[0.5ex]
\ireg^{v_3 \ep}_{[333]} & = \ireg^{v_1 \ep}_{[333]}(k_1 \leftrightarrow k_3).
\end{align}
\end{itemize}

\subsubsection{3-point amplitudes $\jreg_{[\Delta_1 \Delta_2; \Delta_3]}$}

The expansions of the amplitudes $\jreg_{[\Delta_1 \Delta_2; \Delta_3]}$ to order $\ep^1$ are as follows:
\begin{itemize}
\item For $\jreg_{[22;2]}$:
\begin{align}
\jno^{\pm \pm \pm}_{[22;2]} & = 0, \\
\jfin_{[22;2]} & = - \frac{1}{2} \log \left( \frac{\m{12}}{k_t} \right),
 \\
\jreg^{u \ep}_{[22;2]} & = \frac{1}{4} ( \log \m{12} + \gamma_E )^2 - \frac{1}{4} ( \log k_t + \gamma_E )^2, \\[0.5ex]
\jreg^{v_1 \ep}_{[22;2]} & = \frac{1}{4} \Big[ 2 \Li_2 \left( \frac{\m{23}}{k_t} \right) - 2 \Li_2 \left( - \frac{\m{13}}{\m{12}} \right)
\nn\\[0.5ex]
&\qquad\qquad  + ( \log k_t + \gamma_E )^2 - ( \log \m{12} + \gamma_E )^2 \Big], \\[0.5ex]
\jreg^{v_2 \ep}_{[22;2]} & = \jreg^{v_1 \ep}_{[22;2]}( k_1 \leftrightarrow k_2 ), \\[0.5ex]
\jreg^{v_3 \ep}_{[22;2]} & = \frac{1}{2} \left[ 2 \Li_2 \left( \frac{\m{12}}{k_t} \right) - \log \left( \frac{\m{12}}{k_t} \right) ( \log k_t - \gamma_E - 2 \log 2) \right] - \frac{\pi^2}{6}.
\end{align}
\item For $\jreg_{[32;2]}$:
\begin{align}
\jno^{--+}_{[32;2]} & = \frac{k_3}{u - v_1 - v_2 + v_3} \left[ \frac{1}{\ep} + 2 v_3 ( \log k_3 - 1 ) \right.\nn\\
& \qquad\qquad\qquad \left. + 2 v_{3}^2 \ep \, ( \log^2 k_3 - 2 \log k_3 + 2 ) \right], \\[0.5ex]
\jfin_{[32;2]} & = - \frac{k_2 + k_3}{2} (\log k_t + \gamma_E) + \frac{k_2 - k_3}{2} ( \log \m{12} + \gamma_E) + k_3, \\[0.5ex]
\jreg^{u \ep}_{[32;2]} & = \frac{k_2 + k_3}{4} ( \log k_t + \gamma_E - 1 )^2 + \frac{k_3 - k_2}{4} ( \log \m{12} + \gamma_E - 1 )^2 \nn\\
& \qquad + \frac{k_1}{2} \log \left( \frac{\m{12}}{k_t} \right) + \frac{k_3}{2} \left( 1 + \frac{\pi^2}{6} \right), \\[0.5ex]
\jreg^{v_1 \ep}_{[32;2]} & = - \frac{k_2 + k_3}{4} \left[ 2 \Li_2 \left( \frac{\m{23}}{k_t} \right) + ( \log k_t + \gamma_E - 1 )^2 \right] \nn\\
& \qquad + \frac{k_2 - k_3}{4} \left[ 2 \Li_2 \left( - \frac{\m{13}}{\m{12}} \right) + ( \log \m{12} + \gamma_E - 1 )^2  \right] \nn\\
& \qquad + \frac{k_1}{2} \log \left( \frac{\m{12}}{k_t} \right) + \frac{k_3}{2} \left( -1 + \frac{\pi^2}{6} \right), \\[0.5ex]
\jreg^{v_2 \ep}_{[32;2]} & = \frac{k_2 - k_3}{4} \left[ 2 \Li_2 \left( \frac{\m{13}}{k_t} \right) + ( \log k_t + \gamma_E - 1)^2 \right] \nn\\
& \qquad - \frac{k_2 + k_3}{4} \left[ 2 \Li_2 \left( - \frac{\m{23}}{\m{12}} \right) + ( \log \m{12} + \gamma_E - 1)^2 \right] \nn\\
& \qquad + \log \left( \frac{\m{12}}{k_t} \right) \left[ - \frac{k_1}{2} + k_2 ( \log k_2 + \gamma_E + \log 2 - 1 ) \right] \nn\\
& \qquad - \frac{k_3}{2} \left( 1 - \frac{\pi^2}{6} \right), \\[0.5ex]
\jreg^{v_3 \ep}_{[32;2]} & = - \frac{k_2}{2} \left[ 2 \Li_2 \left( \frac{\m{12}}{k_t} \right) + ( \log k_t + \gamma_E - 1)^2 \right] \nn\\
& \qquad + \frac{k_2 + k_3}{2} ( \log k_t + \gamma_E - 1 )( \log \m{12} + \gamma_E - 1 ) \nn\\
& \qquad - k_3 ( \log k_3 - 1) ( \log k_t + \log \m{12} + 2 \gamma_E - 2 ) \nn\\
& \qquad - \left[ \frac{k_1}{2} + k_2 ( \gamma_E + \log 2) \right] \log \left( \frac{\m{12}}{k_t} \right) 
- \frac{3}{2} \left(k_3 - \frac{\pi^2}{18}(2 k_2 + k_3) \right).
\end{align}
\item For $\jreg_{[22;3]}$:
\begin{align}
\jno^{\pm \pm \pm}_{[22;3]} & = 0, \\
\jfin_{[22;3]} & = - \frac{k_1 + k_2}{2} \log \left( \frac{\m{12}}{k_t} \right) - k_3, \\
\jreg^{u \ep}_{[22;3]} & = - \frac{k_1 + k_2}{4} ( \log k_t + \gamma_E - 1 )^2 + \frac{k_1 + k_2}{4} ( \log \m{12} + \gamma_E - 1 )^2 \nn\\
& \qquad + \frac{k_3}{2} ( \log \m{12} + \log k_t + 2 \gamma_E - 2 ), 
\end{align}
\pagebreak
\begin{align}
\jreg^{v_1 \ep}_{[22;3]} & = \frac{k_2 - k_1}{4} \left[ 2 \Li_2 \left( \frac{\m{23}}{k_t} \right) + ( \log k_t + \gamma_E -1 )^2 \right] \nn\\
& \qquad + \frac{k_1 - k_2}{4}  \left[ 2 \Li_2 \left( - \frac{\m{13}}{\m{12}} \right) + ( \log \m{12} + \gamma_E -1 )^2 \right] \nn\\
& \qquad - \log \left( \frac{\m{12}}{k_t} \right) k_1 ( \log k_1 + \gamma_E + \log 2 - 1) \nn\\
& \qquad - \frac{k_3}{2} \left( \log k_t + \log \m{12} + 2 \gamma_E - 2 \right), \\
\jreg^{v_2 \ep}_{[22;3]} & = \jreg^{v_1 \ep}_{[22;3]}( k_1 \leftrightarrow k_2 ), \\
\jreg^{v_3 \ep}_{[22;3]} & = \frac{k_1 + k_2}{2} \left[ 2 \Li_2 \left( \frac{\m{12}}{k_t} \right) + ( \log k_t - \gamma_E - 2 \log 2 + 3)^2 \right] \nn\\
& \qquad - \frac{k_1 + k_2}{2} ( \log k_t - \gamma_E - 2 \log 2 + 3) ( \log \m{12} - \gamma_E - 2 \log 2 + 3) \nn\\
& \qquad + \frac{k_3}{2} ( \log k_t + \log \m{12} - 4 \log k_3 + 2 \gamma_E - 2 ) - \frac{\pi^2}{6} (k_1 + k_2).
\end{align}
\item For $\jreg_{[33;2]}$:
\begin{align}
\jno^{\pm \pm \pm}_{[33;2]} & = 0, \\
\jfin_{[33;2]} & = \frac{k_1^2 + k_2^2 - k_3^2}{4} \log \left( \frac{\m{12}}{k_t} \right) - \frac{k_3 (k_1 + k_2)}{2}, \\
\jreg^{u \ep}_{[33;2]} & = \frac{k_1^2+k_2^2-k_3^2}{8} ( \log k_t + \gamma_E - \tfrac{1}{2} )^2 - \frac{k_1^2+k_2^2-k_3^2}{8} ( \log \m{12} + \gamma_E - \tfrac{1}{2} )^2 \nn\\
& \qquad + \frac{1}{4} ( k_1 k_2 + k_1 k_3 + k_2 k_3 - k_3^2 ) \log \m{12} + \frac{1}{4} ( - k_1 k_2 + k_1 k_3 + k_2 k_3 + k_3^2 ) \log k_t \nn\\
& \qquad + \frac{-3 + 2 \gamma_E}{4} k_3 (k_1 + k_2), 
\end{align}
\begin{align}
\jreg^{v_1 \ep}_{[33;2]} & = -\frac{k_1^2 + k_2^2 - k_3^2}{8} \left[ 2 \Li_2 \left( \frac{\m{23}}{k_t} \right) + ( \log k_t + \gamma_E - \tfrac{3}{2} )^2 \right] \nn\\
& \qquad  + \frac{k_1^2 + k_2^2 - k_3^2}{8} \left[ 2 \Li_2 \left( - \frac{\m{13}}{\m{12}} \right) + ( \log \m{12} + \gamma_E - \tfrac{3}{2} )^2 \right] \nn\\
& \qquad + \frac{-k_1^2 + k_2^2 + k_1 k_2}{4} \log \left( \frac{\m{12}}{k_t} \right) \nn\\
& \qquad + \frac{k_3(k_1 - k_2)}{4} ( \log k_t + \log \m{12} + 2 \gamma_E - 3 ) \nn\\
& \qquad - k_1 k_3 ( \log k_1 + \gamma_E + \log 2 - \tfrac{5}{2} ), \\
\jreg^{v_2 \ep}_{[33;2]} & = \jreg^{v_1 \ep}_{[33;2]}( k_1 \leftrightarrow k_2 ), \\
\jreg^{v_3 \ep}_{[33;2]} & = \frac{-k_1^2 - k_2^2 + k_3^2}{4} \left[ 2 \Li_2 \left( \frac{\m{12}}{k_t} \right) - \log \left( \frac{\m{12}}{k_t} \right) ( \log k_t - \gamma_E - 2 \log 2 - \tfrac{1}{2} ) \right] \nn\\
& \qquad + \frac{-k_1 k_2 + k_1 k_3 + k_2 k_3 - k_3^2}{4} \log \m{12} + \frac{k_1 k_2 + k_1 k_3 + k_2 k_3 + k_3^2}{4} \log k_t \nn\\
& \qquad - \frac{k_3(k_1 + k_2)}{4} ( 4 \log k_3 - 2 \gamma_E + 1) + \frac{\pi^2}{12} ( k_1^2 + k_2^2 - k_3^2 ).
\end{align}
\item For $\jreg_{[32;3]}$:
\begin{align}
\jno^{\pm\pm\pm}_{[32;3]} & = 0, \\
\jfin_{[32;3]} & = - \frac{k_1^2 - k_2^2 + k_3^2}{4} \log \left( \frac{\m{12}}{k_t} \right) + \frac{k_3 (- k_1 + k_2)}{2}, \\
\jreg^{u \ep}_{[32;3]} & = - \frac{k_1^2 - k_2^2 + k_3^2}{8} ( \log k_t + \gamma_E - \tfrac{1}{2} )^2 + \frac{k_1^2 - k_2^2 + k_3^2}{8} ( \log \m{12} + \gamma_E - \tfrac{1}{2} )^2 \nn\\
& \qquad + \frac{1}{4} ( k_1 k_2 + k_1 k_3 - k_2 k_3 + k_2^2 ) \log \m{12} + \frac{1}{4} ( - k_1 k_2 + k_1 k_3 - k_2 k_3 - k_2^2 ) \log k_t \nn\\
& \qquad + \frac{2 \gamma_E - 1}{4} k_1 k_3 - \frac{2 \gamma_E - 3}{4} k_2 k_3, \\
\jreg^{v_1 \ep}_{[32;3]} & = \frac{k_1^2 - k_2^2 + k_3^2}{8} \left[ 2 \Li_2 \left( \frac{\m{23}}{k_t} \right) + ( \log k_t + \gamma_E - \tfrac{1}{2} )^2 \right] \nn\\
& \qquad - \frac{k_1^2 - k_2^2 + k_3^2}{8} \left[ 2 \Li_2 \left( - \frac{\m{13}}{\m{12}} \right) + ( \log \m{12} + \gamma_E - \tfrac{1}{2} )^2 \right] \nn\\
& \qquad - \frac{1}{4} (k_1 + k_2)(2 k_1 - k_2 - k_3) ( \log k_t + \gamma_E - \tfrac{1}{2} ) \nn\\
& \qquad + \frac{1}{4} (k_1 + k_2)(2 k_1 - k_2 + k_3) ( \log \m{12} + \gamma_E - \tfrac{1}{2} ) 
\nn\\& \qquad 
- k_1 k_3 ( \log k_1 + \gamma_E + \log 2 - \tfrac{3}{2} ) - \frac{k_2 k_3}{2}, 
\end{align}
\begin{align}
\jreg^{v_2 \ep}_{[32;3]} & = \frac{k_1^2 - k_2^2 + k_3^2}{8} \left[ 2 \Li_2 \left( \frac{\m{13}}{k_t} \right) + (\log k_t + \gamma_E - \tfrac{1}{2} )^2 \right] \nn\\
& \qquad - \frac{k_1^2 - k_2^2 + k_3^2}{8} \left[ 2 \Li_2 \left( - \frac{\m{23}}{\m{12}} \right) + (\log \m{12} + \gamma_E - \tfrac{1}{2} )^2 \right] \nn\\
& \qquad + \frac{(k_1 + k_2)(k_2 - k_3)}{4} ( \log \m{12} + \gamma_E - \tfrac{1}{2}) \nn\\
& \qquad - \frac{(k_1 + k_2)(k_2 + k_3)}{4} ( \log k_t + \gamma_E - \tfrac{1}{2}) 
+ k_2 k_3 (\log k_2 + \gamma_E + \log 2 ), \\[-1ex]
\jreg^{v_3 \ep}_{[32;3]} & = \frac{k_1^2-k_2^2+k_3^2}{4} \left[ 2 \Li_2 \left( \frac{\m{12}}{k_t} \right) - \log \left( \frac{\m{12}}{k_t} \right) (\log k_t - \gamma_E - 2 \log 2 ) \right] \nn\\
& \qquad - \frac{1}{8} ( 7 k_1^2 + 2 k_1 k_2 - 5 k_2^2 + 3 k_3^2) \log \left( \frac{\m{12}}{k_t} \right) \nn\\
& \qquad + \frac{k_3(k_1 - k_2)}{4} ( \log k_t + \log \m{12} - 4 \log k_3 + 2 \gamma_E - 3 ) 
\nn\\& \qquad 
- \frac{k_2 k_3}{2} - \frac{\pi^2}{12} ( k_1^2 - k_2^2 + k_3^2 ).
\end{align}
\vspace{-5mm}
\item For $\jreg_{[33;3]}$:
\begin{align}
\jno^{--+}_{[33;3]} & = \frac{k_3^3}{u - v_1 - v_2 + v_3} \left[ \frac{1}{3 \ep} + \frac{2}{9} v_3 (3 \log k_3 - 1 ) + \frac{2 v_3^2 \ep}{27} ( 9 \log^2 k_3 - 6 \log k_3 + 2 ) \right], \\[-1ex]
\jfin_{[33;3]} & = \frac{k_1^3+k_2^3+k_3^3}{6} \log \left( \frac{\m{12}}{k_t} \right) - \frac{k_3^3}{3} (\log \m{12} + \gamma_E - \tfrac{4}{3}) 
+ \frac{1}{3} k_3 ( k_1^2 + k_2^2 - k_1 k_2), \\[-2ex]
\jreg^{u \ep}_{[33;3]} & = \frac{k_1^3 + k_2^3 + k_3^3}{12} ( \log k_t + \gamma_E - \tfrac{4}{3} )^2 - \frac{k_1^3 + k_2^3 - k_3^3}{12} ( \log \m{12} + \gamma_E - \tfrac{4}{3} )^2 \nn\\
& \qquad + \frac{1}{6} ( -\s{1}{123} \s{2}{123} + 4 \s{3}{123} ) (\log k_t + \gamma_E - \tfrac{4}{3}) \nn\\
& \qquad + \frac{1}{6} ( \s{1}{123} \s{2}{123} -2 \s{3}{123} - 2 k_3 (k_1^2 + k_2^2) ) (\log \m{12} + \gamma_E - \tfrac{4}{3}) \nn\\
& \qquad + \frac{k_1 k_2 k_3}{3} + \frac{5 k_3^3}{27} \left( 1 + \frac{3 \pi^2}{20} \right), \\[1ex]
\jreg^{v_1 \ep}_{[33;3]} & = \frac{k_1^3 - k_2^3 - k_3^3}{12} \left[ 2 \Li_2 \left( \frac{\m{23}}{k_t} \right) + ( \log k_t + \gamma_E - \tfrac{4}{3} )^2 \right] \nn\\
& \qquad + \frac{- k_1^3 + k_2^3 - k_3^3}{12} \left[ 2 \Li_2 \left( - \frac{\m{13}}{\m{12}} \right) + ( \log \m{12} + \gamma_E - \tfrac{4}{3} )^2 \right] \nn\\
& \qquad - \frac{1}{6} ( \log k_t + \gamma_E - \tfrac{4}{3} ) \left[ 2 k_1^3 ( \log k_1 + \gamma_E + \log 2 - \tfrac{7}{3} ) \right.\nn\\
& \qquad\qquad\qquad \left. - (k_1^2 k_2 + k_1^2 k_3 + k_2^2 k_3 + k_2 k_3^2 - k_1 k_2^2 - k_1 k_3^2 + k_1 k_2 k_3) \right] \nn\\
& \qquad + \frac{1}{6} ( \log \m{12} + \gamma_E - \tfrac{4}{3} ) \left[ 2 k_1^3 ( \log k_1 + \gamma_E + \log 2 - \tfrac{7}{3} ) \right.\nn\\
& \qquad\qquad\qquad \left. + (- k_1^2 k_2 + k_1^2 k_3 + k_2^2 k_3 - k_2 k_3^2 + k_1 k_2^2 + k_1 k_3^2 + k_1 k_2 k_3) \right] \nn\\
& \qquad - \frac{2}{3} k_1 k_2 k_3 ( \log k_1 + \gamma_E + \log 2 - \tfrac{11}{6} ) - \frac{2}{3} k_1^2 k_3 - \frac{5 k_3^3}{27} \left( 1 - \frac{3 \pi^2}{20} \right), 
\end{align}
\begin{align}
\jreg^{v_2 \ep}_{[33;3]} & = \jreg^{v_1 \ep}_{[33;3]}( k_1 \leftrightarrow k_2 ), \\
\jreg^{v_3 \ep}_{[33;3]} & = - \frac{k_1^3 + k_2^3}{6} \left[ 2 \Li_2 \left( \frac{\m{12}}{k_t} \right) - \log \left( \frac{\m{12}}{k_t} \right) (\log k_t - \gamma_E - 2 \log 2 + \tfrac{8}{3} ) \right] \nn\\
& \qquad + \frac{k_3^3}{6} ( \log k_t + \gamma_E - \tfrac{2}{3} ) ( \log \m{12} + \gamma_E - \tfrac{2}{3} ) \nn\\
& \qquad + \frac{1}{6} (\log k_t + \gamma_E - \tfrac{4}{3}) \left[ - 2 k_3^3 \log k_3 + \s{1}{123} \s{2}{123} - 2 \s{3}{123} - 2 k_3(k_1^2+k_2^2) \right] \nn\\
& \qquad + \frac{1}{6} (\log \m{12} + \gamma_E - \tfrac{4}{3}) \left[ - 2 k_3^3 \log k_3 - \s{1}{123} \s{2}{123} + 4 \s{3}{123} \right] \nn\\
& \qquad + \frac{2}{3} k_3 (k_1^2 + k_2^2 - k_1 k_2) ( \log k_3 - \tfrac{1}{3}) \nn\\
& \qquad - \frac{k_1 k_2 k_3}{3} - \frac{17}{27} k_3^3 + \frac{\pi^2}{36} ( 2 k_1^3 + 2 k_2^3 + k_3^3 ).
\end{align}
\end{itemize}

\subsubsection{4-point contact amplitudes}

The 4-point contact amplitudes regulated in the general scheme read
\begin{align}
\ireg_{[2222]} & = \frac{1}{k_T} + O(\ep), \\
\ireg_{[3222]} & = \frac{1}{(2 u - v_T) \, \ep} + \left[ - \log k_T + \frac{k_1}{k_T} - \gamma_E \right] + O(\ep),\\
\label{ireg3322genscheme}
\ireg_{[3322]} & =  - \frac{k_3}{2 u - v_1 - v_2 + v_3 - v_4} \left[ \frac{1}{\ep} + 2  v_3 ( \log k_3 + \gamma_E + \log 2 - 1) \right] \nn\\
& \qquad - \frac{k_4}{2 u - v_1 - v_2 - v_3 + v_4} \left[ \frac{1}{\ep} + 2 v_4 ( \log k_4 + \gamma_E + \log 2 - 1) \right] \nn\\
& \qquad + (k_3 + k_4) (\log k_T + \gamma_E) + \frac{k_1 k_2}{k_T} - k_T + O(\ep),\\[0.5ex]
\ireg_{[3332]} & = - \frac{ k_1^2 + k_2^2 + k_3^2 - k_4^2}{2 \ep \, (2 u - v_T)} + \left[ \frac{k_1^2 + k_2^2 + k_3^2 - k_4^2}{2} ( \log k_T + \gamma_E) \right.\nn\\
& \qquad\qquad \left. + \frac{k_1 k_2 k_3}{k_T} + k_T  \left( k_4 - \frac{k_T}{4} \right) + \frac{1}{2u - v_T} \sum_{j=1}^4 v_j k_j^2 \right] + O(\ep),\\[0.5ex]
\ireg_{[3333]} & = \frac{k_1^3}{2 u + v_1 - v_2 - v_3 - v_4} \left[ \frac{1}{3 \ep} + \frac{2 v_1}{3} ( \log k_1 + \gamma_E + \log 2 - \tfrac{7}{3} ) \right] \nn\\[1ex]
& \qquad + [ (k_1, v_1) \leftrightarrow (k_2, v_2)  ] + [ (k_1, v_1) \leftrightarrow (k_3, v_3)  ] + [ (k_1, v_1) \leftrightarrow (k_4, v_4)  ] \nn\\[1ex]
& \qquad - \frac{k_1^3 + k_2^3 + k_3^3 + k_4^3}{3} ( \log k_T + \gamma_E ) + \frac{k_1 k_2 k_3 k_4}{k_T} + k_T \left( - \s{2}{1234} + \frac{4}{9} k_T^2 \right)  + O(\ep).
\end{align}

\subsubsection{4-point exchange diagrams} \label{sec:div4ptX}

To evaluate the exchange 4-point amplitudes in an arbitrary regularization scheme one uses \eqref{change_scheme_4ptX}. Here, we list the necessary divergent terms:
\begin{align}
&\idiv_{[32,22x2]}  = s^{-1 - 2 \ep v_x} \left[ \frac{\jreg^{2 u - v_3 - v_4 - v_x \{v_1 v_2 v_x\}}_{[32;2]}(k_1, k_2, s)}{\ep \, (-u + v_3 + v_4 + v_x) } + \jreg_{[32;2]}(k_1, k_2, s) \ireg_{[222]}(s, k_3, k_4) \right] \nn\\
& \qquad\qquad \quad + \frac{1}{2 u - v_T} \left[ \frac{1}{\ep} - ( u - v_3 - v_4 + 3 v_x ) \right] + O(\ep),\\[1ex]
&\idiv_{[33,22x2]} \nn\\[-1ex]& = s^{-1 - 2 \ep v_x} \left[ \frac{\jreg^{2 u - v_3 - v_4 - v_x \{v_1 v_2 v_x\}}_{[33;2]}(k_1, k_2, s)}{\ep \, (-u + v_3 + v_4 + v_x) } + \jreg_{[33;2]}(k_1, k_2, s) \ireg_{[222]}(s, k_3, k_4) \right] \nn\\
& \qquad + \frac{k_3}{(2 u - v_1 - v_2 + v_3 - v_4)} \left[ \frac{1}{2 \ep} + v_3 ( \log k_3 + \gamma_E + \log 2 - \tfrac{7}{4} ) + \frac{1}{4} ( -3 u + 3 v_4 + v_x ) \right] \nn\\
& \qquad + \frac{k_4}{(2 u - v_1 - v_2 - v_3 + v_4)} \left[ \frac{1}{2 \ep} + v_4 ( \log k_4 + \gamma_E + \log 2 - \tfrac{7}{4} ) + \frac{1}{4} ( -3 u + 3 v_3 + v_x ) \right] 
\nn\\[0.5ex]&\qquad 
+ O(\ep), 
\end{align}
\pagebreak
\begin{align}
&\idiv_{[32,32x2]} \nn\\[0.5ex]& = s^{-1 - 2 \ep v_x} \jreg_{[32;2]}(k_1, k_2, s) \ireg_{[232]}(s, k_3, k_4) \nn\\
& \qquad - \frac{2^{-1-2 \ep v_4} \Gamma( - \tfrac{1}{2} - \ep v_4 ) }{\ep \, (u - v_3 + v_4 - v_x) \Gamma( \tfrac{1}{2} + \ep v_4 ) } s^{-1 - 2 \ep v_x} k_4^{1 + 2 \ep v_4} \jreg_{[32;2]}^{2u - v_3 + v_4 - v_x \{ v_1 v_2 v_x \}}(k_1, k_2, s) \nn\\
& \qquad - \frac{2^{-1-2 \ep v_x} \Gamma( -\tfrac{1}{2} - \ep v_x )}{\ep \, (u - v_3 - v_4 + v_x) \Gamma( \tfrac{1}{2} + \ep v_x )} \jreg_{[32;2]}^{2u - v_3 - v_4 + v_x \{v_1 v_2 v_x\}}(k_1, k_2, s) \nn\\
& \qquad + \frac{\Gamma( \tfrac{1}{2} + \ep v_x )}{2 \ep \, (u - v_3 - v_4 + v_x) \Gamma (\tfrac{3}{2} + \ep v_x )} \ireg_{[322]}^{2u - v_3 - v_4 + v_x \{v_1 v_2 v_x\}}(k_1, k_2, s) \nn\\
& \qquad - \frac{k_2}{2u - v_1 + v_2 - v_3 - v_4} \left[ \frac{1}{\ep} + 2 v_2 ( \log k_2 + \gamma_E + \log 2 - 1 ) - ( -u + v_3 + v_4 + 3 v_x ) \right] \nn\\
& \qquad - \frac{k_4}{2u - v_1 - v_2 - v_3 + v_4} \left[ \frac{1}{\ep} + 2 v_4 ( \log k_4 + \gamma_E + \log 2 - 1 ) + ( -u + v_3 - v_4 - 3 v_x ) \right], \\
&\idiv_{[32,33x2]}  = s^{-1 - 2 \ep v_x} \left[ \jreg_{[32;2]}(k_1, k_2, s) \ireg_{[233]}(s, k_3, k_4) \right.\nn\\
& \qquad\qquad\qquad + \frac{1}{2 \ep \, (u - v_3 - v_4 - v_x)} \left( \frac{k_3^2}{1 + 2 \ep v_3} + \frac{k_4^2}{1 + 2 \ep v_4} - \frac{s^2}{1 - 2 \ep v_x}  \right) \times\nn\\
& \qquad\qquad\qquad\qquad\qquad \left. \times \jreg^{2u - v_3 - v_4 - v_x \{v_1 v_2 v_x\}}_{[32;2]}(k_1, k_2, s) \right] \nn\\
& \qquad + \frac{1}{4(2u - v_T)} \left[ \frac{k_1^2 - k_2^2 - 2 k_3^2 - 2 k_4^2 - s^2}{\ep} + \tfrac{1}{2} s^2 ( 5 (-u + v_3 + v_4) - 11 v_x ) \right.\nn\\
& \qquad\qquad\quad  + k_1^2 ( \tfrac{3}{2} u - 2 v_1 - \tfrac{3}{2} v_3 - \tfrac{3}{2} v_4 + \tfrac{1}{2} v_x ) + k_2^2 ( - \tfrac{3}{2} u - 2 v_2 + \tfrac{3}{2} v_3 + \tfrac{3}{2} v_4 - \tfrac{1}{2} v_x ) \nn\\
& \qquad\qquad\quad \left. + 2 k_3^2(u + v_3 - v_4 + 3 v_x) + 2 k_4^2(u - v_3 + v_4 + 3 v_x)  \right], \\
&\idiv_{[33,33x2]}\nn\\ & = s^{-1 - 2 \ep v_x} \left[ \jreg_{[33;2]}(k_1, k_2, s) \ireg_{[233]}(s, k_3, k_4) \right.\nn\\
& \qquad\qquad\qquad + \frac{1}{2 \ep \, (u - v_3 - v_4 - v_x)} \left( \frac{k_3^2}{1 + 2 \ep v_3} + \frac{k_4^2}{1 + 2 \ep v_4} - \frac{s^2}{1 - 2 \ep v_x}  \right) \times\nn\\
& \qquad\qquad\qquad\qquad\qquad \left. \times \jreg^{2u - v_3 - v_4 - v_x \{v_1 v_2 v_x\}}_{[33;2]}(k_1, k_2, s) \right] \nn\\
& \qquad - \frac{k_1^3}{3 (2 u + v_1 - v_2 - v_3 - v_4)} \left[ \frac{1}{2 \ep} + v_1( \log k_1 + \gamma_E + \log 2 - \tfrac{7}{3} ) + \tfrac{1}{4} (3 u - 3 v_3 - 3 v_4 + v_x)  \right] \nn\\
& \qquad - \frac{k_2^3}{3 (2 u - v_1 + v_2 - v_3 - v_4)} \left[ \frac{1}{2 \ep} + v_2( \log k_2 + \gamma_E + \log 2 - \tfrac{7}{3} ) + \tfrac{1}{4} (3 u - 3 v_3 - 3 v_4 + v_x)  \right] \nn\\
& \qquad - \frac{k_3^3}{3 (2 u - v_1 - v_2 + v_3 - v_4)} \left[ \frac{1}{2 \ep} + v_3( \log k_3 + \gamma_E + \log 2 - \tfrac{7}{3} ) - \tfrac{1}{4} (3 u + 3 v_3 - 3 v_4 - v_x)  \right] \nn\\
& \qquad - \frac{k_4^3}{3 (2 u - v_1 - v_2 - v_3 + v_4)} \left[ \frac{1}{2 \ep} + v_4( \log k_4 + \gamma_E + \log 2 - \tfrac{7}{3} ) - \tfrac{1}{4} (3 u - 3 v_3 + 3 v_4 - v_x)  \right].
\end{align}
\pagebreak
\begin{align}
&\idiv_{[32,22x3]} = s^{-3 - 2 \ep v_x} \left[ \jreg_{[32;3]}(k_1, k_2, s) \ireg_{[322]}(s, k_3, k_4) \right.\nn\\
& \qquad\qquad\qquad - \frac{2^{-1-2 \ep v_3} \Gamma( -\tfrac{1}{2} - \ep v_3 )}{\ep \, (u + v_3 - v_4 - v_x) \Gamma( \tfrac{1}{2} + \ep v_3 )} k_3^{1 + 2 \ep v_3} \jreg_{[32;3]}^{2u+v_3-v_4-v_x \{v_1 v_2 v_x\}}(k_1, k_2, s) \nn\\
& \qquad\qquad\qquad \left. - \frac{2^{-1-2 \ep v_4} \Gamma( -\tfrac{1}{2} - \ep v_4 )}{\ep \, (u - v_3 + v_4 - v_x) \Gamma( \tfrac{1}{2} + \ep v_4 )} k_4^{1 + 2 \ep v_4} \jreg_{[32;3]}^{2u-v_3+v_4-v_x \{v_1 v_2 v_x\}}(k_1, k_2, s) \right] \nn\\
& \qquad + \frac{1}{2(2u - v_T)} \left[ \frac{1}{\ep} + \frac{1}{2}(u - v_3 - v_4 - 3 v_x) \right] + O(\ep),\\[3ex]
&\idiv_{[33,22x3]}\nn\\ & = s^{-3 - 2 \ep v_x} \left[ \jreg_{[33;3]}(k_1, k_2, s) \ireg_{[322]}(s, k_3, k_4) \right.\nn\\
& \qquad\qquad\qquad - \frac{2^{-1-2 \ep v_3} \Gamma( -\tfrac{1}{2} - \ep v_3 )}{\ep \, (u + v_3 - v_4 - v_x) \Gamma( \tfrac{1}{2} + \ep v_3 )} k_3^{1 + 2 \ep v_3} \jreg_{[33;3]}^{2u+v_3-v_4-v_x \{v_1 v_2 v_x\}}(k_1, k_2, s) \nn\\
& \qquad\qquad\qquad \left. - \frac{2^{-1-2 \ep v_4} \Gamma( -\tfrac{1}{2} - \ep v_4 )}{\ep \, (u - v_3 + v_4 - v_x) \Gamma( \tfrac{1}{2} + \ep v_4 )} k_4^{1 + 2 \ep v_4} \jreg_{[33;3]}^{2u-v_3+v_4-v_x \{v_1 v_2 v_x\}}(k_1, k_2, s) \right] \nn\\
& \qquad - \frac{k_3}{9(2u-v_1-v_2+v_3-v_4)} \left[ \frac{1}{\ep} + 2 v_3 ( \log k_3 + \gamma_E + \log 2 - \tfrac{7}{6} ) - \frac{1}{3} (u - v_4 + 3 v_x) \right] \nn\\
& \qquad - \frac{k_4}{9(2u-v_1-v_2-v_3+v_4)} \left[ \frac{1}{\ep} + 2 v_4 ( \log k_4 + \gamma_E + \log 2 - \tfrac{7}{6} ) - \frac{1}{3} (u - v_3 + 3 v_x) \right]\nn\\[0.5ex]&\qquad + O(\ep), \\[3ex]
&\idiv_{[32,32x3]} \nn\\& = s^{-3 - 2 \ep v_x} \left[ \jreg_{[32;3]}(k_1, k_2, s) \ireg_{[332]}(s, k_3, k_4) \right. \nn\\
& \qquad + \frac{1}{2 \ep \, (u - v_3 - v_4 - v_x) } \left( \frac{k_3^2}{1 + 2 \ep v_3} - \frac{k_4^2}{1 - 2 \ep v_4} + \frac{s^2}{1 + 2 \ep v_x} \right) \times \nn\\
& \qquad\qquad\qquad \left. \times \jreg_{[32;3]}^{2u - v_3 - v_4 - v_x \{v_1 v_2 v_x\}}(k_1, k_2, s) \right] \nn\\
& \qquad - \frac{k_2}{2(2u - v_1 + v_2 - v_3 - v_4)} \left[ \frac{1}{\ep} + 2 v_2 ( \log k_2 + \gamma_E + \log 2 - 1 ) - \frac{1}{2} ( u - v_3 - v_4 + 3 v_x ) \right] \nn\\
& \qquad - \frac{k_4}{2(2u - v_1 - v_2 - v_3 + v_4)} \left[ \frac{1}{\ep} + 2 v_4 ( \log k_4 + \gamma_E + \log 2 - 1 ) + \frac{1}{2} ( u - v_3 + v_4 - 3 v_x ) \right] \nn\\&\qquad+ O(\ep).
\end{align}
\pagebreak
\begin{align}
&\idiv_{[32,33x3]}\nn\\[1ex] & = s^{-3 - 2 \ep v_x} \jreg_{[32;3]}(k_1, k_2, s) \ireg_{[333]}(s, k_3, k_4) \nn\\
& \qquad - \frac{s^{-3 - 2 \ep v_x}}{8 \ep} \left[ \frac{4^{-\ep v_3} \Gamma (-\tfrac{3}{2} - \ep v_3)}{(u + v_3 - v_4 - v_x) \Gamma(\tfrac{3}{2} + \ep v_3)} k_3^{3 + 2 \ep v_3} \jreg_{[32;3]}^{2u + v_3 - v_4 - v_x \{ v_1 v_2 v_x \}}(k_1, k_2, s) \right.\nn\\
& \qquad\qquad\qquad \left. + \frac{4^{-\ep v_4} \Gamma (-\tfrac{3}{2} - \ep v_4)}{(u - v_3 + v_4 - v_x) \Gamma(\tfrac{3}{2} + \ep v_4)} k_4^{3 + 2 \ep v_4} \jreg_{[32;3]}^{2u - v_3 + v_4 - v_x \{ v_1 v_2 v_x \}}(k_1, k_2, s) \right] \nn\\
& \qquad + \frac{1}{8 \ep \, (u - v_3 - v_4 + v_x)} \left[ \frac{8}{3 + 2 \ep v_x} \ireg_{[323]}^{2u - v_3 - v_4 + v_x \{v_1 v_2 v_x\}}(k_1, k_2, s) \right. \nn\\
& \qquad\qquad\qquad \left. - \frac{4^{-\ep v_x} \Gamma(-\tfrac{3}{2} - \ep v_x)}{\Gamma(\tfrac{3}{2} + \ep v_x)} \jreg_{[32;3]}^{2u-v_3-v_4+v_x \{v_1 v_2 v_x\}}(k_1, k_2, s) \right] \nn\\
& \qquad + \frac{1}{36(2u - v_T)} \left[ \frac{-5 s^2-2k_1^2+2k_2^2-9k_3^2-9k_4^2}{\ep} + \frac{s^2}{6}(37(-u+v_3+v_4)+123 v_x) \right.\nn\\
& \qquad\qquad\qquad + \frac{2}{3} k_1^2 (-u + 6 v_1 + v_3 + v_4 + 3 v_x) + \frac{2}{3} k_2^2 (u + 6 v_2 - v_3 - v_4 - 3 v_x) \nn\\
& \qquad\qquad\qquad \left. + \frac{9}{2} k_3^2 (-u+5 v_3 + v_4 + 3 v_x) + \frac{9}{2} k_4^2(-u+v_3+5v_4+3v_x) \right] + O(\ep).
\end{align}
\begin{align}
&\idiv_{[33,33x3]}\nn\\[1ex] & = s^{-3 - 2 \ep v_x} \jreg_{[33;3]}(k_1, k_2, s) \ireg_{[333]}(s, k_3, k_4) \nn\\
& \qquad - \frac{s^{-3 - 2 \ep v_x}}{8 \ep} \left[ \frac{4^{-\ep v_3} \Gamma (-\tfrac{3}{2} - \ep v_3)}{(u + v_3 - v_4 - v_x) \Gamma(\tfrac{3}{2} + \ep v_3)} k_3^{3 + 2 \ep v_3} \jreg_{[33;3]}^{2u + v_3 - v_4 - v_x \{ v_1 v_2 v_x \}}(k_1, k_2, s) \right.\nn\\
& \qquad\qquad\qquad \left. + \frac{4^{-\ep v_4} \Gamma (-\tfrac{3}{2} - \ep v_4)}{(u - v_3 + v_4 - v_x) \Gamma(\tfrac{3}{2} + \ep v_4)} k_4^{3 + 2 \ep v_4} \jreg_{[33;3]}^{2u - v_3 + v_4 - v_x \{ v_1 v_2 v_x \}}(k_1, k_2, s) \right] \nn\\
& \qquad + \frac{1}{8 \ep \, (u - v_3 - v_4 + v_x)} \left[ \frac{8}{3 + 2 \ep v_x} \ireg_{[333]}^{2u - v_3 - v_4 + v_x \{v_1 v_2 v_x\}}(k_1, k_2, s) \right.\nn\\
& \qquad\qquad\qquad \left. - \frac{4^{-\ep v_x} \Gamma(-\tfrac{3}{2} - \ep v_x)}{\Gamma(\tfrac{3}{2} + \ep v_x)} \jreg_{[33;3]}^{2u-v_3-v_4+v_x \{v_1 v_2 v_x\}}(k_1, k_2, s) \right] \nn\\
& \qquad + \frac{k_1^3}{27 (2 u + v_1 - v_2 - v_3 - v_4)} \left[ \frac{1}{\ep} + 2 v_1( \log k_1 + \gamma_E + \log 2 - \tfrac{7}{3} ) + \tfrac{1}{3} (u - v_3 - v_4 - 3 v_x )  \right] \nn\\
& \qquad + \frac{k_2^3}{3 (2 u - v_1 + v_2 - v_3 - v_4)} \left[ \frac{1}{\ep} + 2 v_2( \log k_2 + \gamma_E + \log 2 - \tfrac{7}{3} ) + \tfrac{1}{3} (u - v_3 - v_4 - 3 v_x )  \right] \nn\\
& \qquad + \frac{k_3^3}{3 (2 u - v_1 - v_2 + v_3 - v_4)} \left[ \frac{1}{\ep} + 2 v_3( \log k_3 + \gamma_E + \log 2 - \tfrac{7}{3} ) - \tfrac{1}{3} (u + v_3 - v_4 + 3 v_x )  \right] \nn\\
& \qquad + \frac{k_4^3}{3 (2 u - v_1 - v_2 - v_3 + v_4)} \left[ \frac{1}{\ep} + 2 v_4( \log k_4 + \gamma_E + \log 2 - \tfrac{7}{3} ) - \tfrac{1}{3} (u - v_3 + v_4 + 3 v_x )  \right] \nn\\&\qquad+ O(\ep).
\end{align}

\section{Weight-shifting operators}\label{sec:weightshift}

In this section, we discuss the application of weight-shifting operators to relate correlators of different scaling dimensions.
In principle, such operators offer an elegant construction of the various correlators of interest starting from a smaller set of known `seed'  integrals.  
In the cosmological context, the operators of \cite{Karateev:2017jgd} were applied in \cite{Arkani-Hamed:2018kmz,Baumann:2019oyu} to construct 4-point correlators of conformally coupled and massless scalars in four-dimensional de Sitter.  Holographically,  these are dual to the three-dimensional CFT correlators of dimensions $\Delta=2$, $3$ studied here.
However, there are several important subtleties associated to the use of weight-shifting operators that require careful consideration:

\begin{itemize}
\item While weight-shifting operators can  be applied to 
connect amplitudes in the dimensionally regulated theory,  their action 
on  exchange diagrams generates 
a specific {\it linear combination} of shifted exchange and shifted contact diagrams as given in \eqref{Wpplincombsigma}.
To arrive  purely at a shifted exchange diagram, this shifted contact contribution must therefore be evaluated and subtracted off. 
Sometimes, as we discuss, this shifted contact contribution can itself be constructed by acting with differential operators and symmetrizations.    
Alternatively, when the shifted exchange diagram has a derivative vertex,  the contact contribution is absent for certain special values of the operator dimensions.  This occurs for the cases studied in \cite{Arkani-Hamed:2018kmz,Baumann:2019oyu} but is not a general phenomenon.

\item For other special values of the  operator and spacetime dimensions, the coefficients parametrising the linear combination of shifted exchange and shifted contact diagrams can themselves vanish.  In particular, this must happen whenever a shift operator connects a finite amplitude to a divergent amplitude.  Clearly the result of acting with a  differential operator on any finite amplitude must also be finite, and so any divergent shifted amplitude thus produced must be accompanied by a vanishing coefficient.
When this happens, knowledge of the finite amplitude only allows us to obtain the leading divergence of the shifted amplitude and not its subleading finite part.  

\item After renormalization, the application of weight-shifting operators can no longer reliably connect different renormalized amplitudes.  This is because the weight-shifting operators map only {\it homogeneous} solutions of the conformal Ward identities to shifted homogeneous solutions,  
whereas the renormalized correlators obey {\it inhomogeneous} Ward identities containing anomalies and beta functions reflecting the breaking of conformal invariance.
As a specific example, it is impossible to obtain the amplitude $\ino_{[33,22x2]}$ from $\ino_{[22,22x2]}$ via the standard weight-shifting operators. This follows simply from the fact that $\ino_{[22,22x2]}$ is anomaly-free whereas $\ino_{[33,22x2]}$ is not, and  the anomaly cannot be introduced through differential operators in the external momenta.
Thus, weight-shifting operators can only reliably be applied when working at the level of the regulated theory.

\end{itemize}

In  this section we present a new momentum-space derivation of the  relevant weight-shifting operator for scalar correlators.  We then discuss their application to construct a reduction scheme providing an alternative route to the correlators we evaluated earlier by direct calculation of Witten diagrams.  This shows how, with due care, the obstacles above can be avoided and the regulated amplitudes constructed, although the resulting scheme is not much simpler than evaluating the diagrams directly.

\subsection{Shift operators from the shadow transform}

The weight-shifting operators of principal interest  are those shifting two of the conformal dimensions  either up or down by one unit while preserving the spacetime dimension $d$.
Denoting these operators collectively as $\mathcal{W}_{ij}^{\sigma_i\sigma_j}$, their action is to send 
\[
\Delta_i\rightarrow \Delta_i+\sigma_i,\qquad \Delta_j\rightarrow \Delta_j+\sigma_j,
\] 
where $\sigma_i,\sigma_j\in \pm 1$ represents any independent choice of signs and $(i, j)$ are any pair of legs.  
In momentum space, and acting on the first two legs,  they take the form 
\begin{empheq}[box=\nicebox]{align}
\mathcal{W}^{--}_{12} &=\frac{1}{2}\, \partial_{12}^2\label{Wmm0}\\
\mathcal{W}^{+-}_{12}&= k_1^{2(\beta_1+1)}\mathcal{W}^{--}_{12} k_1^{-2\beta_1}\label{Wpm0}\\
\mathcal{W}^{-+}_{12} &= k_2^{2(\beta_2+1)}\mathcal{W}^{--}_{12} k_2^{-2\beta_2} \label{Wmp0}\\
\mathcal{W}_{12}^{++}  &=
k_1^{2(\beta_1+1)}k_2^{2(\beta_2+1)} \mathcal{W}_{12}^{--}k_1^{-2\beta_1}k_2^{-2\beta_2}\label{Wpp0}
\end{empheq}
where $\partial_{12}^2 = \partial_{12}^\mu\partial_{12\mu}$ with $\partial_{12\mu}=\partial/\partial k_1^\mu-\partial/\partial k_2^\mu$ and $\beta_i=\Delta_i-d/2$ as usual.  The analogous expressions for other pairs of legs follow  trivially through permutations. 
These operators were introduced in position space in \cite{Karateev:2017jgd}, and later applied in momentum space in \cite{Arkani-Hamed:2018kmz,Baumann:2019oyu}.
However, the compact  expressions \eqref{Wpm0} - \eqref{Wpp0} are new and follow directly from the momentum-space shadow transform as we explain below.   Upon multiplying out,  one can verify that \eqref{Wpm0} - \eqref{Wpp0} 
 are equivalent to the expressions used in \cite{Baumann:2019oyu}, namely 
\begin{align}
\mathcal{W}^{-+}_{12}&=k_2^2 \mathcal{W}_{12}^{--} +2\beta_2\Big(\beta_2+1-\frac{d}{2}+k_2^\mu\partial_{12\mu}\Big)\\
\mathcal{W}^{+-}_{12}&=k_1^2 \mathcal{W}_{12}^{--} +2\beta_1\Big(\beta_1+1-\frac{d}{2}-k_1^\mu\partial_{12\mu}\Big),\\
\mathcal{W}_{12}^{++} 
&=k_1^2k_2^2\mathcal{W}_{12}^{--}+2\beta_1\beta_2(k_1^2+k_2^2-s^2)\nn\\[0ex]&\qquad
+2\beta_1 k_2^2 \Big(\beta_1+1-\frac{d}{2}-k_1^\mu\partial_{12\mu}\Big)
+2\beta_2 k_1^2 \Big(\beta_2+1-\frac{d}{2}+k_2^\mu\partial_{12\mu}\Big).\label{oldWpp}
\end{align}
The weight-shifting action of these   operators has been analyzed from an embedding space perspective in \cite{Karateev:2017jgd, Baumann:2019oyu}.   Here, we offer a simple alternative derivation based primarily in momentum space.   Our analysis is restricted to the regulated theory to avoid divergences.  

The action of the lowering operator $\mathcal{W}_{ij}^{--}$ follows immediately since it is simply the Fourier transform (modulo a factor of one half) of $x_{ij}^2$.   The latter acts as a lowering operator in position space since the 
general position-space $n$-point function can be parametrised as
\[
\<\O(\bs{x}_1)\ldots \O(\bs{x}_n)\> = \prod_{1\le i<j\le n}x_{ij}^{2\alpha_{ij}}f(\bs{u}),
\]
where the $f$ is an arbitrary function of the independent cross ratios $\bs{u}$.  In particular, inversions constrain the $\alpha_{ij}$ to satisfy 
\[
\Delta_i = -\sum_{j=1}^n\alpha_{ij},\qquad i=1,2,\ldots, n
\]
where $\alpha_{ij}=\alpha_{ji}$ and $\alpha_{ii}=0$.
Multiplying by $x_{ij}^2$ thus shifts $\alpha_{ij}\rightarrow \alpha_{ij}+1$ and hence we obtain a new solution with $\Delta_i\rightarrow \Delta_i-1$ and $\Delta_j\rightarrow \Delta_j-1$ but the same  $f(\bs{u})$ and spacetime dimension $d$.  

Returning to momentum space, the action of this operator can also be seen from its intertwining relation with the conformal Ward identities.
By direct computation
\begin{align}
\big(\mathcal{K}_{1}^\mu(\Delta_1-1)+\mathcal{K}_{2}^\mu(\Delta_2-1)\big) \mathcal{W}^{--}_{12} = \mathcal{W}^{--}_{12}\big(\mathcal{K}_{1}^\mu(\Delta_1)+\mathcal{K}_{2}^\mu(\Delta_2)\big), 
\end{align}
where 
\[
\mathcal{K}^\mu_i(\Delta_i) = k_i^\mu \frac{\partial}{\partial k_i^\nu}\frac{\partial}{\partial k_{i\nu}}-2k_i^\nu\frac{\partial}{\partial k_i^\nu}\frac{\partial}{\partial k_{i\mu}}+2(\Delta_i-d)\frac{\partial}{\partial k_{i\mu}}
\] 
is the  special conformal generator in momentum space.  One can further check that $\mathcal{W}_{12}^{--}$ commutes with all remaining generators $\mathcal{K}^\mu_i(\Delta_i)$ for $i\neq 1,2$.
Given some solution $\mathcal{I}_{\{\Delta_i\}}$ of the special conformal Ward identities,
\[\label{cwisolnI}
0= \sum_{i=1}^{n-1}\mathcal{K}_{i}^\mu(\Delta_i) \mathcal{I}_{\{\Delta_i\}}, 
\]
by acting with $\mathcal{W}_{12}^{--}$ to the left it then follows that 
\[
0= \sum_{i=1}^{n-1}\mathcal{K}_{i}^\mu(\Delta_i-\delta_{i1}-\delta_{i2}) \mathcal{W}_{12}^{--}\mathcal{I}_{\{\Delta_i\}}.
\]
Thus, $\mathcal{W}_{12}^{--}\mathcal{I}_{\{\Delta_i\}}$ satisfies the  shifted special conformal Ward identities obtained by replacing  $\Delta_i\rightarrow \Delta_i-\delta_{i1}-\delta_{i2}$.
The corresponding shifted dilatation Ward identity is also satisfied as can be seen by noting that $\mathcal{W}_{12}^{--}$ is homogeneous with dimension $-2$. 

The remaining operators \eqref{Wmp0} - \eqref{Wpp0} can then be constructed by applying the shadow transform.  In momentum space, this takes the form of  the operator identity 
\[
\mathcal{K}_{i}^\mu(d-\Delta_i) k_i^{-2\beta_i} = k^{-2\beta_i}\mathcal{K}_{i}^\mu(\Delta_i). 
\]
Multiplying a given solution of the special conformal Ward identities by $k_i^{-2\beta_i}$ thus generates a new solution with dimension $\Delta_i\rightarrow d-\Delta_i$.  The corresponding dilatation Ward identity is also satisfied as can be verified by power counting.
To construct raising operators we then simply  shadow transform  a selection of momenta, apply the lowering operator $\mathcal{W}_{12}^{--}$, and then invert the shadow transform.
For example,
\begin{align}
\mathcal{W}_{12}^{++}  =
k_1^{2(\beta_1+1)}k_2^{2(\beta_2+1)} \mathcal{W}_{12}^{--}k_1^{-2\beta_1}k_2^{-2\beta_2},
\end{align}
since multiplying by $k_1^{-2\beta_1}k_2^{-2\beta_2}$ sends $\Delta_{i}\rightarrow \Delta'_i=d-\Delta_i$ for $i=1,2$, after which $\mathcal{W}_{12}^{--}$ sends $\Delta'_i\rightarrow\Delta_i''= \Delta'_i-1=d-1-\Delta_i$.  The shadow transform  is then inverted by applying $k_1^{-2\beta_1''}k_1^{-2\beta_2''}=k_1^{-2\Delta_1''+d}k_2^{-2\Delta_2''+d} = k_1^{2(\beta_1+1)}k_2^{2(\beta_2+1)}$ sending $\Delta_i'' \rightarrow\Delta_i''' = d-\Delta_i'' = \Delta_i+1$.  Thus, the net effect is to shift $\Delta_i\rightarrow\Delta_i+1$ for $i=1,2$ leaving all other dimensions intact.  If instead we had shadow transformed with respect to only one of the momenta, say the first, then we would have obtained $\mathcal{W}_{12}^{+-}$ since the action on $\Delta_2$ would remain that of $\mathcal{W}_{12}^{--}$.

As for $\mathcal{W}_{12}^{--}$, we can prove the action of these operators more formally by verifying the intertwining relations.
For example, by direct calculation
\[
\big(\mathcal{K}_{1}^\mu(\Delta_1+1)+\mathcal{K}_{2}^\mu(\Delta_2+1)\big) \mathcal{W}^{++}_{12} = \mathcal{W}^{++}_{12}\big(\mathcal{K}_{1}^\mu(\Delta_1)+\mathcal{K}_{2}^\mu(\Delta_2)\big)
\]
while all other $\mathcal{K}_{i}^\mu(\Delta_i)$ with $i\neq 1,2$ commute with $\mathcal{W}_{12}^{++}$.
Thus, given some solution of the special conformal Ward identities \eqref{cwisolnI}, by acting to the left with $\mathcal{W}_{12}^{++}$ we find  
\[
0= \sum_{i=1}^{n-1}\mathcal{K}_{i}^\mu(\Delta_i+\delta_{i1}+\delta_{i2}) \mathcal{W}_{12}^{++}\mathcal{I}_{\{\Delta_i\}},
\]
and so $\mathcal{W}_{12}^{++}$ indeed generates a new solution with raised $\Delta_1$ and $\Delta_2$.  Again, the dilatation Ward identity is automatically satisfied since  $\mathcal{W}_{12}^{++}$ is homogeneous of dimension $+2$.   A similar analysis applies for $\mathcal{W}_{12}^{-+}$ and $\mathcal{W}_{12}^{+-}$.

\subsection{Action on exchange diagrams}  

To analyze the action of the shift operators \eqref{Wmm0} - \eqref{Wpp0} on the various amplitudes we study, it is convenient to convert to Mandelstam variables.  In fact, it will suffice to consider their action on contact diagrams and $s$-channel exchanges only.  
 Since these amplitudes are all independent of $t$, we can then further simplify the shift operators by setting all $t$-derivatives to zero.
The operator $\mathcal{W}^{--}_{12}$ then takes the form 
\begin{align} \label{Wmm12}
\mathcal{W}^{--}_{12} &=\frac{1}{2}\Big[\partial_1^2+\partial_2^2+\frac{(d-1)}{k_1}\partial_1+\frac{(d-1)}{k_2}\partial_2+\frac{1}{k_1k_2}(k_1^2+k_2^2-s^2)\partial_1\partial_2\Big] + O(\partial_t),
\end{align}
where $\partial_i = \partial/\partial k_i$  while the remaining operators follow from \eqref{Wpm0} - \eqref{Wpp0}.

Let us now apply $\mathcal{W}_{12}^{--}$ to an $s$-channel exchange diagram \eqref{amp4x}. 
Acting on a product of bulk-boundary propagators \eqref{KPropagator}, we find the identity
\begin{align}
&\mathcal{W}^{--}_{12}\Big(\mathcal{K}_{d,\Delta_1}(z,k_1) \mathcal{K}_{d,\Delta_2}(z,k_2) \Big)\nn\\&\quad
=\frac{1}{8(\beta_1-1)(\beta_2-1)}\Big(\Box_z-(\beta_1+\beta_2-2)(\beta_1+\beta_2+d-2)\Big)\mathcal{K}_{d,\Delta_1-1}(z,k_1) \mathcal{K}_{d,\Delta_2-1}(z,k_2) \label{actionWmm}
\end{align}
where the box operator for $(d+1)$-dimensional AdS is 
\[\label{boxs}
\Box_z = z^2\partial_z^2+(1-d)z\partial_z-s^2 z^2.
\]
Thus, if we act with $\mathcal{W}^{--}_{12}$ on an $s$-channel exchange diagram, we obtain a box operator which can be integrated by parts
 in $z$ so as to act on the bulk-bulk propagator.\footnote{The boundary terms from $z\rightarrow \infty$ vanish since all propagators decay exponentially.
The lower limit behaves as $\int \D z\,\partial_z z^{\omega+\ep}\sim \int\D z\, z^{\omega-1+\ep}$ 
in the half-integer regularization scheme, where  
$\omega = d+\Delta_x-\tilde{\Delta}_1-\tilde{\Delta}_2$ and $\tilde{\Delta}_1=\Delta_1+\sigma_1$ and $\tilde{\Delta}_2=\Delta_2+\sigma_2$ are the shifted dimensions. 
For $\omega\neq -2m$  with $m\in \mathbb{Z}^+$ this is zero by analytic continuation in $\ep$ from the large-$\ep$ region where it converges, see \cite{Bzowski:2015pba}.  Otherwise, we obtain a divergence which can be removed through a boundary counterterm.
}   This produces a delta function and a constant term, since the bulk-bulk propagator \eqref{GPropagator} obeys
\[
(-\Box_z+m_x^2)\mathcal{G}_{d,\Delta_x}(z,s;\zeta)=z^{d+1}
\delta(z-\zeta), \qquad m_x^2 = \Delta_x(\Delta_x-d)
=\beta_x^2-\frac{d^2}{4}.
\label{Greensfn}
\]
In the regulated theory then, the action of $\mathcal{W}^{--}_{12}$ on an exchange diagram yields a linear combination of a shifted exchange and a shifted contact diagram, 
\[
\label{Wmmexchtocont}
\mathcal{W}^{--}_{12}\,\hat{i}_{[\Delta_1,\Delta_2,\Delta_3,\Delta_4\,x\,\Delta_x]} = 
\mathcal{N}^{--}_{exch.}\,
\hat{i}_{[\Delta_1-1,\Delta_2-1,\Delta_3,\Delta_4 \,x\,\Delta_x]} +\mathcal{N}^{--}_{cont.}\,\hat{i}_{[\Delta_1-1,\Delta_2-1,\Delta_3,\Delta_4]} 
\]
where
\begin{align}\label{Wmmlincomb}
\mathcal{N}^{--}_{exch.} &= \frac{\beta_x^2-d^2/4-(\beta_1+\beta_2-2)(\beta_1+\beta_2+d-2)}{8(\beta_1-1)(\beta_2-1)}, 
\\
\mathcal{N}^{--}_{cont.} &=-\frac{1}{8(\beta_1-1)(\beta_2-1)}.
\end{align}
If our goal is to obtain just the shifted exchange diagram,  we must therefore evaluate and subtract off this contribution from the shifted contact diagram.  Sometimes this can be achieved through the further use of  weight-shifting operators as discussed in  Section \ref{contactsec}.

Let us next consider the action of the raising operator $\mathcal{W}_{12}^{++}$ on an $s$-channel exchange.  This can easily be understood through use of \eqref{actionWmm} in combination with the shadow relation for the bulk-boundary propagator,
\[
k^{-2\beta}\mathcal{K}_{d,\Delta}(z,k) =\Lambda_\beta \,\mathcal{K}_{d,d-\Delta}(z,k) ,\qquad \Lambda_\beta =\frac{\Gamma(-\beta)}{4^\beta\Gamma(\beta)}.\label{shadowK}
\]
First, we write
\begin{align}
&\mathcal{W}_{12}^{++}\Big(\mathcal{K}_{d,\Delta_1}(z,k_1) \mathcal{K}_{d,\Delta_2}(z,k_2) \Big)\nn\\ &\quad
=k_1^{2(\beta_1+1)}k_2^{2(\beta_2+1)} \mathcal{W}_{12}^{--}k_1^{-2\beta_1}k_2^{-2\beta_2}\Big(\mathcal{K}_{d,\Delta_1}(z,k_1) \mathcal{K}_{d,\Delta_2}(z,k_2) \Big)\nn\\&\quad
=\Lambda_{\beta_1}\Lambda_{\beta_2}k_1^{2(\beta_1+1)}k_2^{2(\beta_2+1)} \mathcal{W}_{12}^{--}\Big(\mathcal{K}_{d,d-\Delta_1}(z,k_1) \mathcal{K}_{d,d-\Delta_2}(z,k_2) \Big)\nn\\&\quad
=\frac{\Lambda_{\beta_1}\Lambda_{\beta_2}k_1^{2(\beta_1+1)}k_2^{2(\beta_2+1)} }{8(\beta_1+1)(\beta_2+1)}\nn\\&\qquad \times\Big(\Box_z-(\beta_1+\beta_2+2)(\beta_1+\beta_2-d+2)\Big)\mathcal{K}_{d,d-\Delta_1-1}(z,k_1) \mathcal{K}_{d,d-\Delta_2-1}(z,k_2).
\end{align}
We now commute the factor $k_1^{2(\beta_1+1)}k_2^{2(\beta_2+1)}$ through the differential operator and use again the shadow identity \eqref{shadowK} (replacing $\Delta\rightarrow d-\Delta-1$) to find 
\begin{align}
&k_1^{2(\beta_1+1)}k_2^{2(\beta_2+1)}\mathcal{K}_{d,d-\Delta_1-1}(z,k_1) \mathcal{K}_{d,d-\Delta_2-1}(z,k_2)\nn\\
&\qquad = \Lambda_{-\beta_1-1}\Lambda_{-\beta_2-1}
\mathcal{K}_{d,\Delta_1+1}(z,k_1) \mathcal{K}_{d,\Delta_2+1}(z,k_2).
\end{align}
After simplifying the prefactor 
\begin{align}
\frac{\Lambda_{\beta_1}\Lambda_{\beta_2} }{8(\beta_1+1)(\beta_2+1)}\Lambda_{-\beta_1-1}\Lambda_{-\beta_2-1}
= 2\beta_1\beta_2,
\end{align}
we arrive the desired identity 
\begin{align}
&\mathcal{W}_{12}^{++}\Big(\mathcal{K}_{d,\Delta_1}(z,k_1) \mathcal{K}_{d,\Delta_2}(z,k_2) \Big)\nn\\ &\quad
=2\beta_1\beta_2\Big(\Box_z-(\beta_1+\beta_2+2)(\beta_1+\beta_2-d+2)\Big)\mathcal{K}_{d,\Delta_1+1}(z,k_1) \mathcal{K}_{d,\Delta_2+1}(z,k_2). \label{actionWpp}
\end{align}
This identity is the counterpart to \eqref{actionWmm} and can be used in the same way to evaluate the action of $\mathcal{W}_{12}^{++}$ on an $s$-channel exchange diagram.  After integration by parts in $z$ and use of \eqref{Greensfn},
we obtain the result
\[\label{Wpplincomb}
\mathcal{W}^{++}_{12}\,\hat{i}_{[\Delta_1,\Delta_2,\Delta_3,\Delta_4\,x\,\Delta_x]} = 
\mathcal{N}^{++}_{exch.}\,
\hat{i}_{[\Delta_1+1,\Delta_2+1,\Delta_3,\Delta_4\,x\,\Delta_x]} +\mathcal{N}^{++}_{cont.}\, \hat{i}_{[\Delta_1+1,\Delta_2+1,\Delta_3,\Delta_4]} 
\]
where
\begin{align}
\mathcal{N}^{++}_{exch.} =2\beta_1\beta_2\Big(\beta_x^2-\frac{d^2}{4}-(\beta_1+\beta_2+2)(\beta_1+\beta_2-d+2)\Big), \qquad
\mathcal{N}^{++}_{cont.} &= -2\beta_1\beta_2.
\end{align}

The steps above can similarly be repeated for the remaining operators $\mathcal{W}_{12}^{+-}$ and $\mathcal{W}_{12}^{-+}$.  
Examining the form of these results, and those above, we find they can all be collected into a common form:
\begin{empheq}[box=\nicebox]{align}
&\mathcal{W}_{12}^{\sigma_1 \sigma_2}\Big(\mathcal{K}_{d,\Delta_1}(z,k_1) \mathcal{K}_{d,\Delta_2}(z,k_2) \Big)
=\mathcal{N}^{\sigma_1\sigma_2} \Big(\Box_z-m^2_{(\sigma_1,\sigma_2)}\Big)\mathcal{K}_{d,\Delta_1+\sigma_1}(z,k_1) \mathcal{K}_{d,\Delta_2+\sigma_2}(z,k_2). \label{actionWsigma}
\end{empheq}
where $\sigma_1, \sigma_2\in\pm1$ and 
\begin{align}\label{Nsigmadef}
\mathcal{N}^{\sigma_1\sigma_2} &= 2^{\sigma_1+\sigma_2-1}(\beta_1)_{\sigma_1}(\beta_2)_{\sigma_2},\\
m^2_{(\sigma_1,\sigma_2)} &=(\sigma_1\beta_1+\sigma_2\beta_2+2)(\sigma_1\beta_1+\sigma_2\beta_2-d+2).\label{msigmadef}
\end{align}
Here the Pochhammer symbol $(\beta)_\sigma = \Gamma(\beta+\sigma)/\Gamma(\beta)$ is just a convenient way of writing $(\beta)_1 = \beta$ and $(\beta)_{-1}=(\beta-1)^{-1}$.
Acting on an $s$-channel exchange, we find
\begin{empheq}[box=\nicebox]{align}\label{Wpplincombsigma}
\mathcal{W}^{\sigma_1 \sigma_2}_{12}\,\hat{i}_{[\Delta_1,\Delta_2,\Delta_3,\Delta_4\,x\,\Delta_x]} = 
\mathcal{N}^{\sigma_1\sigma_2}_{exch.}\,
\hat{i}_{[\Delta_1+\sigma_1,\Delta_2+\sigma_2,\Delta_3,\Delta_4\,x\,\Delta_x]} +\mathcal{N}^{\sigma_1\sigma_2}_{cont.}\, \hat{i}_{[\Delta_1+\sigma_1,\Delta_2+\sigma_2,\Delta_3,\Delta_4]} 
\end{empheq}
where
\begin{align}
\mathcal{N}^{\sigma_1\sigma_2}_{exch.} &=\mathcal{N}^{\sigma_1\sigma_2} (m_x^2-m_{(\sigma_1,\sigma_2)}^2),
\qquad
\mathcal{N}^{\sigma_1\sigma_2}_{cont.} =-\mathcal{N}^{\sigma_1\sigma_2}.
\end{align}
Where divergences are present, this identity again applies only in the  regulated theory.

Using the definition of $m_x^2$ from \eqref{Greensfn}, we can rewrite
\[
\mathcal{N}^{\sigma_1\sigma_2}_{exch.} =-\mathcal{N}^{\sigma_1\sigma_2} \Big(\frac{d}{2}-\sigma_1\tilde{\beta}_1
-\sigma_2\tilde{\beta}_2+\beta_x\Big)\Big(\frac{d}{2}-\sigma_1\tilde{\beta}_1
-\sigma_2\tilde{\beta}_2-\beta_x\Big)
\]
where the $\tilde{\beta}_i = \beta_i+\sigma_i$ are the shifted values of the $\beta_i$ parameters under $\Delta_i\rightarrow\Delta_i+\sigma_i$.  From this expression, we see that $\mathcal{N}^{\sigma_1\sigma_2}_{exch.}$ has a zero whenever the condition
\[\label{Nexchzero}
0 =\frac{d}{2}-\sigma_1\tilde{\beta}_1
-\sigma_2\tilde{\beta}_2-\sigma_x\beta_x
\]
is satisfied for some choice of the signs $(\sigma_1,\sigma_2,\sigma_x) = (\pm, \pm, \pm)$, where each sign can be chosen independently of the others.
This makes sense since whenever this condition is satisfied the shifted exchange diagram is  singular\footnote{The divergence of the shifted exchange diagram derives from the region where the vertex corresponding to the shifted operators approaches the boundary of AdS: effectively, this is the lower limit of an integral of three Bessel $K$ functions, two of which come from the bulk-boundary propagators and the third from the exterior part of the bulk-bulk propagator.  The singularities of such triple-$K$ integrals, which also arise for the momentum-space 3-point function $\lla \O_{\Delta_1+\sigma_1}\O_{\Delta_2+\sigma_2}\O_{\Delta_x}\rra$,  have been analyzed in \cite{Bzowski:2015pba} leading to precisely the condition \eqref{Nexchzero}.}, and a differential operator such as $\mathcal{W}^{\sigma_1 \sigma_2}_{12}$ cannot map a finite exchange diagram to a singular one.  Rather, acting with $\mathcal{W}^{\sigma_1 \sigma_2}_{12}$ yields a  finite result which, in the regulated theory, derives from the product of a vanishing coefficient $\mathcal{N}^{\sigma_1\sigma_2}_{exch.}$ and a divergent shifted exchange diagram, plus a contact contribution.  In such situations where $\mathcal{N}^{\sigma_1\sigma_2}_{exch.}$ has a zero, the application of $\mathcal{W}^{\sigma_1 \sigma_2}_{12}$ to a finite diagram therefore only allows us to extract the leading divergence of the shifted exchange diagram: to additionally compute its finite piece requires knowing the finite unshifted exchange diagram to subleading orders in the regulator $\ep$.  This is generally harder to compute since the corresponding Bessel function indices are no longer half-integer.

For this reason, such divergences and their corresponding zeros in the exchange coefficients $\mathcal{N}^{\sigma_i\sigma_j}_{exch.}$ have important consequences for the construction of reduction schemes relating different exchange diagrams.  Fortunately, as we will see in Section \ref{sec:shiftrencorr}, it turns out that for the correlators of interest in this paper, all such zeros of $\mathcal{N}^{\sigma_i\sigma_j}_{exch.}$ can  be avoided through making appropriate use of shadow transformations in addition to the  $\mathcal{W}^{\sigma_i \sigma_j}_{ij}$ operators.

\subsection{Action on contact diagrams}

As we saw above, the action of the $\mathcal{W}_{12}^{\sigma_1\sigma_2}$ on an exchange diagram generates a linear combination of a shifted exchange diagram and a shifted contact diagram.  The action of $\mathcal{W}_{12}^{\sigma_1\sigma_2}$ on a contact diagram produces a linear combination of a shifted contact diagram and a shifted contract diagram containing an AdS box vertex.
In this section we discuss how this box contribution, which also follows from to the action of the Casimir on the shifted contact diagram, can in some cases be removed through an appropriate symmetrization.  Further shift relations for contact diagrams are  discussed in appendix \ref{app_shift_4K}.
Once all shifted contact diagrams are known, either via weight-shifting operators or from direct evaluation, their contribution can be subtracted enabling all shifted exchange diagrams to be constructed.

\subsubsection{Exchanges to contacts}
\label{contactsec}

As an intermediate step, one can always pass from an exchange diagram to the corresponding contact diagram ({\it i.e.,} with no shift to the operator dimensions) through the well-known action of the quadratic Casimir operator  
\begin{align}
\mathcal{C}_{12}& = \big(2k_{1\mu}k_{2\nu}-k_1\cdot k_2\,\delta_{\mu\nu}\big)\partial_{12}^\mu\partial_{12}^\nu+2\big((\Delta_2-d)k_{1\mu}-(\Delta_1-d)k_{2\mu}\big)\partial_{12}^\mu\nn\\&\quad 
-(\Delta_1+\Delta_2-2d)(\Delta_1+\Delta_2-d).
\end{align}
In Mandelstam variables,
\begin{align}\label{Cas12form1}
\mathcal{C}_{12}& = \frac{1}{2}(s^2+k_1^2-k_2^2)K_1+\frac{1}{2}(s^2+k_2^2-k_1^2)K_2-\big(L_1+L_2+\frac{3d}{2}\big)^2+\frac{d^2}{4}+O(\partial_t)
\end{align}
where
\[
K_i = \partial_i^2 +\frac{1-2\beta_i}{k_i}\partial_i,\qquad
L_i = k_i\partial_i-\Delta_i
\]
and we have omitted $t$-derivatives since $s$-channel scalar exchanges are independent of $t$. 
Indeed, 
we can simplify this further since for $s$-channel scalar exchanges 
\[\label{K12onex}
(K_1-K_2) \hat{i}_{[\Delta_1,\Delta_2,\Delta_3,\Delta_4\,x\,\Delta_x]} =0.
\]
The action of the Casimir on such diagrams is thus equal to that of the reduced operator 
\[\label{redCasdef}
\tilde{\mathcal{C}}_{12}=\frac{s^2}{2}(K_1+K_2)-\big(L_1+L_2+\frac{3d}{2}\big)^2 + \frac{d^2}{4}.
\]
We can also use this reduced Casimir operator when acting on scalar contact diagrams, since these are also independent of $t$ (and $s$) and  satisfy $(K_1-K_2)\hat{i}_{[\Delta_1,\Delta_2,\Delta_3,\Delta_4]}=0$.

One can now verify directly the identity
\begin{align}
\tilde{\mathcal{C}}_{12}\Big(\mathcal{K}_{d,\Delta_1}(z,k_1)\mathcal{K}_{d,\Delta_2}(z,k_2) \Big) =-\Box_z\Big(\mathcal{K}_{d,\Delta_1}(z,k_1)\mathcal{K}_{d,\Delta_2}(z,k_2) \Big).\label{Casaction}
\end{align}
Acting on an $s$-channel exchange diagram with $(\tilde{\mathcal{C}}_{12}+m_x^2)$ and integrating by parts with respect to $z$, we  obtain a delta function from \eqref{Greensfn}.  This collapses the exchange diagram to a contact diagram of the same operator dimensions:
\begin{empheq}[box=\nicebox]{align} 
\big(\tilde{\mathcal{C}}_{12}+m_x^2\big)\,\hat{i}_{[\Delta_1,\Delta_2,\Delta_3,\Delta_4\,x\,\Delta_x]}  = \hat{i}_{[\Delta_1,\Delta_2,\Delta_3,\Delta_4]} \label{Casonex} 
\end{empheq}
One can alternatively obtain this result by showing that
\begin{align}\label{Wcomm}
[\mathcal{W}_{12}^{--},\mathcal{W}_{12}^{++}] = 2(\beta_1+\beta_2)\Big(\mathcal{C}_{12}+(\beta_1+\beta_2)^2-\frac{d}{2}(d-2)\Big),\\
[\mathcal{W}_{12}^{-+},\mathcal{W}_{12}^{+-}] = 2(\beta_1-\beta_2)\Big(\mathcal{C}_{12}+(\beta_1-\beta_2)^2-\frac{d}{2}(d-2)\Big),
\end{align}
then making two applications of \eqref{Wpplincombsigma}.

\subsubsection{Generating exchange diagrams with derivative vertices}\label{sec:deriv_vertices}

Rather than integrating by parts the right-hand side of \eqref{Casaction}, we can also simply expand out the action of the AdS box \eqref{boxs} on the two bulk-boundary propagators.  This produces a cross-term plus two terms where the box acts on a single bulk-boundary propagator.
The latter can be replaced via the equation of motion
\[
(z^2\partial_z^2+(1-d)z\partial_z)\mathcal{K}_{d,\Delta_i}(z,k_i) = (m_i^2+k_i^2) \mathcal{K}_{d,\Delta_i}(z,k_i),\qquad m_i^2 = \Delta_i(\Delta_i-d)
\]
leading to the identity
\begin{align}
&-\frac{1}{2}\Big(\tilde{\mathcal{C}}_{12}+m_1^2+m_2^2\Big)\Big(\mathcal{K}_{d,\Delta_1}(z,k_1)\mathcal{K}_{d,\Delta_2}(z,k_2) \Big) \\&\quad
=z^2\Big( \partial_z \mathcal{K}_{d,\Delta_1}(z,k_1)\partial_z \mathcal{K}_{d,\Delta_2}(z,k_2) +\frac{1}{2}(k_1^2+k_2^2-s^2)\mathcal{K}_{d,\Delta_1}(z,k_1) \mathcal{K}_{d,\Delta_2}(z,k_2) \Big).\nn
\end{align}
Since $(k_1^2+k_2^2-s^2)/2 = -\bs{k}_1\cdot\bs{k}_2$, the right-hand side here is precisely the vertex factor appearing in an exchange diagram with a $\Phi_x\partial_\mu\Phi_1\partial^\mu\Phi_2$ derivative interaction, namely
\begin{align}
\ino^{deriv}_{[\Delta_1\Delta_2,\Delta_3\Delta_4x\Delta_x]}&=\int_0^\infty\frac{\D z}{z^{d+1}} z^2\Big( \partial_z \K_{d,\Delta_1}(z,k_1)\partial_z \K_{d,\Delta_2}(z,k_2)\nn\\&\qquad \qquad\qquad+\frac{1}{2}(k_1^2+k_2^2-s^2)\K_{d,\Delta_1}(z,k_1) \K_{d,\Delta_2}(z,k_2)\Big)\nn\\&\quad\times
\int_0^\infty\frac{\D \zeta}{\zeta^{d+1}} \G_{d,\Delta_x}(z,s;\zeta)\K_{d,\Delta_3}(\zeta,k_3)\K_{d,\Delta_4}(\zeta,k_4).
\end{align}
In the regulated theory, we can therefore generate an exchange diagram with a  $\Phi_x\partial_\mu\Phi_1\partial^\mu\Phi_2$ derivative vertex by acting on an ordinary exchange diagram with the Casimir operator:
\begin{align}\label{Casderivrel}
\ireg^{deriv}_{[\Delta_1\Delta_2,\Delta_3\Delta_4x\Delta_x]} &= -\frac{1}{2} (\tilde{\mathcal{C}}_{12}+m_1^2+m_2^2)\ireg_{[\Delta_1\Delta_2,\Delta_3\Delta_4x\Delta_x]}.
\end{align}
Using \eqref{Casonex}, this exchange diagram with a derivative vertex can also be expressed as a sum of an ordinary exchange and a contact diagram,
\begin{align}
\ireg^{deriv}_{[\Delta_1\Delta_2,\Delta_3\Delta_4x\Delta_x]}&= \frac{1}{2} (m_x^2-m_1^2-m_2^2)\ireg_{[\Delta_1\Delta_2,\Delta_3\Delta_4x\Delta_x]}-\frac{1}{2}\ireg_{[\Delta_1\Delta_2\Delta_3\Delta_4]}.
\end{align}
In certain special cases, this linear combination coincides with that produced by the action of the operator $\mathcal{W}_{12}^{\sigma_1\sigma_2}$ in \eqref{Wpplincombsigma} leading to the relation
\begin{align}
\mathcal{W}^{\sigma_1\sigma_2}_{12}\ireg_{[\Delta_1,\Delta_2,\Delta_3,\Delta_4x\Delta_x]} \stackrel{*}{=} 2\mathcal{N}^{\sigma_1\sigma_2}
\,\ireg^{deriv}_{[\Delta_1+\sigma_1,\Delta_2+\sigma_2,\Delta_3,\Delta_4x\Delta_x]}, 
\end{align}
where the coefficient $\mathcal{N}^{\sigma_1\sigma_2}$ is that in  \eqref{Nsigmadef}.
The asterisk indicates that  this equation is only valid when the masses  satisfy the condition 
\[\label{specialm}
m_{\Delta_1+\sigma_1}^2+m_{\Delta_2+\sigma_2}^2 = m_{(\sigma_1,\sigma_2)}^2,
\]
where $m_{\Delta_i+\sigma_i}^2 = (\Delta_i+\sigma_i)(\Delta_i+\sigma_i-d)$ is the mass corresponding to the shifted dimension and
$m_{(\sigma_1,\sigma_2)}^2$ is given in \eqref{msigmadef}.  This  is equivalent to requiring 
\[
(\beta_1+\sigma_1)^2+(\beta_2+\sigma_2)^2-\frac{d^2}{2} = (\sigma_1
\beta_1+\sigma_2\beta_2+2)(\sigma_1\beta_1+\sigma_2\beta_2+2-d).
\]
In particular, this condition holds when $\mathcal{W}_{12}^{++}$ is applied to an exchange diagram of conformal scalars with $\Delta_1=\Delta_2=2$ in $d=3$.  In this case $\beta_1=\beta_2=1/2$, $\sigma_1=\sigma_2=1$ and all the masses in \eqref{specialm} vanish individually.  The action of $\mathcal{W}_{12}^{++}$ then leads to an exchange diagram where two massless scalars are derivatively coupled to a 
 conformal scalar \cite{Arkani-Hamed:2018kmz,Baumann:2019oyu}.  

\subsubsection{Shifted contact diagrams from symmetrization}  

Let us now return to consider the action of $\mathcal{W}_{12}^{\sigma_1,\sigma_2}$ on the contact diagram $\hat{i}_{[\Delta_1,\Delta_2,\Delta_3,\Delta_4]}$.  From \eqref{actionWsigma}, we obtain a linear combination of the shifted contact diagram and a shifted contact diagram with an AdS box vertex.  Via \eqref{Casaction}, the latter can be re-expressed as the Casimir operator acting on the shifted contact diagram giving
\begin{empheq}[box=\nicebox]{align} 
\label{contshiftbyW}
\mathcal{W}_{12}^{\sigma_1,\sigma_2}\ireg_{[\Delta_1,\Delta_2,\Delta_3,\Delta_4]} &=-\mathcal{N}^{\sigma_1\sigma_2}\Big(\tilde{\mathcal{C}}_{12}^{\Delta_1+\sigma_1,\Delta_2+\sigma_2}+m_{(\sigma_1,\sigma_2)}^2\Big)
\ireg_{[\Delta_1+\sigma_1,\Delta_2+\sigma_2,\Delta_3,\Delta_4]}, 
\end{empheq}
where $\mathcal{N}^{\sigma_1\sigma_2}$ and $m_{(\sigma_1,\sigma_2)}^2$ are given in \eqref{Nsigmadef} and \eqref{msigmadef} but  the dimensions appearing inside the Casimir operator are the shifted ones as indicated by the superscript indices.

In the special case where at least three of the shifted operator dimensions are equal,  the Casimir operator on the right-hand side of this relation can now be eliminated through a simple symmetrization procedure. 
Here, it is useful to recall that from an embedding space perspective, the $d$-dimensional conformal Ward identities originate from invariance under the action of the $SO(d+1,1)$ Lorentz generators
\[
0 = \sum_{i=1}^4 L^{AB}_i
\] 
where $L_{i}^{AB}=X_{i}^{A}(\partial/\partial X_{iB})-X_{i}^{B}(\partial/\partial X_{iA})$.
Squaring this relation and using the identities
\[
\mathcal{C}_{ij}^{\Delta_i,\Delta_j} =\frac{1}{2}(L_i^{AB}+L_j^{AB})(L_{iAB}+L_{jAB}), \qquad 
\frac{1}{2}L_i^{AB}L_{iAB}=-\Delta_i(\Delta_i-d),
\]
along with 
\[
\mathcal{C}_{12}^{\Delta_1,\Delta_2}=\mathcal{C}_{34}^{\Delta_3,\Delta_4}, \qquad \mathcal{C}_{14}^{\Delta_1,\Delta_4}=\mathcal{C}_{23}^{\Delta_2,\Delta_3}, \qquad\mathcal{C}_{13}^{\Delta_1,\Delta_3}=\mathcal{C}_{24}^{\Delta_2,\Delta_4},
\]  
as follows from squaring, {\it e.g.,} $L_1^{AB}+L_2^{AB}=-(L_3^{AB}+L_4^{AB})$, 
we obtain
\[
\mathcal{C}_{12}^{\Delta_1,\Delta_2}+\mathcal{C}_{13}^{\Delta_1,\Delta_3}+\mathcal{C}_{23}^{\Delta_2,\Delta_3}=\frac{1}{2}\sum_{i<j}^{4}\mathcal{C}_{ij}^{\Delta_i,\Delta_j} = -\sum_{i=1}^{4}\Delta_i(\Delta_i-d).
\]
If the three operator dimensions on the left-hand side are equal, this reduces to
\begin{align}\label{CasthreeDelta}
\Big(\mathcal{C}_{12}^{\Delta,\Delta}+\mathcal{C}_{13}^{\Delta,\Delta}+\mathcal{C}_{23}^{\Delta,\Delta}\Big)\ireg_{[\Delta,\Delta,\Delta,\Delta_4]}
=-\Big(3\Delta(\Delta-d)+\Delta_4(\Delta_4-d)\Big)\,\ireg_{[\Delta,\Delta,\Delta,\Delta_4]}.
\end{align}
Moreover, we  know from \eqref{contshiftbyW} that
\begin{align}
\mathcal{W}_{12}^{\sigma_1,\sigma_2}\ireg_{[\Delta-\sigma_1,\Delta-\sigma_2,\Delta,\Delta_4]} &=-\mathcal{N}^{\sigma_1\sigma_2}\Big(\mathcal{C}_{12}^{\Delta,\Delta}+m_{(\sigma_1,\sigma_2)}^2\Big)
\ireg_{[\Delta,\Delta,\Delta,\Delta_4]}. 
\end{align}
By permuting momenta and exploiting the fact that the three shifted dimensions are equal, we can now construct all the remaining terms on the left-hand side of \eqref{CasthreeDelta}.  For example,
\begin{align}
\Big(\mathcal{W}_{12}^{\sigma_1,\sigma_2}\ireg_{[\Delta-\sigma_1,\Delta-\sigma_2,\Delta,\Delta_4]}\Big)_{\bs{k}_2\leftrightarrow \bs{k}_3} &=-\mathcal{N}^{\sigma_1\sigma_2}\Big(\mathcal{C}_{13}^{\Delta,\Delta}+m_{(\sigma_1,\sigma_2)}^2\Big)
\ireg_{[\Delta,\Delta,\Delta,\Delta_4]}.
\end{align}
Summing over permutations, the Casimir contributions can  now be eliminated 
\begin{align}
\label{eqshift3}
&\Big(\mathcal{W}_{12}^{\sigma_1,\sigma_2}\ireg_{[\Delta-\sigma_1,\Delta-\sigma_2,\Delta,\Delta_4]} \Big)+(\bs{k}_2\leftrightarrow \bs{k}_3) +(\bs{k}_1\leftrightarrow \bs{k}_3) \nn\\[1ex]&\qquad  =
\mathcal{N}^{\sigma_1\sigma_2}\Big(3\Delta(\Delta-d)+\Delta_4(\Delta_4-d)-3m_{(\sigma_1,\sigma_2)}^2\Big)\ireg_{[\Delta,\Delta,\Delta,\Delta_4]}\nn\\[1ex]
&\qquad =2^{\sigma_1+\sigma_2-1}(\beta-\sigma_1)_{\sigma_1} (\beta-\sigma_2)_{\sigma_2}\nn\\&\qquad \qquad \times
\Big(3\beta^2+\beta_4^2-d^2
-3\beta^2(\sigma_1+\sigma_2)^2 +3\beta d(\sigma_1+\sigma_2)\Big)\ireg_{[\Delta,\Delta,\Delta,\Delta_4]}
\end{align} 
where the Pochammer symbols are simply $(\beta-1)_1=\beta-1$ and $(\beta+1)_{-1}=\beta^{-1}$.  
This relation (and its permutations) thus enable the construction of shifted contact diagrams containing three equal dimensions if the appropriate initial integral is known.  Some further shift relations for contact diagrams can be found in appendix \ref{app_shift_4K}.

\subsection{Reduction scheme}

\label{sec:shiftrencorr}

Equation \eqref{Wpplincombsigma} can be used to set up a reduction scheme. Starting from a smaller set of simple master integrals, one can use the weight-changing operators to obtain other diagrams. Since the lower dimensions generally yield simpler expressions, one would like to start with the lowest dimensional diagrams such as $\ino_{[22,22x\Delta_x]}$ and apply the weight-raising operators $\mathcal{W}^{++}_{ij}$. To see if it is possible, we begin by investigating all available relations due to \eqref{Wpplincombsigma} between the amplitudes.

\subsubsection{Available relations}

First, notice that one can either use the weight-changing operators directly or in a combination with the shadow transform. The shadow transform is an observation that the bulk-to-boundary propagators for the AdS fields with $\Delta$ and $d - \Delta$ are related according to,
\begin{align}\label{shadowK2}
\K_{d, d-\Delta}(z, k) = \frac{\Gamma \left( \Delta - \tfrac{d}{2} \right)}{\Gamma \left( \tfrac{d}{2} - \Delta \right)} \left( \frac{k}{2} \right)^{d - 2 \Delta} \times \K_{d, \Delta}(z, k).
\end{align}
Thus, we can define the shadow transform $\mathcal{S}_j$ acting on the $j$-th external leg of an amplitude as the multiplication by the above constant. This yields, for example,
\begin{align}
\ino_{[d - \Delta_1, \Delta_2; \Delta_3 \Delta_4 x \Delta_x]} & = \mathcal{S}_1 \ino_{[\Delta_1, \Delta_2; \Delta_3 \Delta_4 x \Delta_x]} \nn\\
& = \frac{\Gamma \left( \Delta_1 - \tfrac{d}{2} \right)}{\Gamma \left( \tfrac{d}{2} - \Delta_1 \right)} \left( \frac{k_1}{2} \right)^{d - 2 \Delta_1} \times \ino_{[\Delta_1, \Delta_2; \Delta_3 \Delta_4 x \Delta_x]}.
\end{align}
It is important to use regulated dimensions $\dreg$ and $\Dreg_1$ when dealing with regulated amplitudes.

While the weight-shifting operators $\mathcal{W}^{++}_{ij}$ and $\mathcal{W}^{--}_{ij}$ always raise or lower a pair of dimensions, we can compose them with the shadow transform to raise or lower a single dimension only. As an example, consider $\ino_{[22,22x2]}$. By acting with $\mathcal{W}^{++}_{12}$, according to \eqref{Wpplincombsigma} one should obtain a combination of $\ino_{[33,22x2]}$ and $\ino_{[3322]}$. On the other hand, we can use the shadow transform to flip $\Delta_2 = 2$ to $d - \Delta_2 = 1$. Thus, the combination $\mathcal{W}^{++}_{12} \circ \mathcal{S}_2$ should map $\ino_{[22,22x2]}$ to a combination of $\ino_{[32,22x2]}$ and $\ino_{[3222]}$. In general, we can carry out the following operations on a pair of dimensions:
\begin{equation} \label{red_ops}
\begin{tikzpicture}[scale=1.5]
\node at (0,0) {$(22)$};
\node at (-2,0) {$(33)$};
\node at (2,0) {$(32)$};
\draw[->] (0.3,0.1) -- (1.7,0.1);
\draw[<-] (0.3,-0.1) -- (1.7,-0.1);
\draw[->] (-0.3,0.1) -- (-1.7,0.1);
\draw[<-] (-0.3,-0.1) -- (-1.7,-0.1);
\node[above] at (-1,0.1) {$\mathcal{W}^{++}$};
\node[below] at (-1,-0.1) {$\mathcal{W}^{--}$};
\node[above] at (1,0.1) {$\mathcal{W}^{++} \mathcal{S}_2$};
\node[below] at (1,-0.1) {$\mathcal{S}_2 \mathcal{W}^{--}$};
\end{tikzpicture}
\end{equation}
It is always important to remember that the action of the weight-shifting operators on an exchange diagram results in a combination of shifted exchange and contact diagrams.

\subsubsection{The scheme}

At first glance, all exchange amplitudes with a fixed exchange dimension $\Delta_x$ are related through repeated use of the operations in \eqref{red_ops}. On practical level, however, this turns out not to be the case. In order for the reduction scheme to work, three conditions must be satisfied:
\begin{itemize}
\item[{\bf 1.}] One must know all contact diagrams featuring on the right hand side of \eqref{Wpplincombsigma}.  These are easy to calculate directly and thus pose no difficulty. 

Alternatively,  one can attempt to construct them via the shift relation \eqref{eqshift3} starting from $\ireg_{[2323]}$. This contact diagram can itself be obtained from either $\ireg_{[2323x2]}$ or $\ireg_{[2323x3]}$ via \eqref{Casonex}. 
First, we evaluate
 \begin{align}
 (\mathcal{W}_{24}^{--}\ireg_{[2323]}) + (\bs{k}_2\leftrightarrow \bs{k}_3)+ (\bs{k}_3\leftrightarrow \bs{k}_4)&=-(2+\ep)(5+2\ep)\ireg_{[2222]},\\[1ex]
 (\mathcal{W}_{13}^{++}\ireg_{[2323]}) + (\bs{k}_2\leftrightarrow \bs{k}_3)+ (\bs{k}_3\leftrightarrow \bs{k}_4)&=\ep(3-2\ep)\ireg_{[3333]}.
 \end{align}
In this last equation, $\mathcal{W}^{++}_{13}$ projects out the leading $\ep^{-1}$ divergence of $\ireg_{[2323]}$ meaning the left-hand side is of order $\ep^0$.  We can therefore only construct the leading $\ep^{-1}$ divergence of $\ireg_{[3333]}$ from this relation.
Next, via the shadow transform  \eqref{shadowK2}, we find 
\[
\ireg_{[1323]}= \mathcal{S}_1 \ireg_{[2323]} =\frac{4^{\beta_1} \Gamma(\beta_1)}{\Gamma(-\beta_1)}k_1^{-2\beta_1}\ireg_{[2323]}= -\frac{1}{k_1}\ireg_{[2323]}.
\]
We now construct
  \begin{align}
 (\mathcal{W}_{12}^{+-}\ireg_{[1323]}) + (\bs{k}_2\leftrightarrow \bs{k}_3)+ (\bs{k}_1\leftrightarrow \bs{k}_3)&=(3+6\ep+2\ep^2)\ireg_{[2223]},\\[1ex]
(\mathcal{W}_{23}^{++}\ireg_{[2223]}) + (\bs{k}_2\leftrightarrow \bs{k}_4)+ (\bs{k}_3\leftrightarrow \bs{k}_4)&=(-1+3\ep-2\ep^2)\ireg_{[2333]}.
 \end{align}
From $\ireg_{[2323]}$, we can thus obtain all the contact diagrams of weight two and three operators to finite order, with the exception of $\ireg_{[3333]}$ which we only know to leading $\ep^{-1}$ order.
The finite part of this integral can be found either by direction evaluation or through other shift operators as discussed in appendix \ref{app_shift_4K}.
 

\item[{\bf 2.}] Regularization must be taken care of. For example, consider $\mathcal{W}^{++}_{12} \ino_{[22,22x2]}$, which we denote by $\ino_{[\tilde{3} \tilde{3}, 22 x 2]}$. Since $\ino_{[22,22x2]}$ is finite, so is $\ino_{[\tilde{3} \tilde{3}, 22 x 2]}$. However, we see that both the exchange diagram $\ino_{[33,22x2]}$ in \eqref{idiv3322x2beta} as well as the contact diagram $\ino_{[3322]}$ in \eqref{idiv3322beta} are divergent. Nevertheless, their combination on the right-hand side of \eqref{Wpplincombsigma} is finite in any regularization scheme and
\begin{align}
& \left[ -1 + \frac{\ep}{2} ( 3u - 7 v_1 - 7 v_2 + v_x ) + O(\ep^2) \right] \ireg_{[33,22x2]} \nn\\
& \qquad\qquad\qquad + \left[ -\frac{1}{2} - \ep ( v_1 + v_2 ) + O(\ep^2) \right] \ireg_{[3322]} = \ino_{[\tilde{3} \tilde{3}, 22 x 2]}.
\end{align}
Thus, one can derive $\ireg_{[33,22x2]}$ regulated in an arbitrary scheme once the regulated contact diagram $\ireg_{[3322]}$ is known in that scheme. It is important to emphasize that, whenever regularization is required, \eqref{Wpplincombsigma} holds for regulated amplitudes. If the terms of order $\epsilon$ in the above expression were dropped, the equality would fail.

Indeed, if we naively apply weight-shifting operators to renormalized diagrams, due to the presence of counterterm contributions the two sides of  \eqref{Wpplincombsigma} will generally only agree up to some local terms. 
 For example,
\begin{align}
\mathcal{W}^{++}_{12} \iren_{[22,22x2]} & = - \iren_{[33,22x2]} - \frac{1}{2} \iren_{[3322]} + \frac{k_3 + k_4}{8} \left(3 + 2 \act_{[3322]}^{(1)} - 2 \act_{[33,22x2]}^{(1)} \right), \\
\mathcal{W}^{++}_{12} \iren_{[22,33x2]} & = - \iren_{[33,33x2]} - \frac{1}{2} \iren_{[3333]} + \frac{k_1^3 + k_2^3}{24} \left( 3 - 2 \act_{[3333]}^{(1)} + 2 \act_{[33,33x2]}^{(1)} \right) \nn\\
& \qquad\qquad + \frac{k_3^3 + k_4^3}{24} \left( - 3 - 2 \act_{[3333]}^{(1)} + 2 \act_{[33,33x2]}^{(1)} \right).
\end{align}
The moral is that
\eqref{Wpplincombsigma} holds for regulated diagrams and one should be wary of 
extrapolating this relation to renormalized diagrams.

\item[{\bf 3.}] The value of the constant $\mathcal{N}^{\sigma_1 \sigma_2}_{exch.}$ on the right hand side of \eqref{Wpplincombsigma} cannot vanish when $\ep \rightarrow 0$. The constant vanishes when \eqref{Nexchzero} is satisfied for unregulated parameters. In such a case the finite part of $\mathcal{W}^{\sigma_1 \sigma_2}_{12} \ireg_{[\Delta_1 \Delta_2, \Delta_3 \Delta_4 x \Delta_x]}$ does not allow for the determination of the finite part of the exchange diagram $\ireg_{[\Delta_1 + 1, \Delta_2 + 1, \Delta_3 \Delta_4 x \Delta_x]}$. In cases of $\Delta_j = 2,3$, $j=1,2,3,4,x$ considered in this paper and $d = 3$ this precludes the use of half of the operations listed in \eqref{red_ops}. The operations that still can be used in the scheme, depending on the value of the exchange dimension $\Delta_x$ are presented in the following table:
\begin{center}
\begin{tabular}{|c|c|c|c|c|}
\hline
& $\mathcal{W}_{ij}$ & $\mathcal{W}^{--}$ & $\mathcal{W}^{++} \mathcal{S}$ & $\mathcal{S}\mathcal{W}^{--}$ \\ \hline
$\Delta_x = 2$ & yes & no & no & yes \\ \hline
$\Delta_x = 3$ & no & yes & yes & no \\ \hline
\end{tabular}
\end{center}
This means, for example, that the exchange diagram $\ireg_{[33,22x3]}$ cannot be obtained from $\ireg_{[22,22x3]}$ by the application of \eqref{Wpplincombsigma}. Indeed, in such a case $\mathcal{N}^{++}_{exch.} = \tfrac{3 \ep}{2} (u - v_1 - v_2 - v_x) + O(\ep^2)$ and vanishes in the $\ep \rightarrow 0$ limit.
\end{itemize}

Taking into account the above considerations, a workable reduction scheme connecting the various amplitudes is presented in Figures \ref{fig:reduction2} and \ref{fig:reduction3}.
Given the seed integral $\ireg_{[32,32x2]}$ we can construct all the integrals with $\Delta_x=2$, and from $\ireg_{[33,33x3]}$ we obtain all the integrals with $\Delta_x=3$.

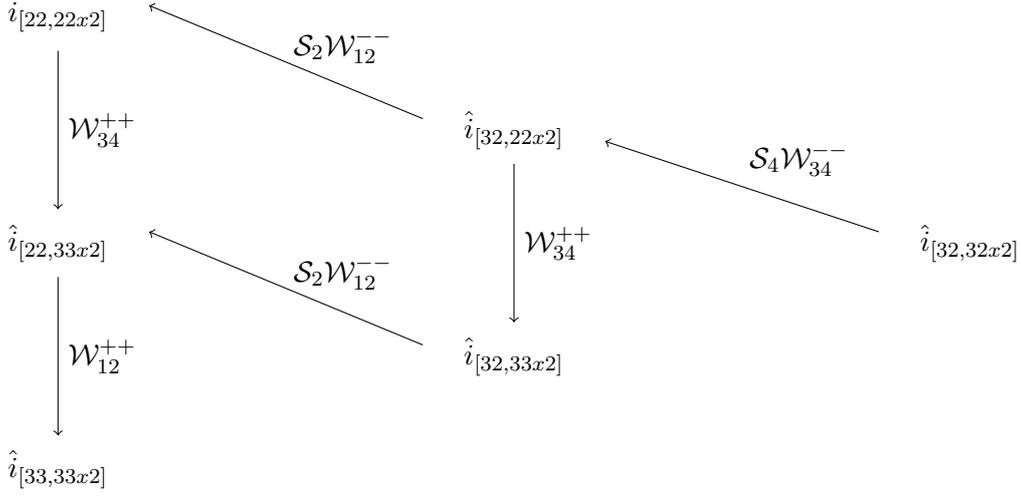
\begin{figure}[t]
\begin{tikzpicture}[scale=1.5]
\node at (-4,2) {$\ino_{[22,22x2]}$};
\draw[->] (-4,1.7) -- (-4,0.3);
\node[right] at (-4,1) {$\mathcal{W}^{++}_{34}$};
\node at (-4,0) {$\ireg_{[22,33x2]}$};
\draw[->] (-4,-0.3) -- (-4,-1.7);
\node[right] at (-4,-1) {$\mathcal{W}^{++}_{12}$};
\node at (-4,-2) {$\ireg_{[33,33x2]}$};
\node at (0,1) {$\ireg_{[32,22x2]}$};
\draw[->] (0,0.7) -- (0,-0.7);
\node[right] at (0,0) {$\mathcal{W}^{++}_{34}$};
\node at (0,-1) {$\ireg_{[32,33x2]}$};
\draw[<-] (-3.2,2.1) -- (-0.8,1.1);
\node[above] at (-1.5,1.5) {$\mathcal{S}_2 \mathcal{W}^{--}_{12}$};
\draw[<-] (-3.2,0.1) -- (-0.8,-0.9);
\node[above] at (-1.5,-0.5) {$\mathcal{S}_2 \mathcal{W}^{--}_{12}$};
\node at (4,0) {$\ireg_{[32,32x2]}$};
\draw[<-] (0.8,0.9) -- (3.2,0.1);
\node[above] at (2.5,0.5) {$\mathcal{S}_4 \mathcal{W}^{--}_{34}$};
\end{tikzpicture}
\centering
\caption{Reduction scheme for the amplitudes with $\Delta_x = 2$.\label{fig:reduction2}}
\end{figure}

\begin{figure}[t]
\begin{tikzpicture}[scale=1.5]
\node at (-4,2) {$\ino_{[22,22x3]}$};
\draw[<-] (-4,1.7) -- (-4,0.3);
\node[right] at (-4,1) {$\mathcal{W}^{--}_{34}$};
\node at (-4,0) {$\ireg_{[22,33x3]}$};
\draw[<-] (-4,-0.3) -- (-4,-1.7);
\node[right] at (-4,-1) {$\mathcal{W}^{--}_{12}$};
\node at (-4,-2) {$\ireg_{[33,33x3]}$};
\node at (0,1) {$\ireg_{[32,22x3]}$};
\draw[<-] (0,0.7) -- (0,-0.7);
\node[right] at (0,0) {$\mathcal{W}^{--}_{34}$};
\node at (0,-1) {$\ireg_{[32,33x3]}$};
\draw[->] (-3.2,2.1) -- (-0.8,1.1);
\node[above] at (-1.5,1.5) {$\mathcal{W}^{++}_{12} \mathcal{S}_2$};
\draw[->] (-3.2,0.1) -- (-0.8,-0.9);
\node[above] at (-1.5,-0.5) {$\mathcal{W}^{++}_{12} \mathcal{S}_2$};
\node at (4,0) {$\ireg_{[32,32x3]}$};
\draw[->] (0.8,0.9) -- (3.2,0.1);
\node[above] at (2.5,0.5) {$\mathcal{W}^{++}_{34} \mathcal{S}_4$};
\end{tikzpicture}
\centering
\caption{Reduction scheme for the amplitudes with $\Delta_x = 3$.\label{fig:reduction3}}
\end{figure}
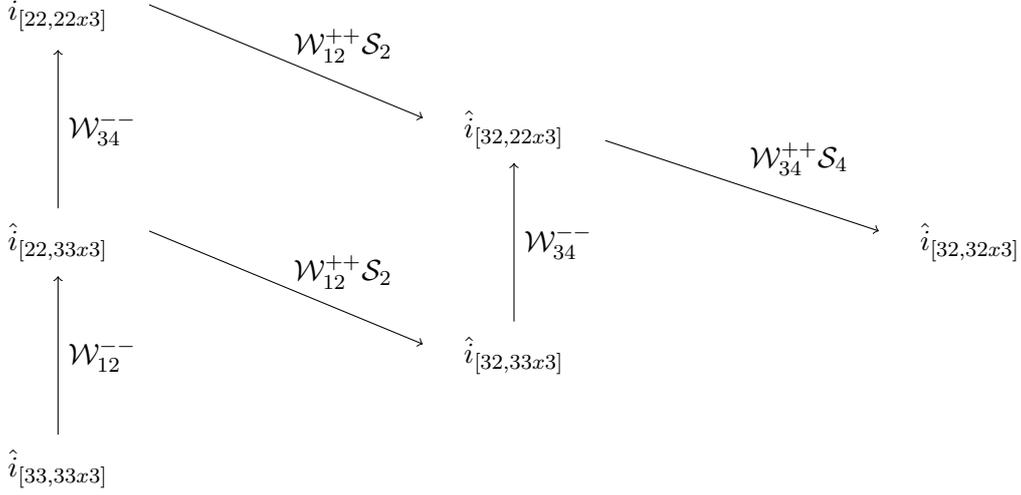

\subsubsection{Shifting the dimension of the exchanged operator}

To complete our analysis, we now seek a weight-shifting operator capable of connecting the two schemes in Figures \ref{fig:reduction2} and \ref{fig:reduction3}, so that only a single exchange diagram is required as a seed integral.  This requires shifting the dimension $\Delta_x$ of the exchanged operator. 

For exchange diagrams with all external dimensions equal to two, a shift operator sending $\Delta_x\rightarrow \Delta_x+1$  was given in \cite{Arkani-Hamed:2018kmz}.  A detailed derivation of this operator \eqref{2222shiftx} is given in  appendix \ref{cosmobootstrapops}.  Here, however, this operator is not suitable since the direction of the arrows in Figure \ref{fig:reduction3} is such that connecting $\ireg_{[22,22x2]}$ to $\ireg_{[22,22x3]}$ does not allow us to recover the three `upstream' integrals $\ireg_{[22,33x3]}$, $\ireg_{[33,33x3]}$ and $\ireg_{[32,33x3]}$.  We will therefore follow a different route, deriving a new shift operator based on the properties of Bessel functions that will allow us to recover $\ireg_{[32,32x2]}$, and hence all the $\Delta_x=2$ integrals, starting from the $\Delta_x=3$ integrals.

In the analysis to follow, it will be clearer to write the AdS propagators as functions of the spacetime dimension $d$ and the parameter $\beta_i=\Delta_i-d/2$.  To this end, we re-define
\begin{align}
\tilde{\mathcal{K}}_{d,\beta_i}(z,k_i) &=\frac{z^{d/2}k_i^\beta K_{\beta_i}(k_iz)}{2^{\beta_i-1}\Gamma(\beta_i)}, \\
\tilde{\mathcal{G}}_{d,\beta_x}(z,s;\zeta) &= (z\zeta)^{d/2}\Big(K_{\beta_x}(sz)I_{\beta_x}(s\zeta)\Theta(z-\zeta)+I_{\beta_x}(sz)K_{\beta_x}(s\zeta)\Theta(\zeta-z)\Big)
\end{align}
where the tildes remind us of the change of argument from $(d,\Delta_i)$ to $(d,\beta_i)$.
From the elementary properties of Bessel functions, the bulk-boundary propagator obeys the relation
\[
\tilde{\mathcal{K}}_{d-2,\, \frac{1}{2}}(z,k_i) =\tilde{\mathcal{K}}_{d, \,\frac{3}{2}}(z,k_i)-k_i \tilde{\mathcal{K}}_{d,\, \frac{1}{2}}(z,k_i). 
\]
Writing the exchange diagram as
\vspace{2mm}
\begin{align}
&\mathcal{I}^{(d)}_{[\beta_1,\beta_2;\beta_3,\beta_4;x\beta_x]} \nn\\[1ex]&\quad = \int_0^\infty \frac{\D z}{z^{d+1}}\int_0^\infty \frac{\D \zeta}{\zeta^{d+1}}\tilde{\mathcal{K}}_{d,\beta_1}(z,k_1)\tilde{\mathcal{K}}_{d,\beta_2}(z,k_2)\tilde{G}_{d,\beta_x}(z,s;\zeta)\tilde{\mathcal{K}}_{d,\beta_3}(\zeta,k_3)\tilde{\mathcal{K}}_{d,\beta_4}(\zeta,k_4),
\end{align}
and using this relation twice, we then find that 
\[\label{diffdrel1}
\mathcal{I}^{(d-2)}_{[\frac{3}{2},\frac{1}{2};\frac{3}{2},\frac{1}{2}; \,x\frac{3}{2}]} = \mathcal{I}^{(d)}_{[\frac{3}{2},\frac{3}{2};\frac{3}{2},\frac{3}{2}; \,x\frac{3}{2}]} -k_2\mathcal{I}^{(d)}_{[\frac{3}{2},\frac{1}{2};\frac{3}{2},\frac{3}{2}; \,x\frac{3}{2}]}-k_4 \mathcal{I}^{(d)}_{[\frac{3}{2},\frac{3}{2};\frac{3}{2},\frac{1}{2}; \,x\frac{3}{2}]}+k_2k_4 \mathcal{I}^{(d)}_{[\frac{3}{2},\frac{1}{2};\frac{3}{2},\frac{1}{2}; \,x\frac{3}{2}]}.
\]
Here, we exploited the fact that the factor of $(z\zeta)^{-d}$ from the two integration measures  cancels with the factors of $z^{d/2}$ and $\zeta^{d/2}$ from the first and third  bulk-boundary propagators and the factor of $(z\zeta)^{d/2}$ from the bulk-bulk propagator.  A shift $d\rightarrow d-2$ on the second and fourth bulk-boundary propagators thus extends to the entire exchange diagram.

Meanwhile,  the bulk-bulk propagator obeys the identity
\[
\frac{1}{2}\Big(z^2+\zeta^2-\partial_s^2+(2\sigma_x \beta_x-1)\frac{1}{s}\partial_s\Big)\tilde{\mathcal{G}}_{d-2,\beta_x}(z,s;\zeta) = \tilde{\mathcal{G}}_{d,\beta_x+\sigma_x}(z,s;\zeta),\quad \sigma_x=\pm 1,
\]
which shifts the spacetime dimension up by two while increasing or decreasing $\beta_x$ by one. 
To apply this operator to $s$-channel exchange diagrams, we further replace the factors of $z^2$ and $\zeta^2$ with  Bessel operators acting on the external legs,
\[
\Big(\partial_{i}^2 + \frac{1-2\beta_i}{k_i}\partial_{i}\Big) \tilde{\mathcal{K}}_{d,\beta_i}(z,k_i) = z^2 \tilde{\mathcal{K}}_{d,\beta_i}(z,k_i).
\]
So long as we choose one external leg corresponding to each vertex the result is the same.  Choosing legs 2 and 4, this gives
\begin{align}
&\mathcal{I}^{(d)}_{[\beta_1,\beta_2;\beta_3,\beta_4;x\beta_x+\sigma_x]}  \nn\\[1ex]&\quad =
\frac{1}{2}\Big(\partial_{2}^2 + \frac{1-2\beta_2}{k_2}\partial_{2}+\partial_{4}^2 + \frac{1-2\beta_4}{k_4}\partial_{4}-\partial_s^2+(2\sigma_x \beta_x-1)\frac{1}{s}\partial_s\Big)\mathcal{I}^{(d-2)}_{[\beta_1,\beta_2;\beta_3,\beta_4;x\beta_x]} 
\end{align}
Thus, applying this operator with $\sigma_x=-1$ to the left-hand side of \eqref{diffdrel1}, we find
\[\label{diffdrel2}
\mathcal{I}^{(d)}_{[\frac{3}{2},\frac{1}{2};\frac{3}{2},\frac{1}{2}; \,x\frac{1}{2}]}  =
\frac{1}{2}\Big(\partial_{2}^2 +\partial_{4}^2 -\partial_s^2-\frac{4}{s}\partial_s\Big)\mathcal{I}^{(d-2)}_{[\frac{3}{2},\frac{1}{2};\frac{3}{2},\frac{1}{2}; \,x\frac{3}{2}]}. 
\]

Using \eqref{diffdrel1} and \eqref{diffdrel2}, we can now obtain the $\Delta_x=2$ master integral $\ireg_{[32,32x2]}$ starting from the $\Delta_x=3$ master integral $\ireg_{[33,33x3]}$. 
In the first step, we compute all the $\Delta_x=3$ integrals using the reduction scheme illustrated in Figure \ref{fig:reduction3}.  The right-hand side of \eqref{diffdrel1} can then be constructed in the half-integer scheme:
\begin{align}
\mathcal{I}^{(1+2\ep)}_{[\frac{3}{2},\frac{1}{2};\frac{3}{2},\frac{1}{2}; \,x\frac{3}{2}]} &=\ireg_{[33,33x3]}-k_2 \ireg_{[32,33x3]}-k_4\ireg_{[33,32x3]}+k_2k_4 \ireg_{[32,32x3]}.
\end{align}
Equation \eqref{diffdrel2} now gives the desired result
\begin{empheq}[box=\nicebox]{align}
\ireg_{[32,32x2]} = \frac{1}{2}\Big(\partial_{2}^2 +\partial_{4}^2 -\partial_s^2-\frac{4}{s}\partial_s\Big)\Big(\ireg_{[33,33x3]}-k_2 \ireg_{[32,33x3]}-k_4\ireg_{[33,32x3]}+k_2k_4 \ireg_{[32,32x3]}\Big).
\end{empheq}
Using this relation in combination with the reduction schemes in Figures \ref{fig:reduction2} and \ref{fig:reduction3} we can thus obtain all exchange diagrams starting from only the exchange diagram $\ireg_{[33,33x3]}$ and contact diagrams.  As the most complicated integral, $\ireg_{[33,33x3]}$ is the natural starting point since the action of differential operators results in a reduction in complexity ({\it i.e.,} a lowering of the degree of transcendality and/or a reduction in the order of polynomial coefficients).

\section{Mathematica notebooks} \label{sec:Math}

All our results and their derivations can be found in the Mathematica notebooks included in the arXiv submission of this paper.\footnote{If any discrepancies exist as a result of typos, the expressions in these notebooks are definitive.}   Explicit expressions for the regulated and renormalized amplitudes are stored in the Mathematica package \verb|HandbooK.wl|, whose contents are described 
in Section \ref{sec:package} below.  
The package is accompanied by a number of  notebooks where details of the remaining calculations described in this paper can be found. We summarize the contents of these notebooks in Section \ref{sec:notebooks}.

\subsection{The package} \label{sec:package}

The package \verb|HandbooK.wl| provides  explicit expressions for both regulated and renormalized 2-, 3-, and 4-point functions.

\begin{itemize}
\item The package does not contain a dedicated installer and is loaded instead through Mathematica's \verb|Get| command:

\includegraphics*[scale=0.9, trim=2.0cm 24.8cm 0cm 2.3cm]{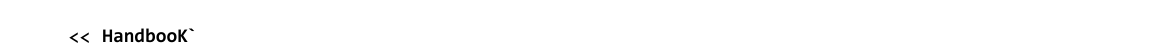}

\item All the regulated and renormalized expressions listed in the previous sections can be accessed through the following commands:

\hspace{-1.1cm}\begin{tabular}{|c|c|c|}
\hline
Amplitude & Regulated & Renormalized \\ \hline
$\ino_{[\Delta_1 \Delta_2]}$ & \verb|iReg2pt[|$\dreg, \{ \Dreg_1, \Dreg_2 \}$\verb|]| & \verb|iRen2pt[|$d, \{ \Delta_1, \Delta_2 \}$\verb|]| \\ \hline
$\ino_{[\Delta_1 \Delta_2 \Delta_3]}$ & \verb|iReg3pt[|$\dreg, \{ \Dreg_1, \Dreg_2, \Dreg_3 \}$\verb|]| & \verb|iRen3pt[|$d, \{ \Delta_1, \Delta_2, \Delta_3 \}$\verb|]| \\ \hline
$\jno_{[\Delta_1 \Delta_2; \Delta_3]}$ & \verb|jReg3pt[|$\dreg, \{ \Dreg_1, \Dreg_2, \Dreg_3 \}$\verb|]| & --- \\ \hline
$\ino_{[\Delta_1 \Delta_2 \Delta_3 \Delta_4]}$ & \verb|iReg4ptC[|$\dreg, \{ \Dreg_1, \Dreg_2, \Dreg_3, \Dreg_4 \}$\verb|]| & \verb|iRen4ptC[|$d, \{ \Delta_1, \Delta_2, \Delta_3, \Delta_4 \}$\verb|]| \\ \hline
$\ino_{[\Delta_1 \Delta_2, \Delta_3 \Delta_4 x \Delta_x]}$ & \verb|iReg4ptX[|$\dreg, \{ \Dreg_1, \Dreg_2, \Dreg_3, \Dreg_4, \Dreg_x \}$\verb|]| & \verb|iRen4ptX[|$d, \{ \Delta_1, \Delta_2, \Delta_3, \Delta_4, \Delta_x \}$\verb|]| \\ \hline
\end{tabular}

\item For example, the amplitude $\ireg_{[222]}$ regulated in the half-integer scheme with $u = 1$ and $v_j = 0$ can be obtained by the command

\includegraphics*[scale=0.9, trim=2.0cm 23.8cm 0cm 2.3cm]{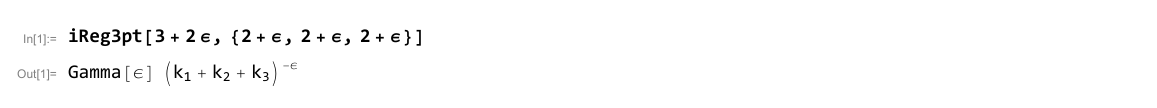}

By default, the regulator is represented by $\ep$ while the momentum magnitudes are $k_1, k_2, k_3, k_4$ and $s$ for the Mandelstam variable.

If no regulator is specified, the half-integer scheme is assumed by default. Thus, for example, \verb|iReg3pt[3, {2,2,2}]| is equivalent to \verb|iReg3pt[|$3 + 2 \ep, \{ 2 + \ep, 2 + \ep, 2 + \ep \}$\verb|]| evaluated above and returns the amplitude in the half-integer scheme.

\item By default the amplitude is expanded to the highest available order in the regulator. Since the 3-point amplitudes regulated in the half-integer scheme are exact, the expression above is not expanded in $\ep$ at all. In a general regularization scheme, the expressions are expanded to order $\ep^1$:

\includegraphics*[scale=0.9, trim=2.0cm 19.3cm 0cm 2.3cm]{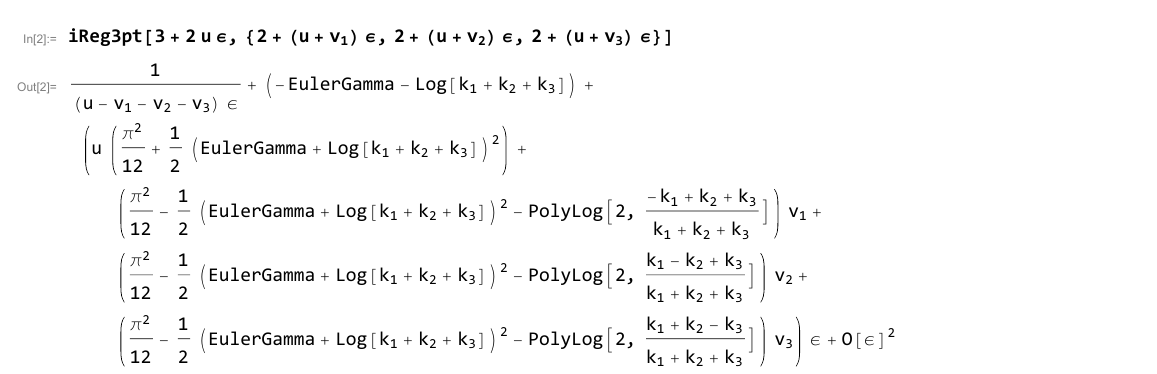}

\item The default settings can be adjusted by passing the following options to \verb|iReg3pt|, \verb|jReg3pt|, \verb|iReg4ptC| and \verb|iReg4ptX|:
\begin{center}
\begin{tabular}{|l|l|}
\hline
Option & Description \\ \hline
\verb|Regulator| & Symbol for the regulator \\ \hline
\verb|Momenta| & A list of symbols for the momentum magnitudes \\ \hline
\verb|ExpansionOrder| & The expansion order in the regulator \\ \hline
\verb|DefaultRegularization| & Default regularization when no regulator is specified \\ \hline
\end{tabular}
\end{center}

Alternatively, say we are interested in the amplitude $\ireg_{[222]}$ regulated in the scheme $\dreg = 3 + \delta$ and $\Dreg_j = 2$ for $j=1,2,3$, where $\delta$ is the regulator.  Moreover, we want the expansion to include finite and divergent terms, \textit{i.e.}, terms up to order $\ep^0$, and with momentum magnitudes denoted by $p_1, p_2$ and $p_3$. This can be achieved by:

\includegraphics*[scale=0.9, trim=2.0cm 23.5cm 0cm 2.3cm]{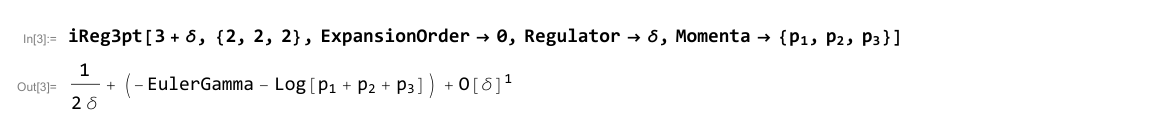}

\item The explicit expressions for 4-point exchange  amplitudes can be quite involved, particularly in a general regularization scheme. This is due to the use of the equation \eqref{change_scheme_4} and the complicated structure of the scheme changing terms $\idiv(u, v_j) - \idiv(1, 0)$. As an example, consider the amplitude $\ireg_{[33,22x2]}$ regulated in the scheme where $\dreg = 3 - \ep$ while conformal dimension $\Delta_j$ are not regulated for all $j=1,2,3,4,x$. The amplitude can be accessed by the command

\includegraphics*[scale=0.9, trim=2.0cm 24.5cm 0cm 2.3cm]{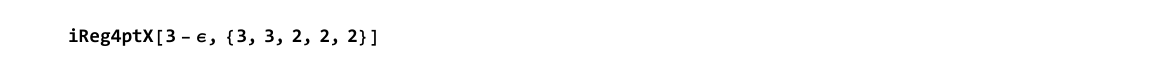}

By passing the option \verb|KeepHeld->True| one can prevent the  scheme-changing terms from expanding.  This produces results similar in presentation to those given in Section \ref{sec:scheme_change_results}, as illustrated in Figure \ref{KeepHeldFig}.
\begin{figure}[ht]
\includegraphics*[scale=0.9, trim=2.0cm 12.5cm 0cm 2.3cm]{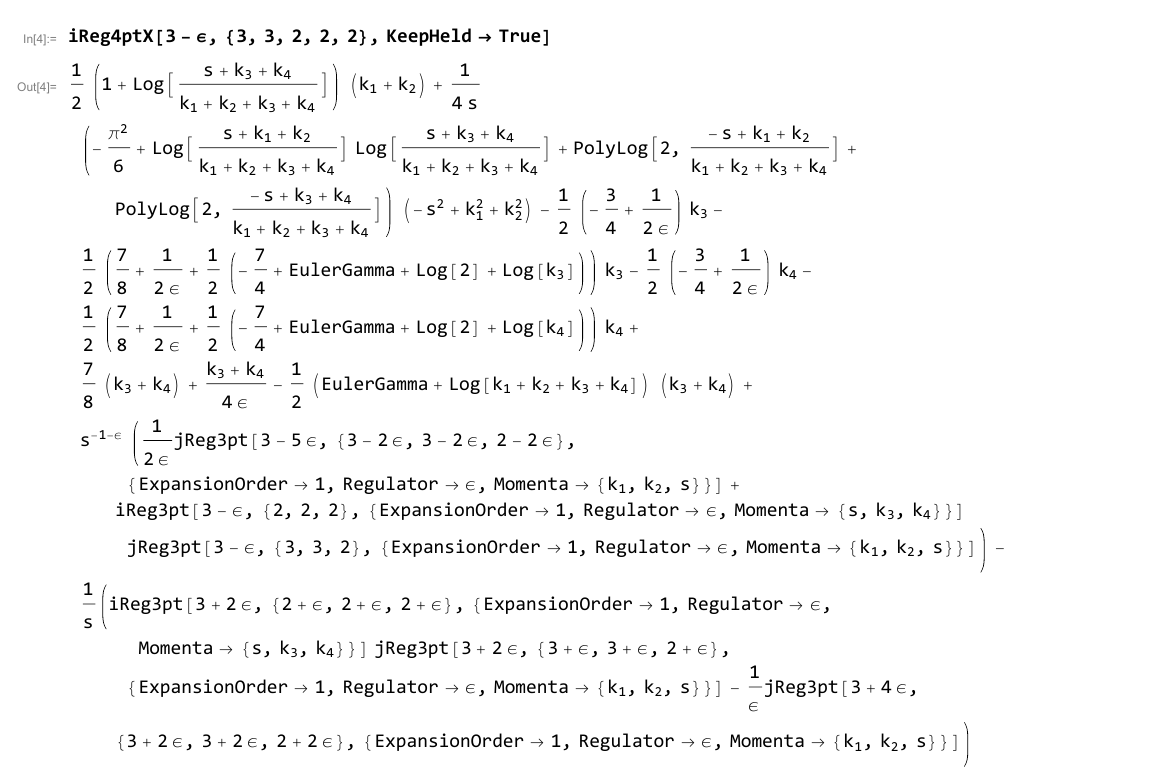}
\caption{Suppressing the expansion of scheme-changing terms. \label{KeepHeldFig}
}
\end{figure} 

To fully expand these expressions, use  Mathematica's \verb|ReleaseHold| command.

\item Renormalized amplitudes are accessed in a similar fashion, but no regulator is present. For $\iren_{[222]}$, for example,  we get

\includegraphics*[scale=0.9, trim=2.0cm 23.5cm 0cm 2.3cm]{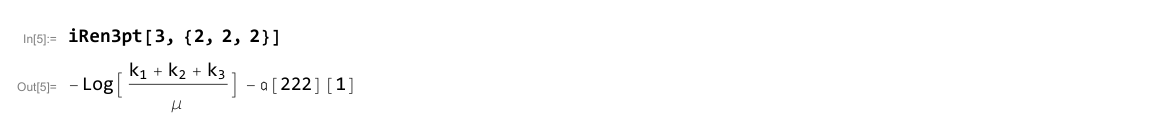}

\item In renormalized amplitudes, by default  the renormalization scale is denoted by $\mu$ while renormalization constants are denoted by $\mathfrak{a}$. These can be changed by  the following options passed to the functions:
\begin{center}
\begin{tabular}{|l|l|}
\hline
Option & Description \\ \hline
\verb|RenormalizationScale| & Symbol for the renormalization scale \\ \hline
\verb|RenormalizationConstant| & Symbol for the renormalization constants \\ \hline
\verb|Momenta| & A list of symbols for the momentum magnitudes \\ \hline
\end{tabular}
\end{center}

\end{itemize}

\subsection{Notebooks} \label{sec:notebooks}

The package is accompanied by a number of Mathematica notebooks which derive the results
presented in this paper.

\begin{itemize}
\item \verb|BetaScheme.nb| contains calculations of 2-, 3-, and 4-point functions regulated in the half-integer scheme \eqref{special}. The notebook requires the \verb|HypExp| package \cite{Huber:2005yg}, which is included in the packet. The raw results of the calculations are then saved to \verb|Results_BetaScheme.nb|.

\item  \verb|GeneralScheme.nb| contains calculations of 2-, 3-, and 4-point functions regulated in a general scheme. It also contains calculations of the 3-point amplitudes $\ireg_{[\Delta_1 \Delta_2 \Delta_3]}$ and $\jreg_{[\Delta_1 \Delta_2 \Delta_3]}$ including terms of order $\ep^1$ in the regulator. The notebook requires the \verb|HypExp| package as well as the \verb|TripleK| package  \cite{Bzowski:2020lip}, both of which are included in the packet. The results of the calculations are saved to \verb|Results_GeneralScheme.nb|.

\item \verb|Checks.nb| provides checks on the results contained in \verb|Results_BetaScheme.nb| and \verb|Results_GeneralScheme.nb|. Furthermore, it compares these raw results with the expressions presented in Sections \ref{sec:reg_amp} and \ref{sec:scheme_change_results} and stored in the package file \verb|HandbooK.wl|.

\item  \verb|Renormalization.nb| contains the results presented in Section \ref{sec:renormalization}. First, we check that the regulated expressions are renormalized by the counterterms listed, and that they produce the renormalized amplitudes presented in Section \ref{sec:ren_amp} and stored in \verb|HandbooK.wl|. Next, we calculate the beta functions of the symmetric theory from Section \ref{sec:beta}, and finally, we check the Callan-Symanzik equation for all 3- and 4-point functions in the symmetric model.

\end{itemize}

\section{Outlook}\label{sec:outlook}

This handbook has presented explicit closed-form expressions for all renormalized holographic $4$-point functions derived from contact and exchange diagrams of scalar operators with dimensions $\Delta=2,3$ in $d=3$.  The results were obtained using dimensional regularization. We saw that the  half-integer scheme \eqref{special}  is very convenient for carrying out explicit computations, and we also discussed how to change scheme and presented our results both in the half-integer  scheme and the general scheme \eqref{genreg}.
We believe this is the most extensive set of renormalized correlators of this kind currently available.
Besides the listings given in the text, all our results are available in the accompanying Mathematica notebooks as explained in the previous section.

There are many natural extensions of our work:
\begin{itemize}
\item Generalizations to cases involving both external and internal operators with spin.
\item Analysis of even-dimensional spacetimes including $d=4$.
\item Computations for  more general operator dimensions. 
\end{itemize} 
As we have seen, there are also subtleties in using the OPE in momentum space and it would be interesting to understand how to set up  the corresponding bootstrap program.
Finally, as we hope to report soon \cite{toappear}, our results have immediate implications for the computation of wavefunction coefficients and correlators in de Sitter spacetime.

\subsection*{Acknowledgments}
AB is supported by the NCN POLS grant No.~2020/37/K/ST2/02768 financed from the Norwegian Financial Mechanism 2014-2021 \includegraphics[width=12pt]{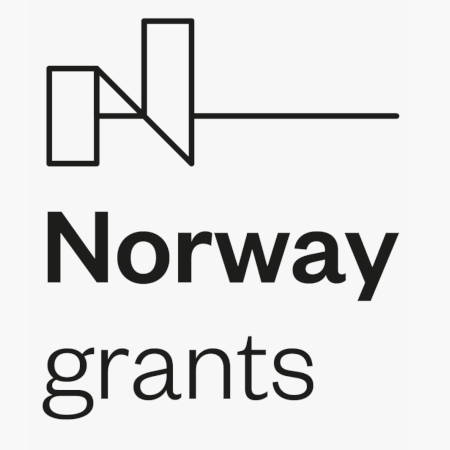}. PM is supported by the Science and Technology Facilities Council through an Ernest Rutherford Fellowship (ST/P004326/2).  KS is supported in part by the Science and Technology Facilities Council (Consolidated Grant ST/T000775/1).

\appendix

\section{Conventions and definitions}\label{sec:conventions}

\subsection{QFT conventions}

\begin{itemize}
\item We work exclusively in $d = 3$ Euclidean spacetime dimensions. 
\item Scaling dimensions are denoted throughout via square brackets, including where they appear as superscripts on operators and sources.  We assign the dimensions $[\bs{x}] = -1$ and  $[\partial_\mu] = 1$ for coordinates and their derivatives, $[\bs{k}] = 1$ for momenta,  $[\mu] = 1$ for the RG scale, $[\O_j] = \Delta$ for operators and $[\phi_j] = d - \Delta$ for their sources.

\item Our main focus will be scalar operators $\O^{[2]}$ and  $\O^{[3]}$ of dimensions $\Delta=2,3$,  their sources $\phi^{[1]}, \phi^{[0]}$ and the holographically dual bulk fields $\Phi^{[2]}$ and  $\Phi^{[3]}$. 
\item Regarding the relation of the generating functional of connected diagrams $W$ to the action $S$, we follow the conventions of \cite{Skenderis:2002wp} and \cite{Bzowski:2016kni} and define $W$ as
\begin{align}
Z = e^W = \< e^{-S} \>
\end{align}
from which it follows that
\begin{align}
\< \O_1(\bs{x}_1) \ldots \O_n(\bs{x}_n) \> = (-1)^n \frac{\delta^n W}{\delta \phi_1(\bs{x}_1) \ldots \delta \phi_n (\bs{x}_n)}.
\end{align}

\item The most general regularization scheme we use is \eqref{genreg}.  
The parameterization is such that the natural parameters arising in holographic calculations \cite{Bzowski:2015pba},
\begin{align} \label{alphabeta}
& \alpha = \frac{d}{2} - 1, && \beta_j = \Delta_j - \frac{d}{2}
\end{align}
are regulated according to
\begin{align}
& \alpha \longmapsto \areg = \alpha + u \ep, && \beta_j \longmapsto \breg_j = \beta_j + v_j \ep.
\end{align}
Here, $\beta_j$ is the index associated with the Bessel function representing the bulk-boundary propagator for external leg $j$.

Any quantity $f = f(d, \Delta_j)$ depending on the dimensions $d$ and $\Delta_j$ is regulated according to the selected scheme in \eqref{genreg}. We denote the regulated version of $f$ as $\reg{f}$, \textit{i.e.}, $\reg{f} = f(\dreg, \Dreg_j)$.
\item Mostly, though, we work in the special `half-integer' scheme \eqref{special}.  This  corresponds to setting $u = 1$ and $v_j = 0$ for all  $j = 1,2,3,4,x$ in the general scheme \eqref{genreg}.
In this scheme the value of the $\beta_j$ parameters 
do not change:
\begin{align}
& \alpha \longmapsto \areg = \alpha + \ep, && \beta_j \longmapsto \breg_j = \beta_j.
\end{align}
This is a good renormalization scheme in $d = 3$ for operators of dimension $\Delta = 2, 3$ in the sense that it regulates all correlation functions of such operators.

\item Divergences and scale-violating terms are parametrized by
\begin{align} \label{def:div}
\Div_a(k) & = \Gamma(a \ep) k^{-a \ep} \nn\\
& = \frac{1}{a \ep} - ( \log k + \gamma_E) + \frac{a \ep}{2} \left( \log k^2 + 2 \gamma_E \log k + \gamma_E^2 + \frac{\pi^2}{6} \right) + O(\ep^2),
\end{align}
where the momentum magnitude $k$ can also be replaced by the RG scale $\mu$.  

$\Div_a(k)$ satisfies the identity
\begin{align}
\Div_a(k) - \Div_a(\mu) & = - \log \left( \frac{k}{\mu} \right) + O(\ep).
\end{align}
\end{itemize}

\subsection{Definitions for momenta} \label{sec:notation}

\begin{itemize}
\item External momenta are denoted $\bs{k}_j$, with lengths or magnitudes $k_j = |\bs{k}_j|$, where $j=1,2,\ldots$. The Mandelstam variables are
\begin{align} \label{def:stu}
& s = | \bs{k}_1 + \bs{k}_2 |, && t = | \bs{k}_1 + \bs{k}_3 |, && u = | \bs{k}_2 + \bs{k}_3 |
\end{align}
without squares. For convenience, we also adopt the convention $k_s = s$.
\item The $3$- and $4$-point total magnitudes are denoted
\begin{align} \label{total}
& k_t = k_1 + k_2 + k_3, && k_T = k_1 + k_2 + k_3 + k_4.
\end{align}
\item A contact diagram with $n$ identical particles has symmetry group $S_n$ corresponding to permutations of the external momenta: $\bs{k}_j \mapsto \bs{k}_{\sigma(j)}$ for any $\sigma \in S_4$.  

\item We use $\s{m}{J}$ to denote  the corresponding  $m$-th symmetric polynomial on the set of indices $J$. 
 To be precise, let $J$ be an ordered set of indices and let $m$ be an integer such that $1 \leq m \leq |J|$. Then,
\begin{align} \label{def:sigma}
\s{m}{J} = \sum_{\substack{L \subseteq J\\|L|=m}} k_{L_1} \ldots k_{L_m},
\end{align}
where the sum is taken over all ordered subsets $L \subseteq J$ of cardinality $m$. In particular
\begin{align}
\s{1}{12} & = k_1 + k_2, & \s{1}{123} & = k_1 + k_2 + k_3, \\
\s{2}{12} & = k_1 k_2, & \s{2}{123} & = k_1 k_2 + k_1 k_3 + k_2 k_3, \\
&& \s{3}{123} & = k_1 k_2 k_3,
\end{align}
\begin{align}
\s{1}{1234} & = k_1 + k_2 + k_3 + k_4, \\
\s{2}{1234} & = k_1 k_2 + k_1 k_3 + k_1 k_4 + k_2 k_3 + k_2 k_4 + k_3 k_4, \label{def:sigma2} \\
\s{3}{1234} & = k_1 k_2 k_3 + k_1 k_2 k_4 + k_1 k_3 k_4 + k_2 k_3 k_4, \\
\s{4}{1234} & = k_1 k_2 k_3 k_4.
\end{align}
We also allow for the indices to take the value $s$, so that, for example, $\sigma_{(1)12s} = k_1 + k_2 + s$ and so on.

\item For 4-point exchange diagrams we define the following variables,
\begin{align}\label{mij}
& \m{ij} = k_i + k_j - | \bs{k}_i + \bs{k}_j|, && \p{ij} = k_i + k_j + | \bs{k}_i + \bs{k}_j|.
\end{align} 
In particular all 4-point exchange diagrams will contain the following combinations,
\begin{align}
& \m{12} = k_1 + k_2 - s, && \p{12} = k_1 + k_2 + s, \\
& \m{34} = k_3 + k_4 - s, && \p{34} = k_3 + k_4 + s,
\end{align}
which we will use in addition to the standard momentum variables. 

Note that all $\m{ij}$ and $\p{ij}$ are non-negative. Furthermore, $\p{ij} = 0$ corresponds to $\bs{k}_i = \bs{k}_j = 0$, while $\m{ij} = 0$ indicates the collinear limit, $\bs{k}_i \parallel \bs{k}_j$, when the two momenta are parallel.

\item The highest possible symmetry group of an exchange diagram $12 \mapsto 34$, arising when all external particles are identical, is the dihedral group $D_4$.  This contains the eight permutations generated by swapping the numbers within each pair, $12;34 \mapsto 21;34$ and $12;34 \mapsto 12;43$, as well as exchanging the pairs, $12;34 \mapsto 34;12$. 

The following dilogarithmic quantities then arise in exchange diagrams:
\begin{align}
\PLp & = \Li_2 \left( \frac{\m{34}}{k_T} \right) + \Li_2 \left( \frac{\m{12}}{k_T} \right) + \log \left( \frac{\p{12}}{k_T} \right) \log \left( \frac{\p{34}}{k_T} \right) - \frac{\pi^2}{6}, \label{defD+} \\
\PLm & = \Li_2 \left( \frac{\m{34}}{k_T} \right) - \Li_2 \left( \frac{\m{12}}{k_T} \right) + \frac{1}{2} \log^2 \left( \frac{\p{12}}{k_T} \right) - \frac{1}{2} \log^2 \left( \frac{\p{34}}{k_T} \right). \label{defD-}
\end{align}
$\PLp$ is invariant under the group $D_4$,
\begin{align}
\PLp(\bs{k}_1, \bs{k}_2; \bs{k}_3, \bs{k}_4) = \PLp(\bs{k}_2, \bs{k}_1; \bs{k}_3, \bs{k}_4) = \PLp(\bs{k}_3, \bs{k}_4; \bs{k}_1, \bs{k}_2),
\end{align}
while $\PLm$ acquires a sign when any two pairs of indices are exchanged,
\begin{align}
\PLm(\bs{k}_1, \bs{k}_2; \bs{k}_3, \bs{k}_4) = \PLm(\bs{k}_2, \bs{k}_1; \bs{k}_3, \bs{k}_4) = - \PLm(\bs{k}_3, \bs{k}_4; \bs{k}_1, \bs{k}_2).
\end{align}

\end{itemize}

\section{Conformal Ward identities}\label{CWIapp}

Dilatation and special conformal Ward identities for scalar operators $\O_1, \ldots, \O_n$ in momentum space can be found in \cite{Bzowski:2013sza} and read
\begin{align}
0 & = \left[ \sum_{j=1}^n \Delta_j - (n - 1) d - \sum_{j=1}^{n-1} k_j^\alpha \frac{\partial}{\partial k_j^\alpha} \right] \lla \mathcal{O}_1(\bs{k}_1) \ldots \mathcal{O}_n(\bs{k}_n) \rra, \label{DWI} \\
0 & = \left[ \sum_{j=1}^{n-1} \left( 2 (\Delta_j - d) \frac{\partial}{\partial k_j^\kappa} - 2 k_j^\alpha \frac{\partial}{\partial k_j^\alpha} \frac{\partial}{\partial k_j^\kappa} + (k_j)_\kappa \frac{\partial}{\partial k_j^\alpha} \frac{\partial}{\partial k_{j \alpha}} \right) \right] \lla \mathcal{O}_1(\bs{k}_1) \ldots \mathcal{O}_n(\bs{k}_n) \rra. \label{SCWI}
\end{align}
In the absence of renormalization effects, the dilatation Ward identity fixes the total dimension of the $n$-point function to $\sum_{j=1}^n \Delta_j - (n - 1) d$.

Defining the operators
\begin{align}
\K_{\beta}(k) & = \frac{\partial^2}{\partial k^2} + \frac{1- 2\beta}{k} \frac{\partial}{\partial k}, & L_{\Delta}(k) & = k \frac{\partial}{\partial k} - \Delta,
\end{align}
an independent set of special conformal Ward identities for 2-, 3-, and 4-point functions can be written as follows:
\begin{itemize}
\item For 2-point functions, we have the single equation
\begin{align}
0 & = \K_{\Delta - \frac{d}{2}}(k) \lla \O_{[\Delta]}(\bs{k}) \O_{[\Delta]}(-\bs{k}) \rra. \label{scwi2}
\end{align}
\item For 3-point functions, we have two independent equations
\begin{align}
0 & = \left[ \K_{\Delta_i - \frac{d}{2}}(k_i) - \K_{\Delta_j - \frac{d}{2}}(k_j) \right] \lla \O_{[\Delta_1]}(\bs{k}_1) \O_{[\Delta_2]}(\bs{k}_2) \O_{[\Delta_3]}(\bs{k}_3) \rra, \label{scwi3}
\end{align}
with $i,j=1,2,3$, where we treat the 3-point function as a function of three independent magnitudes $k_1$, $k_2$, and $k_3$.
\item For 4-point functions there are three independent equations. In \cite{Arkani-Hamed:2018kmz}, these are chosen as
\begin{align}
0 & = (D_m - D_n) \lla \O_{[\Delta_1]}(\bs{k}_1) \O_{[\Delta_2]}(\bs{k}_2) \O_{[\Delta_3]}(\bs{k}_3) \O_{[\Delta_4]}(\bs{k}_4) \rra, \label{scwi4a}
\end{align}
where the 4-point function is treated as the function of six scalar parameters: four magnitudes $k_1, k_2, k_3, k_4$ as well as the Mandelstam variables $s$ and $u$ and
\begin{align}
D_1 & = \frac{\partial^2}{\partial k_1^2} + \frac{1}{s} \frac{\partial}{\partial s} \left( k_1 \frac{\partial}{\partial k_1} + k_2 \frac{\partial}{\partial k_2} \right) + \frac{1}{u} \frac{\partial}{\partial u} \left( k_1 \frac{\partial}{\partial k_1} + k_4 \frac{\partial}{\partial k_4} \right) - \frac{k_3^2}{s u} \frac{\partial^2}{\partial s \partial u} \nn\\
& \qquad - 2 \left( \Delta_1 - \frac{d}{2} - \frac{1}{2} \right) \frac{1}{k_1} \frac{\partial}{\partial k_1} - ( \Delta_1 + \Delta_2) \frac{1}{s} \frac{\partial}{\partial s} - (\Delta_1 + \Delta_4) \frac{1}{u} \frac{\partial}{\partial u}.
\end{align}
The operators $D_2, D_3$ and $D_4$ are obtained by cyclic permutation of the momenta $\bs{k}_1, \bs{k}_2, \bs{k}_3, \bs{k}_4$ as well as the conformal dimensions $\Delta_1, \Delta_2, \Delta_3, \Delta_4$. Note that under a single cyclic permutation $s \leftrightarrow u$.

Another useful representation of the special conformal Ward identities is
\begin{align}
0 & = D_{ij} \lla \O_{[\Delta_1]}(\bs{k}_1) \O_{[\Delta_2]}(\bs{k}_2) \O_{[\Delta_3]}(\bs{k}_3) \O_{[\Delta_4]}(\bs{k}_4) \rra, \label{scwi4b}
\end{align}
where
\begin{align}
D_{12} & = \K_{\Delta_1 - \frac{d}{2}}(k_1) - \K_{\Delta_2 - \frac{d}{2}}(k_2) + ( - k_3^2 + k_4^2 ) \frac{1}{s t} \frac{\partial^2}{\partial s \partial t} \nn\\
& \qquad + \frac{1}{t} \frac{\partial}{\partial t} \left[ L_{\Delta_1}(k_1) - L_{\Delta_2}(k_2) - L_{\Delta_3}(k_3) + L_{\Delta_4}(k_4) \right].
\end{align}
Here we treat the 4-point function as the function of 4 magnitudes $k_1, k_2, k_3, k_4$ and the Mandelstam variables $s$ and $t$.
The other operators $D_{ij}$ are obtained by permuting the momenta and dimensions. Only three are independent, which can be chosen as $D_{12}, D_{23}, D_{34}$ corresponding to cyclic permutations.
\end{itemize}

When deriving the simplified expressions for special conformal Ward identities from \eqref{SCWI}, one uses repeatedly the dilatation Ward identity \eqref{DWI} and its derivatives. This is the reason why different (equivalent) representations of the special conformal Ward identities exist in the literature.

\section{Useful formulae} \label{sec:form}

The series expansion of the Bessel-$I$ function is
\begin{align} \label{I_Bessel}
I_{\beta}(z) & = \sum_{n=0}^{\infty} a_n(\beta) z^{\beta + 2n}, \qquad a_n(\beta)  = \frac{1}{2^{\beta + 2 n} n! \Gamma(\beta + n + 1)}.
\end{align}
Defining the Bessel-$K$ function via 
\begin{align} \label{defBesselK}
K_{\beta}(z) & = \frac{\pi}{2} \frac{I_{-\beta}(z) - I_{\beta}(z)}{\sin (\beta \pi)},
\end{align}
we can concentrate on $\beta \geq 0$ since $K_{\beta}(z) = K_{-\beta}(z)$. Thus, the series expansion of the bulk-boundary propagator \eqref{KPropagator} around $z = 0$ equals
\begin{align} \label{Iexp}
\K_{d, \Delta}(z, k) & = \sum_{\sigma = \pm 1} \sum_{n=0}^{\infty} b_{n}(\sigma \beta) z^{\sigma \beta + 2 n}, \\ 
b_n(\nu) & = \frac{(-1)^n \Gamma(-\nu - n)}{n! \Gamma(|\nu|)} \left( \frac{p}{2} \right)^{2 n + \nu + |\nu|}, \label{bcf}
\end{align}
where $\beta = \Delta - \frac{d}{2}$ as usual and $\beta \notin \Z$.

The derivatives of the Bessel functions with respect to the order read
\begin{align}
\left. \frac{\partial K_{\beta}(z)}{\partial \beta} \right|_{\beta = \frac{1}{2}} & = \sqrt{\frac{\pi}{2 z}} E_1(2 z) e^z, \label{dKdb12} \\ 
\left. \frac{\partial I_{\beta}(z)}{\partial \beta} \right|_{\beta = \frac{1}{2}} & = - \frac{1}{\sqrt{2 \pi z}} \left[ E_1(2 z) e^z + \Ei(2 z) e^{-z} \right], \label{dIdb12}
\end{align}
and
\begin{align}
\left. \frac{\partial K_{\beta}(z)}{\partial \beta} \right|_{\beta = \frac{3}{2}} & = \sqrt{\frac{\pi}{2 z^3}} \left[ 2 e^{-z} + e^z (1 - z) E_1(2z) \right], \label{dKdb32} \\
\left. \frac{\partial I_{\beta}(z)}{\partial \beta} \right|_{\beta = \frac{3}{2}} & = - \frac{1}{\sqrt{2 \pi z^3}} \left[ 4 \sinh z + e^z (z - 1) E_1(2z) - e^{-z} (z + 1) \Ei(2z) \right], \label{dIdb32}
\end{align}
where\footnote{See, {\it e.g.,} functions.wolfram.com.}
\begin{align}
& E_1(z) = \int_z^{\infty} \D t\,\frac{e^{-t}}{t}, && \Ei(z) = - P.V. \int_{-z}^{\infty}\D t\, \frac{e^{-t}}{t}.
\end{align}
 The two exponential integrals are related, but their continuations to negative arguments differ. For $x > 0$, one has the relations
\begin{align}
\Ei(-x) & = - E_1(x), & E_1(-x \pm \I 0) = - \Ei(x) \mp \I \pi.
\end{align}

Finally, we have the integrals
\begin{align}
\int_0^{\infty} \D z \, z^{\alpha - 1} e^{-\mu z} E_1(2 p z) & = \frac{\Gamma(\alpha)}{\alpha (\mu + 2 p)^{\alpha}} \, {}_2 F_1 \left(1, \a; \a + 1; \frac{\mu}{\mu + 2 p} \right), \label{intE1} \\
\int_0^{\infty} \D z \, z^{\alpha - 1} e^{-\mu z} \Ei(- 2 p z) & = - \frac{\Gamma(\alpha)}{\alpha (\mu + 2 p)^{\alpha}} \, {}_2 F_1 \left(1, \a; \a + 1; \frac{\mu}{\mu + 2 p} \right) \nn\\
& \qquad\qquad - \frac{\I \pi \Gamma(\alpha)}{\mu^{\alpha}} \theta(-p). \label{intEi}
\end{align}

\section{Further shift operators} \label{appD}

In this appendix, we collect together some further results on weight-shifting operators supplementary to our discussion in Section \ref{sec:weightshift}.  Section \ref{app_shift_4K} presents additional shift operators for contact diagrams based on their representation as a quadruple Bessel integral.  Section \ref{cosmobootstrapops}
shows  the weight-shifting shift operators employed in the cosmological bootstrap paper \cite{Arkani-Hamed:2018kmz}
are a special case of the $\mathcal{W}^{\sigma_i\sigma_j}_{ij}$ operator discussed here.  We also give an explicit derivation of an additional weight-shifting operator for exchange diagrams proposed in \cite{Arkani-Hamed:2018kmz}.  This operator acts to shifts the exchange dimension provided all external dimensions are equal to two.
 
\subsection{ Contact diagrams of pairwise-equal shifted dimension}
\label{app_shift_4K}

In analogy to the triple-$K$ integrals of \cite{Bzowski:2013sza}, let us define the quadruple-$K$ integrals as
\begin{align}
I_{\alpha \{ \beta_1 \beta_2 \beta_3 \beta_4 \}}(k_1, k_2, k_3, k_4) = \int_0^{\infty} \D x \, x^{\alpha} \prod_{j=1}^4 k_j^{\beta_j} K_{\beta_j}(k_j x).
\end{align}
From \eqref{amp4c} we see that the parameter $\alpha=d-1$ where $d$ is  the spacetime dimension. However, by keeping $\alpha$ (or $d$) general we can use additional identities which relate quadruple-$K$ integrals in different spacetime dimensions. Our 3-dimensional amplitudes are then
\begin{align}
\ino_{[\Delta_1 \Delta_2 \Delta_3 \Delta_4]} = \frac{I_{2 + 2 \ep \{ \beta_1 \beta_2 \beta_3 \beta_4 \}}}{2^{\beta_t - 4} \Gamma(\beta_1)\Gamma(\beta_2)\Gamma(\beta_3)\Gamma(\beta_4)}.
\end{align}

The identities between quadruple-$K$ integrals can be derived from the identities satisfied by the Bessel functions following the same reasoning as in Section 3.2 of \cite{Bzowski:2015yxv}. By employing the identities
\begin{align}
& \frac{\partial}{\partial k} \left[ k^{\nu} K_{\nu}(k x) \right] = - x k^\nu K_{\nu - 1}(k x), && K_{-\nu}(x) = K_{\nu}(x)
\end{align}
we obtain two relations, both raising the value of $\alpha$ by one,
\begin{align} \label{incAdecB}
I_{\alpha + 1 \{ \beta_1 - 1, \beta_2 \beta_3 \beta_4 \}} & = - \frac{1}{k_1} \frac{\partial}{\partial k_1} I_{\alpha \{ \beta_1 \beta_2 \beta_3 \beta_4 \}}, \\
I_{\alpha + 1 \{ \beta_1 + 1, \beta_2 \beta_3 \beta_4 \}} & = \left( 2 \beta_1 - k_1 \frac{\partial}{\partial k_1} \right) I_{\alpha \{ \beta_1 \beta_2 \beta_3 \beta_4 \}}. \label{incAincB}
\end{align}
To derive a relation decreasing the value of $\alpha$ consider the following integral
\begin{align}
& \int_0^{\infty} \D x \, \frac{\partial}{\partial x} \left[ x^{\alpha} \prod_{j=1}^4 k_j^{\beta_j} K_{\beta_j}(k_j x) \right] = (\alpha - \beta_t) I_{\alpha - 1 \{ \beta_1 \beta_2 \beta_3 \beta_4 \}} \nn\\
& \qquad\qquad - \left[ k_1^2 I_{\alpha \{ \beta_1-1, \beta_2 \beta_3 \beta_4 \}} + k_2^2 I_{\alpha \{ \beta_1, \beta_2 - 1, \beta_3 \beta_4 \}} + k_3^2 I_{\alpha \{ \beta_1 \beta_2, \beta_3-1, \beta_4 \}} + k_4^2 I_{\alpha \{ \beta_1 \beta_2 \beta_3, \beta_4 - 1 \}} \right],
\end{align}
where $\beta_t = \beta_1 + \beta_2 + \beta_3 + \beta_4$. For generic (regulated) values of $\alpha$ and $\beta_j$ parameters the left hand side vanishes. Thus we obtain,
\begin{align} \label{decAincB}
I_{\alpha \{ \beta_1 \beta_2 \beta_3 \beta_4 \}} = \frac{1}{\alpha + 1 - \beta_t} \sum_{j=1}^4 k_j^2 I_{\alpha + 1 \{ \beta_i - \delta_{ij} \} } 
\end{align}
We can combine this identity with \eqref{incAdecB} and \eqref{incAincB} in such a way that the value of the parameter $\alpha$ stays the same. Since in general the amplitudes with larger conformal dimensions are more complicated, we will focus on increasing the conformal dimensions. Thus, we will only apply \eqref{incAincB}. Before we do this, however, note that for this identity to be useful, the constant $\alpha + 1 - \beta_t$ should not vanish. If this was the case, its regulated version, $\reg{\alpha} + 1 - \reg{\beta}_t$, would be of order $\ep^1$. This implies that the knowledge of the finite part of order $\ep^0$ of the integrals on the right-hand side of \eqref{decAincB} would only determine the divergent piece of the integral on the left hand side. This happens for $\beta_t = d$.

To proceed, let $\mathcal{N}_{\beta_t}$ be the constant which multiplies the sum in \eqref{decAincB},
\begin{align}
\mathcal{N}_{\beta_t} & = \frac{1}{3 + 2\ep - \beta_t} = \frac{1}{3 - \beta_t} - \frac{2 \ep}{(3 - \beta_t)^2} + O(\ep^2).
\end{align}
Furthermore, define the differential operators $\mathcal{L}_n, \mathcal{M}_n$ as follows,
\begin{align}
\mathcal{L}_n & = - \frac{1}{k_n} \frac{\partial}{\partial k_n}, \\
\mathcal{M}_n^{(\beta_n)} & = 2 \beta_n - k_n \frac{\partial}{\partial k_n}, 
\end{align}
These are the operators featuring on the right hand sides of \eqref{incAdecB} and \eqref{incAincB}. They are related by $k^{2(\beta+1)} \mathcal{L} ( k^{-2 \beta} f ) =  \mathcal{M}(\beta) f$, as follows by conjugating $\mathcal{L}$ with shadow transforms.

A priori the relation \eqref{decAincB} contains four different integrals on its right hand side. This number can be decreased in some special cases. For example, we can use \eqref{incAincB} to express both $I_{\alpha+1 \{ \beta_1 - 1, \beta_2 \beta_3 \beta_4\}}$ and $I_{\alpha +1\{ \beta_1, \beta_2-1, \beta_3 \beta_4\}}$ in terms of the same integral, $I_{\alpha \{ \beta_1 - 1, \beta_2-1 \beta_3 \beta_4\}}$. In this way we obtain
\begin{align} \label{red1}
\ino_{[\Delta_1 \Delta_2 \Delta_3 \Delta_4]} & = \mathcal{N}_{\beta_t} \left[ \frac{k_1^2 \mathcal{M}_2^{(\beta_2 - 1)} + k_2^2 \mathcal{M}_1^{(\beta_1 - 1)}}{4 (\beta_1 - 1)(\beta_2 - 1)} \ino_{[\Delta_1 - 1, \Delta_2 - 1, \Delta_3 \Delta_4]} \right.\nn\\
& \qquad \left. + \frac{k_3^2 \mathcal{M}_4^{(\beta_4 - 1)} + k_4^2 \mathcal{M}_3^{(\beta_3 - 1)}}{4 (\beta_3 - 1)(\beta_4 - 1)} \ino_{[\Delta_1 \Delta_2, \Delta_3 - 1, \Delta_4 - 1]} \right].
\end{align}
This expression contains only two integrals on its right hand side and increases the values of the appropriate dimensions. Assume now that the dimensions are pairwise equal, \textit{i.e.}, $\Delta_1 = \Delta_3$ and $\Delta_2 = \Delta_4$. In such a case
\begin{align}
\ino_{[\Delta_1 \Delta_2, \Delta_3-1, \Delta_4-1]}(k_1, k_2, k_3, k_4) = \ino_{[\Delta_1 - 1, \Delta_2 - 1, \Delta_3 \Delta_4]}(k_3, k_4, k_1, k_2)
\end{align}
and the right hand side of \eqref{red1} contains only a single integral. 

Additional relations are available in other special cases, for example
\begin{align}
\ireg_{[3222]} & = \frac{4}{\pi^2} I_{\reg{2} \{ \frac{3}{2} \frac{1}{2} \frac{1}{2} \frac{1}{2} \}} \nn\\
& = \frac{4}{\pi^2} \mathcal{N}_3 \left[ k_1^2 I_{\reg{3} \{ \frac{1}{2} \frac{1}{2} \frac{1}{2} \frac{1}{2} \}} + k_2^2 I_{\reg{3} \{\frac{3}{2}, -\frac{1}{2}, \frac{1}{2} \frac{1}{2} \}} + k_3^2 I_{\reg{3} \{ \frac{3}{2} \frac{1}{2}, -\frac{1}{2}, \frac{1}{2} \}} + k_4^2 I_{\reg{3} \{ \frac{3}{2} \frac{1}{2} \frac{1}{2}, -\frac{1}{2} \}} \right] \nn\\
& =  \frac{4}{\pi^2} \mathcal{N}_3 \left[ k_1^2 I_{\reg{3} \{ \frac{1}{2} \frac{1}{2} \frac{1}{2} \frac{1}{2} \}} + (k_2 + k_3 + k_4) I_{\reg{3} \{ \frac{3}{2} \frac{1}{2} \frac{1}{2} \frac{1}{2} \}} \right] \nn\\
& = \frac{4}{\pi^2} \mathcal{N}_3 \left[ k_1^2 \mathcal{M}_1^{(-\frac{1}{2})} I_{\reg{2} \{ -\frac{1}{2} \frac{1}{2} \frac{1}{2} \frac{1}{2} \}} + (k_2 + k_3 + k_4) \mathcal{M}_1^{(\frac{1}{2})} I_{\reg{2} \{ \frac{1}{2} \frac{1}{2} \frac{1}{2} \frac{1}{2} \}} \right] \nn\\
& = \frac{4}{\pi^2} \mathcal{N}_3 \left[ k_1^3 \mathcal{L}_1 + (k_2 + k_3 + k_4) \mathcal{M}_1^{(\frac{1}{2})} \right] I_{\reg{2} \{ \frac{1}{2} \frac{1}{2} \frac{1}{2} \frac{1}{2} \}} \nn\\
& = \mathcal{N}_3 \left[ k_1^3 \mathcal{L}_1 + (k_2 + k_3 + k_4) \mathcal{M}_1^{(\frac{1}{2})} \right] \ireg_{[2222]} \nn\\
& = \frac{1}{2 \ep} \left[ - k_1 k_T \frac{\partial}{\partial k_1} + (k_2 + k_3 + k_4) \right] \ireg_{[2222]},
\end{align} 
and if we keep going,
\begin{align}
\ireg_{[3322]} & = \frac{1}{-1 + 2 \ep} \left[ \left( k_1^2 \mathcal{M}_2^{(\frac{1}{2})} + k_2^2 \mathcal{M}_1^{(\frac{1}{2})} \right) \ireg_{[2222]} + (k_3 + k_4) \mathcal{M}_2^{(\frac{1}{2})} \ireg_{[3222]} \right].
\end{align}

\subsection{Relation to `cosmological bootstrap' operators}
\label{cosmobootstrapops}

In this section we discuss briefly the weight-shifting operators for scalar correlators employed in the cosmological bootstrap paper \cite{Arkani-Hamed:2018kmz}.  We confirm that, as expected from   \cite{Baumann:2019oyu}, they are a special case of the more general weight-shifting operators $\mathcal{W}_{ij}^{\sigma_i\sigma_j}$. 
We also provide a detailed derivation of the operator \eqref{2222shiftx} proposed in \cite{Arkani-Hamed:2018kmz} which acts to shift the dimension of the exchanged operator in the  case where all external dimensions are  equal to two.

\begin{itemize}

\item 
Weight-shifting operators acting on three-dimensional CFT correlators are discussed in Section 5 and Appendix E of 
\cite{Arkani-Hamed:2018kmz}.  The relation of these operators to the  $\mathcal{W}_{ij}^{\sigma_i\sigma_j}$ in \eqref{Wmm0} - \eqref{Wpp0} is not immediately apparent, however, since the operators in \cite{Arkani-Hamed:2018kmz} are restricted to act on 4-point functions  of the form
$
f(u,v,s),
$
where
\[\label{uvdef}
u = \frac{s}{k_{12}}, \qquad v = \frac{s}{k_{34}}, \qquad k_{ij} = k_i + k_j.
\]
As a starting point, this ansatz is clearly valid for $s$-channel exchange and contact diagrams with all $\Delta=2$, since propagators with $\beta=1/2$ reduce to plane waves.


Using \eqref{Wmm12}  and the identities
\[
\frac{1}{k_1}+\frac{1}{k_2} = \frac{s}{uk_1k_2},\qquad 
k_1^2+k_2^2 = \frac{s^2}{u^2}-2k_1k_2,\qquad \partial_{k_1} f = \partial_{k_2} f = -\frac{u^2}{s}\partial_u f,
\]
one can show that the action of $\mathcal{W}_{12}^{--}$ on this ansatz is 
\begin{align}
\mathcal{W}_{12}^{--} f(u,v,s) =
\frac{1}{2k_1 k_2}\Big((1-u^2)\partial_u(u^2 \partial_u f)-(d-1)u\partial_u f\Big).
\end{align}
The raising operator $\mathcal{W}_{12}^{++}$ then follows from \eqref{oldWpp}.    
If we explicitly set $\Delta_1=\Delta_2=2$ and $d=3$,   
the result can be written in the form
\begin{align}
\mathcal{W}_{12}^{++}f(u,v,s) = \frac{s^2}{2}\Big(1+\frac{k_1k_2}{s^2} u^3\partial_u\Big)\Big[\Big(\frac{1-u^2}{u^2}\Big)\partial_u(uf)\Big] = s^2 U_{12} f.
\end{align}
where $U_{12}$ is defined in (5.10) of  \cite{Arkani-Hamed:2018kmz}.  For this latter definition, note that the operator $\O_{12}$ given in (5.3) of  \cite{Arkani-Hamed:2018kmz} can be rewritten as
\[
\O_{12} f(u,v,s) = \Big(1-\frac{k_1k_2}{k_{12}}\partial_{k_{12}} \Big)f\Big(\frac{s}{k_{12}},\frac{s}{k_{34}},s\Big) = \Big(1+\frac{k_1k_2}{s^2} u^3\partial_u\Big) f(u,v,s)
\]
so indeed 
\[
U_{12} f(u,v,s) = \frac{1}{2}\O_{12}\Big[\Big(\frac{1-u^2}{u^2}\Big)\partial_u(uf)\Big] 
\]
as per  (5.10)  of  \cite{Arkani-Hamed:2018kmz}.  
As the dilatation Ward identity fixes the 4-point function to be 
$f(u,v,s) = s^{\Delta_t-9} \hat{f}(u,v)$, one then has
\[
\hat{f}(u,v)\Big|_{\Delta_1=\Delta_2=3} = U_{12}\hat{f}(u,v)\Big|_{\Delta_1=\Delta_2=2}.
\]
This is also consistent with (5.11) of \cite{Arkani-Hamed:2018kmz}, since on the right-hand side the 4-point function with all $\Delta_j=2$ for $j=1,\ldots,4$ is written as $f=s^{-1}\hat{f}(u,v)$ and so 
\[
\mathcal{W}_{12}^{++}\mathcal{W}_{34}^{++}f=s^4 U_{12} U_{34}f = s^3 U_{12} U_{34}\hat{f}(u,v).
\]

\item

Acting on the ansatz $f(u,v,s)$,  the Casimir operator \eqref{Cas12form1} with $d=3$ and $\Delta_1=\Delta_2=2$ reduces to 
\begin{align}
\mathcal{C}_{12}\, f(u,v,s) = (2-\Delta_u) f(u,v,s),
\end{align}
where 
\[
\Delta_u = u^2(1-u^2)\partial_u^2-2u^3\partial_u = u^2\partial_u\Big((1-u^2)\partial_u\Big)
\]
as defined in (2.30) of  \cite{Arkani-Hamed:2018kmz}.  
For general $\Delta_x$ and $\Delta_3=\Delta_4=2$, we then have
\begin{align}\label{ECasdef}
E_{12,\Delta_x} &\equiv \mathcal{C}_{12}+\Delta_x(\Delta_x-3)
=-\Delta_u+(\Delta_x-1)(\Delta_x-2),\\
E_{34,\Delta_x}&\equiv \mathcal{C}_{34}+\Delta_x(\Delta_x-3)=-\Delta_v+(\Delta_x-1)(\Delta_x-2).
\end{align}
compatible with, {\it e.g.,} (4.57).  
Let us now define the shift operator
\begin{align}
\label{Sopdef}
S_{u,\Delta_x} \equiv (1-u^2)\partial_u+\frac{1-\Delta_x}{u}
\end{align}
which obeys the intertwining relations
\begin{align}\label{Casintertwiner1}
E_{12,\Delta_x+1}S_{u,\Delta_x} &= S_{u,\Delta_x+2}E_{12,\Delta_x},\\
E_{34,\Delta_x+1}S_{v,\Delta_x} &= S_{v,\Delta_x+2}E_{34,\Delta_x}.
\end{align}
Given a homogeneous solution $h_{\Delta_x} = h_{\Delta_x}(u,v,s)$ of the Casimir equations
\[
E_{12,\Delta_x}h_{\Delta_x}=
E_{34,\Delta_x}h_{\Delta_x}=0,
\] 
we find 
\[
h_{\Delta_x+1}\equiv S_{u,\Delta_x}S_{v,\Delta_x}h_{\Delta_x}
\] 
is a shifted homogeneous solution satisfying 
\[
E_{12,\Delta_x+1}h_{\Delta_x+1}=E_{34,\Delta_x+1}h_{\Delta_x+1}=0.
\]

\item The shift operator $S_{u,\Delta_x}$ can itself be derived from the solution of the homogeneous Casimir equation $E_{12,\Delta_x}=0$.  
Making use of the standard quadratic transformations of the hypergeometric function,\footnote{ See, {\it e.g.,} 15.8.13 in the online {\it Digital Library of Mathematical Functions}.} this can be expressed as 
\[\label{homsolns}
h(u) = c_1 z^a {}_2F_1(a,a,2a;z)+c_2 z^{1-a}{}_2F_1(1-a,1-a,2(1-a);z)
\]
where
\[
z=\frac{2u}{1+u}, \qquad u=\frac{z}{2-z},\qquad a = \Delta_x-1.
\]
Here, the two solutions in \eqref{homsolns} are related by the shadow transformation $\Delta\rightarrow 3-\Delta$ which sends $a\rightarrow 1-a$. 
 In $z$ variables, 
the shift operator reads
\[
S_{u,\Delta_x} = 2(1-z)\partial_z + \frac{a(z-2)}{z}
\]
and one can verify that it acts as expected, namely
\begin{align}\label{2F1shift1}
&S_{u,\Delta_x}\Big(z^a {}_2F_1(a,a,2a;z)\Big) = -\frac{a^2}{2(1+2a)}z^{a+1} {}_2F_1(a+1,a+1,2(a+1);z),\\
&S_{u,\Delta_x}\Big(z^{1-a} {}_2F_1(1-a,1-a,2(1-a);z)\Big) = 2(1-2a) z^{-a} {}_2F_1(-a,-a,1-2a;z).
\end{align}
To construct $S_{u,\Delta_x}$ in the first place, we concatenate the standard ${}_2F_1$ shift relations 
\begin{align}
\partial_z \,{}_2F_1(a,b,c;z)&=\frac{ab}{c}{}_2F_1(a+1,b+1,c+1;z),\\
\big((z-1)\partial_z+a+b-c\big){}_2F_1(a,b,c;z)&=\Big(a+b-c-\frac{ab}{c}\Big){}_2F_1(a,b,c+1;z)
\end{align}
then eliminate the second derivative using the hypergeometric equation.  This gives the first-order shift operator
\[\label{Fshift}
\frac{1}{z}\big(c(1-z)\partial_z-ab){}_2F_1(a,b,c;z)=-\frac{ab(c-a)(c-b)}{c(c+1)}{}_2F_1(a+1,b+1,c+2;z)
\]
which, on specializing to $(a,b,c)=(a,a,2a)$, reads
\[
-\frac{2(2a+1)}{a^2 z}\Big(2(1-z)\partial_z-a\Big){}_2F_1(a,a,2a;z) ={}_2F_1(a+1,a+1,2(a+1);z). 
\]
Multiplying by $z^{1+a}$ and pushing this factor inside the derivative, we recover
\[
-\frac{2(2a+1)}{a^2}\Big(2(1-z)\partial_z+\frac{a(z-2)}{z}\Big)z^a {}_2F_1(a,a,2a;z) = z^{a+1}{}_2F_1(a+1,a+1,2(a+1);z) 
\]
which is precisely \eqref{2F1shift1}.

\item Having solved the homogeneous problem, let us now turn to the inhomogeneous one.  
From \eqref{Casonex}, exchange diagrams with external dimensions $\Delta_j=2$ satisfy 
\[
E_{12,\Delta_x}f_{\Delta_x}=E_{34,\Delta_x}f_{\Delta_x}= \ireg_{[2222]} = \frac{1}{k_T}=\frac{u v}{s(u+v)} \equiv C,
\]
where the same contact term appears on the right-hand side for all $\Delta_x$.

To solve this inhomogeneous equation, let us make the ansatz
\[
f_{\Delta_x+1} = \gamma_{\Delta_x}(S_{u,\Delta_x}S_{v,\Delta_x}f_{\Delta_x} + g),
\]
where $\gamma_{\Delta_x}$ is a constant depending on $\Delta_x$ and $g=g(u,v)$ is a function that we will assume (in order to find a simple solution as below) is independent of $\Delta_x$.   This gives
\begin{align}
C = 
E_{12,\Delta_x+1}f_{\Delta_x+1} &=\gamma_{\Delta_x} E_{12,\Delta_x+1}( S_{u,\Delta_x}S_{v,\Delta_x}f_{\Delta_x} + g)\nn\\&
=\gamma_{\Delta_x}(S_{u,\Delta_x+2}S_{v,\Delta_x}E_{12,\Delta_x}f_{\Delta_x}+E_{12,\Delta_x+1}g)\nn\\
&=\gamma_{\Delta_x}(S_{u,\Delta_x+2}S_{v,\Delta_x}C+E_{12,\Delta_x+1}g).
\end{align}
Thus, $g$ must solve the inhomogeneous Casimir problem
\[\label{geqn}
E_{12,\Delta_x+1}g = (\gamma_{\Delta_x}^{-1}-S_{u,\Delta_x+2}S_{v,\Delta_x})C \equiv G(u,v).
\]
Evaluating the right-hand side, the only terms with dependence on $\Delta_x$ are
\[
G = -\frac{\Delta_x(\Delta_x-1)}{s(u+v)} + \frac{(\gamma_{\Delta_x}^{-1}-\Delta_x)uv}{s(u+v)}+\ldots 
\]
These terms must match those on the left-hand side, which,  since  $g$ is independent of $\Delta_x$ by assumption, are
\[
\Delta_x(\Delta_x-1)g+\ldots
\]
This gives 
\[
\gamma_{\Delta_x}^{-1} = c_1\,\Delta_x(\Delta_x-1)+\Delta_x+c_2,\qquad 
g = \frac{c_1\,uv-1}{s(u+v)}
\]
where $c_1$ and $c_2$ are $\Delta_x$-independent constants.   Here $c_2$ encodes the $\Delta_x$-independent part of $\gamma_{\Delta_x}^{-1}$ which is unconstrained by the above. 
Plugging this trial solution back into \eqref{geqn}, we find $c_1=c_2=-1$ and hence the solution 
\[
\gamma_{\Delta_x}^{-1} = -(\Delta_x-1)^2,\qquad g = -\frac{(uv+1)}{s(u+v)}.
\]
We thus have the shift relation
\begin{empheq}[box=\nicebox]{align}\label{2222shiftx}
\ireg_{[22,22\,x\, \Delta_x+1]} = -\frac{1}{(\Delta_x-1)^2}\Big(S_{u,\Delta_x}S_{v,\Delta_x} \ireg_{[22,22\,x\,\Delta_x]}-\frac{(uv+1)}{s(u+v)}\Big),
\end{empheq}
with $S_{u,\Delta_x}$ as given in \eqref{Sopdef} and $u,v,$ variables as defined in \eqref{uvdef}.
This result is equivalent to (4.58) in \cite{Arkani-Hamed:2018kmz}\footnote{Note the sign of the final inhomogeneous term in (4.58) of  \cite{Arkani-Hamed:2018kmz} is misprinted.}, noting  the $\Delta_\sigma$ there is $\Delta_\sigma=\Delta_x+1$, {\it i.e.,} the final dimension of the exchanged operator {\it after} applying the shift relation.\footnote{This can be seen from (5.6) in \cite{Baumann:2020dch}, correcting the typo in the sign of the first term.}

Note that if we were to drop the simplifying assumption  that $g$ is independent of $\Delta_x$,  we could instead attempt to solve the inhomogeneous Casimir problem \eqref{geqn} through the method of variation of parameters.
Using the two homogeneous solutions $h_1(u)$ and $h_2(u)$ from \eqref{homsolns}, this gives
\[
g = -h_1(u)\int^u\D \tilde{u}\, (1-\tilde{u}^2) G(\tilde{u},v) h_2(\tilde{u})+h_2(u)\int^u\D \tilde{u}\, (1-\tilde{u}^2) G(\tilde{u},v) h_1(\tilde{u})
\]
where the Wronskian is $W(\tilde{u})=(1-\tilde{u}^2)^{-1}$. 
Evaluating these integrals is difficult in general however, and for the special cases that can be evaluated ({\it e.g.,} $\Delta_x=2$) we obtain complicated solutions involving the dilogarithm.  Thus, the assumption that $g$ is independent of $\Delta_x$ is a useful one in that it leads to a simple shift relation that can be obtained without having to evaluate any integrals.

\end{itemize}

\bibliographystyle{JHEP}
\bibliography{exchange}

\end{document}